\documentclass[oneside,letterpaper,11pt]{book}   % one-sided
\usepackage{amsbsy}
\usepackage[usenames]{color}
\usepackage{graphicx}
\usepackage{latexsym}
\usepackage{dissertp}

\usepackage[hyperindex,breaklinks,linktocpage,colorlinks]{hyperref}

\def\A{{\rm \AA}}
\def\apriori{\emph{a priori}}
\def\CRSS{{\rm CRSS}}
\def\d#1{{\rm d}#1}
\def\deg{^\circ}
\def\eps{\varepsilon}
\def\eV{{\rm eV}}
\def\gdir#1{\langle{#1}\rangle}
\def\gplane#1{\{#1\}}
\def\K{{\rm K}}

\def\mat#1{\boldsymbol{#1}}

\def\MPa{{\rm MPa}}

\def\refeq#1{Eq.\,\ref{#1}}
\def\refeqs#1{Eqs.\,\ref{#1}}
\def\reffig#1{Fig.\,\ref{#1}}
\def\reftab#1{Tab.\,\ref{#1}}
\def\s{{\rm s}}

\newcommand\systab[4]{#1 & (#2)[#3] & [#3] & [#2] & [#4]}

\begin{document}

  \setcounter{page}{1}
  \title{DEVELOPMENT OF PHYSICALLY BASED PLASTIC FLOW RULES FOR
  BODY-CENTERED CUBIC METALS WITH TEMPERATURE AND STRAIN RATE
  DEPENDENCIES}

\author{Roman Gr\"oger}

\major{Materials Science and Engineering}

\university{University of Pennsylvania}

\degree{Doctor of Philosophy}

\date{2007}

\supervisor{Professor Vaclav Vitek}

\gradchair{Professor Russell J.~Composto}

\maketitle

\cleardoublepage

  \renewcommand{\thepage}{\roman{page}}

  \ \vskip1cm

\begin{flushright}
  My advice to you is get married: \\
  if you find a good wife you'll be happy;\\
  if not, you'll become a philosopher. \\
  \emph{Socrates}
\end{flushright}

\vskip2cm

\begin{center}
  {\it To Veronika} \\
  for I have not become a philosopher
\end{center}

\cleardoublepage

  \section*{Acknowledgments}

This work would not have been possible without the initial impuls and never-ending encouragement of
my advisor, Vasek Vitek, whom I thank for helping me develop independent thinking, build extensive
research skills in computational Materials Science, and shape my scientific writing. His attention
to my every thought and their subsequent implementations in the theory presented in this Thesis has
never allowed me to go astray. I am similarly grateful to John Bassani, for his explanation and many
fruitful discussions on the non-associated plastic flow model that I would have missed should I not
have come to Penn.

Among my collaborators, I have especially benefited from discussions with Vikranth Racherla on
fitting the parameters of the effective yield criterion and further elaborations on the shapes of
yield and flow surfaces. The most recent part of this Thesis, dealing with the behavior of screw
dislocations in tungsten under stress, is partly the work of Aimee Bailey (now Imperial College,
London) whose rapid progress with atomistic simulations allowed an interesting comparison between
the plastic flow of molybdenum and tungsten. Many thanks are due to Khantha Mahadevan whom I highly
respect for encouraging me to pursue any idea I am interested in, no matter how improbable it may
seem to be, and for her recent help with starting my future career. I am also grateful to Matou\v{s}
Mrov\v{e}c (Fraunhofer Institut f\"ur Werkstoffmechanik) who helped me understand the role of
screening of bond integrals and especially for his continuing support during my construction of the
Bond Order Potentials for niobium and tantalum. Similar thanks are also due to Duc Nguyen-Manh
(UKAEA Fusion/Culham Science Centre) for his countless discussions on the construction of the Bond
Order Potentials, their testing and final implementation in the simulation of extended defects. I am
indebted to Lutz Hollang (Technische Universit\"at Dresden) for providing me with the results of his
careful experiments on single crystals of molybdenum and also for subsequent discussions on the
subject. We also acknowledge Fran\c{c}ois Louchet (Laboratoire Glaciologie et de G\'eophysique,
Grenoble), Ladislas Kubin (CNRS/ONERA, Ch\^{a}tillon) and Drahos Vesely (University of Oxford) for
their comments on our model explaining the discrepancy between the theorerical and experimental
yield stresses in body-centered cubic metals. Finally, I am thankful for the most recent discussions
with David Lassila (Lawrence Livermore National Laboratory) on their six-degrees-of-freedom
microstrain experiments.

There could not have been a more enjoyable beginning at Penn than sharing an apartment with my
Trinidadian friend Darryl Romano, sniffing in the crock pot during his cooking of \emph{pelau} or
\emph{callaloo} and celebrating Christmas together with my wife Veronika as a family. My early years
at Penn were marked by interaction with many new people from which I am mainly grateful to Ana
Claudia Costa whose spontaneity made our office a living place, Marc Cawkwell for balancing an
excessive joy with work, Rado Po\v{r}\'izek and Pavol Juh\'as for being my family at Penn, Michelle
Chen for sharing endless hours on cracking the physics homeworks, and Chang-Yong Nam, Evan Goulet,
John Garra, Chris Rankin, Papot Jaroenapibal and Niti Yongvanich for pulling me out of the lab for
beer. I am delighted to have met Ondrej Hovorka (Drexel University), for our regular refreshing
coffee breaks and not-so-regular Fridays at the Cavanaugh's talking about the Ising model, field
Hamiltonians, renormalization group theory, spin entanglements and other irresistible topics.

Many thanks are also due to Pat Overend, Irene Clements for their help with many administrative
issues and to Fred Hellmig and Alex Radin for never turning blind eye to our hardware issues.

I appreciate my Dissertation committee -- Vasek Vitek, John Bassani, David Pope, Charles McMahon
Jr. and Bill Graham -- for devoting their time to reading this Thesis and mainly for their
suggestions and encouragement throughout my graduate study at Penn. Similar thanks should be
directed towards Avadh Saxena (Los Alamos National Laboratory) whose limitless research interests
serve as a deep source of inspiration for me.

Finally, many thanks should be given to my parents, Marie and Franti\v{s}ek Gr\"oger for their
remote emotional support, and to my brother Milan and his daughters Eva and Monika for always
cheering me up, keeping me busy with building puzzles and watching fairy-tales.

\cleardoublepage

  \centerline{\bf ABSTRACT}

\vskip1cm

\begin{center}
  {\bf DEVELOPMENT OF PHYSICALLY BASED PLASTIC FLOW RULES FOR BODY-CENTERED 
    CUBIC METALS WITH TEMPERATURE AND STRAIN RATE DEPENDENCIES} \\[2em]
  Roman Gr\"oger \\[1em]
  Professor Vaclav Vitek \\
\end{center}
\vskip1cm

Plastic flow of all bcc metals is controlled by the glide of $1/2\langle{111}\rangle$ screw
dislocations since they possess non-planar cores and thus experience high Peierls stress.
Atomistic studies at 0~K determine the Peierls stress and reveal that it is strongly dependent on
non-glide stresses, i.e. components of the stress tensor other than the shear stress in the slip
plane parallel to the Burgers vector. At finite temperatures the corresponding Peierls barrier is
surmounted via the formation of pairs of kinks.  Theoretical description of this thermally activated
process requires knowledge of not only the height and shape of the barrier but also its intrinsic
dependence on the applied stress tensor.  This information is not obtainable from any experimental
data and the atomistic studies at 0~K determine the Peierls stress but not the shape of the Peierls
barrier.

In this Thesis we first show how the shape of the Peierls barrier and its dependence on the applied
loading can be extracted from the data obtained in atomistic studies at 0~K.  We consider the
Peierls barrier as a two-dimensional periodic function of the position of the intersection of the
dislocation line with the perpendicular $\gplane{111}$ plane, with adjustable terms dependent on the
shear stresses parallel and perpendicular to the slip direction.  The functional forms of these
terms are based on the effective yield criterion recently developed on the basis of atomistic
modeling of the glide of screw dislocations at 0~K.  The minimum energy path between two potential
minima, and thus the corresponding activation barrier, is obtained using the Nudged Elastic Band
method.  The constructed Peierls barrier reproduces correctly both the well-known
twinning-antitwinning asymmetry observed for pure shear parallel to the slip direction and the
effect of shear stresses perpendicular to the slip direction.  This advancement introduces for the
first time the effect of both shear stresses parallel and perpendicular to the slip direction into
the model of thermally activated dislocation motion.  Based on this model we formulate a general
yield criterion that includes not only the full stress tensor but also effects of temperature and
strain rate.  This approach forms a basis for multislip yield criteria and flow relations for
continuum analyses in both single and polycrystals the results of which can be compared with
experimental observations.

\vskip2em

This research has been supported by the NSF grant DMR02-19243 and by the U.S. Department of Energy,
BES grant DE-PG02-98ER45702. R.G. acknowledges the Hl\'avka foundation of the Czech Republic for
awarding him the 2002 travel stipend. We have benefited from unrestricted access to the Penn's linux
workstations \emph{presto} on which most of the simulations were performed and acknowledge
professional support of their administrators. The computations on tungsten were performed in part on
the National Science Foundation HP~GS1280 system at the Pittsburgh Supercomputing Center under the
grant SEE060004P.

\cleardoublepage

  \tableofcontents  
  \def\addvspace#1{}  % no extra vertical space in the lot and lof
  \listoftables  
  \listoffigures  
  \def\addvspace#1{\vspace{#1}}  % no extra vertical space in the lot and lof

  \cleardoublepage
  \renewcommand{\thepage}{\arabic{page}} 
  \setcounter{page}{1}

  \chapter{Introduction}

\begin{flushright}
  We say that we will put the sun into a box. The idea is pretty. \\
  The problem is, we don't know how to make the box. \\
  \emph{Pierre-Gilles de Gennes}
\end{flushright}

Refractory metals crystalizing in the body-centered cubic (bcc) structure are materials employed in
many modern applications such as structural components of fusion reactors or kinetic energy
penetrators. Modern nuclear applications seek materials that exihibit large radiographic densities
that allow for shielding or collimation of radiation \citep{zinkle:02}. This requirement
disqualifies most of the usual structural materials because they yield rather bulky
components. Apart from very dense alloys that are often inapplicable due to their prohibitive cost,
the pool of potential candidates comprises pure forms of refractory metals and iron.

One of the fundamental requirements for every structural material is its stability under mechanical
loading in a given range of working conditions. In structural applications, a material is frequently
regarded to loose its functionality upon reaching the yield limit state when a permanent
macroplastic deformation sets in. For a given loading, this limiting state can be predicted
theoretically using a proper yield criterion that is independent of the shape of the structural
component.

%----------------------------------------------------------------------------------------------------
%----------------------------------------------------------------------------------------------------

\section{Historical background}

The experimental studies of plastic flow of crystalline materials date back to the work of
\citet{taylor:25} and \citet{taylor:26a} on face-centered cubic (fcc) aluminum, and to
\citet{elam:26} on fcc copper and gold. These materials were found to deform by sliding of
close-packed atomic planes, $\gplane{111}$, over each other in the direction of densest atomic
packing, $\gdir{110}$. Around the same time, similar experiments on body-centered cubic (bcc)
$\alpha$-iron were carried out by \citet{taylor:26}. In this material, the deformation occured by
crystallographic planes sliding virtually parallel to the $\gdir{111}$ axis which is the direction
of closest atomic packing but, unlike in fcc aluminum, copper and gold, the slip plane was not
well-defined. Instead, the plane of slip appeared to be related to the distribution of stress albeit
coinciding with a plane containing the $\gdir{111}$ slip direction. Perhaps even more surprising was
the observation that the sliding of parallel atomic planes did not occur by a rigid displacement of
one plane over the other but, instead, the particles of the material appeared to ``cling'' together
in lines or rods. Two years later, \citet{taylor:28} performed similar experiments on $\beta$-brass
with B2 structure whose lattice is composed of two interpenetrating simple cubic (sc) lattices of
copper and zinc. For some orientations, the slip plane did not coincide with any well-defined
crystallographic plane, similarly as in $\alpha$-iron, although for other orientations the slip
occured on $\gplane{110}$ planes, as expected. Unlike $\alpha$-iron, the resistance to slipping two
parallel planes over each other in $\beta$-brass depended on the sense of slip. Furthermore, the
magnitudes of axial compressive stresses needed to plastically deform single crystals of
$\alpha$-iron and $\beta$-brass were significantly larger than those employed in earlier experiments
on fcc aluminum \citep{taylor:25, taylor:26a}.

The marked differences in plastic deformation of fcc and bcc metals led \citet{schmid:35} to
investigate hexagonal close-packed (hcp) metals zinc and cadmium. The single crystals of these
metals were observed to deform in a similar way as aluminum, copper and gold. These observations
together with the earlier results of studies on the fcc metals were condensed into the so-called
Schmid law according to which the macroscopic plastic deformation occurs when the shear stress
parallel to the slip direction resolved in the most highly stressed slip system attains its critical
value. Although the Schmid law was originally deemed to be generally valid for any crystalline
material, its deficiency in bcc metals was known already from the works of \citet{taylor:26} on
$\alpha$-iron and of \citet{taylor:28} on $\beta$-brass. This implies that the mechanism governing
the plastic flow of bcc metals is likely to be different from that of fcc and hcp metals.

The observations of \citet{taylor:26} remained puzzling until the birth of the concept of
dislocations established by three classical papers by \citet{orowan:34}, \citet{polanyi:34} and
\citet{taylor:34} that provided the needed microscopic understanding of the plastic flow of
crystalline materials. Dislocations soon became recognized as carriers of plastic flow, which
immediately raised important questions about their behavior under stress and later also the effect
of temperature on their motion. Since the propagation of dislocations requires a collective motion
of many atoms, it was impossible to obtain the atomic positions as a rigorous solution of some
equilibrium conditions. Instead, the periodic structure of the crystal in the glide plane was
represented by a continous function that merely obeyed the periodicity of the lattice
\citep{peierls:40}. Based on this model, the shear stress to move the dislocation was predicted to
be about one-thousandth of the theoretical shear strength of a perfect lattice \citep{nabarro:47}, a
hundred times smaller value than the yield stresses normally encountered in experiments on bcc
metals. The final breakthrough came two decades later, when \citet{hirsch:60} realized that the
screw dislocation core has to obey the underlying three-fold screw symmetry of the $\gdir{111}$
direction in the bcc lattice and may spread into three planes in the zone of this direction. He
postulated that, provided the screw dislocation core spreads on several planes that all contain the
$\gdir{111}$ slip direction, a high stress would be needed to transform this initially sessile core
into a configuration that is glissile in a slip plane. This conjecture not only provided a plausible
explanation of the origin of large Peierls stresses but also implied a strong temperature dependence
and complex slip geometry in bcc metals.

%----------------------------------------------------------------------------------------------------
%----------------------------------------------------------------------------------------------------

\section{Low-temperature experiments}

One of the most notable characteristics of the plastic deformation of bcc metals is a steep increase
of both yield and flow stresses with decreasing temperature, which had been a matter of debate for
many years and attributed by some to the effect of impurities. This hypothesis survived until the
discovery of new purification techniques such as electron beam floating zone method and
ultra-high-vacuum annealing that provided macroscopic single crystals essentially free of
interstitial solutes. Even in these high-purity bcc metals, the plastic flow at temperatures below
$0.1T_m$, where $T_m$ is the melting temperature, displayed remarkable differences with respect to
the behavior of close-packed metals. Among the most important ones were the tendency to cleave at
low temperatures, large strain rate sensitivity, strong influence of interstitial impurities and,
most remarkably, a tension-compression asymmetry detected in uniaxial loading tests of single
crystals of practically all bcc metals \citep{christian:83}. This asymmetry was originally
attributed to the so-called twinning-antitwinning asymmetry of shearing in the $\gdir{111}$ slip
direction along $\gplane{112}$ planes.

The majority of the early experiments have been done on niobium \citep{mitchell:63, duesbery:69,
  bolton:72, louchet:75, reed:76, bowen:77}, tantalum \citep{webb:74, shields:75, nawaz:75,
  takeuchi:77, wasserbach:85}, molybdenum \citep{vesely:68, guiu:69, matsui:76, saka:76,
  kitajima:81, matsui:82, aono:83}, and $\alpha$-iron \citep{allen:56, arsenault:64, keh:65,
  aono:81}. On the other hand, significantly less attention has been devoted to tungsten
\citep{argon:66, tabata:76, brunner:04} owing to the complicated purification process caused by its
extremely high melting temperature. Similar experiments on alkali metals have long been impossible
due to their strong reactivity to water and rapid oxidization in air, and have been limited so far
to potassium \citep{basinski:81, duesbery:93, pichl:97} which retains its bcc structure down to the
lowest temperatures investigated. On the other hand, experiments on bcc sodium and lithium are
rather rare \citep{pichl:97b}, mainly due to their martensitic transformation to close-packed
structures at low temperatures. 

Pure bcc metals at low temperatures exhibit a phenomenon called anomalous slip which was first
observed experimentally on single crystals of niobium \citep{duesbery:69b, bolton:72}, and later in
vanadium \citep{taylor:73}, tantalum \citep{nawaz:75}, molybdenum \citep{matsui:76, kitajima:81} and
tungsten \citep{kaun:68}. In these experiments the slip did not occur on the most highly stressed
slip system, as expected, but typically on the fourth or fifth most highly stressed
$\gplane{110}\gdir{111}$ system. At the same time, the slip traces of the expected slip system with
the highest Schmid stress were discontinuous or sometimes not observed at all. Similar behavior was
also observed in dilute transition metal alloys \citep{jeffcoat:76, taylor:91} and alkali metals
\citep{pichl:97}. In contrast, the low-temperature form of ferromagnetic $\alpha$-iron deforms
exclusively by slip on the most highly stressed slip system \citep{aono:81} and the anomalous slip
has never been observed.

The direct observation of dislocations was first performed on single crystals of gold and aluminum
\citep{hirsch:56}, as recalled by \citet{hirsch:80} in his review of the period 1946-56. The
evolution of high-resolution electron microscopy (HREM) recently allowed to observe the dislocation
core structure in thin foils of bcc metals \citep{sigle:99}. However, since screw dislocations in
few-nanometer thick foils, which are currently employed in the HREM studies, inherently generate the
so-called Eshelby twist \citep{eshelby:51}, the interpretation of these measurements is still a
matter of controversy \citep{mendis:06}.

%----------------------------------------------------------------------------------------------------
%----------------------------------------------------------------------------------------------------

\section{Theoretical studies and computer simulations}

The modern theoretical studies of the motion of screw dislocations in bcc metals were initiated by
the study of internal friction in polycrystalline copper \citep{seeger:56}. At high temperatures and
low applied stresses, screw dislocations were postulated to move by nucleating two non-screw
segments, called kinks, that connected the original straight screw dislocation with its activated
segment lying in the neighboring valley of the Peierls potential. This work was generalized to
finite stresses by \citet{dorn:64} by solving for the stationary shape of the dislocation in the
presence of applied stress. They demonstrated that, at finite applied shear stress parallel to the
slip direction, the dislocation surmounts the Peierls barrier by nucleating a critical bow-out
represented by a smooth local deflection of otherwise straight dislocation line. For a prototypical
Peierls potential whose shape merely obeyed the periodicity of the lattice in the slip plane, these
studies correctly predicted the increasing trend of the yield stress with decreasing temperature for
all refractory metals and $\alpha$-iron. The shape of the Peierls potential and its effect on the
temperature dependence of the yield stress was recently studied by \citet{suzuki:95}, who concluded
that the best agreement with experiment is obtained if the Peierls potential exhibits a flat maximum
or a ``camel-hump'' shape with an intermediate minimum. However, since the Peierls potential was
still defined as a one-dimensional function of the position of the dislocation in the slip plane,
this model could not reproduce the cross-slip in bcc metals and the onset of anomalous slip
frequently encountered in low-temperature experiments. In order to allow for these phenomena to
occur, \citet{edagawa:97} recently suggested that the Peierls potential should be a two-dimensional
function of the position of the intersection of the dislocation line with the $\gplane{111}$ plane.

Computer-based atomistic modeling of the structure and energetics of screw dislocations using simple
pair potentials became possible around 1970 and was mainly pioneered by \citet{duesbery:69},
\citet{vitek:70}, and \citet{basinski:71}. As envisaged already by \citet{hirsch:60}, the unstressed
dislocation cores possessed the three-fold symmetry of the underlying lattice and extended on the
three $\gplane{110}$ planes containing the $\gdir{111}$ slip direction. Because two such
configurations of the same energy were found that were related by a diad symmetry, the core has been
called as degenerate. Subsequent studies of dislocations under stress revealed that the spreading of
the core on the three $\gplane{110}$ planes is indeed responsible for a strong twinning-antitwinning
asymmetry of the shear stress parallel to the slip direction \citep{christian:83}. In the following
years, these simulations were revisited using more accurate potentials such as Finnis-Sinclair (F-S)
or Embedded Atom Method (EAM) that \emph{a~priori} include also the many-body contribution to the total
energy \citep{duesbery:98}. The latter was further extended to incorporate the directional bonding
that gave rise to the Modified Embedded Atom Method (MEAM) \citep{baskes:92}.

Currently the most accurate semi-empirical schemes formulated in real-space are the Modified
Generalized Pseudopotential Theory (MGPT) \citep{moriarty:88, xu:96, xu:98} and the Bond Order
Potential (BOP) \citep{pettifor:95, horsfield:96}. The latter is an $O(N)$, two-center orthogonal
tight binding method that correctly describes the mixed metallic and covalent character of bonding
in $d$-electron transition metals. In BOP, the bonding part is obtained directly from the
first-principles calculations, while the rest of the potential is constructed purely empirically to
reproduce such fundamental parameters as the lattice constant, cohesive energy, and elastic
moduli. The incorporation of directional bonding revealed a new structure of the dislocation core
that spreads symmetrically on the three $\gplane{110}$ planes. This, so-called non-degenerate core
possesses not only the three-fold symmetry dictated by the lattice but also an additional
$[\bar{1}01]$ diad symmetry. Since the number of atoms employed in these simulations is usually of
the order of thousands, they allow extensive atomistic simulations of the onset of plastic flow, the
results of which can be used as a basis for the development of continuum laws of plasticity in bcc
metals.

The most basic simulations of dislocations in bcc metals utilize the first-principles-based Density
Functional Theory (DFT). Since this theory is formulated in k-space, it necessitates the use of
periodic simulation cells whose boundary conditions must eliminate the contribution of the periodic
images of the dislocation in the neighboring cells. One of the most recent achievements in the
simulation of dislocations using DFT is the work of \citet{woodward:01, woodward:02} introducing the
flexible Green's function boundary condition (GFBC) that self-consistently couples the local
dislocation strain field with the long-range elastic field.  This method predicts the existence of
the same non-degenerate dislocation core as obtained by BOP.

Apart from the single dislocation molecular statics simulations at $0~\K$ outlined above that aim at
studying the plasticity of bcc metals ``from the bottom up'', there has been recently a growing interest
in large-scale simulations at finite temperatures and strain rates. The methods of molecular
dynamics (MD) are employed to study the microscopic behavior of dislocations by simulating an entire
ensemble of up to a few million atoms \citep{bulatov:98, zhou:98, marian:04,chaussidon:06} whose
mutual interactions are governed by a given interatomic potential. In these studies, interactions
between the dislocations present in the simulation cell occur intrinsically and, besides the
periodicity of the block, no additional conditions are imposed. These simulations provide the laws
for rates of dislocation motion that are intended to be utilized in mesoscopic-level discrete
dislocation dynamics (DDD) simulations \citep{kubin:98, zbib:00, jonsson:03} in which one studies an
evolution of an ensemble of mutually interacting dislocations without reference to individual
atoms. Unfortunately, due to their inherent computational complexity and limited time scale, both MD
and DDD studies employ strain rates of the order of $10^5~\s^{-1}$, which are about ten orders of
magnitude greater than those normally achieved in engineering applications. Consequently, these
simulations are mainly useful in studies of plastic flow under such extreme conditions as nuclear
explosions or laser-induced shock waves.

%----------------------------------------------------------------------------------------------------
%----------------------------------------------------------------------------------------------------

\section{Continuum description of plastic flow of bcc metals}

Despite the discrete nature of crystalline materials that determines the onset of their microplastic
deformation at short length scales, these features average out at large scales and give rise to the
continuum response of the material. For engineering calculations, it is therefore highly desirable
to work with continuum yield criteria in which the microscopic behavior is represented only
indirectly by a few fundamental parameters. The framework of time-independent single-crystal
plasticity has been investigated through the work of \citet{hill:65} and \citet{rice:71} and applied
in studies of multislip hardening behavior and strain localization. These theories are commonly
based on the validity of the Schmid law and thus provide good models of the plastic flow of
close-packed metals for which this law is well-established. Because the Schmid law is used in
formulation of both the yield surface and the flow potential, the plastic flow predicted from these
models is said to be associated with the yield function. In contrast, none of these models apply to
non-close-packed materials, such as bcc metals and certain intermetallic compounds, in which the
breakdown of the Schmid law has been known since the work of \citet{taylor:26} and
\citet{taylor:28}.

The first systematic work, inspired by the early developments of \citet{hill:72} and
\citet{hill:82}, in which the influence of non-Schmid effects has been explicitly included, is due to
\citet{qin:92b, qin:92} for Ni$_3$Al, an intermetallic compound crystallizing in the close-packed
L1$_2$ structure. In this alloy, the critical resolved shear stress in the primary slip system is a
function of both the orientation of the loading axis and of the sense of shear. Based on this
observation, Qin and Bassani proposed a simple form of an effective yield criterion in which the
yield stress is written as a linear combination of the Schmid stress and other (non-Schmid) stresses
that determine energy dissipation in the crystal due to slip on a given slip system. An important
implication of this work is that the plastic strain rate $\dot\gamma^\alpha$ on system $\alpha$ is
not determined only by the Schmid stress but is also affected by the non-Schmid stress
components. Due to the presence of the non-Schmid stresses the yield and flow surfaces do not
coincide, and the plastic flow is then said to be non-associated with the yield criterion. This
non-associated plastic flow model based on simple linear yield criterion has been shown to correctly
capture the experimentally observed non-Schmid behavior in Ni$_3$Al and also the occurrence of
strain localization in the form of shear bands.

Apart from the yield criteria that do not explicitly involve any thermodynamic parameters, one is
often interested in the rate of plastic flow at a given temperature and applied loading. This is
routinely expressed using a rate equation represented by the Arrhenius formula $\dot\gamma \propto
\exp(-\Delta{H}/kT)$, where $\Delta H$ is the activation enthalpy, $k$ the Boltzmann constant, and
$T$ the absolute temperature. In the phenomenological theory of plastic flow of bcc metals due to
\citet{kocks:75}, the stress dependence of the activation enthalpy is written as a power law with
constant exponents adjusted so that the activation enthalpy and the activation volume agree with
experimental data. In recent years, considerable attention has been devoted to studying the
influence of microstructure on the continuum response of materials. Along these lines, various
approximations of the temperature dependence of the yield stress have been proposed
\citep{meyers:02, zerilli:04, voyiadjis:05}. Although these models have been shown to correctly
reproduce certain macroscopic features of the plastic flow of bcc metals, such as the strong increase
of the yield stress with decreasing temperature, they do not reveal some important manifestations of
the microscopic features governing the onset of the plastic flow. Particularly, none of these models
reproduces the strong orientational dependence of the yield stress at low temperatures and its
gradual decay with increasing temperature.

%----------------------------------------------------------------------------------------------------
%----------------------------------------------------------------------------------------------------

\section{Objectives and organization of this Thesis}

The main objective of this Thesis is to develop a theoretical framework for continuum-level
predictions of plastic flow in bcc metals that would be valid within a broad range of temperatures
and strain rates. To achieve this goal, we developed a multiscale model in which the fundamental
information is obtained by means of $0~\K$ molecular statics simulations of isolated screw
dislocations in molybdenum using the BOP whose formalism is introduced briefly in
Chapter~\ref{chap_BOP}.

Atomistic studies are discussed in detail in Chapter~\ref{chap_MoBOP}. After identifying the stress
components that affect the motion of the dislocation, the obtained dependencies determining the
behavior of the dislocation under stress are generalized to real single crystals that contain mobile
dislocations of all eight Burgers vectors. This provides us with the 0~K version of an effective
yield criterion, constructed in Chapter~\ref{chap_tstarcrit}, where a number of experiments are
invoked to test the accuracy of the criterion at low temperatures.

The incorporation of the effects of temperature and strain rate into the effective yield criterion
is presented in Chapter~\ref{chap_thermalact}, where we first construct the Peierls potential on the
basis of the known relation between the Peierls stress and the maximum slope of the Peierls
barrier. The shape of the Peierls potential obeys the underlying three-fold symmetry of the lattice,
which breaks down as a result of the action of non-Schmid stresses. We will demonstrate that the
obtained Peierls potential, in conjunction with the theory of kink-pair formation \citep{seeger:56,
  dorn:64} and the Nudged Elastic Band method \citep{jonsson:98, henkelman:00}, successfully predicts
not only the twinning-antitwinning asymmetry and the strong increase of the yield stress at low
temperatures, but also the change of the slip plane and the related onset of anomalous slip.

The temperature and strain rate dependence of the effective yield criterion is developed in the
subsequent Chapter~\ref{chap_yieldtemp}, where the activation enthalpy to nucleate a pair of kinks
is treated in an approximate fashion. It is shown that the effective yield stress can be written as
a relatively simple analytical function of both temperature and strain rate or, equivalently, the
plastic strain rate can be expressed as a function of applied loading and temperature. Within the
framework of the proposed theory, the plastic flow at a given temperature and strain rate occurs
when the instantaneous value of the effective stress reaches its critical value, the effective yield
stress.

To demonstrate the general validity of this multiscale approach, the same development as on molybdenum
is repeated in Chapter~\ref{chap_WBOP} for tungsten, where we focus mainly on the differences
between the plastic deformation of these two metals. The approximate equations for the plastic
strain rate and the temperature and strain rate dependence of the yield criterion for molybdenum and
tungsten are found to have the same functional forms and differ merely in the magnitudes of a few
adjustable parameters.

The most important achievements of the theory presented in this Thesis are discussed in
Chapter~\ref{chap_conclusion}.

Recent developments related to the work presented here, together with the topics for future
research, are summarized in Chapter~\ref{chap_future}.

For completeness, we propose in Appendix~\ref{chap_interact} a physical model that provides an
explanation of the origin of the well-known discrepancy between the yield stresses calculated
theoretically and those measured in experiments. This model estimates the impact of interactions
between dislocations that were not included in the atomistic model and provides a justification for
scaling the theoretical yield stresses down to experimental values.

The multiscale approach developed in this Thesis provides, for the first time, a physically-based
yield criterion that is an explicit function of both temperature and strain rate. Within the model,
the microscopic details of slip percolate through many length scales up to the macroscopic level,
where their effect is manifested in a coarse-grained fashion. The observation that this model works
well for two different metals, molybdenum and tungsten, suggests a general validity of the proposed
theory for refractory metals and possibly also for other bcc metals, such as alkali metals.

  \chapter{The Bond Order Potentials for refractory metals}
\label{chap_BOP}

\begin{flushright}
  In physics, you don't have to go around \\
  making trouble for yourself -- nature does it for you.”\\
  \emph{Frank Wilczek}
\end{flushright}

Body-centered cubic transition metals of the groups VB and VIB of the periodic table exhibit strong
directional bonding that is the consequence of partial filling of their $d$-electron bands
\citep{friedel:69, pettifor:95}. Central-force schemes, such as pair potentials and many-body
potentials of Finnis-Sinclair or EAM type, have originally been developed for metals with a filled
$d$-band where the bonding is mediated by $s$ and $p$ electrons. Since directional bonding is not
included, when these schemes are applied to transition metals the results may reveal some generic
features of the given class of materials but not characteristics of specific materials. However, the
structure of the core of extended defects, such as dislocations, may be governed by the directional
character of bonding. For example, it has been generally found that central-force schemes prefer
degenerate screw dislocation cores \citep{duesbery:98, vitek:04} that are in contrast with the
non-degenerate cores obtained from studies based on first principles \citep{ismail-beigi:00,
  woodward:01, woodward:02, frederiksen:03}.

There have been a number of attempts to develop empirical potentials that would capture the
directional character of bonding in bcc metals. The details of bonding are often obtained from first
principles, notably from calculations utilizing the density functional theory (DFT), while the rest
of the potential is constructed empirically by fitting a few fundamental characteristics such as the
lattice parameter, cohesive energy, elastic moduli and, in some cases, also vacancy and interstitial
formation energies. Among the currently most popular schemes explicitly incorporating angular-force
contributions are the modified embedded atom method (MEAM) of \citet{baskes:92}, potentials
developed from tight-binding theory \citep{sutton:88, pettifor:95, carlsson:90, carlsson:90a} and
from first-principles generalized pseudopotential theory \citep{moriarty:88, moriarty:90}. For the
dislocation studies presented in this Thesis, we have adopted the tight-binding-based Bond Order
Potential (BOP) whose formalism has been developed by Pettifor and co-workers \citep{pettifor:95,
  horsfield:96}. A number of papers explaining the details of the development of BOPs for various
metals have already appeared in the literature and in Ph.D. Theses of \citet{girshick:97},
\citet{znam:01}, \citet{mrovec:02} and \citet{cawkwell:05a}, and, therefore, we will present here
only a rather brief overview of the basic aspects of the method. Up to date, the BOPs based on the
approach advanced by Pettifor and co-workers have been constructed for several transition metals and
transition-metal-based alloys: titanium \citep{girshick:98}, TiAl \citep{znam:03}, molybdenum
\citep{mrovec:04}, molybdenum silicides \citep{mrovec:02}, iridium \citep{cawkwell:06} and, most
recently, for tungsten \citep{mrovec:07}. The BOPs for other refractory metals -- niobium, tantalum
and vanadium -- as well as for $\alpha$-iron that incorporates ferromagnetism utilizing the Stoner
model of band magnetism \citep{stoner:38, stoner:39}, are now under development. Similarly, BOPs
have also been developed for a number of $sp$-valent systems \citep{pettifor:02, pettifor:04,
  drautz:05, murdick:06}.

Within the BOP, the binding energy of transition metals and their alloys can be written as
\begin{equation}
  U = U_{bond} + U_{env} + U_{pair} \ ,
  \label{eq_BOP_energy}
\end{equation}
where $U_{bond}$ is the bond energy, $U_{env}$ the repulsive environmentally dependent term that
represents the repulsion due to the valence $s$- and $p$-electrons being squeezed into the ion core
regions under the influence of the large covalent $d$-bonding forces experienced in transition
metals, and $U_{pair}$ is a pair-wise interaction arising nominally from the overlap repulsion and
the electrostatic interaction between the atoms. Each of these terms is constructed independently
and sequentially; this is achieved by a systematic incorporation of various fundamental
characteristics of these materials.

In elemental transition metals the most important quantities determining $U_{bond}$ are the
two-center bond integrals $dd\sigma$, $dd\pi$, $dd\delta$ entering the tight-binding Hamiltonian.
The angular dependence of the intersite Hamiltonian matrix elements takes the usual Slater-Koster
form \citep{slater:54}. The dependence on the separation of atoms is determined using the
first-principles tight-binding linear muffin-tin orbital (TB-LMTO) method \citep{andersen:85,
  andersen:94}. In the original formulation of the so-called \emph{unscreened BOP}, the bond
integrals $dd\tau$, where $\tau$ symbolizes $\sigma$, $\pi$ or $\delta$ orbital, were represented by
a continuous analytical function $\beta_\tau(r_{ij})$ taking the generalized
Goodwin-Skinner-Pettifor (GSP) form \citep{goodwin:89} 
\begin{equation}
  \beta_\tau(r_{ij}) = \beta_\tau(r_0) \left( \frac{r_{ij}}{r_0} \right)^{n_a}
  \exp \left\{ 
    n_b \left[ \left( \frac{r_0}{r_c} \right)^{n_c} - 
      \left( \frac{r_{ij}}{r_c} \right)^{n_c} \right] \right\} \ ,
  \label{eq_GSPfun}
\end{equation}
where $r_{ij}$ is the distance between atoms $i$ and $j$, $r_0$ the equilibrium separation of first
nearest neighbors, and $n_a$, $n_b$, $n_c$ and $r_c$ are fitting parameters used to reproduce the
numerical data. However, in bcc transition metals the bond integrals display a marked discontinuity
between the data corresponding to the first and second nearest neighbors that cannot be reproduced
sufficiently well by the above analytical function. This discontinuity was recognized to result from
different environments of the first and second nearest neighbors and can be captured by considering
different screening of their $d-d$ bonds by $s$ valence electrons on neighboring atoms
\citep{nguyen-manh:00}. The difference between these environments has been accounted for by
introducing the so-called \emph{screening function} $S_\tau(r_{ij})$ that determines the degree of
screening of the bond integral $dd\tau$ for a $d-d$ bond between atoms $i$ and $j$. Within this
modified formulation, called hereafter the \emph{screened BOP}, the bond integrals read
\begin{equation}
  \tilde{\beta}_\tau(r_{ij}) = \beta_\tau(r_{ij}) \left[ 1-S_\tau(r_{ij}) \right] \ ,
\end{equation}
where $\beta_\tau(r_{ij})$ is the GSP function (\ref{eq_GSPfun}). The screening function involves
contributions arising from hopping of electrons between different atoms via both bond and overlap,
and details of this functional form can be found, for example, in \citet{mrovec:04} and
\citet{aoki:07}. In practical calculations the bond part, $U_{bond}$, is evaluated using the Oxford
Order-N (OXON) package that has been modified at the University of Pennsylvania during the earlier
developments of BOPs. This term in \refeq{eq_BOP_energy} is based solely on data obtained from
\emph{ab initio} calculations. It is important to emphasize that $U_{bond}$ represents the bonding
between two atoms and does not include any short-range repulsion. Therefore, $U_{bond}$ alone
does not predict the lattice parameter, elastic moduli, or cohesive energy.

A part of the repulsion is described by the environmental term, $U_{env}$. This term is represented
by the screened Yukawa-type potential and is parameterized such that the experimental Cauchy
pressure, $C_{12}-C_{44}$, is reproduced exactly by $U_{bond}+U_{env}$; $U_{pair}$ does not
contribute to the Cauchy pressure.

Finally, the pair potential part, $U_{pair}$, is added to account for the overlap and electrostatic
interaction between individual atoms. It is repulsive at short separations of atoms but may be
attractive at intermediate distances. It is constructed as a sum of cubic splines whose coefficients
are adjusted such that the sum of the three terms entering the binding energy (\ref{eq_BOP_energy})
reproduces correctly the cohesive energy, lattice parameter, and two elastic moduli that are left
after fixing the Cauchy pressure. This term completes the construction of the BOP. The total
internal energy of a system of atoms is then given according to \refeq{eq_BOP_energy}.

  \chapter{Atomistic simulations of an isolated screw dislocation in Mo under stress}
\label{chap_MoBOP}

\begin{flushright}
  Science is facts; \\
  just as houses are made of stone, so is science made of facts;\\
  but a pile of stones is not a house, and a collection of facts is not necessarily science. \\
  \emph{Jules H. Poincar\'e}
\end{flushright}

Plastic deformation of single crystals of bcc metals is governed by the properties of screw
dislocations. The reason is that screw dislocations in these materials exhibit low mobility owing to
their complex dislocation cores \citep{duesbery:89, duesbery:98}. In the following, we will
investigate the behavior of screw dislocations in molybdenum by carrying out a series of extensive
atomistic simulations in which we study the motion of an isolated $1/2[111]$ screw dislocation under
various loadings. The goal of this modeling is to identify the stress components that affect the
motion of individual screw dislocations and subsequently to quantify their effects on the magnitude
of the Peierls stress, the shear stress parallel to the slip direction that moves the
dislocation. The analysis of the effect of interactions between dislocations is provided in
Appendix~\ref{chap_interact}.

The crystallographic data and elastic moduli of molybdenum are given below.

\begin{table}[!htb]
  \begin{minipage}{0.5\textwidth}
    \centering
    \begin{tabular}{lcl}
      \hline
      $\gdir{100}$ lattice parameter & $a$   & 3.147 \\
      Shortest periodicity in $\gplane{110}\gdir{112}$ & $a_0$ & 2.570 \\
      Magnitude of the Burgers vector & $b$   & 2.726 \\
      \hline
    \end{tabular} \\
    Dimensions:  [\A]
  \end{minipage}
  \hfill
  \begin{minipage}{0.4\textwidth}
    \centering
    \begin{tabular}{lcl}
      \hline
                     & $C_{11}$ & 4.647 \\
      Elastic moduli & $C_{12}$ & 1.615 \\
                     & $C_{44}$ & 1.089 \\
      $\gdir{111}$ shear modulus & $\mu$ & 1.374 \\
      Anisotropy factor & & 0.72 \\
      \hline
    \end{tabular} \\
    Dimensions:  $C_{ij},\mu~[10^{5}~\MPa]$
  \end{minipage}
\end{table}

%----------------------------------------------------------------------------------------------------
%----------------------------------------------------------------------------------------------------

\section{$\gamma$-surface and energy of stacking faults}
\label{sec_gsurf_MoBOP}

It has been known for a long time that dislocations in crystalline materials minimize their energy
either by splitting into partials separated by well-defined stacking faults or by spreading their
cores continuously into certain crystallographic planes \citep{hirth:82}. In the latter case the
fault within the core varies continuously, sampling relative displacements of the two parts of the
crystal shifted with respect to each other along the plane of spreading. The early models of core
spreading are due to \citet{peierls:40} and \citet{nabarro:47} who assumed that the energy
associated with the relative displacement of the two parts of the crystal varies sinusoidally. A
more general analysis of this energy is based on the concept of the generalized stacking fault,
introduced by \citet{vitek:68} in connection with the search for stacking faults in bcc metals.
We shall briefly introduce this approach here.

Consider a perfect crystal that is separated into two parts by a planar cut. The lower part of the
crystal is kept fixed while the upper part is rigidly displaced by an arbitrary vector $\mat{t}$
defined in the plane of the cut. The planar fault created in this way, which is in general unstable,
is a generalized stacking fault. It has a higher energy per unit area than the original perfect
crystal, and this surplus energy will be denoted as $\gamma(\mat{t})$. As the vector $\mat{t}$ spans
the unit cell in the plane of the cut, $\gamma(\mat{t})$ generates a surface that represents
energies of the generalized stacking faults, commonly called the $\gamma$-surface. Usually the
$\gamma$-surfaces are calculated while allowing relaxations of atoms in the direction perpendicular
to the plane of the cut but not parallel to the cut. In the following we call such a
$\gamma$-surface \emph{relaxed}, while if the positions of atoms are not relaxed, i.e. the atoms are
frozen in their displaced positions, the $\gamma$-surface is called \emph{unrelaxed}. It should be
emphasized that relaxed $\gamma$-surfaces have always lower energies per unit area than the
corresponding unrelaxed $\gamma$-surfaces that are calculated by freezing the atoms in their
displaced positions.

\begin{figure}[!htb]
  \centering
  \includegraphics[width=13cm]{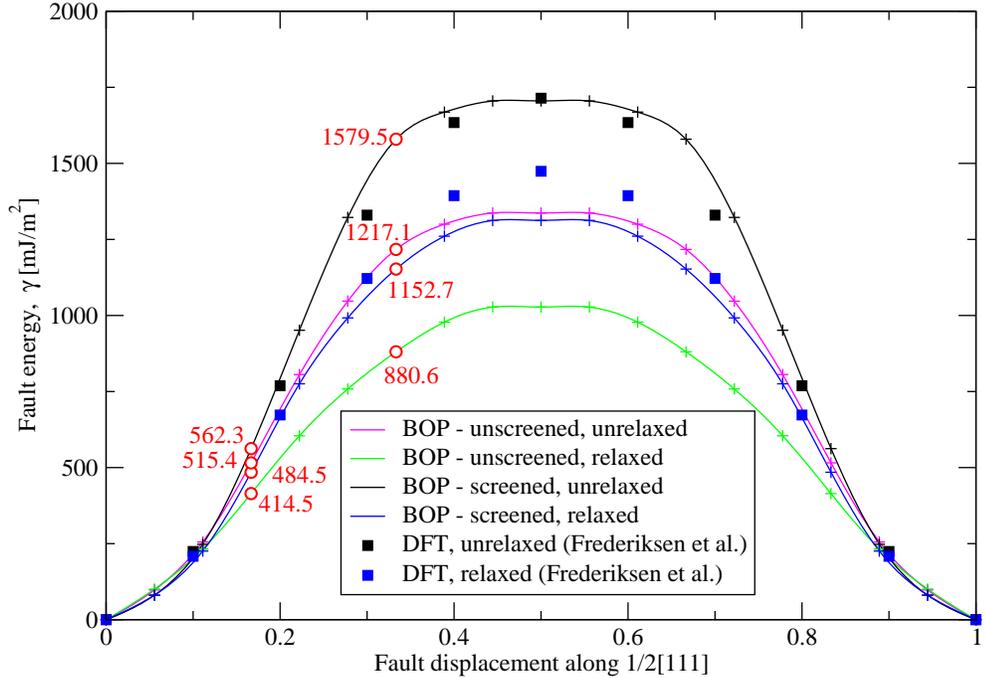}
  \parbox{15cm}{\caption{$\gdir{111}$ cross-section of $\gplane{110}$ $\gamma$-surfaces for molybdenum
  calculated using BOP. The squares correspond to the DFT calculations of \citet{frederiksen:03}.}
    \label{fig_scrMo_110gsurf}}
\end{figure}

The concept of $\gamma$-surfaces is in the first place invaluable in finding possible metastable
stacking faults that play an important role in dislocation dissociation. This can be done by first
recognizing that $\mat{f}=-\nabla\gamma(\mat{t})$ determines the restoring force of the lattice that
tends to destroy the disregistry between the upper and lower part of the crystal. If the
$\gamma$-surface displays intermediate minima, it is favorable for the dislocation to split into
partial dislocations whose Burgers vectors correspond to these minima, in which case the restoring
force would vanish. An example is the splitting of $1/2\gdir{110}$ dislocations into a pair of
$1/6\gdir{112}$ Shockley partials in fcc crystals.

$\gamma$-surfaces can be calculated relatively easily by means of atomistic simulations utilizing
empirical potentials \citep{duesbery:98} or even using the DFT-based methods (e.g. for bcc
transition metals see \citet{frederiksen:03}). For the prediction of the structure of the
dislocation core in bcc metals, it is natural to look at the $\gdir{111}$ cross-section of these
surfaces, since this is the direction of the Burgers vector of dislocations in this lattice. Since
the most densely packed plane in the bcc structure is $\gplane{110}$ and, as shown below, the screw
dislocations spread onto these planes, we have calculated the $[111]$ cross-section of the
$\gamma$-surface on the $(\bar{1}01)$ plane using both the BOP with and without screening, and with
and without relaxation. The results are shown in \reffig{fig_scrMo_110gsurf} together with analogous
DFT-based calculations of \citet{frederiksen:03}. As we see from \reffig{fig_scrMo_110gsurf}, no
minima appear on the calculated $\gamma$-surface, and, therefore, there are no metastable stacking
faults on the $\gplane{110}$ plane. Similarly, no minima are found on the $\gplane{112}$
$\gamma$-surface. The same result was obtained in all previous calculations of $\gamma$-surfaces in
bcc metals (for reviews see e.g. \citet{duesbery:89, duesbery:91, duesbery:02, moriarty:02}), and it
is thus likely that the non-existence of metastable stacking faults is a general feature of the bcc
structure. This implies that $1/2\gdir{111}$ screw dislocations in molybdenum cannot dissociate into
partial dislocations and always preserve their total Burgers vector.

The comparison of the $\gamma$-surfaces calculated using BOP and \emph{ab initio} is complicated by
different characters of the simulation cell in the direction perpendicular to the fault
plane. Unlike BOP whose block is effectively infinite in this direction, the DFT-based studies
employ a fully periodic simulation cell. This makes no difference when comparing the unrelaxed
$\gamma$-surfaces calculated by the two methods, and hence the close agreement between the two shows
that the screened BOP reproduces the \emph{ab initio} results very closely. However, if the atoms
are allowed to relax, the periodicity of the cell used in \emph{ab initio} calculations partially
confines the motion of atoms during the relaxation, while no such restriction is imposed on the
energy minimization using BOP. Consequently, the difference between the relaxed $\gamma$-surfaces
obtained \emph{ab initio} and using BOP, shown in \reffig{fig_scrMo_110gsurf}, can be attributed to
the incompatibility of the simulation cells in the two methods. Finally, the results also
demonstrate that the screened BOP reproduces the \emph{ab initio} data much more closely than the
unscreened BOP, and thus introduction of screening of bond integrals is an essential ingredient of
the BOPs; this was also demonstrated on other examples by \citet{mrovec:04}.

\begin{figure}[!htb]
  \centering
  \includegraphics{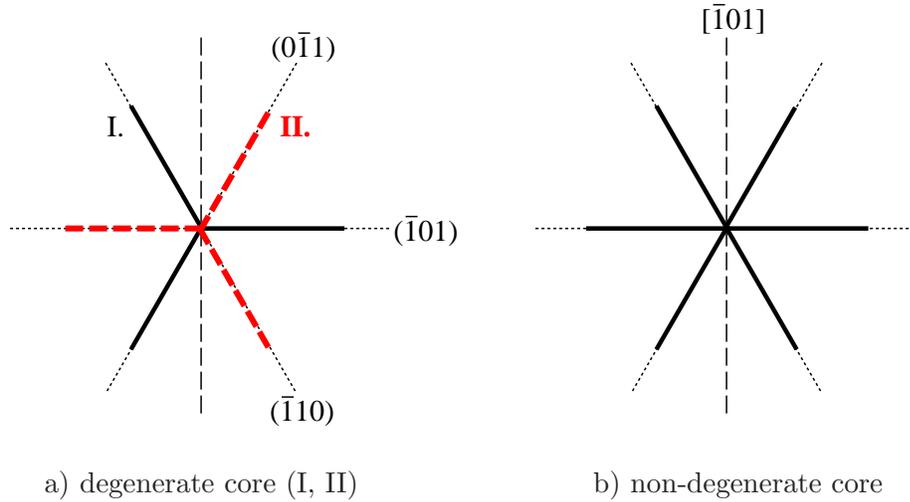} \\
  a) degenerate core (I, II) \hskip3cm b) non-degenerate core \\[1em]
  \parbox{14cm}{\caption{Schematic illustration of the two types of dislocation cores that obey the
      three-fold symmetry dictated by the lattice.}
  \label{fig_core_types}}
\end{figure}

Following von Neumann's principle \citep{neumann:1885}, the symmetry of any physical property of a
crystal must include the symmetry elements of the point group of this crystal. Hence, the three-fold
symmetric character of $\gdir{111}$ axes dictates that the unstressed cores of $1/2\gdir{111}$
screw dislocations must possess at least the three-fold rotational symmetry. This permits two different
types of cores that are shown schematically in \reffig{fig_core_types}. Each of the two variants of
the degenerate core in \reffig{fig_core_types}a can be perceived as a generalized splitting into
three fractional dislocations with Burgers vectors $1/6[111]$ on $(\bar{1}01)$, $(0\bar{1}1)$, and
$(\bar{1}10)$ planes \citep{vitek:04a}. Similarly, the non-degenerate core in
\reffig{fig_core_types}b can be regarded as a generalized splitting into six fractional dislocations
with Burgers vectors $1/12[111]$ on the same planes as above. However, unlike the degenerate core
that possesses the minimal symmetry dictated by the von Neumann's principle, the non-degenerate core
is further invariant with respect to the $[\bar{1}01]$ diad. Which type of core is energetically
favored can be assessed by comparing the values of the $\gamma$-surface corresponding to the two
types of fractional displacements according to the Duesbery-Vitek rule \citep{duesbery:98}. In
particular, the degenerate core will be preferred if $3\gamma(b/3)<6\gamma(b/6)$, while if the
opposite is true the non-degenerate core is favored. From the $\gamma$-surfaces in
\reffig{fig_scrMo_110gsurf}, one can easily verify that both screened and unscreened BOPs predict
the existence of the non-degenerate core (see also \citet{mrovec:02}).

%----------------------------------------------------------------------------------------------------
%----------------------------------------------------------------------------------------------------

\section{Simulation block and structure of the dislocation core}

In the following simulations, we use a block of atoms that is oriented such that the $z$-axis coincides
with the $[111]$ direction (and thus with the Burgers vector and line direction of the dislocation
studied), $y$ is perpendicular to the $(\bar{1}01)$ plane, and $x$ to both $y$ and $z$ such that the
coordinate system is right-handed. To simulate an infinitely long straight screw dislocation, we use
periodic boundary conditions along the $z$-direction. This reduces the number of $(111)$ atomic
layers in the model to three, with the nearest layer-to-layer distance equal to $1/2\sqrt{3}$ in
units of the lattice parameter. In our case, the size of the block in the $xy$ plane was about $20
\times 20$ lattice parameters.

\begin{figure}[!htb]
  \centering
  \includegraphics[width=8cm]{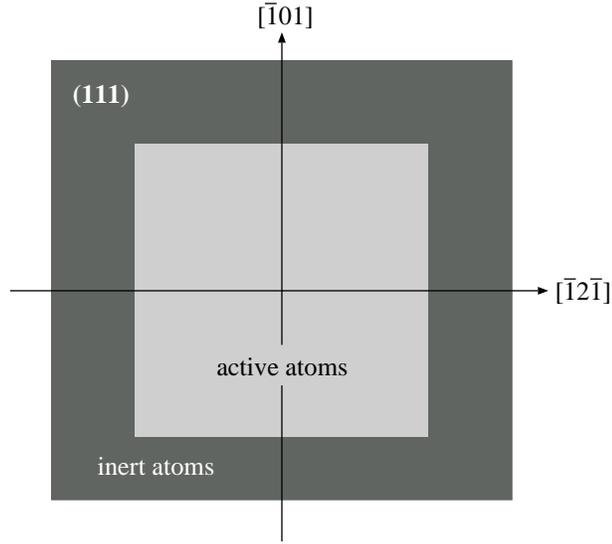}
  \parbox{12cm}{\caption{The simulated block consists of three $\gplane{111}$ planes. The dark region extends
    effectively to infinity.}
  \label{fig_simul_block}}
\end{figure}

\begin{figure}[!htb]
  \centering
  \includegraphics[width=8cm]{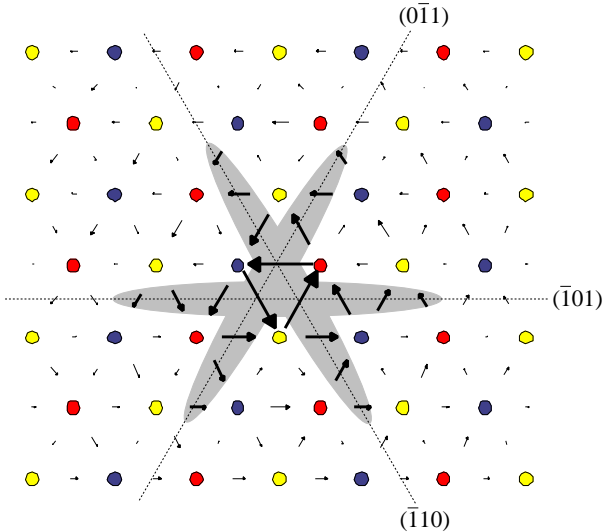}
  \parbox{11cm}{\caption{Structure of the $1/2[111]$ screw dislocation core calculated using the BOP
      for molybdenum.}
  \label{fig_bl_core_MoBOP}}
\end{figure}

Starting with an ideal crystal, we inserted a $1/2[111]$ screw dislocation by displacing all atoms
in the block according to the elastic anisotropic strain field of the dislocation
\citep{hirth:82}. The active atoms of the block, shown in \reffig{fig_simul_block}, were
subsequently relaxed using the BOP for molybdenum \citep{mrovec:04} while holding the atoms in the
inert region fixed. The differential displacement map of the relaxed dislocation core is shown
in \reffig{fig_bl_core_MoBOP}. In this projection, the circles stand for atoms in the three
successive $(111)$ layers. The lengths of the arrows correspond to the displacements of two
neighboring atoms parallel to the Burgers vector, i.e. perpendicular to the plane of the figure,
relative to their distance in the perfect lattice. The three longest arrows close to the center of
the figure, each corresponding to the relative displacement vector $1/6[111]$ in units of the
lattice parameter, define a Burgers circuit that gives $1/2[111]$, the total Burgers vector of the
dislocation. The same net product is obtained when going around the six second-largest arrows in the
figure, each giving a relative displacement equal to $1/12[111]$, or around any other circuit
encompassing the dislocation.

The dislocation core shown in \reffig{fig_bl_core_MoBOP} is non-degenerate, as predicted from the
shape of the $\gamma$-surface using the Duesbery-Vitek rule. The core is spread symmetrically on the
three $\gplane{110}$ planes of the $[111]$ zone, namely on the $(\bar{1}01)$ plane whose trace on
the $(111)$ plane coincides with the $x$-axis, and on $(0\bar{1}1)$ and $(\bar{1}10)$ planes that
both make $60\deg$ with the $(\bar{1}01)$ plane. Due to the non-planar character of the dislocation
core, screw dislocations in molybdenum are less mobile than both edge and mixed dislocations, and
their motion therefore determines the onset of macroscopic plastic deformation. In order to move the
screw dislocation, the applied stress must first transform the core from its initially sessile
configuration (\reffig{fig_bl_core_MoBOP}) into a less symmetric form that can be glissile in the
slip plane. Associated with this transformation is a large energy barrier that must be overcome to
move the dislocation. This is in contrast to fcc metals and to basal slip in hexagonal crystals, in
which screw dislocations, dissociated into Shockley partials, move at low stresses without
significant changes in the dislocation core.

In order to obtain the full information about the motion of the screw dislocation subjected to a
general loading, it is first necessary to identify the components of the stress tensor that
affect the magnitude of the shear stress parallel to the slip direction at which the dislocation
moves. For example, in the case of bcc tantalum, \citet{yang:01a} have shown that ambient
hydrostatic pressure causes only marginal changes of the dislocation core and does not affect the
stress at which the dislocation moves. However, under extreme conditions such as nuclear explosions
or laser-induced shock waves, large hydrostatic stresses may be responsible for significant changes
in the electronic structure, and this may no longer be compatible with the interatomic potential.

%----------------------------------------------------------------------------------------------------
%----------------------------------------------------------------------------------------------------

\section{Loading by shear stress parallel to the slip direction}
\label{sec_spar_MoBOP}

For the application of stress the simulated block is divided into two domains, as shown in
\reffig{fig_simul_block}. The atoms in the outer part, called inert atoms, are displaced
according the anisotropic strain field corresponding to the applied stress tensor and the long range
strain field of the dislocation. The atoms in the inner region, called active atoms, are then
relaxed by minimizing the total energy (\ref{eq_BOP_energy}) of the block. In the first series of
simulations, we study the effect of pure shear parallel to the slip direction that exerts a nonzero
Peach-Koehler force on the dislocation and thus directly drives its motion.

\begin{figure}[!htb]
  \centering 
  \includegraphics[width=8.5cm]{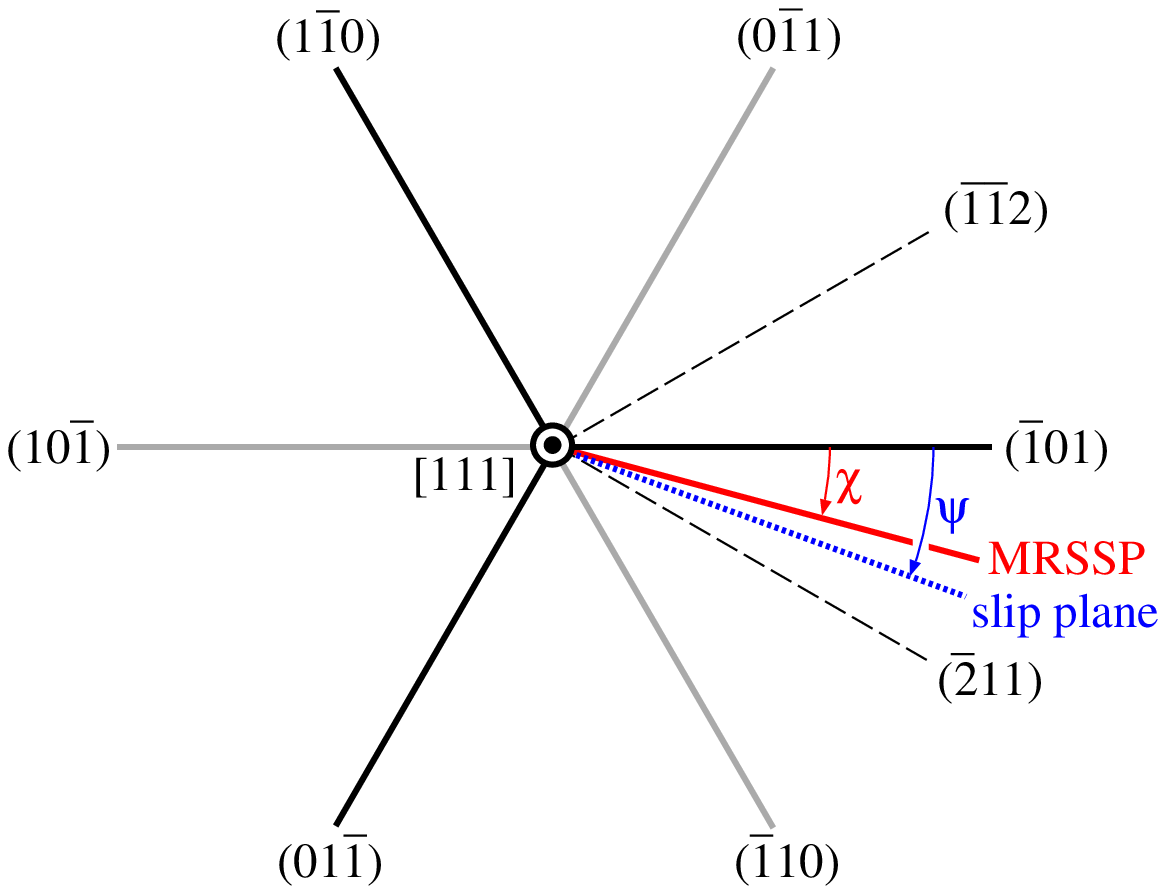} \\
  \parbox{13cm}{\caption{Orientation of the planes in the $[111]$ zone.}
  \label{fig_planes}}
\end{figure}

According to the Schmid law, which is well established in close-packed structures but does not hold
in bcc metals, the shear stress parallel to the slip direction in the plane of the slip at which the
dislocation starts to move is independent of the orientation of the plane of applied loading. In
order to obtain the actual orientation dependence of the critical resolved shear stress (CRSS)
parallel to the slip direction at which the dislocation starts to move, we carried out a series of
atomistic simulations for different orientations of the plane in which the shear stress parallel to
the slip direction is applied. This plane is called the maximum resolved shear stress plane (MRSSP)
and its orientation is defined by the angle $\chi$ which it makes with the $(\bar{1}01)$ plane, as
shown in \reffig{fig_planes}. Due to the crystal symmetry, it is sufficient to consider only the
MRSSPs in the angular region $-30\deg<\chi<+30\deg$. This region is bounded by two $\gplane{112}$
planes that are twinning planes in bcc crystals. For $\chi<0$, the nearest $\gplane{112}$ plane,
i.e. $(\bar{1}\bar{1}2)$, is sheared in the \emph{twinning} sense. On the other hand, for $\chi>0$,
the nearest $\gplane{112}$ plane, i.e. $(\bar{2}11)$, is sheared in the \emph{antitwinning} sense.

The applied stress tensor is given in the right-handed coordinate system with the $y$-axis normal to
the MRSSP and $z$ parallel to the slip direction. In this orientation, the stress tensor for the
applied shear stress, $\sigma$, parallel to the slip direction acting in the MRSSP is
\begin{equation}
  \mathbf{\Sigma} = \left(
    \begin{array}{ccc}
      0 & 0 & 0 \\
      0 & 0 & \sigma \\
      0 & \sigma & 0
    \end{array}
  \right) \quad.
  \label{eq_tensor_sigma}
\end{equation}
In the application of stress, we started with a completely relaxed block with the $1/2[111]$ screw
dislocation in the middle, as shown in \reffig{fig_bl_core_MoBOP}. To ensure that the stress at
which the dislocation starts to move, the critical resolved shear stress (CRSS), is found with
sufficient precision, the shear stress $\sigma$ was built up incrementally in steps of
$0.001C_{44}$, where $C_{44}$ is the elastic modulus. In each loading step, the atomic block was
fully relaxed before increasing $\sigma$. At low stresses, the dislocation core transforms from its
initially sessile, three-fold symmetric non-degenerate core, to a less symmetric form. This
transformation is purely elastic in that the block returns into its original configuration when the
stress is removed. Once the applied shear stress $\sigma$ attains the CRSS, the transformation is
complete, and the dislocation moves through the crystal.

\begin{figure}[!htb]
  \centering
  \includegraphics[width=13cm]{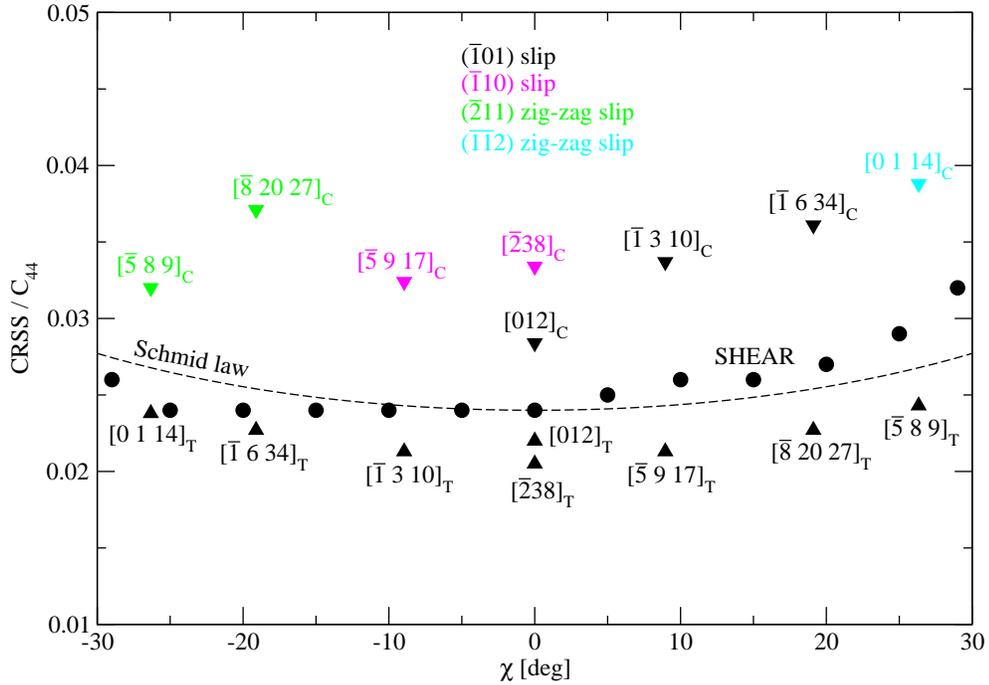}
  \parbox{14cm}{\caption{Orientation dependence of the CRSS for loading by shear stress parallel to
      the slip direction (circles), tension (up-triangles), and compression (down-triangles). The
      circles and up-triangles correspond to slip on the $(\bar{1}01)$ plane; different slip
      planes for loading in compression are labeled separately.}
  \label{fig_CRSS_chi_MoBOP}}
\end{figure}

The obtained dependence of the CRSS on the orientation of the MRSSP, $\chi$, is plotted as circles
in \reffig{fig_CRSS_chi_MoBOP}. For all orientations of the MRSSP, the dislocation moved on the
$(\bar{1}01)$ plane, which coincides here with the most highly stressed $\gplane{110}$ plane of the
$[111]$ zone. If the Schmid law were valid in molybdenum, the CRSS would vary with $\chi$ as
$\cos^{-1}\chi$, showed as the dashed curve in \reffig{fig_CRSS_chi_MoBOP}, and all data points in the
graph would lie on this curve. In our case, however, the CRSS for antitwinning shear ($\chi>0$)
is always higher than the corresponding value for the twinning shear ($\chi<0$). This is the
well-known twinning-antitwinning asymmetry that is commonly observed in experiments on many
bcc metals and is one evidence of the breakdown of the Schmid law in these materials.

%----------------------------------------------------------------------------------------------------

\section{Loading in tension and compression}
\label{sec_tc_MoBOP}

\begin{figure}[!htb]
  \centering
  \includegraphics[width=9cm]{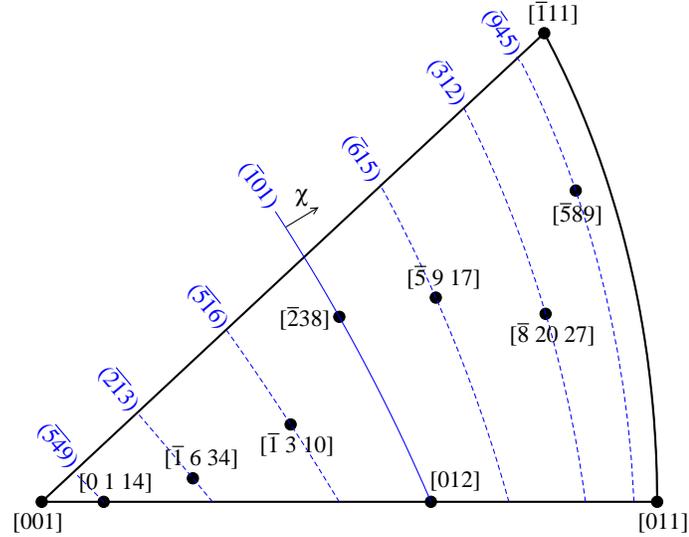}
  \parbox{12cm}{\caption{Stereographic triangle corresponding to the orientation of the block in
      which $(\bar{1}01)$ is the most highly stressed $\gplane{110}$ plane in the zone of the
      $[111]$ slip direction.}
  \label{fig_sgtria_tc}}
\end{figure}

Recall that any uniaxial loading exerts a general triaxial \emph{strain} state at any point of the
loaded body.  Therefore, an important test that reveals whether or not the shear stress parallel to
the slip direction is the only stress component that affects the dislocation movement is to apply
uniaxial loadings in several different orientations and calculate the corresponding CRSS at which
the dislocation moves. Due to the symmetry of bcc crystals, the complete set of loading axes can be
found within the stereographic triangle shown in \reffig{fig_sgtria_tc}. For our orientation of the
crystal, where $(\bar{1}01)$ is the most highly stressed $\gplane{110}$ plane of the $[111]$ zone,
the three corners of the stereographic triangle of interest coincide with axes $[001]$, $[011]$,
$[\bar{1}11]$. 

The orientations of the loading axes in tension and compression that were studied are plotted in
\reffig{fig_sgtria_tc}. For any loading axis, one can always find the orientation of the MRSSP in
the zone of the $[111]$ slip direction. For the set of chosen orientations, the corresponding MRSSPs
are shown in \reffig{fig_sgtria_tc} as dashed curves. The shear stress parallel to the slip
direction resolved in the MRSSP, which moves the dislocation in tension/compression, is the CRSS
that can be directly compared with the value calculated for the same $\chi$ by applying the pure
shear stress parallel to the slip direction. In \reffig{fig_CRSS_chi_MoBOP}, the values of the CRSS
for tension are plotted as up-triangles and for compressions as down-triangles. If, for the same
$\chi$, the CRSS for tension/compression (triangles) were the same, or very close, to those
calculated for pure shear stress parallel to the slip direction (circles), we would conclude that
only the shear stress parallel to the slip direction controls the plastic flow of single crystals of
bcc molybdenum. However, the observed large deviation of the CRSS for tension/compression from the
CRSS for pure shear parallel to the slip direction implies that the motion of the screw dislocation
is also affected by other stress component(s).

The analysis described above was performed originally by \citet{ito:01} using the Finnis-Sinclair
potential for molybdenum. They concluded that the shear stresses perpendicular to the slip direction
also play role in the dislocation motion and thus in the plastic flow of bcc metals. The analysis in
\citet{ito:01} was confined only to $\chi=0$ and to exactly $\chi=\pm30\deg$ for which the MRSSP
coincides with the highly symmetrical $\gplane{211}$ planes. In the following section, our objective
is to investigate in detail the effect of the shear stress perpendicular to the slip direction and
its impact on the magnitude of the CRSS for slip for a number of orientations of the MRSSP.

%----------------------------------------------------------------------------------------------------
%----------------------------------------------------------------------------------------------------

\section{Effect of the shear stress perpendicular to the slip direction}
\label{sec_sperp_section}

Shear stress perpendicular to the slip direction does not exert any Peach-Koehler force
\citep{peach:50} on the dislocation and, therefore, cannot cause its movement. However, it plays an
important role in the transformation of the dislocation core \citep{duesbery:89, vitek:92, ito:01}.

%----------------------------------------------------------------------------------------------------

\subsection{Transformation of the dislocation core}
\label{sec_sperp_MoBOP}

The first obvious question is how the dislocation core changes upon applying a pure shear stress
perpendicular to the slip direction. Since no shear stress parallel to the slip direction is applied
here, it is not appropriate to use the term MRSSP when referring to the plane that defines the
orientation of applied loading. The stress tensor applied in the coordinate system where the
$y$-axis is normal to the plane defined by the angle $\chi$, and $z$ is parallel to the dislocation
line (and the slip direction) now has the following structure:
\begin{equation}
  \mathbf{\Sigma} = \left(
    \begin{array}{ccc}
      -\tau & 0 & 0 \\
      0 & \tau & 0 \\
      0 & 0 & 0
    \end{array}
  \right) \quad.
  \label{eq_tensor_tau}
\end{equation}
Here, $\tau$ is the magnitude of the shear stress perpendicular to the slip direction that is
resolved in this orientation as a combination of two normal stresses. In these simulations, the
applied stress was built up in steps of $0.005C_{44}$.

\begin{figure}[!htb]
  \centering
  \includegraphics[width=7.5cm]{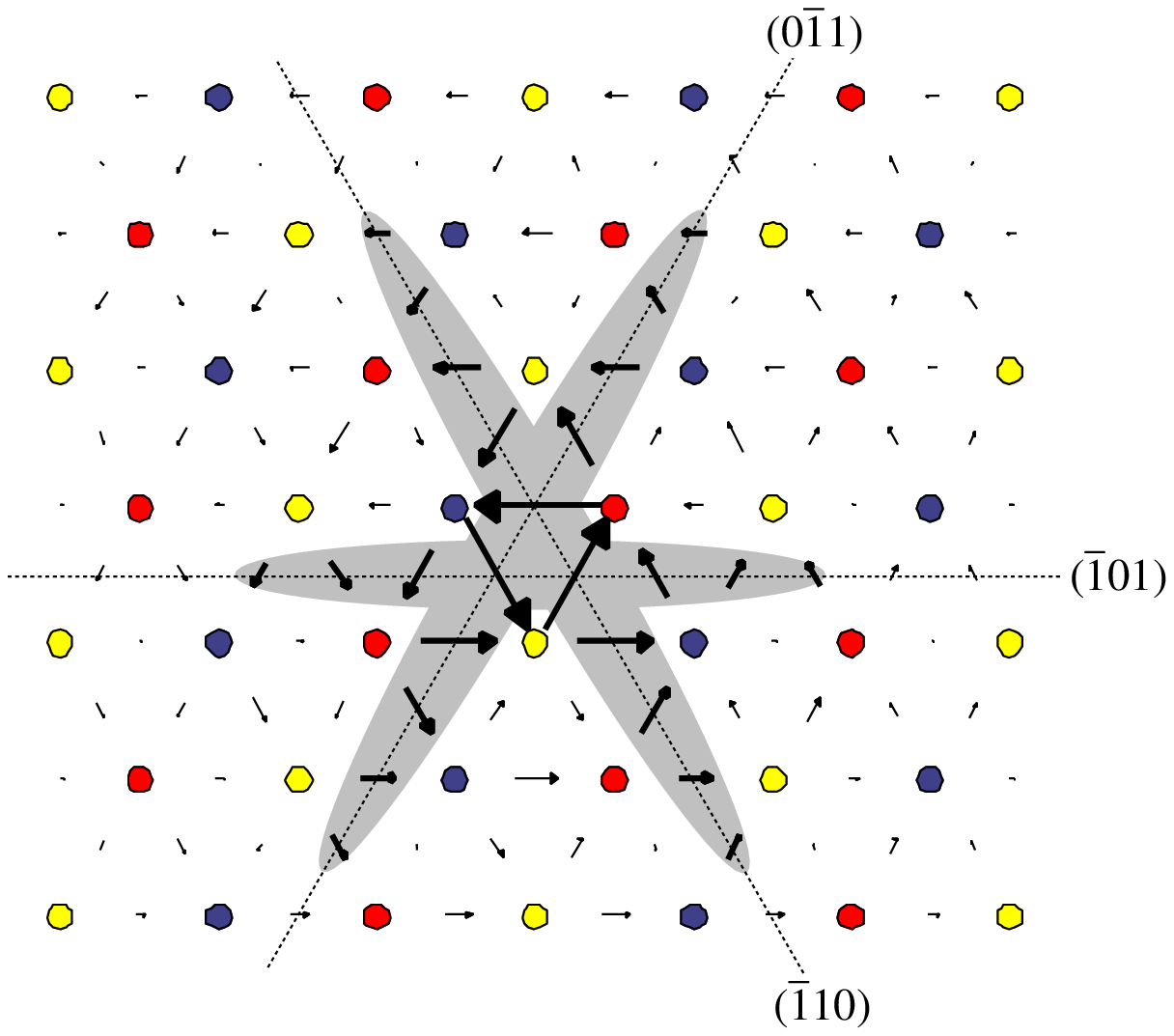} \hfill
    \includegraphics[width=7.5cm]{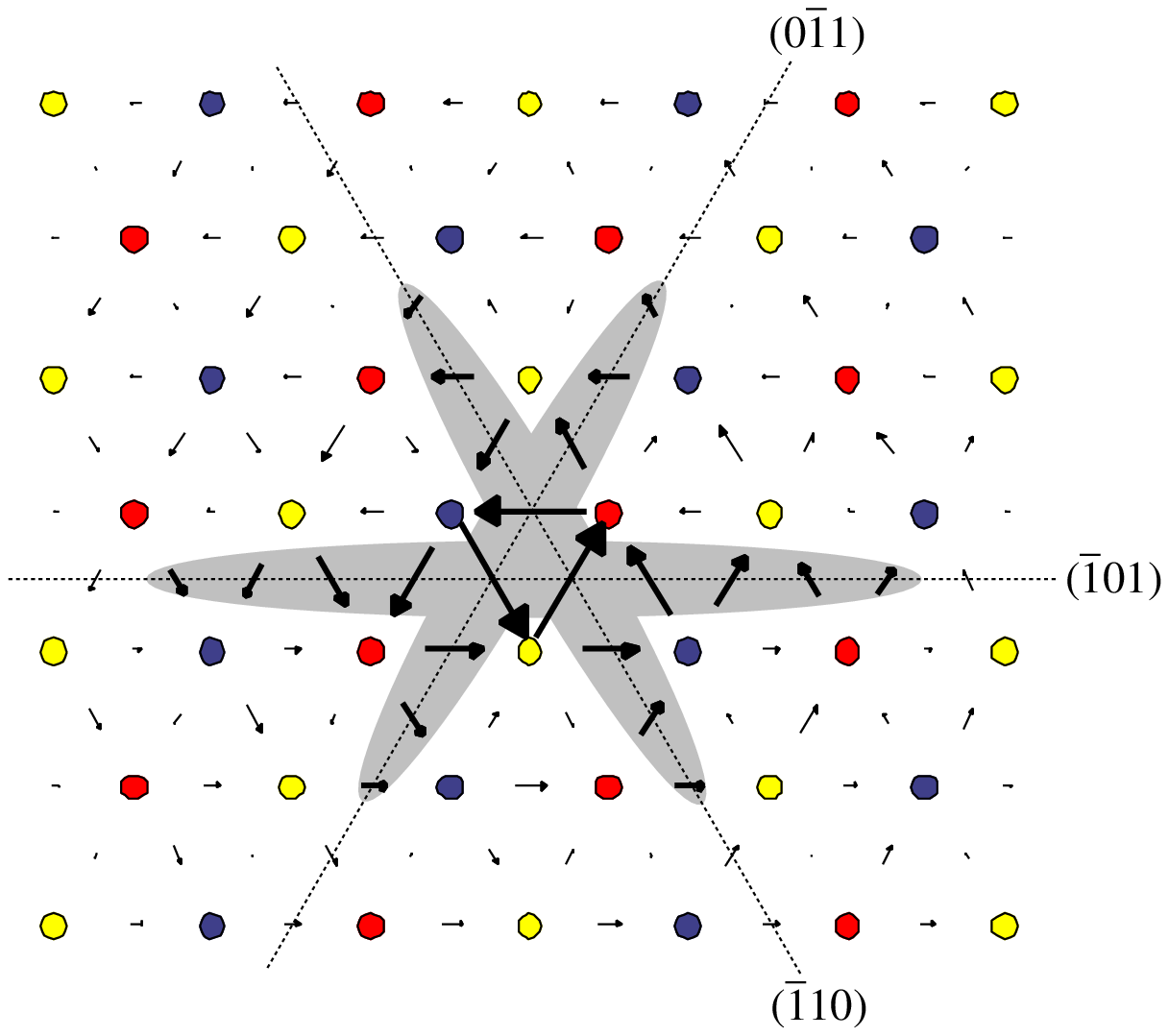}\\
  \hskip-1cm a) $\tau/C_{44}=-0.05$  \hskip5cm b) $\tau/C_{44}=+0.05$ \\
  \caption{Structure of the dislocation core upon applying (a) negative and (b)
  positive shear stress perpendicular to the slip direction in the coordinate system for which
  $\chi=0$.
  \label{fig_bl_tau0.05_MoBOP}}
\end{figure}

Because the shear stress perpendicular to the slip direction cannot move the dislocation, the
deformation exerted by this stress in the crystal is purely elastic. In the following, we will focus
on the case $\chi=0$, but the same analysis holds also for other angles $\chi$. The final structure
of the dislocation core obtained by relaxing the simulated block at $\tau=\pm 0.05C_{44}$ is shown
in \reffig{fig_bl_tau0.05_MoBOP}. For the negative $\tau$, the core constricts on the $(\bar{1}01)$
plane and extends on both $(0\bar{1}1)$ and $(\bar{1}10)$ planes. Due to the larger spreading of the
core on the two low-stressed $\{110\}$ planes, this suggests that the dislocation will be easier to
move in $(0\bar{1}1)$ or $(\bar{1}10)$ planes than in the $(\bar{1}01)$ plane. On the other hand,
for the positive $\tau$, the dislocation core extends on the $(\bar{1}01)$ plane and constricts on
both $(0\bar{1}1)$ and $(\bar{1}10)$ planes which suggests that the dislocation will move most
easily on the $(\bar{1}01)$ plane. Hence, one can expect that the subsequent loading by the shear
stress parallel to the slip direction will move the dislocation in the $(\bar{1}01)$ plane for
$\tau>0$, while $(0\bar{1}1)$ and $(\bar{1}10)$ may be the slip planes for $\tau<0$.

Besides the preference for slip on a particular $\{110\}$ plane of the $[111]$ zone, the changes in
the structure of the dislocation core also suggest how large a CRSS is needed to drive the dislocation
glide. For example, assume again that the crystal is loaded by the stress tensor $\mathbf{\Sigma}$
(Eq. \ref{eq_tensor_tau}) defined in the coordinate system where the $y$-axis coincides with the
normal to the $(\bar{1}01)$ plane. If we apply positive $\tau$, the core extension on the
$(\bar{1}01)$ plane makes the glide on this plane easier, as compared to the three-fold symmetric
core at zero $\tau$. Hence, one may expect that the $\CRSS$ at $\tau>0$ will decrease with
increasing $\tau$. On the other hand, applying a negative $\tau$ makes the glide on the
$(\bar{1}01)$ plane increasingly more difficult and, at larger negative $\tau$, the $(\bar{1}01)$
glide may be suppressed completely. Instead, the dislocation glide may proceed exclusively on one of
the other two $\gplane{110}$ planes of the $[111]$ zone. However, because the shear stress parallel
to the slip direction resolved into the planes with $\chi=\pm 60 \deg$ is only $\CRSS/2$, an
appreciably larger CRSS for slip of the dislocation can be expected at larger negative $\tau$.

%----------------------------------------------------------------------------------------------------

\subsection{Dependence of the CRSS on the shear stress perpendicular to the slip direction}
\label{sec_sperp_dep}

From the previous text we know that the shear stress perpendicular to the slip direction can make
the slip either easier (the core extends further onto the $(\bar{1}01)$ plane) or more difficult
(the core extends out of the $(\bar{1}01)$ plane). In order to investigate the dependence of the
CRSS on the magnitude of the shear stress $\tau$ perpendicular to the slip direction, we carried out
a number of atomistic simulations of the $1/2[111]$ screw dislocation subjected to simultaneous
loading by shear stresses perpendicular and parallel to the slip direction. The shear stress
parallel to the slip direction was applied in the seven differently oriented MRSSPs for which we
have performed the simulations in tension/compression. In each case the shear stress perpendicular
to the slip direction (\ref{eq_tensor_tau}) was first applied in steps, as explained in
Section~\ref{sec_sperp_MoBOP}. Keeping the shear stress $\tau$ fixed, we then incrementally built up
the shear stress parallel to the slip direction, i.e. the stress tensor (\ref{eq_tensor_sigma}),
until the CRSS was attained and the dislocation moved into the next equivalent site. At this point
the applied stress tensor reads
\begin{equation}
  \mathbf{\Sigma} = \left(
    \begin{array}{ccc}
      -\tau & 0 & 0 \\
      0 & \tau & \CRSS \\
      0 & \CRSS & 0
    \end{array}
  \right) \quad.
  \label{eq_tensor_full}
\end{equation}
Note, that the plane of the maximum shear stress \emph{perpendicular} to the slip direction is
inclined with respect to the MRSSP by the angle $-45\deg$.

For each orientation of the MRSSP, the procedure outlined above was implemented for a number of
values of $\tau$, both positive and negative.  Without the loss of generality, we will discuss here
in detail only three orientations of the MRSSP, namely $(\bar{1}01)$ at $\chi=0$, and
$(\bar{3}12)$, $(\bar{2}\bar{1}3)$ at $\chi \approx \pm19\deg$. However, the complete set of
the $\CRSS-\tau$ dependencies calculated for the seven MRSSPs considered is shown in
\reffig{fig_CRSS_tau_MoBOP}. If the Schmid law were valid, the CRSS would be independent of $\tau$,
and for $\chi=0$ the CRSS would be the same as in the case of pure shear parallel to the Burgers
vector, specifically $0.024C_{44}$. In contrast, one can observe a strong dependence of the CRSS on
the shear stress perpendicular to the slip direction, in particular for negative $\tau$.

\begin{figure}[p]
  \centering
  \includegraphics[width=12cm]{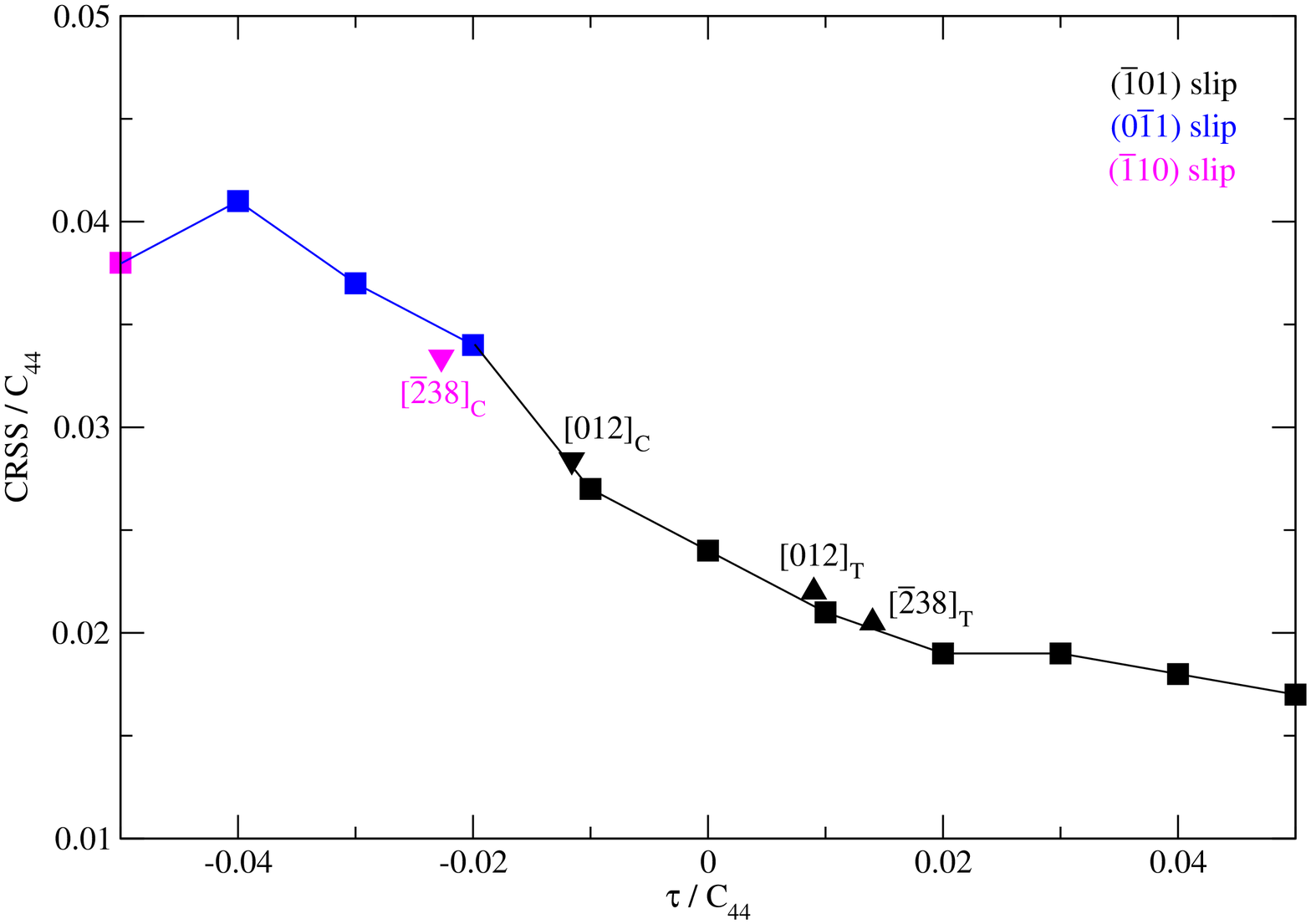} \\
  a) MRSSP $(\bar{1}01)$, $\chi=0$
\vskip2em
  \includegraphics[width=12cm]{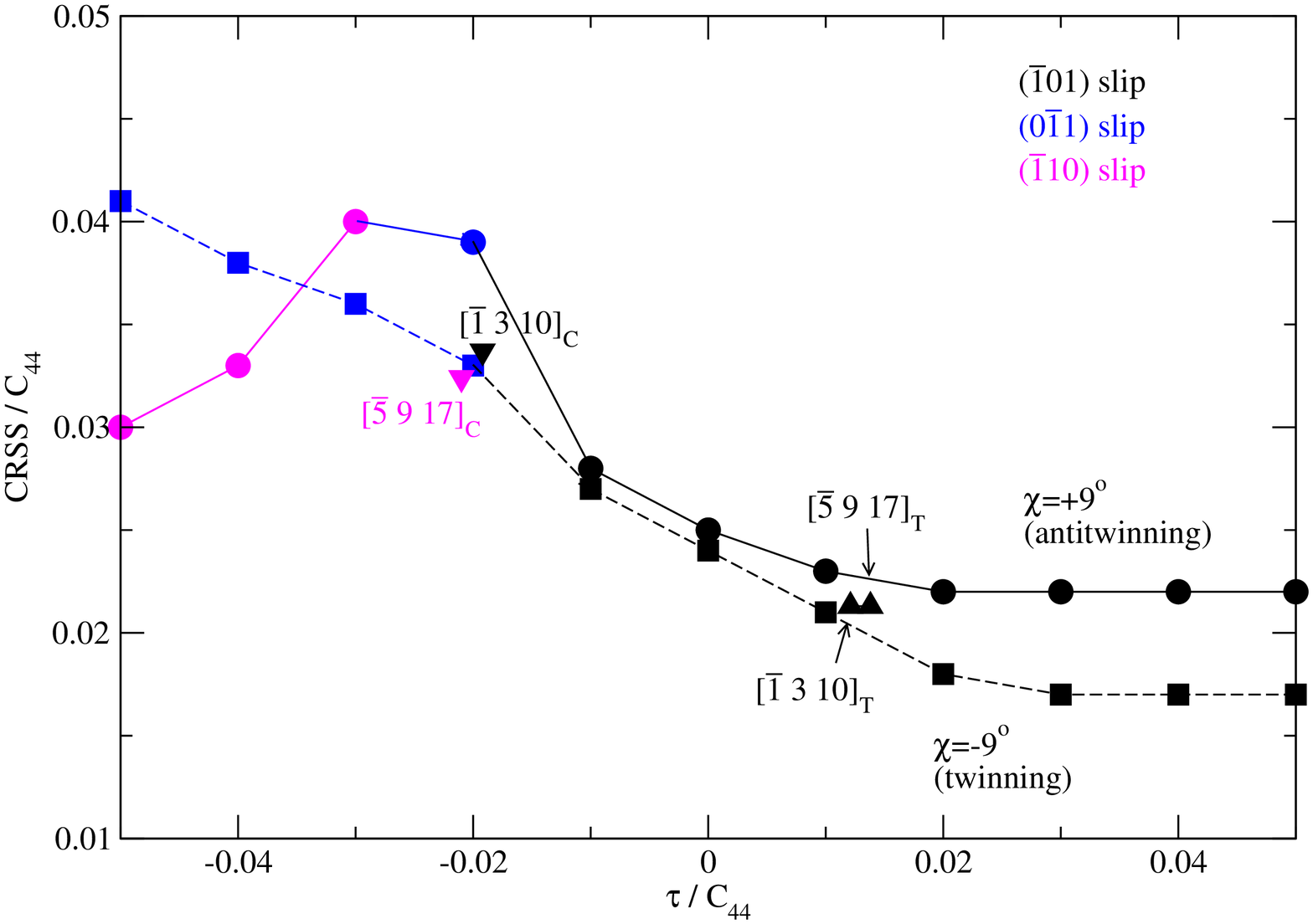} \\
  b) MRSSP $(\bar{6}15)$, $\chi=+9\deg$ (circles) and 
    $(\bar{5}\bar{1}6)$, $\chi=-9\deg$ (squares)
\end{figure}
\begin{figure}[p]
  \centering
  \includegraphics[width=12cm]{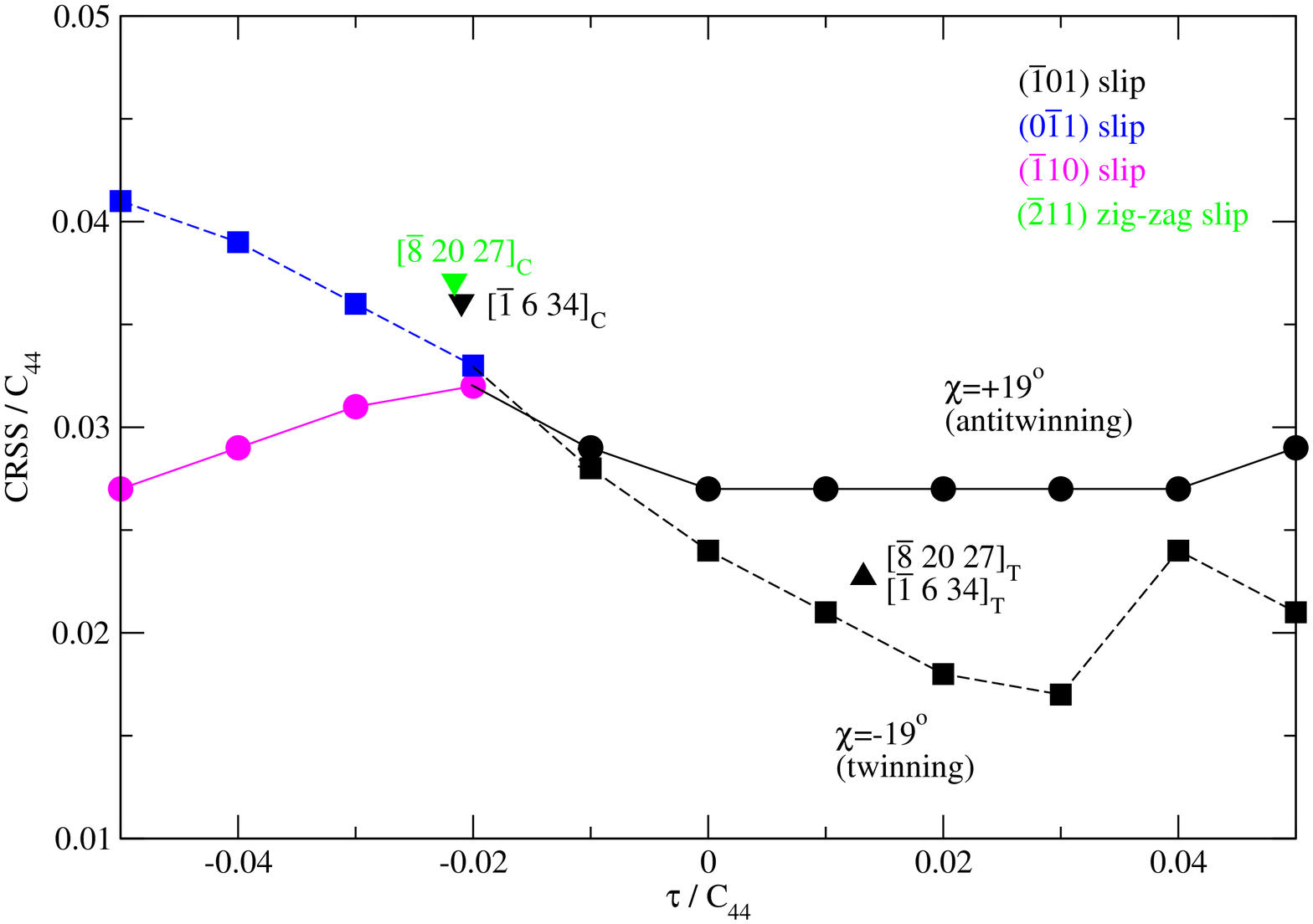} \\
  c) MRSSP $(\bar{3}12)$, $\chi=+19\deg$ (circles) and 
    $(\bar{2}\bar{1}3)$, $\chi=-19\deg$ (squares)
\vskip2em
  \includegraphics[width=12cm]{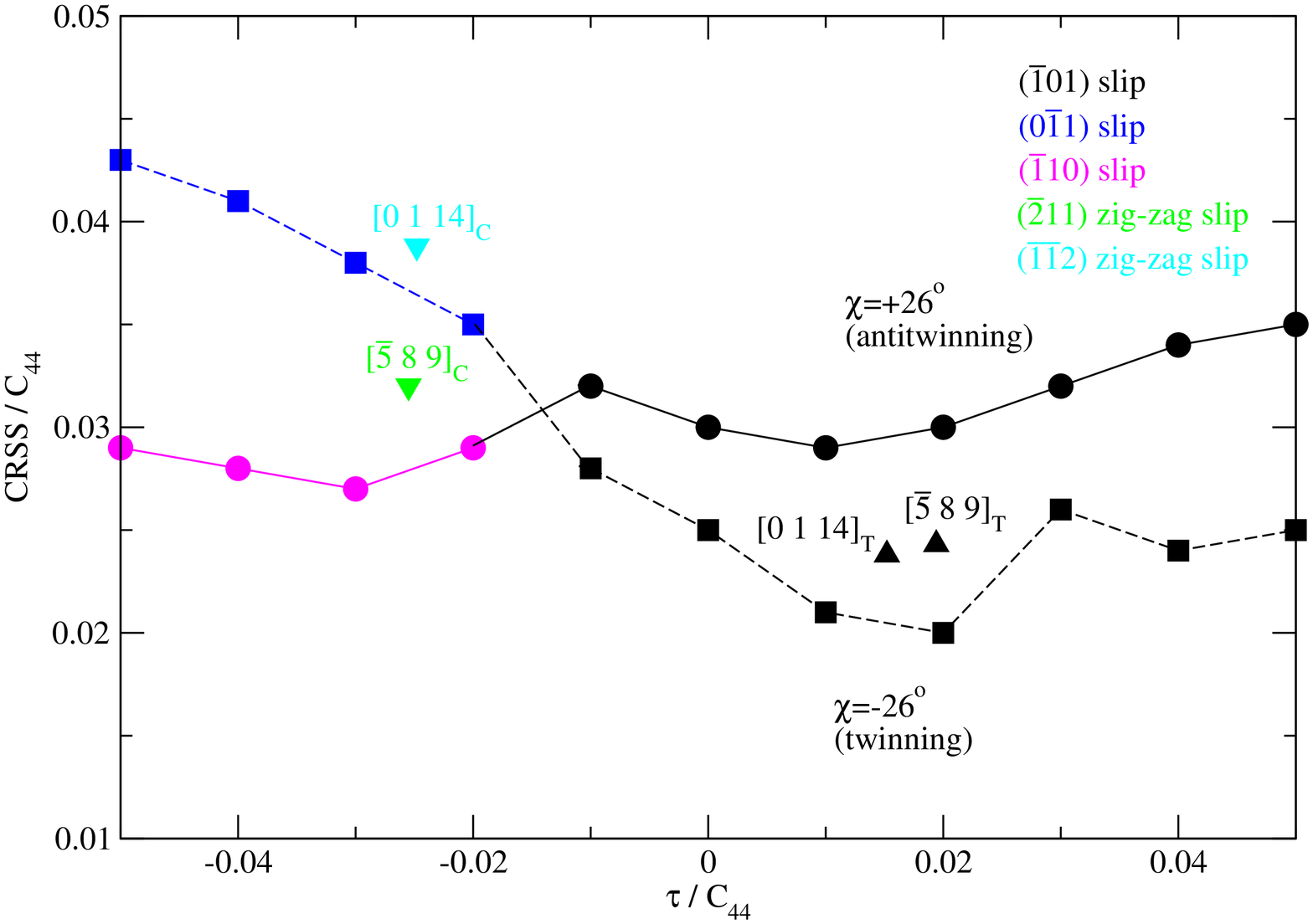} \\
  d) MRSSP $(\bar{9}45)$, $\chi=+26\deg$ (circles) and 
    $(\bar{5}\bar{4}9)$, $\chi=-26\deg$ (squares)
  \caption{Dependence of the CRSS on the shear stress perpendicular to the slip direction, $\tau$,
  for seven studied orientations of the MRSSP.}
  \label{fig_CRSS_tau_MoBOP}
\end{figure}

At positive $\tau$, the CRSS is lower when compared to $\tau=0$, and the slip always occurs on the
expected $(\bar{1}01)$ plane with the highest Schmid factor. Because the dislocation core
corresponding to positive $\tau$ spreads predominantly on the $(\bar{1}01)$ plane, see
\reffig{fig_bl_tau0.05_MoBOP}b, the shear stress perpendicular to the slip direction effectively
promotes the dislocation glide by lowering the CRSS required to overcome the barrier for slip on the
$(\bar{1}01)$ plane.

In contrast, negative $\tau$ extends the dislocation core out of the $(\bar{1}01)$ plane and thereby
makes the slip on this plane more difficult. For values of $\tau/C_{44}>-0.02$, the constriction of
the core on the $(\bar{1}01)$ plane is less than its extension on the $(0\bar{1}1)$ or the
$(\bar{1}10)$ plane. In this case the slip still proceeds on the expected $(\bar{1}01)$
plane. However, for larger negative values of $\tau$ the extension of the core into the
$(0\bar{1}1)$ or the $(\bar{1}10)$ plane becomes significant, and the slip then occurs preferentially
on one of these planes. It should be noted that the Schmid factors corresponding to both the
$(0\bar{1}1)[111]$ and the $(\bar{1}10)[111]$ slip systems are typically a half of that for the most
highly stressed $(\bar{1}01)[111]$ slip system. The occurrence of the slip on these planes is thus
reminiscent of the experimentally observed \emph{anomalous slip} that takes place on the slip
systems with very low Schmid factors \citep{bolton:72, creten:77, jeffcoat:76, matsui:76, reed:76}.

One can see by examining the $\CRSS-\tau$ dependencies shown in \reffig{fig_CRSS_tau_MoBOP}c, that
for $\chi=-19\deg$, which is in the regime of the twinning shear on the nearest $\gplane{112}$
plane, the behavior is similar to that found for $\chi=0$. However, for $\chi=+19\deg$, when
shearing on the nearest $\gplane{112}$ plane is in the antitwinning sense, the $\CRSS$ is almost
independent of $\tau$. The origin in this behavior must be hidden in the effect of $\tau$ on the
structure of the dislocation core. Let us consider application of the stress tensor
(\ref{eq_tensor_tau}) defined in the right-handed coordinate system where the $z$-axis is $[111]$
and $y$ is perpendicular to the $(\bar{3}12)$ plane, the MRSSP for $\chi=+19\deg$. Projecting the
stress tensor $\mathbf{\Sigma}$ into the coordinate system in which the $y$-axis is perpendicular to
$(\bar{1}01)$, $(0\bar{1}1)$ and $(\bar{1}10)$, respectively, one finds that the $(\bar{1}01)$ plane
is subjected to the shear stress perpendicular to the slip direction with the magnitude $+0.8\tau$, the
$(0\bar{1}1)$ plane, inclined by $-60\deg$ with respect to the $(\bar{1}01)$ plane, to $-0.9\tau$,
and the $(\bar{1}10)$ plane, inclined by $+60\deg$ with respect to the $(\bar{1}01)$ plane, to
$+0.1\tau$. Note that a positive projected shear stress perpendicular to the slip direction results
in an extension of the core in a given $\gplane{110}$ plane and a negative projected shear to a
constriction of the core. Clearly, the dislocation core extends with increasing $\tau$ on both
$(\bar{1}01)$ and $(\bar{1}10)$ planes. This competing extension of the core on two different
$\{110\}$ planes then effectively inhibits the sessile $\rightarrow$ glissile transformation of the
core in the $(\bar{1}01)$ slip plane and results in the observed independence of the CRSS on $\tau$.

Superimposed in \reffig{fig_CRSS_tau_MoBOP} are also results for the uniaxial loadings examined
earlier in Section \ref{sec_tc_MoBOP}. For loading in tension, the resolved shear stress
perpendicular to the slip direction is always positive, while for compression it is always
negative. For consistency, we consider here only the deviatoric part of the stress tensor. One can
observe a close agreement of the $\CRSS-\tau$ data calculated for tension/compression (triangles)
with the data for combined loading by the shear stress perpendicular and parallel to the slip
direction (squares/circles). Because the same agreement is obtained also for all other orientations
of the MRSSP, one can conclude that we have identified all stress components that affect screw
dislocation glide and thus the plastic deformation of single crystals of molybdenum. These stresses
are: (i) shear stress perpendicular to the slip direction that merely changes the structure of the
dislocation core but does not drive the glide process, and (ii) shear stress parallel to the slip
direction whose critical value (CRSS) determines the onset of the dislocation glide. The concomitant
effect of these stress components on the onset of glide of an isolated $1/2[111]$ screw dislocation
is given by the calculated $\CRSS-\tau$ dependencies.

%----------------------------------------------------------------------------------------------------
%----------------------------------------------------------------------------------------------------

\section{Prediction of the macroscopic slip plane from atomistics}

A number of plastic deformation experiments on pure single crystals of molybdenum and also other
refractory metals reveal that, for certain orientations of tensile/compressive axes, the macroscopic
slip does not proceed on any of the available $\gplane{110}$ planes but, instead, on planes of lower
symmetry. At first glance, the observation of the slip on a high-index plane appears to disagree
with the results of our atomistic simulations in which the $1/2[111]$ screw dislocation always moves
by elementary steps on one of the three $\gplane{110}$ planes of the zone of the slip direction. In
the following, we will demonstrate how the elementary microscopic steps of the dislocation on the
three $\gplane{110}$ planes can lead naturally to macroscopic slip on any plane containing the
$[111]$ slip direction.

\begin{figure}[!htb]
  \centering
  \includegraphics[width=6cm]{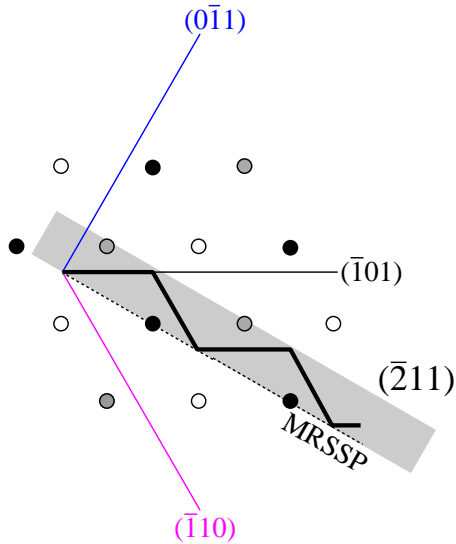} \\
  \parbox{12cm}{\caption{Schematic illustration of the $\gplane{112}$ zig-zag slip by elementary
      steps of the dislocation on the two adjacent $\gplane{110}$ planes. The circles represent the
      positions of atoms and the filled band the macroscopically observed slip trace on the
      $(\bar{2}11)$ plane.}
  \label{fig_slip112plane}}
\end{figure}

Slip on $\gplane{112}$ planes is often observed in experiments if the orientation of the MRSSP is
close to $\pm30\deg$, in which case two $\gplane{110}$ planes become subjected to almost identical
shear stresses parallel to the slip direction. In view of the results of atomistic modeling
presented earlier, this macroscopically observed $\gplane{112}$ slip can be understood as a
consequence of dislocations moving in a zig-zag fashion by elementary steps on the two most highly
stressed $\gplane{110}$ planes; see \reffig{fig_slip112plane}. Due to the limited resolution in
experiments, the individual $\gplane{110}$ steps cannot be directly observed, and the onset of
plastic deformation reveals itself on the macroscopic level by slip traces on the intermediate
$\gplane{112}$ plane.

If the MRSSP does not coincide exactly with any $\pm30\deg$ plane, one of the two adjacent
$\gplane{110}$ planes will be more stressed than the other, which may result in slip on any
high-index plane bounded by the two $\gplane{110}$ planes. On the microscopic level, this motion
corresponds to a particular type of the zig-zag slip in which the dislocation makes more steps on
one $\gplane{110}$ plane than on the other. This slip mode was observed in molybdenum under
compression for $\chi=+25\deg$ at the temperature $77~\K$ by \citet{jeffcoat:76}. Two groups of fine
slip traces appeared on the surface, one at $\psi\approx+10\deg$ and the other at
$\psi\approx+51\deg$, which are close to the $(\bar{1}01)$ and $(\bar{1}10)$ planes that are
subjected to the highest shear stresses parallel to the slip direction. A more thorough analysis of
the observations of slip traces in single crystals of molybdenum at low temperatures and their
correlation with the results of our atomistic studies will be presented in Section
\ref{sec_expt_correl}.

  \chapter{The 0~K effective yield criterion}
\label{chap_tstarcrit}

\begin{flushright}
  The opposite of a correct statement is a false statement.\\
  The opposite of a profound truth may well be another profound truth.\\
  \emph{Niels Bohr}
\end{flushright}

Based on the atomistic results presented in the previous chapter, we will now formulate a model that
generalizes these data to real single crystals containing dislocations of all possible Burgers
vectors. We will then proceed to formulate the 0~K effective yield criterion that explicitly
involves the effect of shear stresses perpendicular and parallel to the slip direction. The
functional form of this criterion follows from the non-associated plastic flow model
\citep{qin:92b,qin:92} proposed originally for Ni$_3$Al. For an exhaustive explanation of the
effects of non-Schmid stresses and the related differences between the yield function and the flow
potential, refer to \citet{bassani:94}.

%----------------------------------------------------------------------------------------------------
%----------------------------------------------------------------------------------------------------

\section{Model of plastic flow of real single crystals}
\label{sec_realcryst_ex}

\begin{figure}
  \centering
  \includegraphics[width=15cm]{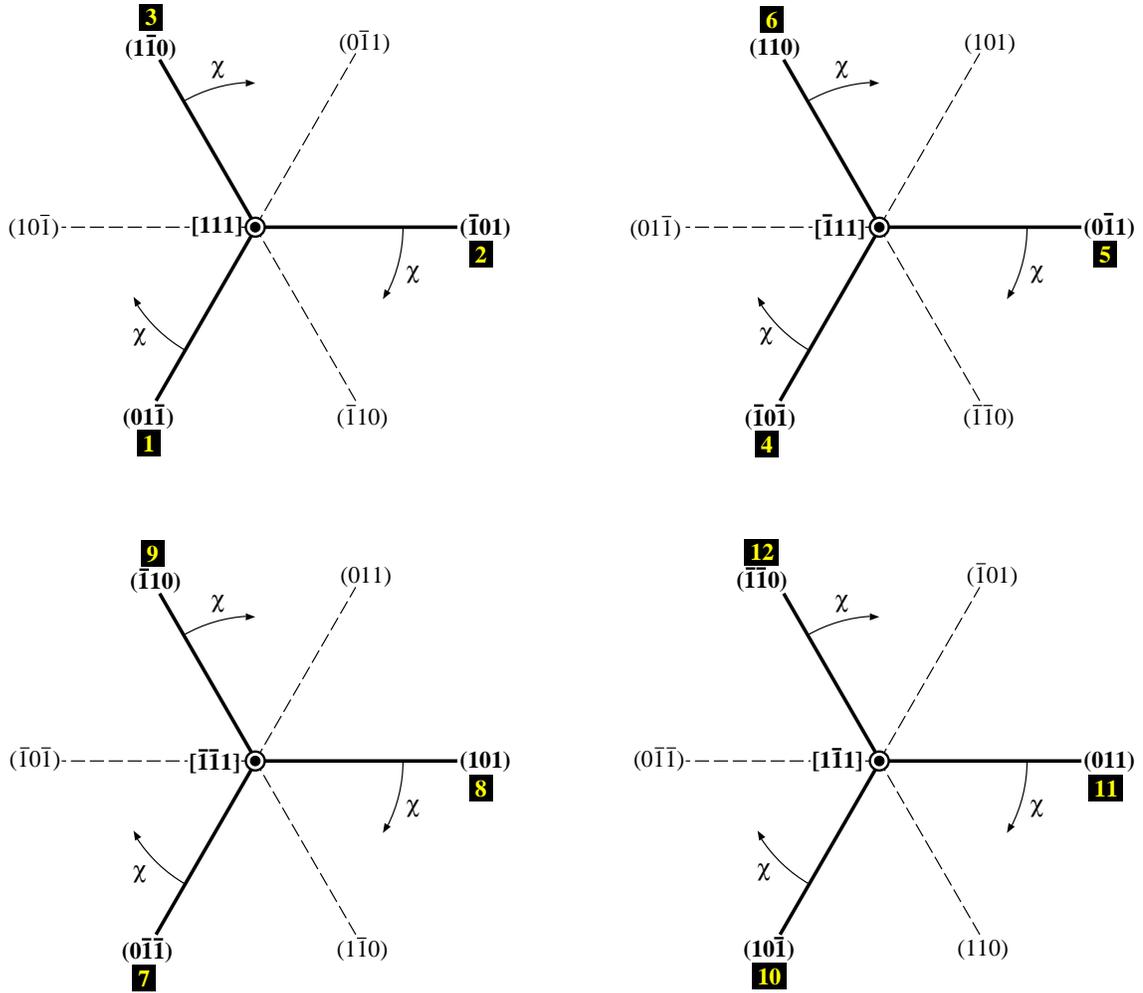}
  \caption{Orientations of $\gplane{110}$ reference planes in the zones of the four $\gdir{111}$
    slip directions in bcc crystals corresponding to systems 1 to 12 in \reftab{tab_bcc24sys}. The
    positive sense of $\chi$ in each system is marked relative to the solid lines representing the
    reference planes. The orientations of MRSSPs in the zones of the opposite slip directions,
    corresponding to the conjugate systems 13 to 24, can be obtained simply by reversing the sense
    of $\chi$.}
  \label{fig_8slipdir}
\end{figure}

In order to extend the calculated $\CRSS-\chi$ and $\CRSS-\tau$ dependencies to real single crystals
at $0~\K$, we will consider that mobile screw dislocations populate all $\gplane{110}\gdir{111}$
systems. We have seen earlier that the actual slip plane does not necessarily coincide with the most
highly stressed $\gplane{110}$ plane in the zone of the slip direction. For the sake of clarity, we
thus define a \emph{reference plane} as a particular $\gplane{110}$ plane in the zone of the slip
direction from which the angle of the MRSSP, $\chi$, and the angle of the slip plane, $\psi$, are
measured. From symmetry, 24 reference systems can be recognized in bcc crystals, each defined by a
\emph{reference plane} and a slip direction, and these are shown in \reffig{fig_8slipdir}. For
example, there are three reference planes in the zone of the $[111]$ slip direction, namely
$(01\bar{1})$, $(\bar{1}01)$ and $(1\bar{1}0)$.

If we neglect the interactions between dislocations, the motion of each individual dislocation is
governed by the same $\CRSS-\chi$ and $\CRSS-\tau$ dependencies as those for an isolated $1/2[111]$
dislocation. To apply these dependencies to any reference system $\alpha$, it is first necessary to
find the angle $\chi_\alpha$ of the MRSSP in the zone of the corresponding $\gdir{111}$ slip
direction that lies within the $\pm 30\deg$ angular region measured from the respective
$\gplane{110}$ reference plane. For instance, the MRSSP of the $(\bar{1}01)[111]$ reference system
always lies between the $(\bar{1}\bar{1}2)$ plane at $\chi=-30\deg$ and the $(\bar{2}11)$ plane at
$\chi=+30\deg$ measured relative to the $(\bar{1}01)$ reference plane, as shown in
\reffig{fig_planes}. Furthermore, because the CRSS is always assumed to be positive, we will require
that the shear stress parallel to the slip direction resolved in each of these MRSSPs,
$\sigma_\alpha$, is also positive. It is then a simple task to show that only 4 out of the total of
24 reference systems satisfy both requirements, i.e. $-30\deg<\chi_\alpha<+30\deg$ and
$\sigma_\alpha>0$, and thus only these four systems can be activated for slip by the applied
stress. For the opposite sense of loading, the four reference systems are sheared in the
\emph{opposite} sense and thus the relevant systems change to those with the same reference planes
as above but \emph{opposite} slip directions.

In each loading step and for each of the four reference systems $\alpha$, one can determine the
shear stresses perpendicular and parallel to the slip direction applied in each of the four MRSSPs
at $\chi_\alpha$. This then yields the stress state $(\tau,\sigma)_\alpha$ associated with the MRSSP
of the system $\alpha$. Since all $1/2\gdir{111}$ dislocations are equivalent, the shear stresses
$(\tau,\sigma)_\alpha$ for the four $\gplane{110}\gdir{111}$ systems can now be directly compared
with atomistic results for $\chi\approx \chi_\alpha$ obtained for the isolated $1/2[111]$
dislocation. Within the framework of this model, a particular system $\alpha$ is activated for slip
when $\sigma_\alpha$ reaches the CRSS for the given $\tau_\alpha$, as determined by the calculated
$\CRSS-\tau$ dependence for the MRSSP defined by angle $\chi_\alpha$.  From our atomistic
simulations, these dependencies are available only for seven orientations of the MRSSP and,
therefore, the comparison will be made in the $\CRSS-\tau$ graph for $\chi_\alpha$ that is closest
to one of the values from the set $\chi\approx\{0\deg,\pm 9\deg,\pm 19\deg,\pm 26\deg\}$.

\begin{figure}[!htb]
  \centering
  \includegraphics[width=13cm]{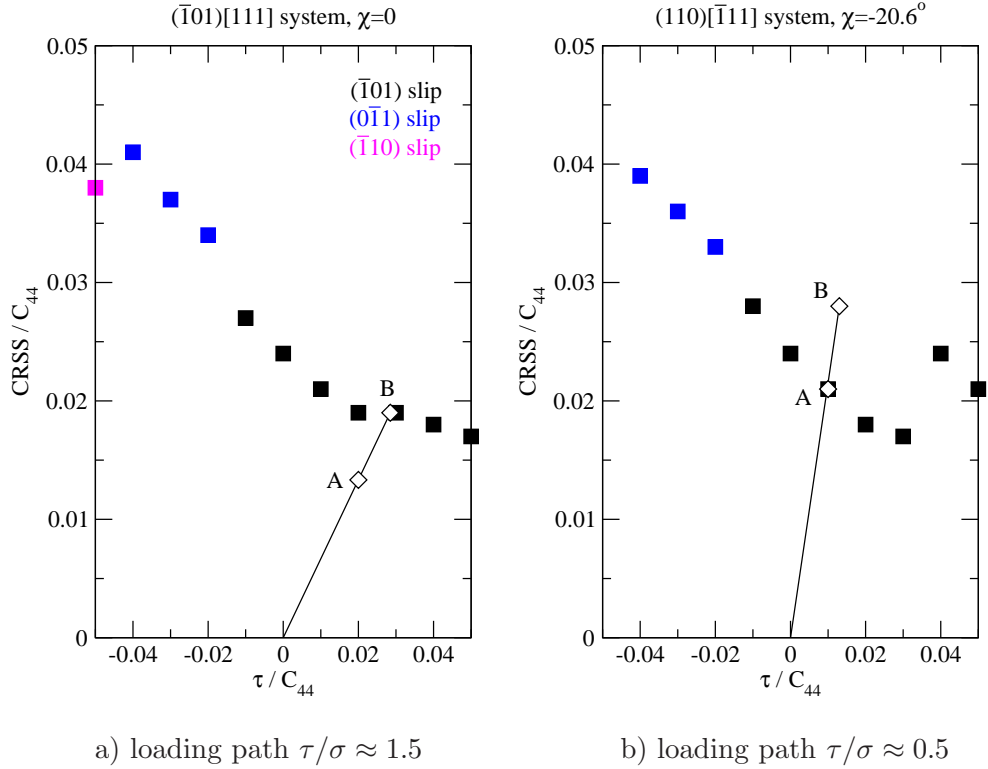} \\
  \ \hskip5mm a) loading path $\tau/\sigma\approx1.5$ \hskip2.5cm 
  b) loading path $\tau/\sigma\approx0.5$ \\[1em]
  \parbox{15cm}{\caption{Evolution of loading in two different $\gplane{110}\gdir{111}$ systems
      (lines) induced by shear stresses perpendicular and parallel to the slip direction applied in
      the $(\bar{1}01)[111]$ system. Squares correspond to the atomistic data calculated for a
      single $1/2[111]$ dislocation in Section \ref{sec_sperp_dep}. The points marked "A" and "B" in
      the two panels correspond to the same applied loading.}
    \label{fig_chi0_real_MoBOP}}
\end{figure}

In order to demonstrate this transfer of the atomistic results to real single crystals, we will now
consider a combined loading by the shear stresses perpendicular and parallel to the $[111]$ slip
direction, applied in the MRSSP that coincides with the $(\bar{1}01)$ plane. As an example we choose
one particular orientation of the loading axis for which the shear stresses perpendicular and
parallel to the slip direction resolved in the MRSSP $(\bar{1}01)$ vary along a straight loading
path with the slope $\tau/\sigma\approx1.5$, see \reffig{fig_chi0_real_MoBOP}a. If the crystal
contained only $1/2[111]$ dislocations, the slip would occur when this loading path reaches the CRSS
given by the atomistically calculated $\CRSS-\tau$ dependence, specifically the point
``B''. However, in real crystals, all slip systems are stressed simultaneously and the activation of
a particular system depends on the interplay between the shear stresses perpendicular and parallel
to the slip direction resolved in the corresponding MRSSP. Hence, while the loading evolves along
the path in \reffig{fig_chi0_real_MoBOP}a, the two shear stresses resolved in the MRSSP of the
$(110)[\bar{1}11]$ reference system follow the path drawn in \reffig{fig_chi0_real_MoBOP}b with the
slope $\tau/\sigma\approx0.5$. Note that the slope of the loading path in the MRSSP is generally
different for the two reference systems. As we increase the load, the stress states in the two
MRSSPs, in zones of $[111]$ and $[\bar{1}11]$ slip directions, follow their respective loading
paths. When the applied loading in \reffig{fig_chi0_real_MoBOP}b reaches the point marked ``A'', the
$(110)[\bar{1}11]$ system becomes operative. However, at this stress state, the resolved loading in
the MRSSP of the $(\bar{1}01)[111]$ system is subcritical in that the corresponding stress state is
still well below the calculated CRSS for this system (see
\reffig{fig_chi0_real_MoBOP}a). Consequently, the $(110)[\bar{1}11]$ system will become operative
first, followed by the operation of the $(\bar{1}01)[111]$ system.

This simple model demonstrates that the onset of slip in real single crystals can be associated with
the stress for which the loading path of any of the four potentially active reference systems, drawn
in their respective MRSSPs, reaches the CRSS found in atomistic calculations. Even though
$(\bar{1}01)[111]$ may be the most highly stressed system in the sense of the Schmid law, it does
not automatically follow that no other slip system can be activated prior to this system. This would
only be the case if shear stresses perpendicular to the slip direction did not affect the magnitude
of the CRSS, which is clearly not the case in bcc molybdenum. Indeed, even pure shear stress
perpendicular to the slip direction of one system can be resolved as shear stress parallel to the
slip direction in the MRSSP of a \emph{different} slip direction, and consequently it can induce
glide of dislocations in this system.

%----------------------------------------------------------------------------------------------------
%----------------------------------------------------------------------------------------------------

\section{The 0~K effective yield criterion}
\label{sec_yieldcrit_MoBOP}

The theory of non-associated plastic flow in bcc metals, put forward by \citet{qin:92b,qin:92}, aims
at reformulating the plasticity of bcc metals by introducing an effective yield criterion that
captures the breakdown of the Schmid law in these materials. As we have seen previously, this is
inherently associated with the behavior of individual screw dislocations and, therefore, the results
of atomistic simulations presented earlier can be used to obtain the actual form of the effective
yield stress. This development has been carried out jointly with V.~Racherla and J.~Bassani and is
explained in more detail in \citet{vitek:04}.

%----------------------------------------------------------------------------------------------------

\subsection{Restricted form}

For the sake of clarity, we first construct the so-called restricted form of the effective yield
criterion that merely reproduces the dependence of the CRSS on $\chi$ for the case of pure shear
stress parallel to the slip direction (circles in \reffig{fig_CRSS_chi_MoBOP}), i.e. without
considering the effect of perpendicular shear stresses. This can be achieved by writing the
effective stress, $\tau^*$, as a linear combination of shear stresses parallel to the slip direction
acting in two distinct planes chosen to be of the $\gplane{110}$ type,
\begin{equation}
  \tau^* = \sigma^{(\bar{1}01)} + a_1\sigma^{(0\bar{1}1)} \leq \tau^*_{cr} \ ,
  \label{eq_tstar_restr}
\end{equation}
where $a_1$ and $\tau_{cr}^*$ are parameters that can be determined by fitting the $\CRSS-\chi$
atomistic results. If $\tau^*<\tau^*_{cr}$, screw dislocations do not move, and the deformation is
purely elastic. However, when $\tau^*=\tau^*_{cr}$, the applied stress induces the glide of screw
dislocations and the crystal is said to yield or flow. In this model, $\tau^*_{cr}$ is an effective
yield stress that represents the critical value of $\tau^*$. Since the $(\bar{1}01)$ plane is the
actual slip plane for any orientation of the applied shear stress, $\sigma^{(\bar{1}01)}$ is the
Schmid stress. The shear stress $\sigma^{(0\bar{1}1)}$, parallel to the slip direction and acting in
the $(0\bar{1}1)$ plane, deforms the dislocation core and thus affects the magnitude of the CRSS at
which the dislocation starts to move. This core transformation gives rise to the
twinning-antitwinning asymmetry of the CRSS for shear stress parallel to the slip direction that is
shown in \reffig{fig_CRSS_chi_MoBOP}. Without this additional term, \refeq{eq_tstar_restr} reduces
to the Schmid law. Note that the choice of the auxiliary stress component, in our case
$\sigma^{(0\bar{1}1)}$, is arbitrary, and different choices result in different constants $a_1$ but
identical overall reproduction of the atomistic data.

\begin{figure}[!htb]
  \centering
  \includegraphics[width=12cm]{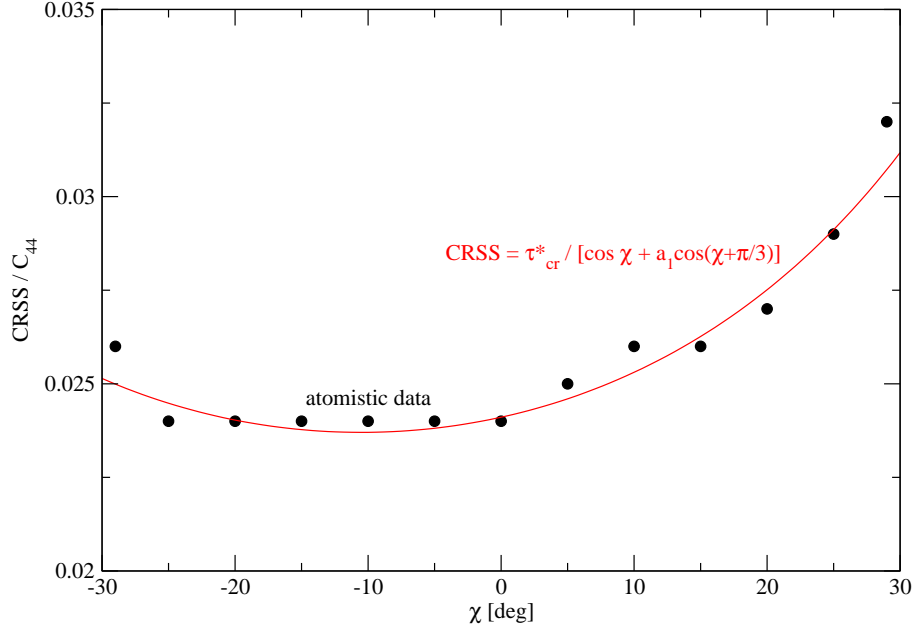}
  \parbox{14cm}{\caption{Atomistically calculated CRSS for pure shear stress parallel to the slip
      direction (circles) and the prediction of the effective yield criterion (curve).}
  \label{fig_CRSS_chi_fit_MoBOP}}
\end{figure}

The parameters $a_1$ and $\tau_{cr}^*$ in \refeq{eq_tstar_restr} are determined by fitting the
atomistic data for pure shear stress parallel to the slip direction, see
\reffig{fig_CRSS_chi_MoBOP}. For this purpose, it is convenient to rewrite (\ref{eq_tstar_restr}) in
terms of the angle $\chi$ between the MRSSP and the $(\bar{1}01)$ plane. When the applied stress
$\sigma$ reaches its critical value, $\CRSS$, the magnitude of $\tau^*$ becomes equal to the
effective yield stress, $\tau^*_{cr}$, and the criterion reads
\begin{equation}
  \CRSS \left[ \cos\chi + a_1 \cos\left(\chi+\frac{\pi}{3}\right) \right] = \tau^*_{cr} \ .
  \label{eq_tstar_restr_angle}
\end{equation}
Each data point $(\chi,\CRSS)$ from \reffig{fig_CRSS_chi_MoBOP} thus provides one equation of the
type (\ref{eq_tstar_restr_angle}). The optimal values of $a_1$ and $\tau_{cr}^*$ are then obtained
by the least-squares procedure. In our case, the best fit corresponds to $a_1=0.24$ and
$\tau_{cr}^*/C_{44}=0.027$, where the nonzero value of $a_1$ quantifies the degree of the
twinning-antitwinning asymmetry observed in atomistic simulations. The quality of this fit is seen
in \reffig{fig_CRSS_chi_fit_MoBOP}.

%----------------------------------------------------------------------------------------------------

\subsection{Full form}

The restricted form of the effective yield criterion (\ref{eq_tstar_restr}) does not capture the
effect of shear stresses perpendicular to the slip direction which, as we now know, play an
important role in the glide of screw dislocations. In order to reproduce the $\CRSS-\tau$ dependence
found in atomistic calculations, we have to incorporate this dependence into $\tau^*$. This can be
achieved by complementing $\tau^*$ with two additional terms whose coefficients have to be fitted so
as to reproduce the $\CRSS-\tau$ dependence. The full form of the effective yield criterion then
reads
\begin{equation}
  \tau^* = \sigma^{(\bar{1}01)} + a_1\sigma^{(0\bar{1}1)} + 
  a_2 \tau^{(\bar{1}01)} + a_3\tau^{(0\bar{1}1)} \leq \tau_{cr}^* \ ,
  \label{eq_tstar_full}
\end{equation}
where $\sigma$ are shear stresses parallel and $\tau$ shear stresses perpendicular to the slip
direction. In this equation, $a_1$ and $\tau^*_{cr}$ are taken to be the same as the values obtained
from the restricted model and, therefore, only $a_2$ and $a_3$ have to be determined by fitting the
$\CRSS-\tau$ data for the combination of the shear stresses perpendicular and parallel to the slip
direction. Similarly as above, it is convenient to rewrite \refeq{eq_tstar_full} for the case when
the applied stress $\sigma$ reaches its critical value, CRSS, and thus $\tau^*=\tau^*_{cr}$:
\begin{equation}
  \CRSS \left[ \cos\chi + a_1 \cos\left(\chi+\frac{\pi}{3}\right) \right] + 
  \tau \left[ a_2 \sin2\chi + a_3 \cos\left(2\chi+\frac{\pi}{6}\right) \right] = 
  \tau^*_{cr} \ .
  \label{eq_tstar_full_angle}
\end{equation}

For each point in the $\CRSS-\tau$ graph corresponding to an angle $\chi$, we thus obtain an
equation for two unknowns, $a_2$ and $a_3$. Since the number of points in all $\CRSS-\tau$ graphs
greatly exceeds the number of unknowns, the corresponding least-squares fitting is far from
unique. However, for all orientations of \emph{uniaxial} loading studied here, the corresponding
shear stresses perpendicular to the slip direction are within $-0.02<\tau/C_{44}<+0.02$ when the
CRSS has been attained. Therefore, we have determined $a_2$ and $a_3$ by fitting only to those
points in the $\CRSS-\tau$ graphs that are within this interval of $\tau$ values, where the
$\CRSS-\tau$ dependence is approximately linear.

We have found that the best fit is obtained when considering only three orientations of the MRSSP,
namely $\chi=0$ and $\chi\approx\pm 9\deg$. In each of these planes, only the data for
$\tau/C_{44}=\pm 0.01$ were taken into account, since only two points are necessary to specify a
slope of the straight line.  However, even in this case, the system of 6 algebraic equations for 2
unknowns is overdetermined, and, therefore, the values of $a_2$ and $a_3$ have been found by
least-squares fitting of \refeq{eq_tstar_full_angle}. Finally, note that, since $\tau^*$ is
\emph{linearly} dependent on the shear stress perpendicular to the slip direction, the criterion is
not capable of reproducing the $\CRSS-\tau$ dependencies at $|\tau/C_{44}|>0.02$. However, this is not
a problem when applying the criterion to study plastic behavior of real single crystals, since for
large values of $\tau$ another $\gplane{110}\gdir{111}$ system always becomes operative before
activating the $(\bar{1}01)[111]$ system (see e.g. \reffig{fig_chi0_real_MoBOP}). In other words,
straight loading paths for \emph{large} ratios of $|\tau/\sigma|$ always correspond to orientations
that lie outside the stereographic triangle for the $(\bar{1}01)[111]$ system, and, therefore,
another $\gplane{110}\gdir{111}$ system becomes active.

The best fit to the atomistic data obtained by the procedure outlined above yields $a_2\approx 0$
and $a_3=0.35$. The fact that $a_2$ is zero implies that $\tau^*$ is independent of the shear stress
perpendicular to the slip direction applied in the $(\bar{1}01)$ plane by the stress tensor
(\ref{eq_tensor_tau}). This is clearly a coincidence and not any basic rule, since one can choose a
different combination of the two shear stresses perpendicular to the slip direction for which both
$a_2$ and $a_3$ will be nonzero. In summary, the effective yield criterion for molybdenum obtained
by fitting the 0~K atomistic data is given by \refeq{eq_tstar_full} with the parameters summarized
in \reftab{tab_tstar_params}.

\begin{table}[!htb]
  \centering
  \parbox{10cm}{\caption{Coefficients of the effective yield criterion (\ref{eq_tstar_full}) 
      for molybdenum.}
  \label{tab_tstar_params}}\\[1em]

  \begin{tabular}{cccc}
    \hline
    $a_1$ & $a_2$ & $a_3$ & $\tau^*_{cr}/C_{44}$ \\
    \hline
    0.24 & 0 & 0.35 & 0.027 \\
    \hline
  \end{tabular}
\end{table}

% why do we want the linear tau* criterion ?

If the $\tau^*$ criterion is to be employed in large-scale simulations, such as finite-element
calculations, it has to be sufficiently simple so that efficient calculations of $\tau^*$ at every
node of the finite-element mesh can be performed. Moreover, yield criteria are used not only to
check whether the plastic deformation takes place but often also to predict the magnitude of the
stress (e.g. CRSS) at which the yielding occurs. It then follows that $\tau^*$ has to be written in
a form that can be easily inverted to get the CRSS as a function of other parameters. We have
demonstrated that both requirements above are easily satisfied by the linear $\tau^*$ criterion
(\ref{eq_tstar_full}), which provides a relatively simple expression for the CRSS that follows from
\refeq{eq_tstar_full_angle}.

%----------------------------------------------------------------------------------------------------
%----------------------------------------------------------------------------------------------------

\section{Onset of plastic flow in real single crystals}
\label{sec_4slipsys}

The full form of the effective yield criterion (\ref{eq_tstar_full}) can be directly employed to
obtain the $\CRSS-\tau$ dependencies for real single crystals of molybdenum in which any
$\gplane{110}\gdir{111}$ system may be potentially operative. These dependencies will generalize the
$\CRSS-\tau$ data displayed in \reffig{fig_CRSS_tau_MoBOP} that correspond to the glide of a
dislocation with a specific Burgers vector, $1/2[111]$.

For any loading, the primary system and the corresponding magnitude of critical loading can be found
by: (i) identifying the four reference systems $\alpha$ that can become operative under this
loading, (ii) calculating the orientations $\chi_\alpha$ of the MRSSPs in zones of these slip
directions, and (iii) resolving the shear stresses perpendicular ($\tau_\alpha$) and parallel
($\sigma_\alpha$) to the slip direction in the MRSSP at $\chi_\alpha$. Similarly as in
\reffig{fig_chi0_real_MoBOP}, these shear stresses then define a straight loading path in the
$\CRSS-\tau$ graph of the MRSSP at $\chi_\alpha$, whose slope is
$\eta_\alpha=\tau_\alpha/\sigma_\alpha$. Hence, one can express the CRSS in this system from
\refeq{eq_tstar_full_angle} as
\begin{equation}
  \CRSS_\alpha = \frac{\tau^*_{cr}}{\cos\chi_\alpha + a_1 \cos(\chi_\alpha+\pi/3) + 
    \eta_\alpha [ a_2 \sin2\chi_\alpha + a_3 \cos(2\chi_\alpha+\pi/6) ] } \ ,
  \label{eq_tstar_CRSS}
\end{equation}
where the term with $a_2$ is retained for generality but vanishes for molybdenum. These calculations
yield a set of four values of $\CRSS_\alpha$ in four MRSSPs at $\chi_\alpha$ corresponding to
\emph{different} reference systems. In order to compare the prominence of these systems, we project
the three stresses $\CRSS_\alpha$ for systems \emph{other} than $(\bar{1}01)[111]$ into the
$\CRSS-\tau$ graph that corresponds to the MRSSP for the $(\bar{1}01)[111]$ reference system. This
yields four critical points that correspond to the activation of slip in the four reference
systems. The point on the straight loading path emanating from the origin of the $\CRSS-\tau$ graph
that is closest to the origin marks the stress that causes operation of the primary
$\gplane{110}\gdir{111}$ system or, equivalently, the stress at which a real crystal starts to
deform plastically. This analysis can be carried out for many different loading paths in a number of
MRSSPs, which yields a set of critical points corresponding to the activation of individual
systems. The inner envelope of these lines is then a projection of the 0~K yield surface.

\begin{figure}[!p]
  \centering
  \includegraphics[width=12cm]{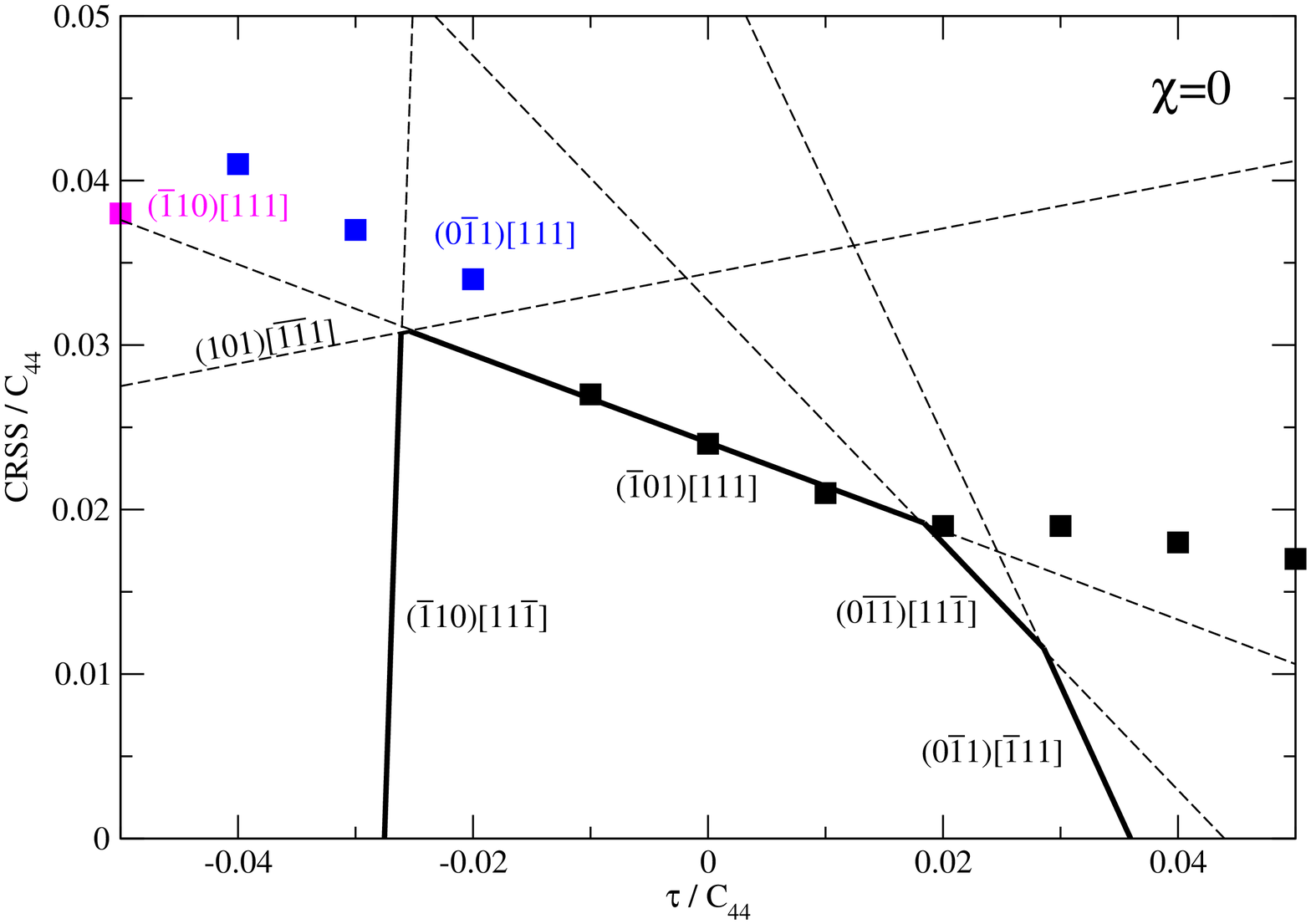} \\[2em]
  \includegraphics[width=12cm]{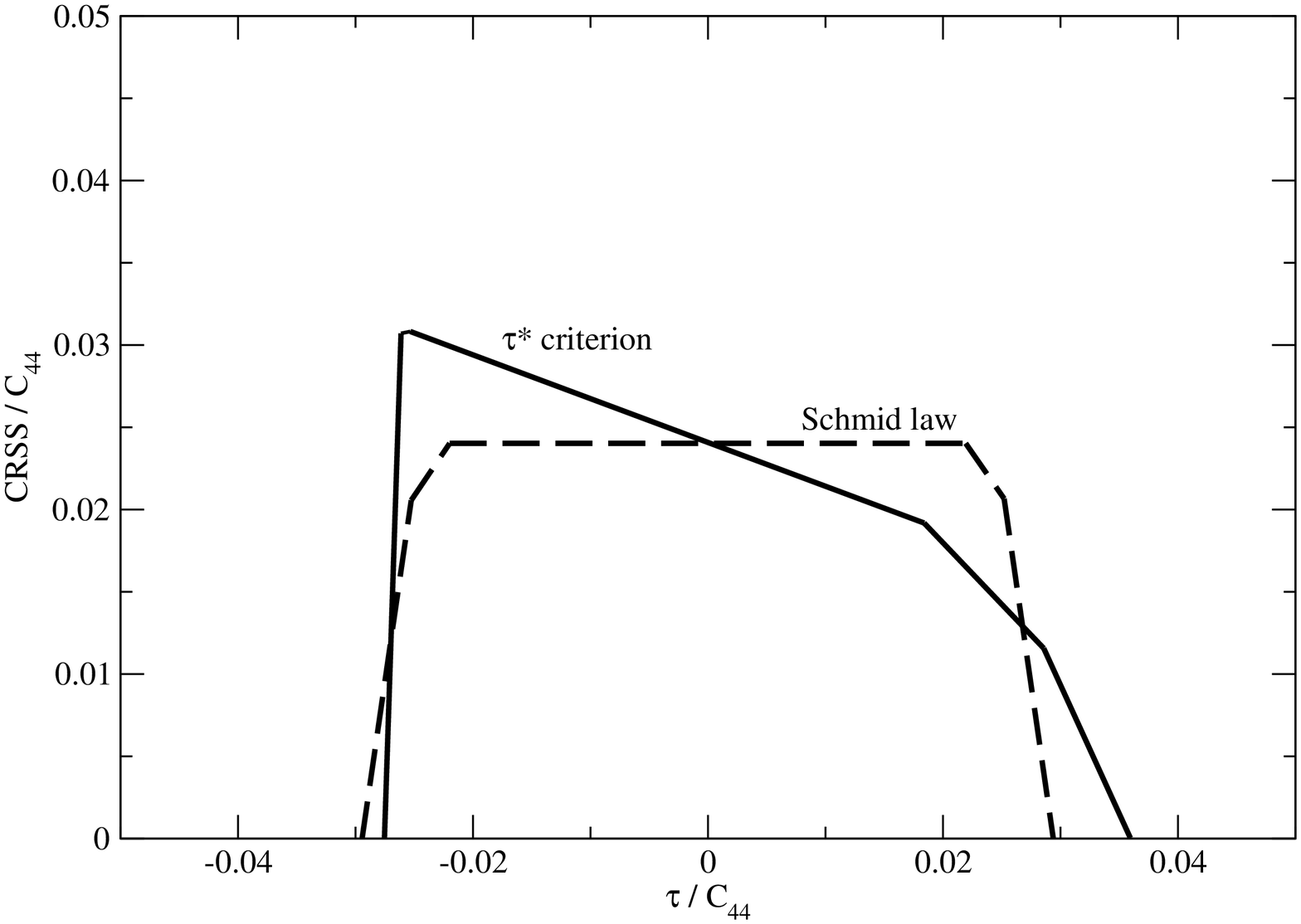} \\
  a) $\chi=0$, MRSSP $(\bar{1}01)$
\end{figure}

\begin{figure}[!p]
  \centering
  \includegraphics[width=12cm]{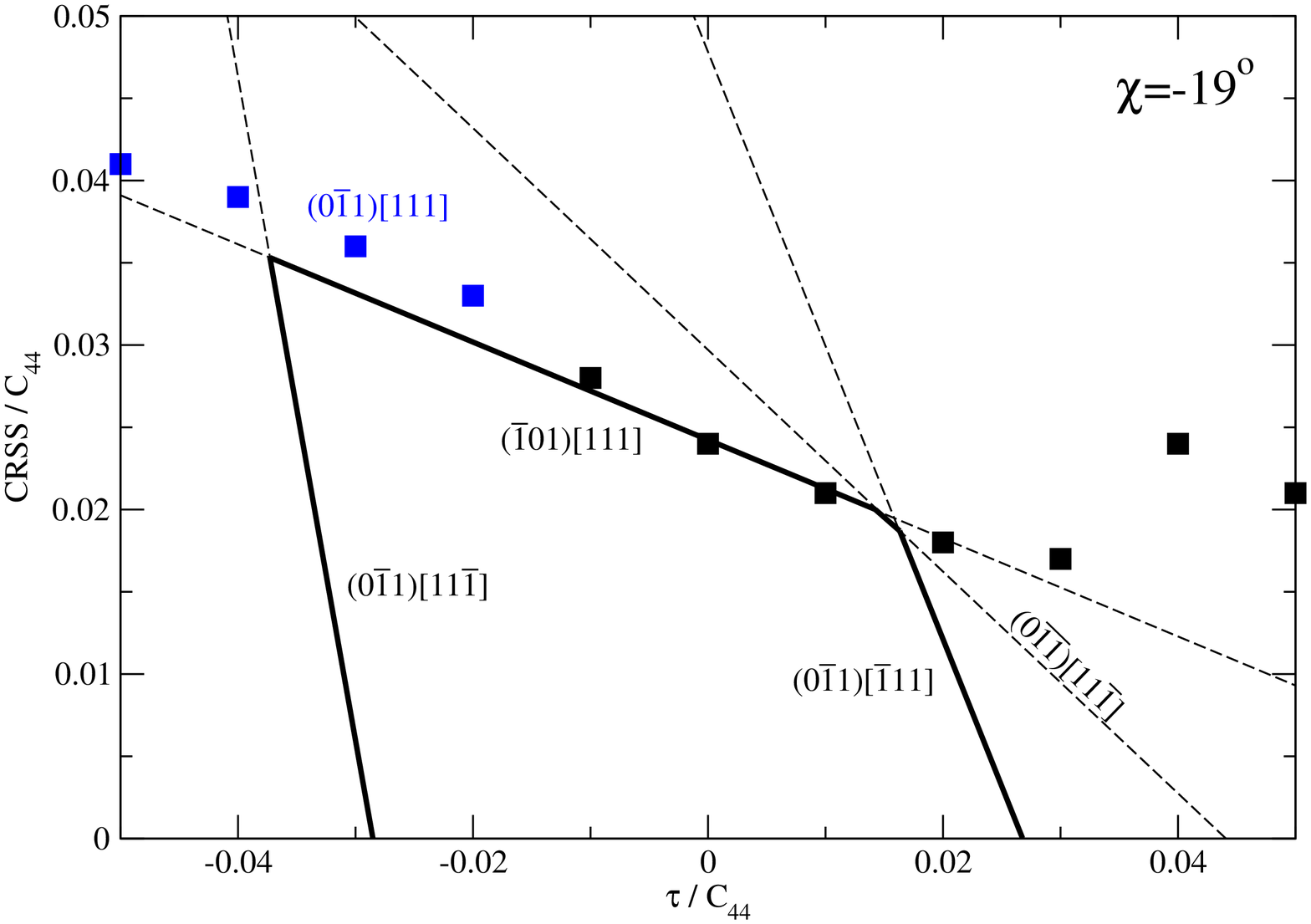} \\[2em]
  \includegraphics[width=12cm]{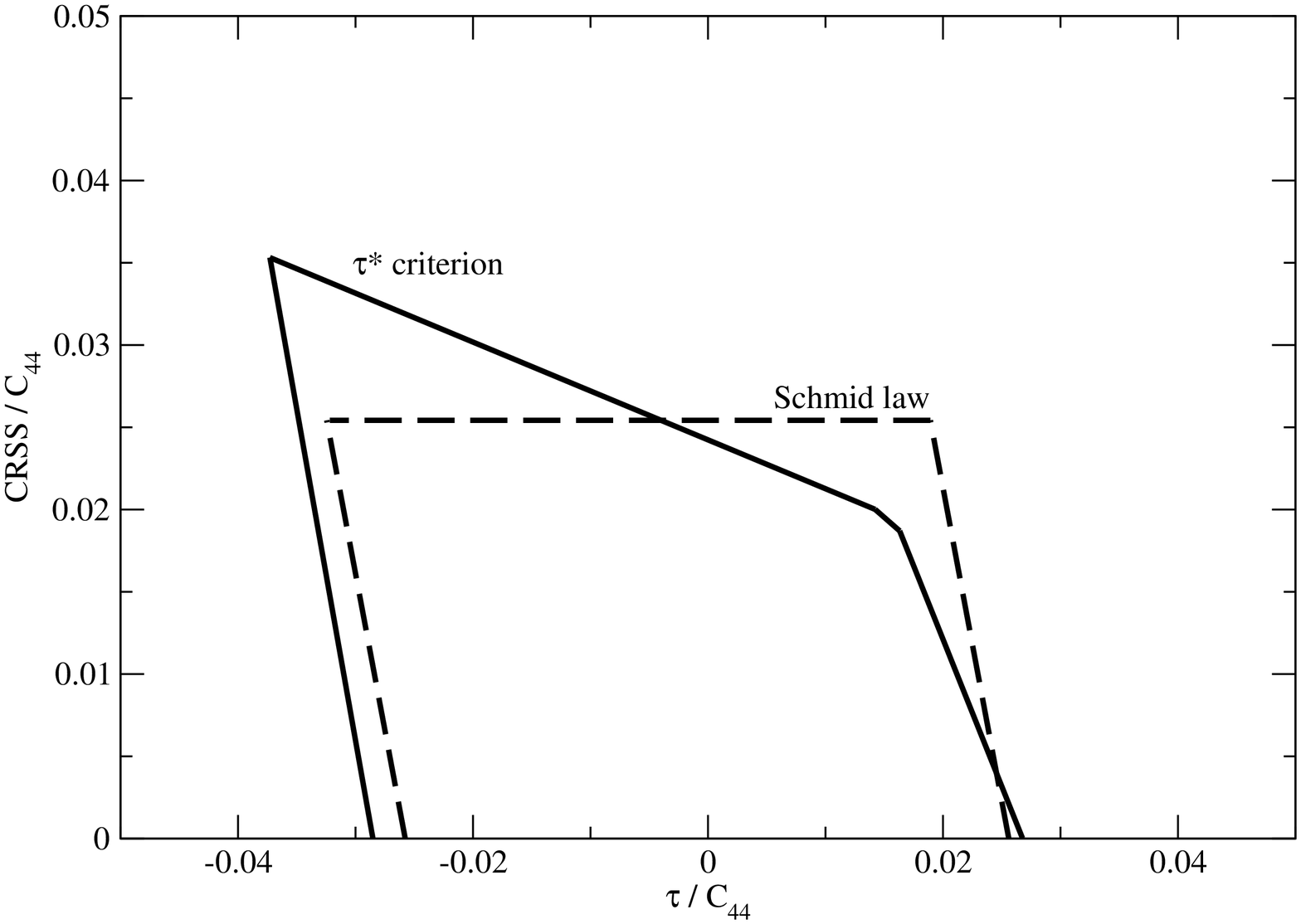} \\
  b) $\chi=-19\deg$, MRSSP $(\bar{2}\bar{1}3)$
\end{figure}

\begin{figure}[!p]
  \centering
  \includegraphics[width=12cm]{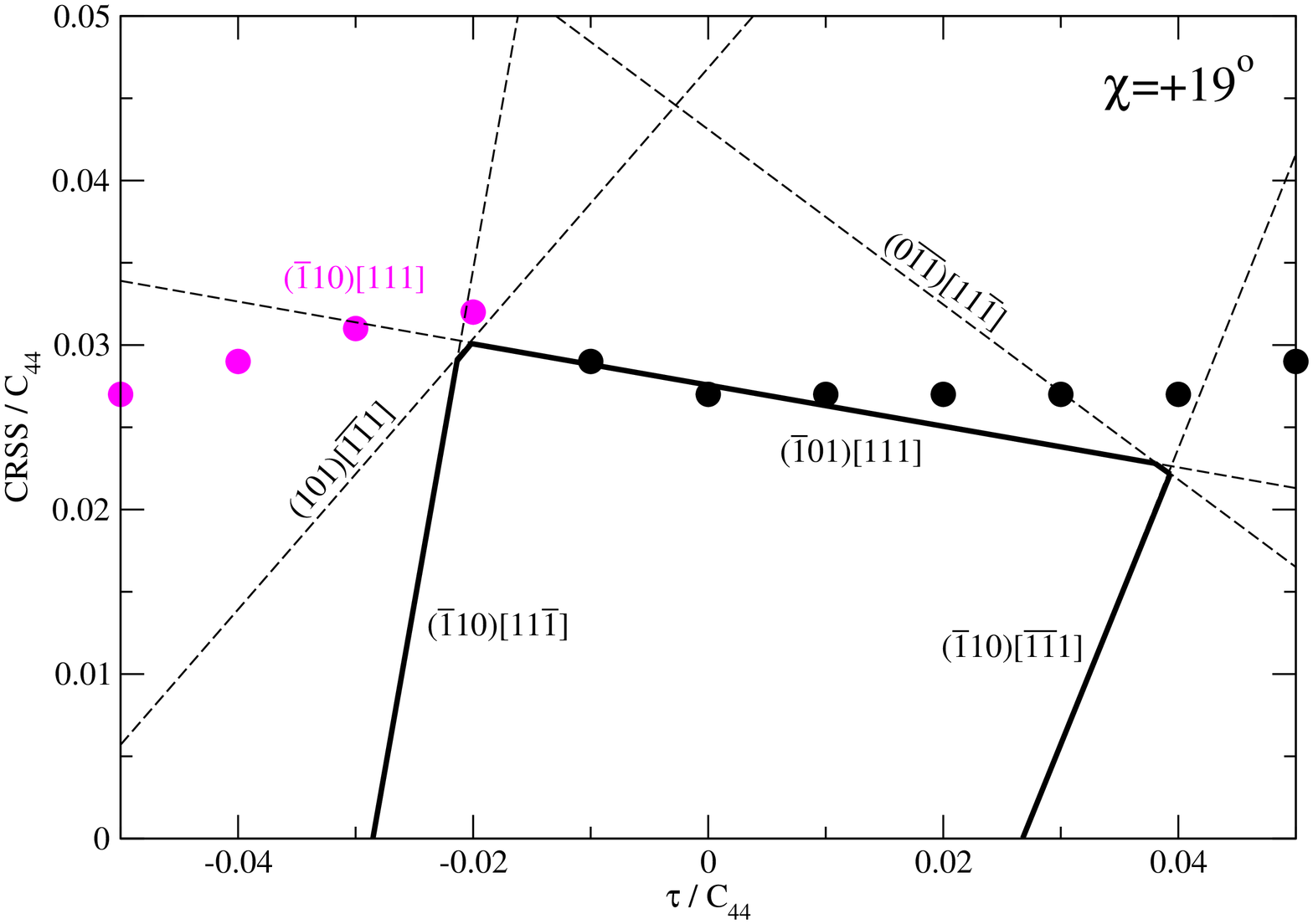} \\[2em]
  \includegraphics[width=12cm]{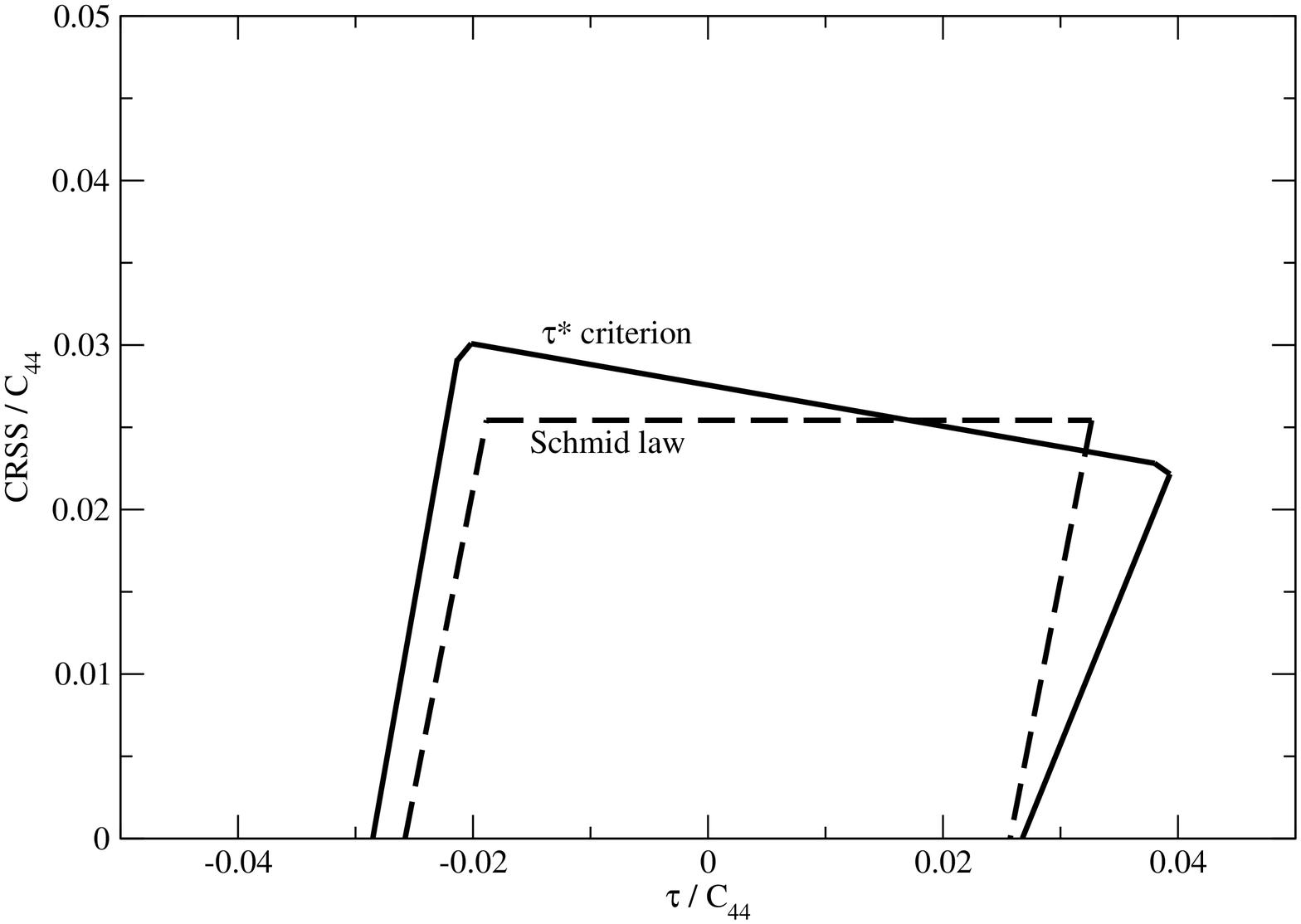} \\
  c) $\chi=+19\deg$, MRSSP $(\bar{3}12)$ \\
  \caption{Critical lines for various MRSSPs of the $(\bar{1}01)[111]$ reference system, calculated
    from the $\tau^*$ criterion. The lower panels show the comparison of the shape of the yield
    surface obtained from the $\tau^*$ criterion and that predicted by the Schmid law.}
  \label{fig_CRSS_tau_fit_MoBOP}
\end{figure}

The critical lines marking the onset of activation of the four reference systems $\alpha$, predicted
using the procedure described above, are plotted for $\chi=0$ in \reffig{fig_CRSS_tau_fit_MoBOP}a
and for $\chi=\pm19\deg$ in \reffig{fig_CRSS_tau_fit_MoBOP}b,c. If the magnitude of the shear stress
perpendicular to the slip direction is small, roughly $|\tau/C_{44}|<0.02$, the primary slip system
coincides with the most highly stressed $(\bar{1}01)[111]$ system. However, as $|\tau|$ becomes
larger, another $\gplane{110}\gdir{111}$ system becomes dominant when the loading path intersects
the corresponding critical line. Since the values of $\tau$ at which the plastic deformation of real
crystals takes place are bounded by the yield polygon, $|\tau/C_{44}|$ can never be larger than
about $0.03$. If the loading path reaches the corner of the inner polygon generated by the
critical lines for the primary slip systems, more systems become activated simultaneously and a
multiple slip occurs.

For illustration, we also show by dashed lines in the lower plots of
\reffig{fig_CRSS_tau_fit_MoBOP}a-c how the projection of the yield surface looks if the $\tau^*$
criterion reduces to the Schmid law. In this case, the CRSS for the most highly stressed
$(\bar{1}01)[111]$ system is clearly independent of $\tau$, and its magnitude follows the Schmid law,
i.e. it varies as $1/\cos\chi$. At larger $\tau$, the yield polygon is bounded by the critical lines
that correspond to \emph{different} reference systems than $(\bar{1}01)[111]$. However, it is very
important to emphasize that this does \emph{not} mean that the CRSS is a function of $\tau$ at large
$\tau$. The slope of the critical lines for systems other than $(\bar{1}01)[111]$ is obtained purely
by the projection of the CRSS from the MRSSPs of these other systems. Without the influence of
non-glide stresses, the yield polygons for positive and negative $\chi$ are mirror images of each
other. Moreover, since the $(\bar{1}01)$ plane is a mirror plane in bcc crystals, the yield surface
projected in the $\CRSS-\tau$ plot for $\chi=0$ is completely symmetrical with respect to
$\tau=0$. All these symmetries are clearly broken in molybdenum as a result of twinning-antitwinning
asymmetry and the strong effect of the shear stresses perpendicular to the slip direction.

%----------------------------------------------------------------------------------------------------
%----------------------------------------------------------------------------------------------------

\section{Tensorial form of the effective yield criterion}

The full form of the effective yield criterion (\ref{eq_tstar_full}) can be written in tensorial
representation, where one uses directly the crystallographic indices and the stress tensor of
applied loading. In a matrix notation, the $\tau^*$ criterion reads
\begin{equation}
  \tau^{*\alpha} = {\bf m}^\alpha {\bf \Sigma}^c {\bf n}^\alpha + 
     a_1 {\bf m}^\alpha {\bf \Sigma}^c {\bf n}_1^\alpha + 
     a_2 ({\bf n}^\alpha \times {\bf m}^\alpha) {\bf \Sigma}^c {\bf n}^\alpha + 
     a_3 ({\bf n}_1^\alpha \times {\bf m}^\alpha) {\bf \Sigma}^c {\bf n}_1^\alpha \leq \tau_{cr}^* \ ,
  \label{eq_tstar_tensor}
\end{equation}
where ${\bf \Sigma}^c$ is the stress tensor expressed in the $[100], [010], [001]$ cube coordinate
system, and $\alpha$ is one of the 24 reference systems in bcc crystals listed in
\reftab{tab_bcc24sys}. 

For a particular $\gplane{110}\gdir{111}$ reference system, ${\bf m}^\alpha$ is a unit vector
defining the slip direction, ${\bf n}^\alpha$ a unit vector perpendicular to the reference plane,
and ${\bf n}_1^\alpha$ a unit vector perpendicular to the $\gplane{110}$ plane in the zone of
${\bf m}^\alpha$ that makes angle $-60\deg$ with the reference plane. For example, if the reference
system considered is $(\bar{1}01)[111]$, the three unit vectors are ${\bf m}^\alpha=[111]/\sqrt{3}$,
${\bf n}^\alpha=[\bar{1}01]/\sqrt{2}$, ${\bf n}_1^\alpha=[0\bar{1}1]/\sqrt{2}$. The complete list of
the three crystallographic vectors of all $\gplane{110}\gdir{111}$ systems is given in
\reftab{tab_bcc24sys}. Notice that the systems labeled 13 to 24 are conjugate to the systems
numbered 1 to 12. A pair of systems $\alpha$ and $\alpha+12$ have identical normal of the reference
plane, ${\bf n}^\alpha$, opposite slip directions ${\bf m}^\alpha$, and complementary non-glide
plane normals ${\bf n}_1^\alpha$.

For any applied loading described by the stress tensor ${\bf \Sigma}^c$, one can determine the
activity of individual reference systems by calculating $\tau^{*\alpha}$ for $\alpha$ from 1 to
24. The slip is activated on a particular system $\alpha$ when the magnitude of $\tau^{*\alpha}$
reaches the effective yield stress, $\tau_{cr}^*$. In the framework of the $\tau^*$ criterion, the
plastic deformation of a single crystal of molybdenum at 0~K commences when the largest of the 24
values of $\tau^{*\alpha}$ reaches the effective yield stress. The corresponding system $\alpha$ is
then called the primary system.

\begin{table}[!htb]
  \centering
  \parbox{14cm}{\caption{The 24 slip systems in bcc crystals. Note, that the crystallographic
  vectors ${\bf m}^\alpha$, ${\bf n}^\alpha$, ${\bf n}_1^\alpha$ have to be normalized before their
  use in \refeq{eq_tstar_tensor}.}
  \label{tab_bcc24sys}}\\[1em]
  \begin{tabular}{ccccc||ccccc}
    \hline
    $\alpha$ & $\rm{ref.\ system}$ & ${\bf m^\alpha}$ & ${\bf n^\alpha}$ & ${\bf n_1^\alpha}$ &
    $\alpha$ & $\rm{ref.\ system}$ & ${\bf m^\alpha}$ & ${\bf n^\alpha}$ & ${\bf n_1^\alpha}$ \\
    \hline
    \systab {1} {$01\bar{1}$} {$111$} {$\bar{1}10$} & 
      \systab {13} {$01\bar{1}$} {$\bar{1}\bar{1}\bar{1}$} {$10\bar{1}$} \\
    \systab {2} {$\bar{1}01$} {$111$} {$0\bar{1}1$} & 
      \systab {14} {$\bar{1}01$} {$\bar{1}\bar{1}\bar{1}$} {$\bar{1}10$} \\
    \systab {3} {$1\bar{1}0$} {$111$} {$10\bar{1}$} & 
      \systab {15} {$1\bar{1}0$} {$\bar{1}\bar{1}\bar{1}$} {$0\bar{1}1$} \\
    \systab {4} {$\bar{1}0\bar{1}$} {$\bar{1}11$} {$\bar{1}\bar{1}0$} & 
      \systab {16} {$\bar{1}0\bar{1}$} {$1\bar{1}\bar{1}$} {$01\bar{1}$} \\
    \systab {5} {$0\bar{1}1$} {$\bar{1}11$} {$101$} & 
      \systab {17} {$0\bar{1}1$} {$1\bar{1}\bar{1}$} {$\bar{1}\bar{1}0$} \\
    \systab {6} {$110$} {$\bar{1}11$} {$01\bar{1}$} & 
      \systab {18} {$110$} {$1\bar{1}\bar{1}$} {$101$} \\
    \systab {7} {$0\bar{1}\bar{1}$} {$\bar{1}\bar{1}1$} {$1\bar{1}0$} & 
      \systab {19} {$0\bar{1}\bar{1}$} {$11\bar{1}$} {$\bar{1}0\bar{1}$} \\
    \systab {8} {$101$} {$\bar{1}\bar{1}1$} {$011$} & 
      \systab {20} {$101$} {$11\bar{1}$} {$1\bar{1}0$} \\
    \systab {9} {$\bar{1}10$} {$\bar{1}\bar{1}1$} {$\bar{1}0\bar{1}$} & 
      \systab {21} {$\bar{1}10$} {$11\bar{1}$} {$011$} \\
    \systab {10} {$10\bar{1}$} {$1\bar{1}1$} {$110$} & 
      \systab {22} {$10\bar{1}$} {$\bar{1}1\bar{1}$} {$0\bar{1}\bar{1}$} \\
    \systab {11} {$011$} {$1\bar{1}1$} {$\bar{1}01$} & 
      \systab{23} {$011$} {$\bar{1}1\bar{1}$} {$110$} \\
    \systab {12} {$\bar{1}\bar{1}0$} {$1\bar{1}1$} {$0\bar{1}\bar{1}$} & 
      \systab{24} {$\bar{1}\bar{1}0$} {$\bar{1}1\bar{1}$} {$\bar{1}01$} \\
    \hline
  \end{tabular}
\end{table}

%----------------------------------------------------------------------------------------------------
%----------------------------------------------------------------------------------------------------

\section{The yield surface and effect of non-glide stresses}

Any stress state can be represented in the so-called principal space in which the orientations of
the three principal axes are the eigenvectors of the stress tensor of the applied loading. The
eigenvalues of this stress tensor then give a triad of the so-called principal stresses that define
a position in the principal space corresponding to a particular stress state exerted at a point in
the loaded body. The locus of the stress states that induce irreversible plastic deformation then
characterizes the so-called \emph{yield surface} which is of a fundamental importance in predictions
of the reliability of structural components.

\begin{figure}[!htb]
  \centering
  \includegraphics[width=10cm]{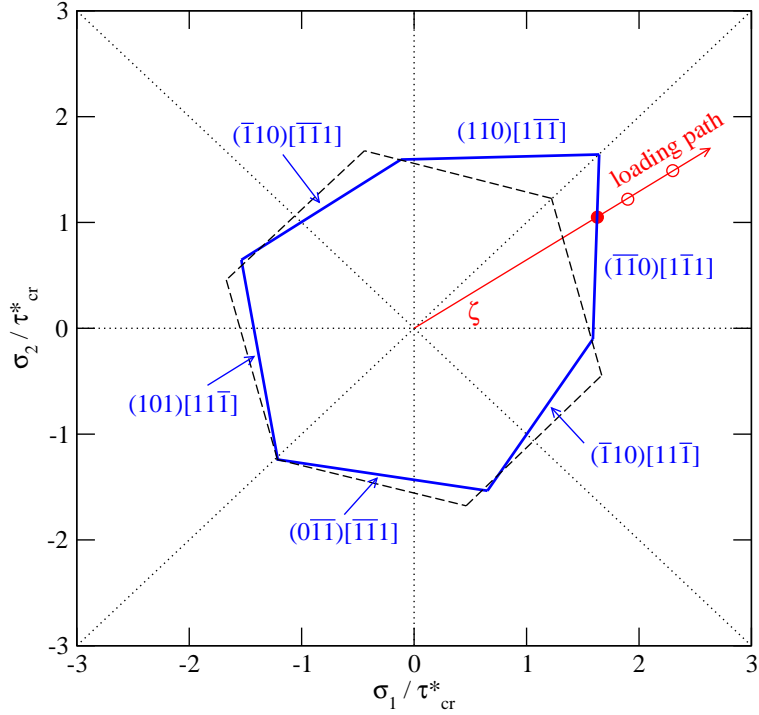}
  \parbox{13cm}{\caption{Projection of the yield surface generated by the $\tau^*$ criterion
      (\ref{eq_tstar_tensor}) for $a_1=0.24$, $a_2=0$, $a_3=0.35$, $\tau^*_{cr}/C_{44}=0.027$ (solid
      line). Only the deviatoric component of the stress tensor is considered. The
      $\gplane{110}\gdir{111}$ primary slip systems that become activated when the loading path
      reaches the corresponding edge of the solid yield polygon are marked. The dashed Tresca
      hexagon corresponds to the purely Schmid behavior, i.e. $a_1=a_2=a_3=0$.}
  \label{fig_yieldsurf_MoBOP}}
\end{figure}

For the sake of clarity, the yield surface is often plotted as a projection of its three-dimensional
variant onto the so-called $\pi$-plane whose normal is defined by the three Euler angles%
\footnote{These Euler angles are given in the so-called $x$-convention which is the most commonly
used scheme. The transformation from the principal coordinate system is accomplished by first
rotating about $\sigma_3$, followed by a rotation about $\sigma_1$, and finally again about
$\sigma_3$.}
$\phi=\psi=0.7854$, $\theta=0.9553$. Because small hydrostatic stresses encountered in engineering
applications do not affect the magnitude of the yield stress, we consider the applied stress tensor
to be a pure deviator 
\begin{equation}
  \mat{\Sigma^\pi} = \left[
    \begin{array}{ccc}
      \sigma_1 & 0 & 0 \\
      0 & \sigma_2 & 0 \\
      0 & 0 & -\sigma_1-\sigma_2
    \end{array}
  \right] 
  \label{eq_tensor_s1s2}
\end{equation}
that is applied in a particular orientation in the principal space that corresponds to the three
Euler angles $(\phi,\theta,\psi)$. In order to calculate the yield surface, we choose a set of
loading paths described by angles $-\pi\leq\zeta\leq\pi$, emanating from the origin in
\reffig{fig_yieldsurf_MoBOP}. Each of these paths constitutes a well-defined orientation of loading
for which we can write the two principal stresses as $\sigma_1=\cos\zeta$ and
$\sigma_2=\sin\zeta$. This determines the applied stress tensor (\ref{eq_tensor_s1s2}) that is
further transformed into the cube coordinate system to arrive at $\mat{\Sigma}^c$, used in
\refeq{eq_tstar_tensor}. For every reference system $\alpha$ from \reftab{tab_bcc24sys} and the
corresponding unit vectors $\mat{m}^\alpha$, $\mat{n}^\alpha$ and $\mat{n}_1^\alpha$,
\refeq{eq_tstar_tensor} provides the effective stress $\tau^{*\alpha}$. From the $\tau^*$ criterion,
the reference system $\alpha$ becomes activated when $\tau^{*\alpha}$ reaches $\tau^*_{cr}$,
i.e. when the applied loading increases $k^\alpha=\tau^*_{cr}/\tau^{*\alpha}$ times. This
corresponds to loading by principal stresses $\sigma_1^\alpha=k^\alpha\sigma_1$ and
$\sigma_2^\alpha=k^\alpha\sigma_2$ that provides a set of yield points
$(\sigma_1^\alpha,\sigma_2^\alpha)$ marked along the loading path in \reffig{fig_yieldsurf_MoBOP} by
circles, which correspond to the activation of individual systems $\alpha$. Note that the point
closest to the origin (filled circle) determines a stress state that causes activation of the
primary system. When the procedure outlined above is performed for a number of angles $\zeta$, the
inner envelope of the yield points generates the $\pi$-plane projection of the yield surface for
molybdenum that is shown in \reffig{fig_yieldsurf_MoBOP} as a solid polygon.

For comparison, we also plot in \reffig{fig_yieldsurf_MoBOP} by dashed lines the projection of the
yield surface that is generated by the Schmid law, i.e. when $a_1=a_2=a_3=0$. In this case, the
yield criterion reduces to that of Tresca whose $\pi$-plane projection is the regular hexagon. The
marked difference between the shapes of the two polygons in \reffig{fig_yieldsurf_MoBOP} is the
result of the effect of non-glide stresses that are clearly responsible for the asymmetry between
the positive and negative loading along a given direction. This asymmetry is most noticeable when
$\sigma_1=\sigma_2$ (i.e. $\zeta=45\deg$) in which case the yield locus at negative stresses
coincides with a corner of the Tresca hexagon, while at positive stresses larger elastic deformation
is permitted prior to yielding than that predicted by the Schmid law.

\begin{figure}[!tb]
  \centering
  \includegraphics[width=10cm]{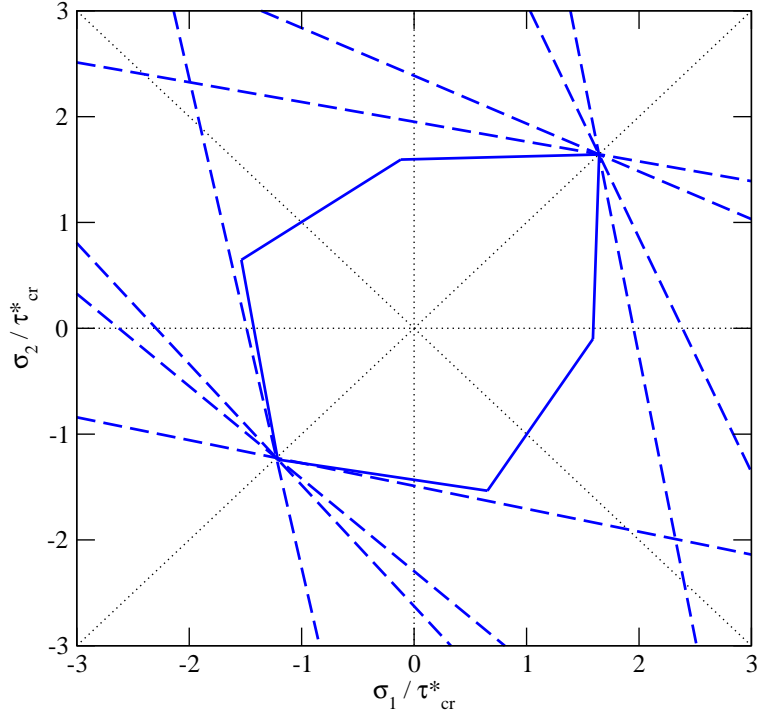}
  \parbox{13cm}{\caption{$\pi$-plane projection of the yield polygon for molybdenum showing all
      potentially active slip systems. The dashed lines correspond to various slip systems that
      become operative if the loading path intersects the yield polygon at its corner.}
  \label{fig_yieldsurf_allsys_MoBOP}}
\end{figure}

If the loading path, defined by a pair of principal stresses $(\sigma_1,\sigma_2)$, intersects one
of the edges of the polygon, the plastic deformation proceeds by \emph{single slip} on the plane
that is marked in the figure. In contrast, if the loading path reaches the yield polygon at one of
its corners at which $\sigma_1=\sigma_2$, many slip systems are typically activated at once, and the
plastic deformation proceeds by so-called \emph{multislip}. This is shown graphically in
\reffig{fig_yieldsurf_allsys_MoBOP}, where the dashed lines correspond to all slip systems that can
become activated for slip if the loading path reaches the yield surface at its corner. For
comparison, if the plastic deformation of bcc molybdenum were governed by the Schmid law, all edges
of the yield polygon that are \emph{not} parallel to the diagonal ($\sigma_1=\sigma_2$) would be
shared by two different slip systems. One can thus clearly see from
\reffig{fig_yieldsurf_allsys_MoBOP}, that non-glide stresses not only modify the shape of the yield
surface, but they are also responsible for the confinement of the multislip to the corners of the
yield polygon.

%----------------------------------------------------------------------------------------------------
%----------------------------------------------------------------------------------------------------

\section{Primary slip systems in uniaxial loading}
\label{sec_predact_MoBOP}

We are now in the position to investigate the activity of all $\gplane{110}\gdir{111}$ systems for
any orientation within any stereographic triangle, e.g. that shown in \reffig{fig_sgtria_tc}
corresponding to the $(\bar{1}01)[111]$ system. This requires identification of the four slip
systems $\alpha$ that can become operative and subsequent calculations of the orientations of their
MRSSPs, $\chi_\alpha$, with the corresponding shear stresses perpendicular and parallel to the slip
direction, $\tau_\alpha$ and $\sigma_\alpha$, respectively. The tensorial expression of the yield
criterion (\ref{eq_tstar_tensor}) provides a way of deciding the activity of individual systems and
can be used directly to predict the primary slip system for various orientations of uniaxial
loading. If one considers a unit loading $\hat\sigma_{ax}$ (+1 for tension, -1 for compression) with
the axis along a chosen direction and determines the corresponding magnitude of $\tau^{*\alpha}$ for
every system $\alpha$, the magnitude of the uniaxial stress for which a system $\alpha$ is activated
can be determined as $\sigma_{ax}^\alpha = k^\alpha \hat\sigma_{ax}$ where
$k^\alpha=\tau^*_{cr}/\tau^{*\alpha}$. The stress $\sigma^\alpha_{ax}$ corresponding to the highest
$\tau^{*\alpha}$, calculated among the four systems, defines the actual uniaxial yield stress and
$\alpha$ the corresponding primary system.

\begin{figure}[!htb]
  \centering
  \includegraphics[width=15cm]{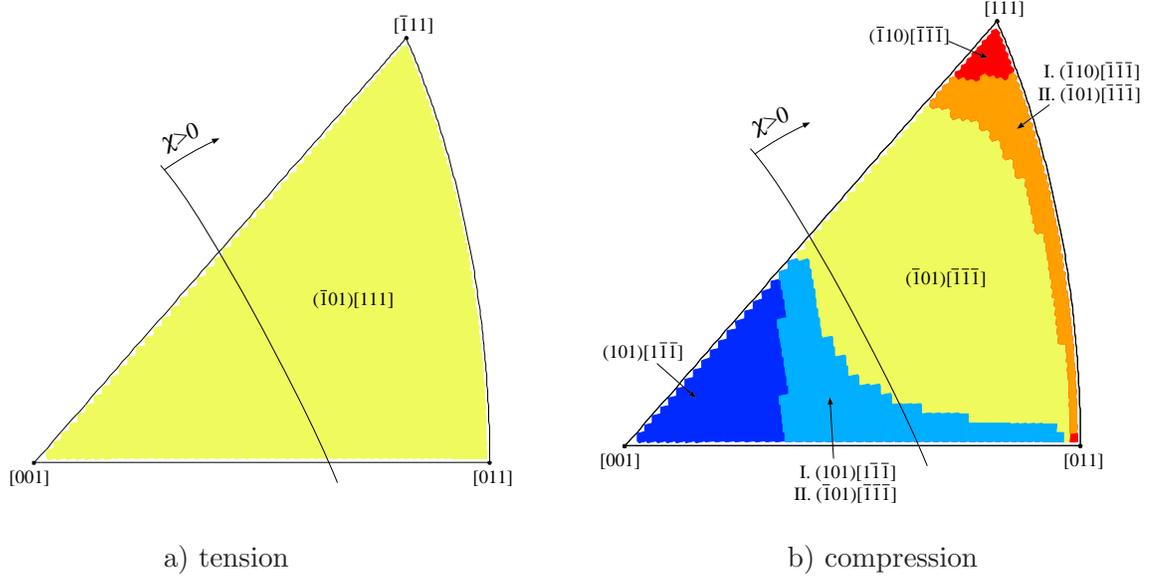} \\
  a) tension \hskip6.5cm b) compression \\
  \caption{Primary slip systems for uniaxial loading predicted from the effective yield criterion
    (\ref{eq_tstar_tensor}). In the regions of simultaneous activity of two slip systems, the
    primary slip system with lower yield stress is marked as ``I'' and the secondary, whose yield
    stress is at most 2\% higher than I, as ``II''.}
  \label{fig_sgtria_uniax}
\end{figure}

The primary slip systems in both tension and compression, identified by the $\tau^*$ criterion, are
shown in \reffig{fig_sgtria_uniax}. In tension, $(\bar{1}01)[111]$ is the primary slip system for
any orientation of the loading axis within the standard stereographic triangle. This can be easily
understood by looking at the distribution of the critical lines plotted in
\reffig{fig_CRSS_tau_fit_MoBOP}. In tension (positive $\tau$) the ratio of the shear stresses
perpendicular and parallel to the slip direction resolved in the MRSSP, $\eta=\tau/\sigma$, is
always positive and small enough such that the loading path plotted in the $\CRSS-\tau$ projection
first reaches the critical line for the operation of the $(\bar{1}01)[111]$ system. The slip on
other systems is also possible, but at larger stresses than those for which the primary
$(\bar{1}01)[111]$ system becomes operative.

In compression (negative $\tau$), the map of primary slip systems for different loading directions
is more complex. Because the sense of shearing is now reversed, the slip system with the highest
\emph{positive} Schmid stress is not $(\bar{1}01)[111]$, as in tension, but its conjugate
$(\bar{1}01)[\bar{1}\bar{1}\bar{1}]$. In the central region of the triangle, the most operative slip
system is, indeed, $(\bar{1}01)[\bar{1}\bar{1}\bar{1}]$. As the loading axis deviates towards the
$[011]-[\bar{1}11]$ edge, the $(\bar{1}10)[\bar{1}\bar{1}\bar{1}]$ system becomes increasingly more
prominent. There exists a region in which the two systems are equally operative, i.e. the uniaxial
yield stress of the secondary system (marked as II) is within 2\% from that of the primary system
(denoted as I). In this case, both systems operate simultaneously, which may give rise to the
macroscopic slip on the $(\bar{2}11)$ plane. For orientations close to the $[\bar{1}11]$ corner, the
dislocation moves by single slip on the $(\bar{1}10)[\bar{1}\bar{1}\bar{1}]$ system. A similar
situation arises on the other side of the triangle for orientations closer to the $[001]$
corner. Two slip systems have again the same prominence, namely $(\bar{1}01)[\bar{1}\bar{1}\bar{1}]$
and $(101)[1\bar{1}\bar{1}]$. However, since the slip directions are now Burgers vectors of two
different dislocations, $1/2[111]$ and $1/2[\bar{1}11]$, these dislocations move
simultaneously. This so-called multislip mechanism of plastic deformation was frequently observed in
low-temperature experiments, not only on molybdenum but also on all other bcc refractory metals of
high purity. For loading axes close to the $[001]$ corner of the standard triangle, the plastic
deformation is dominated by the $(101)[1\bar{1}\bar{1}]$ system and single slip on this system
ensues.

\begin{figure}[!htb]
  \centering
  \includegraphics[width=8.5cm]{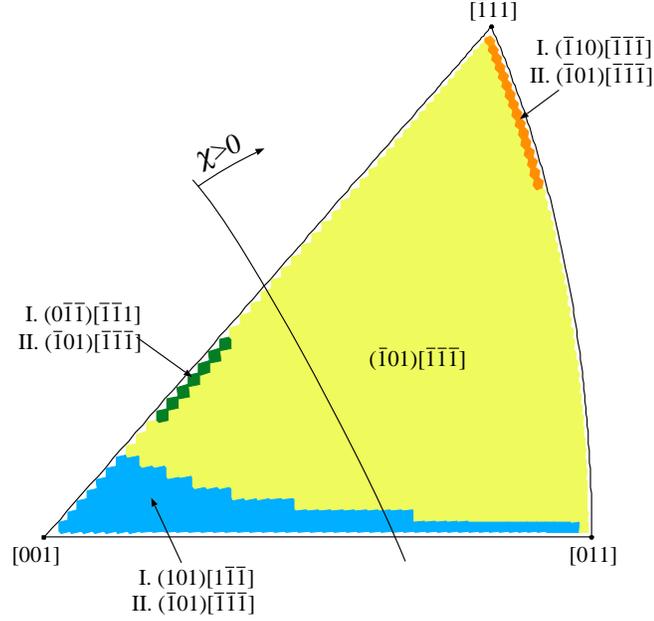} \\
  \parbox{12cm}{\caption{Primary slip systems for loading in \emph{compression} predicted from the
      Schmid law that arises if the $a_1$,$a_2$,$a_3$ in the effective yield criterion
      (\ref{eq_tstar_tensor}) are all zero. In the regions of simultaneous activity of two slip
      systems, the primary slip system with lower yield stress is marked as ``I'' and the
      secondary, whose yield stress is at most 2\% higher than I, as ``II''.}
  \label{fig_sgtria_uniax_Schmid}}
\end{figure}

For comparison, we show in \reffig{fig_sgtria_uniax_Schmid} the hypothetical distribution of the
most operative slip systems for loading in compression dictated by the Schmid law, i.e. when
$\tau^*$ is equal to the Schmid stress and $a_1=a_2=a_3=0$. In this case, the region of single slip
on the $(\bar{1}01)[\bar{1}\bar{1}\bar{1}]$ system occupies most of the stereographic triangle, and
only in the three narrow domains close to the corners or edges of the triangle this system operates
together with another $\gplane{110}\gdir{111}$ system. Comparing \reffig{fig_sgtria_uniax_Schmid}
with \reffig{fig_sgtria_uniax}b, we can conclude that for compression the non-glide stresses are
responsible for a substantial broadening of the regions of multiple slip with the single
$(\bar{1}01)[\bar{1}\bar{1}\bar{1}]$ system confined to the center of the stereographic
triangle. 

%----------------------------------------------------------------------------------------------------
%----------------------------------------------------------------------------------------------------

\section{Comparisons with low-temperature experiments}
\label{sec_expt_correl}

In the following, we will show that the effective yield criterion (\ref{eq_tstar_tensor}) correctly
reproduces the activity of the slip systems whose traces are observed in low-temperature plastic
deformation experiments. We will focus on a characteristic sample of experiments on ultra-high
purity single crystals and dilute alloys done at temperatures up to 77~K.

%----------------------------------------------------------------------------------------------------

\subsection{Experiments in pure shear}

Due to their inherent experimental complexities, the plastic flow experiments in which the loading
is applied as a pure shear are rather rare. A classical study was carried out by \citet{guiu:69}, who
applied the shear at temperatures ranging from $77~\K$ to room temperature such that the MRSSPs
were $\gplane{110}$ and/or $\gplane{112}$ planes. At 77~K, the experiments revealed a significant
twinning-antitwinning asymmetry for shearing on $(\bar{2}11)$ and $(\bar{1}\bar{1}2)$ planes,
respectively. At higher temperatures, the difference between the twinning and antitwinning shear
diminished. For $\gplane{110}$ and $\gplane{112}$ as the MRSSPs, the primary slip plane was always
$(\bar{1}01)$. This observation is in a good agreement with the results of atomistic simulations,
particularly with \reffig{fig_CRSS_chi_MoBOP}, in which the slip plane corresponding to pure shear
parallel to the slip direction is always the $(\bar{1}01)$ plane with the highest Schmid stress.

%----------------------------------------------------------------------------------------------------

\subsection{Experiments in tension}

The prediction of the $\tau^*$ criterion that $(\bar{1}01)[111]$ is the primary slip system for any
orientation of tensile loading (see \reffig{fig_sgtria_uniax}a) agrees well with many experimental
observations of slip traces in macroscopic specimens. In their tensile experiments performed at
4.2~K, \citet{kitajima:81} observed that the most prominent slip system for any orientation of the
loading axis is $(\bar{1}01)[111]$. For orientations for which $\chi$ was close to $+30\deg$, the
$(\bar{1}10)[111]$ slip was observed. At the same time secondary slip on the $(101)[\bar{1}11]$
system with much weaker activity was observed for all orientations corresponding to $\chi<30\deg$.

\citet{matsui:76} performed tensile experiments with the loading axis close to the center of the
stereographic triangle. At low strains, anomalous $(0\bar{1}1)[111]$ slip dominated the plastic
flow. At strains larger than $1.57\%$, long straight slip traces of the primary $(\bar{1}01)[111]$
system were observed with only a minor contribution from the $(0\bar{1}1)[111]$ system. The
suppression of the activity of the anomalous $(0\bar{1}1)[111]$ system at high strains implies that
this system is not responsible for the macroscopic plastic deformation and, instead, the primary
$(\bar{1}01)[111]$ system plays the dominant role. This view is also supported by the tensile
experiments of \citet{lau:70} that reveal the dominant slip on the most highly stressed
$(\bar{1}01)[111]$ system within a broad range of orientations. Nevertheless, the occurrence of the
$(0\bar{1}1)[111]$ slip at very low strains remains an unexplained phenomenon. 

%----------------------------------------------------------------------------------------------------

\subsection{Experiments in compression}
\label{sec_expt_jeffcoat}

Compression experiments that are particularly suitable for comparison with the predictions of the
$\tau^*$ criterion are those of \citet{jeffcoat:76} on Mo-5\,at.\%\,Nb and Mo-5\,at.\%\,Re. The
reason is that they provide explicit information about the macroscopic slip planes corresponding to
the most prominent slip traces. The orientation of loading is given by the angle $\lambda$ between
the loading axis and the $[111]$ direction, and the angle $\chi$ between the $(\bar{1}01)$ plane and
the MRSSP in which the resolved shear stress parallel to the slip direction is at maximum. Three
orientations used by \citet{jeffcoat:76} will be discussed here, namely: (i)
$\lambda=50\deg,\chi=0$, (ii) $\lambda=50\deg,\chi=-15\deg$, and (iii) $\lambda=50\deg,\chi=+25\deg$,
which correspond approximately to $[\bar{5}\ 11\ 27]$, $[\bar{1}\ 3\ 14]$ and $[\overline{40}\ 101\
116]$ axes, respectively.

For $\lambda=50\deg,\chi=+25\deg$, single slip of $1/2[111]$ dislocations on planes corresponding to
$\psi=+10\deg$ and $\psi=+51\deg$, termed by \citet{jeffcoat:76} as an irrational slip, was
observed. This observation can be compared directly with the distribution of the primary slip
systems calculated from the $\tau^*$ criterion and plotted in \reffig{fig_sgtria_uniax}b. For
$\chi=25\deg$, the $\tau^*$ criterion predicts that the $(\bar{1}01)[\bar{1}\bar{1}\bar{1}]$ and
$(\bar{1}10)[\bar{1}\bar{1}\bar{1}]$ systems will be activated simultaneously. Since both systems
share the same slip direction, a $1/2[111]$ screw dislocation can glide with the same propensity
on both $(\bar{1}10)$ and $(\bar{1}01)$ planes whose orientations correspond to $\psi=+60\deg$ and
$\psi=0$, respectively. Consequently, the macroscopic slip plane composed of $(\bar{1}01)$ and
$(\bar{1}10)$ segments can be any plane in $0<\psi<+60\deg$, which agrees well with the experimental
observations.

For the other two orientations, i.e. for $\lambda=50\deg$ and $\chi=0$ or $\chi=-15\deg$, the
most prominent slip traces corresponded to the $(0\bar{1}1)$ plane that completely dominates the
plastic deformation at larger negative $\chi$. Besides this system, traces of the most highly
stressed $(\bar{1}01)[\bar{1}\bar{1}\bar{1}]$ system and the $(101)[1\bar{1}\bar{1}]$
system were also observed%
\footnote{The slip planes observed in the experiments of \citet{jeffcoat:76} were originally written
  with opposite slip directions. For consistency, we always use the Miller indices for which the
  shear stress parallel to the slip direction resolved in the MRSSP is positive. In compression,
  this means that instead of $(\bar{1}01)[111]$, we write $(\bar{1}01)[\bar{1}\bar{1}\bar{1}]$.},
the latter of which became more prominent at negative angles $\chi$. From the stereographic triangle in
\reffig{fig_sgtria_uniax}b, one again observes that the $\tau^*$ criterion predicts a simultaneous
operation of the $(\bar{1}01)[\bar{1}\bar{1}\bar{1}]$ and $(101)[1\bar{1}\bar{1}]$ systems at
negative $\chi$, as observed experimentally.

\begin{figure}[!htb]
  \centering
  \includegraphics[width=12cm]{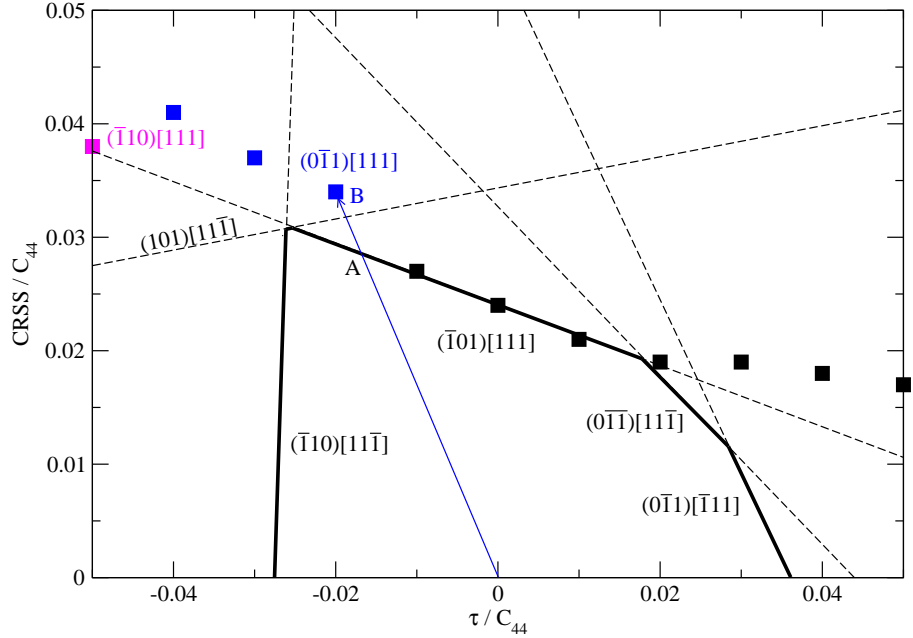}
  \parbox{14cm}{\caption{Loading path for compression at $\lambda=50\deg$, $\chi=0$ superimposed on
      the $\CRSS-\tau$ dependence for the MRSSP $(\bar{1}01)$. The dashed lines are the critical
      lines calculated from the $\tau^*$ criterion with the inner envelope (solid line)
      corresponding to the yield polygon.}
    \label{fig_chi0_0-11slip_MoBOP}}
\end{figure}

% why is the (0-11) slip not predicted by the tau* criterion

We have seen that the results of atomistic simulations, plotted in \reffig{fig_CRSS_tau_MoBOP},
correctly predict the $(0\bar{1}1)$ slip at negative shear stresses perpendicular to the slip
direction that correspond to loading in compression. At the same time, the proposed $\tau^*$
criterion correctly predicts the onset of slip when $-0.02<\tau/C_{44}<+0.02$ for which the CRSS
varies linearly with $\tau$. However, at large negative $\tau$, the atomistically calculated
$\CRSS-\tau$ dependence is nonlinear, and the linear $\tau^*$ criterion does not reproduce the
results of atomistic simulations sufficiently well. To illustrate this, we plot in
\reffig{fig_chi0_0-11slip_MoBOP} the loading path corresponding to compression along
$\lambda=50\deg$, $\chi=0$. From the $\tau^*$ criterion, slip occurs when the loading path reaches
the yield polygon, particularly point A at which the $(\bar{1}01)[111]$ system becomes
activated. However, using directly the atomistic results, one may argue that the slip occurs when
the loading path reaches point B, which leads to the anomalous slip on the $(0\bar{1}1)[111]$
system. Ideally, both these predictions should give the same answer, and this agreement would
certainly be achieved if $\tau^*$ were a nonlinear function of the applied stress. However, the
nonlinear $\tau^*$ criterion would clearly introduce serious complications when predicting the
$\CRSS-\chi$ and $\CRSS-\tau$ dependencies, and future finite-element simulations would become
significantly more complicated and less efficient. Quoting \citet{strogatz:03}: "No insight is
gained if the model is as perplexing as the phenomena it's supposed to describe". The fact that the
linear $\tau^*$ criterion does not directly predict the occurrence of the anomalous slip at negative
$\tau$ is thus relevant only when dealing with microscopic details of the slip and does not affect
the accuracy of the $\tau^*$ criterion in continuum analyses.

\begin{table}[!p]
  \caption{The most operative slip systems predicted by the Schmid law and from the $\tau^*$ 
    criterion, respectively, listed in the order of descending $\tau^*$. The Schmid factors and the
    values of $\tau^*$ are normalized by the values corresponding to the system with the highest
    $\tau^*$. The experimentally observed systems are written in bold.}
  \label{tab_expt_jeffcoat_act}
  \vskip2em
  % 
  % The following is calculated as
  %   tstar12sys_tc([-40 101 116],-2.2387,0.24,0,0.35);
  % 
  \begin{minipage}{0.4\textwidth}
    \centering
    a) $\lambda=50\deg$, $\chi=+25\deg$ \\
    \begin{tabular}{cccc}
      \hline
      order & ref. system & Schmid f. & $\tau^*$ \\
      \hline
      \bf 1 & $\bf (\bar{1}01)[\bar{1}\bar{1}\bar{1}]$ & \bf 1.00 & \bf 1.00 \\
      \bf 2 & $\bf (\bar{1}10)[\bar{1}\bar{1}\bar{1}]$ & \bf 0.90 & \bf 0.97 \\
      3 & $(101)[1\bar{1}\bar{1}]$                     & 0.71 & 0.79 \\
      4 & $(110)[1\bar{1}\bar{1}]$                     & 0.57 & 0.71 \\
      5 & $(101)[11\bar{1}]$                           & 0.15 & 0.45 \\
      6 & $(\bar{1}\bar{1}0)[\bar{1}1\bar{1}]$         & 0.06 & 0.41 \\
      7 & $(\bar{1}10)[11\bar{1}]$                     & 0.28 & 0.26 \\
      8 & $(0\bar{1}\bar{1})[\bar{1}\bar{1}1]$         & 0.43 & 0.13 \\
      9 & $(0\bar{1}1)[\bar{1}\bar{1}\bar{1}]$         & 0.10 & 0.10 \\
      $\vdots$ & $\vdots$ & $\vdots$ & $\vdots$ \\
      \hline
    \end{tabular}
  \end{minipage}
  \hskip2cm
  %
  % The following is calculated as
  %   tstar12sys_tc([-5 11 27],-2.1592,0.24,0,0.35);
  %
  \begin{minipage}{0.4\textwidth}
    \centering
    b) $\lambda=50\deg$, $\chi=0$
    \begin{tabular}{cccc}
      \hline
      order & ref. system & Schmid f. & $\tau^*$ \\
      \hline
      \bf 1 & $\bf (\bar{1}01)[\bar{1}\bar{1}\bar{1}]$ & \bf 1.00 & \bf 1.00 \\
      \bf 2 & $\bf (101)[1\bar{1}\bar{1}]$             & \bf 0.90 & \bf 0.96 \\
      3 & $(\bar{1}10)[\bar{1}\bar{1}\bar{1}]$ & 0.50 & 0.79 \\
      4 & $(0\bar{1}1)[1\bar{1}\bar{1}]$      & 0.65 & 0.69 \\
      5 & $(0\bar{1}\bar{1})[\bar{1}\bar{1}1]$ & 0.76 & 0.65 \\
      6 & $(101)[11\bar{1}]$                   & 0.44 & 0.62 \\
      \bf 7 & $\bf (0\bar{1}1)[\bar{1}\bar{1}\bar{1}]$ & \bf 0.50 & \bf 0.60 \\
      $\vdots$ & $\vdots$ & $\vdots$ & $\vdots$ \\
      \hline
    \end{tabular}
  \end{minipage}
  \vskip1cm
  % 
  % The following is calculated as
  %   tstar12sys_tc([-1 3 14],-2.1867,0.24,0,0.35);
  % 
  \centering
  c) $\lambda=50\deg$, $\chi=-15\deg$ \\
  \begin{tabular}{cccc}
    \hline
    order & ref. system & Schmid f. & $\tau^*$ \\
    \hline
    \bf 1 & $\bf (101)[1\bar{1}\bar{1}]$ & \bf 0.98 & \bf 1.00 \\
    \bf 2 & $\bf (\bar{1}01)[\bar{1}\bar{1}\bar{1}]$ & \bf 1.00 & \bf 1.00 \\
    3 & $(0\bar{1}1)[1\bar{1}\bar{1}]$ & 0.89 & 0.87 \\
    \bf 4 & $\bf (0\bar{1}1)[\bar{1}\bar{1}\bar{1}]$ & \bf 0.73 & \bf 0.81 \\
    $\vdots$ & $\vdots$ & $\vdots$ & $\vdots$ \\
    \hline
  \end{tabular}
\end{table}

The discussion above holds also for the other loading direction, specifically $\lambda=50\deg$,
$\chi=-15\deg$. Very briefly, the ratio $\eta$ of the two shear stresses resolved in the MRSSP at
$\chi=-15\deg$ is again negative, and thus the loading path in the $\CRSS-\tau$ projection passes
between the atomistic data corresponding to slip on the $(\bar{1}01)$ and the $(0\bar{1}1)$
planes. Hence, the dislocation can again move on either of these two planes, which was, indeed,
observed in experiments of \citet{jeffcoat:76} for orientations corresponding to zero and negative
$\chi$. Thus, the plastic deformation occurs via multiple slip of dislocations with Burgers vectors
$1/2[111]$ and $1/2[\bar{1}11]$.

It is also important to check the \emph{degree} of prominence of the most operative slip
systems. This can be done either by comparing the Schmid factors or the $\tau^{*\alpha}$ values
calculated for each system $\alpha$. For the three orientations of applied loading,
\reftab{tab_expt_jeffcoat_act} shows the Schmid factors and the $\tau^*$ values that are both
normalized by the values corresponding to the system with the highest $\tau^*$. One can clearly see
that the $\tau^*$ criterion predicts the increasing prominence of the
$(0\bar{1}1)[\bar{1}\bar{1}\bar{1}]$ system for MRSSPs at zero and negative $\chi$. In particular,
for $\chi=0$, this system is only the seventh most highly stressed one, but its activity
increases at more negative $\chi$. For $\chi=-15\deg$, this system is already the fourth most
prominent one. A similar mechanism of slip, called hereafter ``anomalous slip'', has been
frequently encountered in low-temperature experiments on high purity single crystals of all bcc
metals and thus seems to be a universal feature of slip of the group of VB and VIB metals.

\begin{figure}[!htb]
  \centering
  \includegraphics[width=8cm]{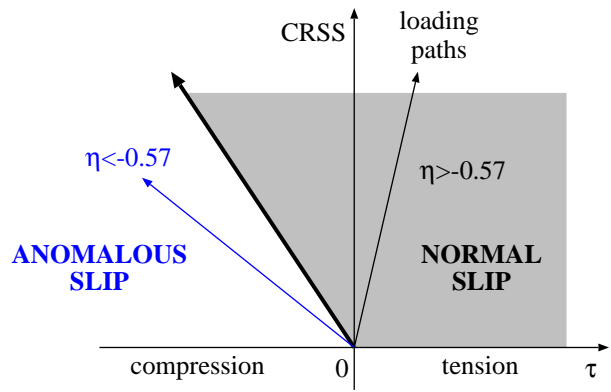}
  \parbox{13cm}{\caption{The two regions representing two different modes of slip.}
    \label{fig_eta_regions}}
\end{figure}

Although the onset of anomalous slip under loading in compression is not directly predicted by the
$\tau^*$ criterion, it can be directly predicted using the results of atomistic simulations
performed at 0~K. The crucial observation is that the slip on the most highly stressed $(\bar{1}01)$
plane occurs when $\eta>-0.57$, where $\eta=\tau/\sigma$. In contrast, the slip on the anomalous
$(0\bar{1}1)$ or $(\bar{1}10)$ plane takes place at larger negative $\tau$ or, more specifically,
when $\eta<-0.57$. The orientations of loading paths leading to these two modes of slip are
illustrated schematically in \reffig{fig_eta_regions}. The crossover value, $\eta\approx-0.57$, is
independent of the orientation of the MRSSP and is thus a characteristic parameter of
molybdenum. This suggests the possibility of finding a set of orientations of loading axes for which
the atomistic results predict the onset of anomalous slip in compression. For any orientation of
loading axis, this can be accomplished by resolving the shear stresses $\tau$ and $\sigma$ in the
MRSSP and calculating their ratio $\eta$.

\begin{figure}[!htb]
  \centering
  \includegraphics[width=10cm]{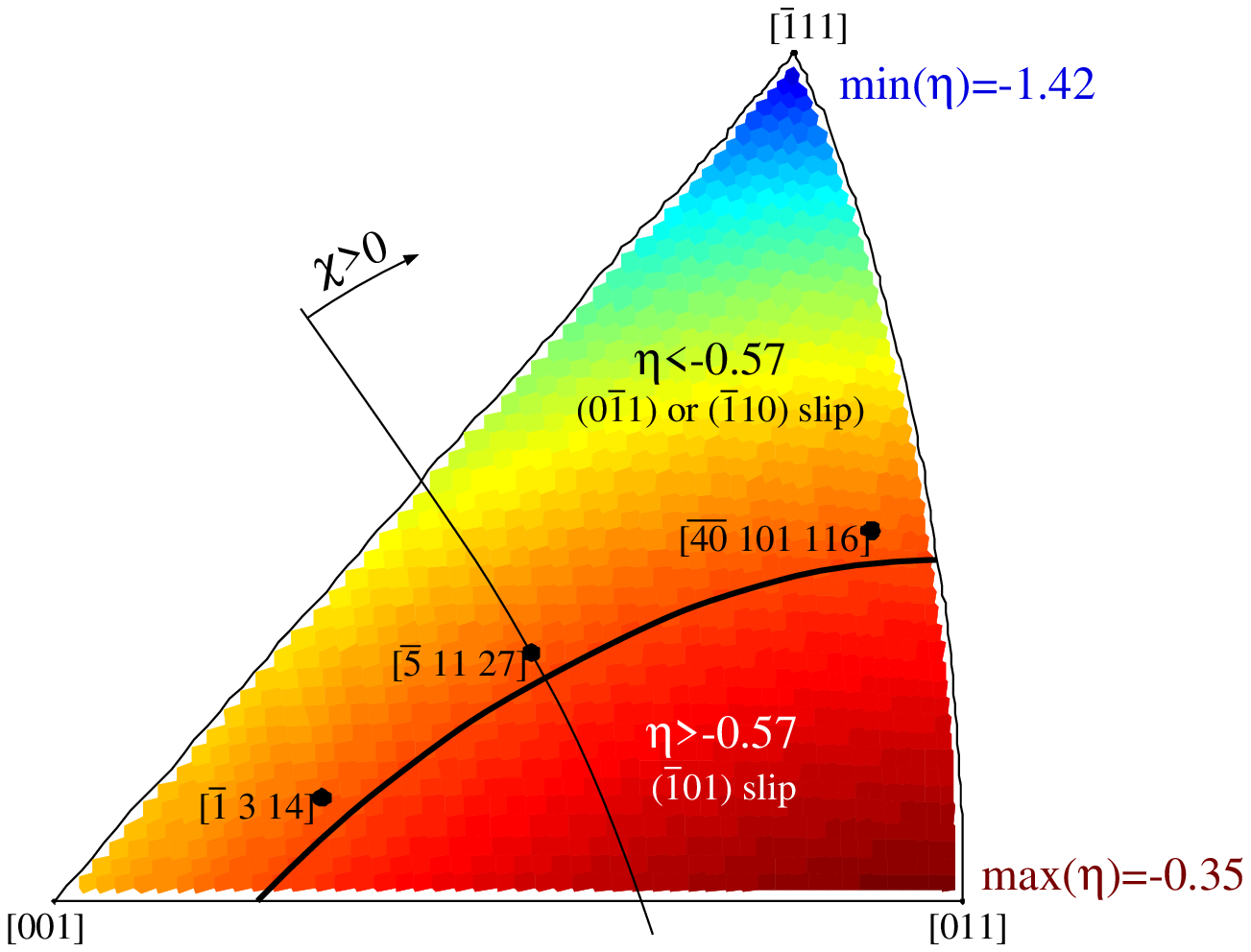} \qquad 
  \includegraphics[height=8cm]{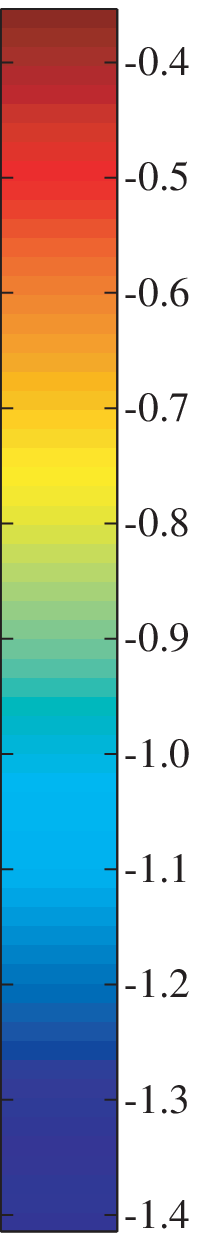}
  \parbox{14cm}{\caption{Distribution of the values of $\eta=\tau/\sigma$. The map is calculated for
      loading in \emph{compression} in all possible directions inside the stereographic
      triangle. The solid line, $\eta\approx-0.57$, represents orientations for which the slip on
      the $(\bar{1}01)$ and either $(0\bar{1}1)$ or $(\bar{1}10)$ plane is equally likely.}
  \label{fig_sgtria_eta}}
\end{figure}

\reffig{fig_sgtria_eta} shows the distribution of the values of $\eta$ for loading in
compression. As explained above, the orientations for which a $1/2[111]$ dislocation moves on the
$(\bar{1}01)$ plane correspond to the region where $\eta>-0.57$, which is represented by the loading
axes close to the $[011]$ corner of the triangle. However, for a large number of orientations closer
to the $[001]-[\bar{1}11]$ edge, $\eta<-0.57$ and the slip proceeds on one of the low-stressed
$\gplane{110}$ planes, as observed in atomistic simulations. All orientations of loading axes used
in experiments of \citet{jeffcoat:76} and plotted in \reffig{fig_sgtria_eta} correspond to
$\eta<-0.57$, which predicts the activation of slip on one of the two low-stressed $\gplane{110}$
planes.

%----------------------------------------------------------------------------------------------------

\subsection{Yield stress asymmetry in tension and compression}
\label{sec_ysasym}

In order to investigate the plastic deformation of ultra-high purity single crystals of molybdenum
under tension and compression, \citet{seeger:00} recently performed a series of experiments for
three different orientations of applied loading and temperatures between $123~\K$ and $460~\K$. At
high temperatures, approaching $460~\K$, the tensile and compressive yield stress of all crystals
were practically the same. However, at temperatures close to $123~\K$ an interesting yield stress
asymmetry arose for orientations $[\bar{1}49]$ and $[\bar{5}79]$ of tensile/compressive axes, which
correspond to $\chi=0$ and $\chi=+21\deg$, respectively, in that the yield stress in compression
was appreciably higher than that obtained for the same orientation in tension.

\begin{figure}[!p]
  \centering
  \includegraphics[width=9cm]{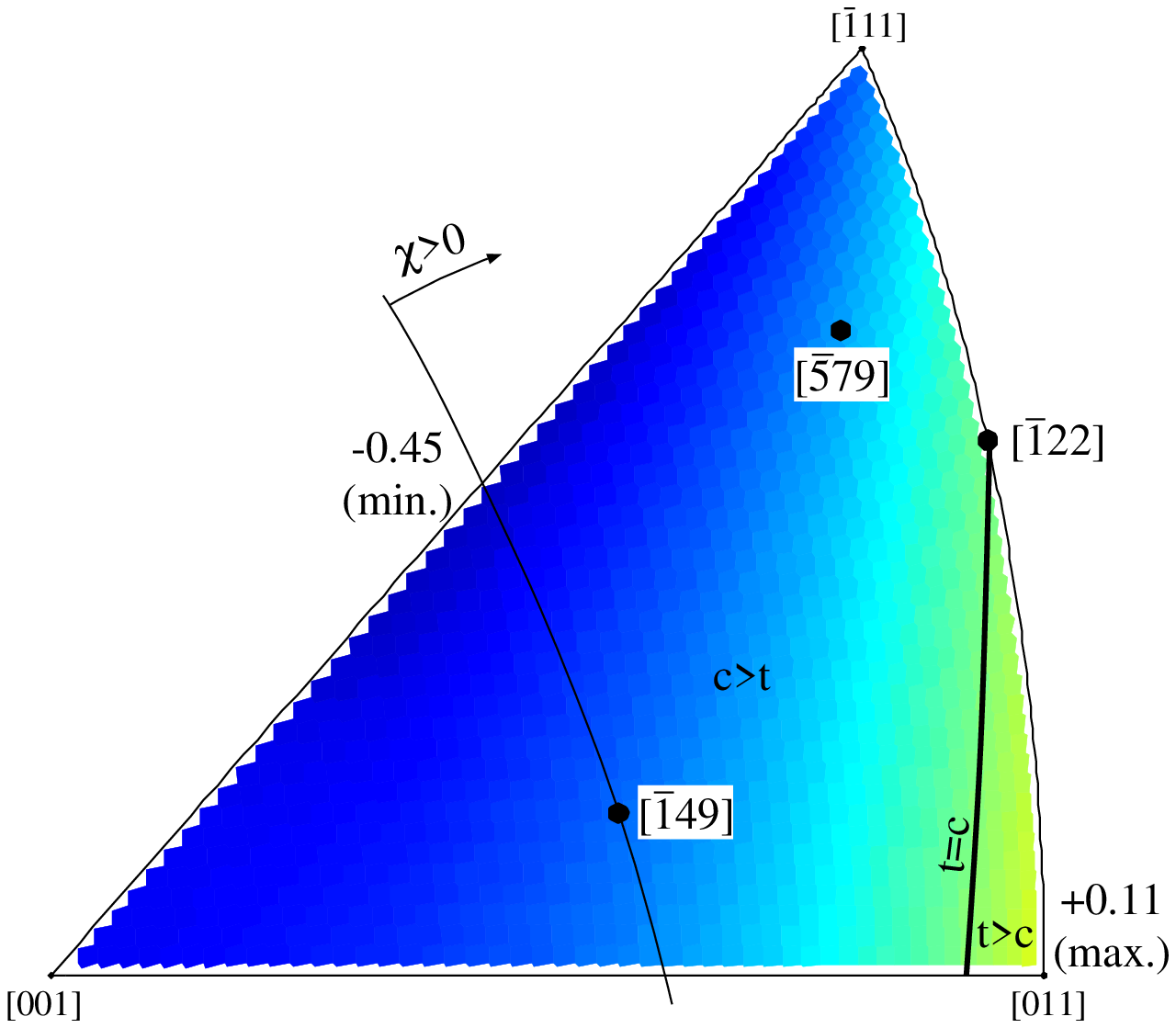} \qquad
  \includegraphics[height=8cm]{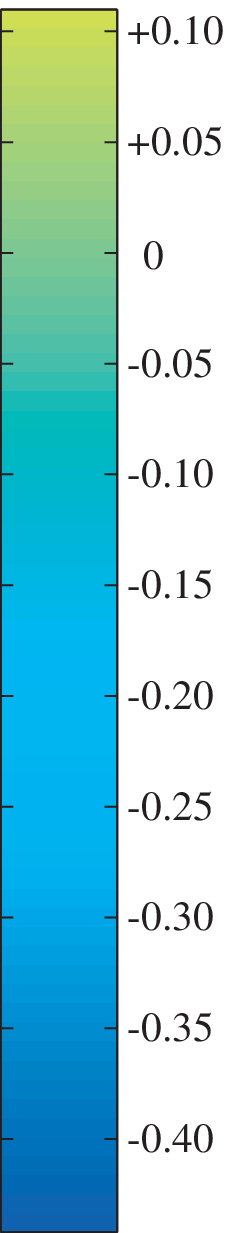} \\[2em]
  a) both shear stresses parallel and perpendicular to the slip direction considered \\[2em]
  \includegraphics[width=9cm]{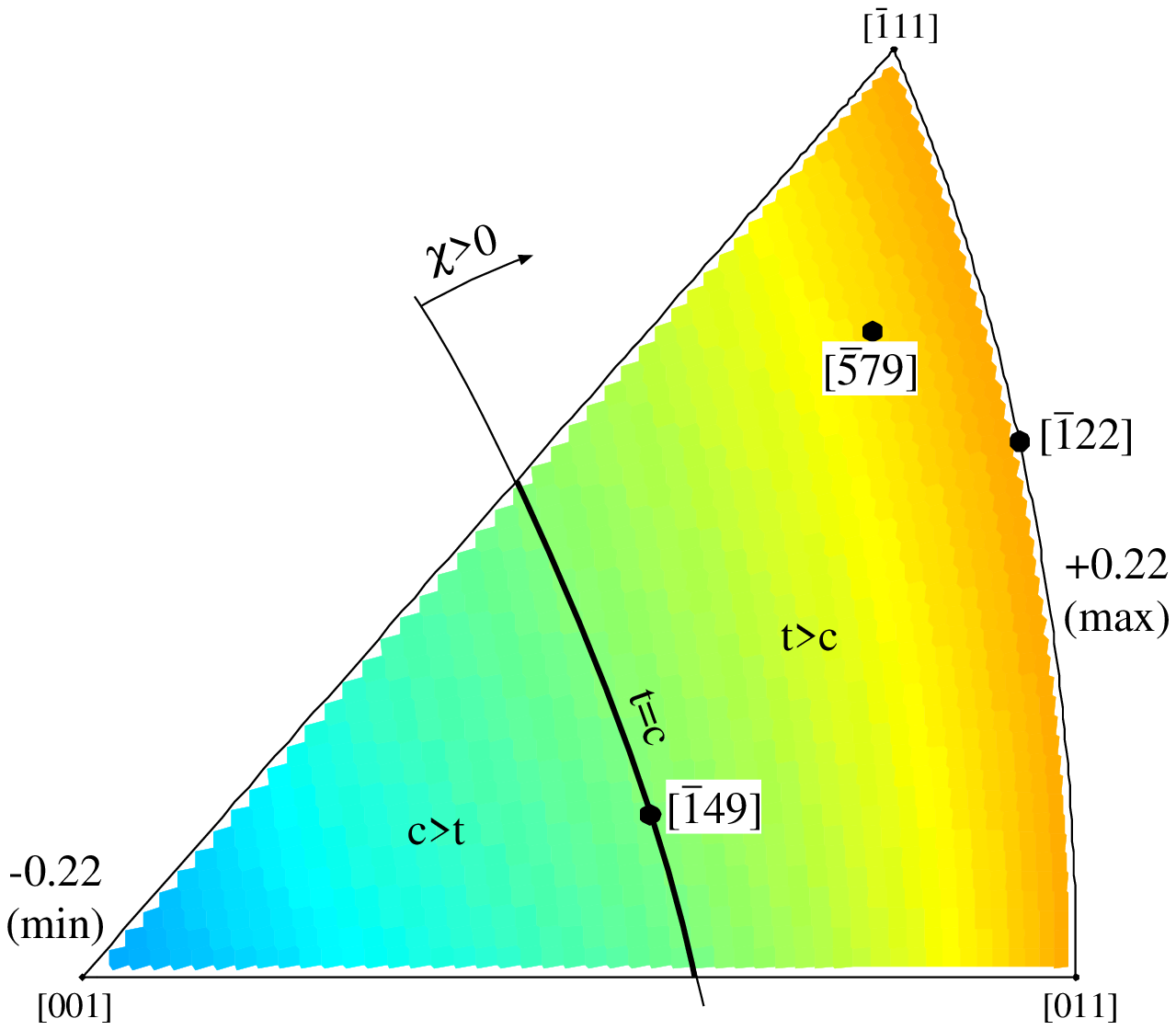} \qquad\hskip-1mm
  \includegraphics[height=8cm]{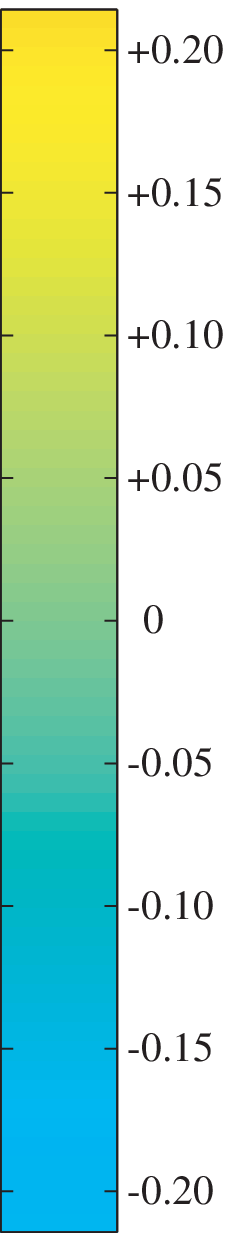} \\
  b) only shear stress parallel to the slip direction considered ($a_2=a_3=0$) \\
  \caption{Maps of the stress differential $\Delta\sigma_{t,c}$ calculated from the $\tau^*$
  criterion. Notice the antisymmetric character of $\Delta\sigma_{t,c}$ in b) but a more complex
  distribution in a).}
  \label{fig_sgtria_tcasym}
\end{figure}

In order to test whether our effective yield criterion is capable of predicting correctly this
asymmetry between tension and compression, we will first introduce a stress differential
\begin{equation}
  \Delta\sigma_{t,c} = \frac{\sigma_t-\sigma_c}{(\sigma_t+\sigma_c)/2} \ ,
  \label{eq_stressdiff}
\end{equation}
where $\sigma_t$ and $\sigma_c$ are the uniaxial yield stresses in tension and compression,
respectively. For any orientation of applied loading, these yield stresses can be easily calculated
from the tensorial form of the $\tau^*$ criterion (\ref{eq_tstar_tensor}). If one applies unit
loading (+1 for tension, -1 for compression) along the given direction and calculates
$\tau^{*\alpha}$ for each of the 24 slip systems from \reftab{tab_bcc24sys}, the ratio
$\tau_{cr}^*/\tau^{*\alpha}$ determines the magnitude of the uniaxial yield stress for each system
$\alpha$. For a given loading axis, the actual yield stress that induces plastic flow in the
crystal corresponds to the minimum value among the 24 possibilities, and the $\gplane{110}\gdir{111}$
system corresponding to this value is the primary slip system that becomes operative first. By
performing these calculations for all orientations in the stereographic triangle, one obtains a
distribution of the stress differential (\ref{eq_stressdiff}) that is plotted in
\reffig{fig_sgtria_tcasym}a. The effective yield criterion predicts that the yield stress in
compression is larger than that in tension within a large area of the stereographic triangle. As the
loading axis deviates from the center of the triangle towards the $[011]-[\bar{1}11]$ edge,
$\sigma_t$ increases significantly and becomes equal to $\sigma_c$ for orientations along the solid
line in \reffig{fig_sgtria_tcasym}a. For orientations closer to the $[011]$ corner of the triangle,
$\sigma_t>\sigma_c$ and the sign of $\Delta\sigma_{t,c}$ becomes positive. The maximum asymmetry
between the yield stress in tension and compression is reached for exactly the $[011]$ orientation
in which case $\max(\Delta\sigma_{t,c})\approx+0.11$. For comparison, the minimum value of the
stress differential, corresponding to the loading axis in the middle of the $[001]-[\bar{1}11]$
edge, is $\min(\Delta\sigma_{t,c})\approx-0.45$.

If the twinning-antitwinning asymmetry were the only reason for the tension-compression asymmetry in
molybdenum, the stress differential $\Delta\sigma_{t,c}$ would be \emph{antisymmetric} with respect
to $\chi$ and thus would necessarily vanish at $\chi=0$. This is shown in
\reffig{fig_sgtria_tcasym}b, which corresponds to the restricted model of the $\tau^*$ criterion
(\ref{eq_tstar_restr}), i.e. when no shear stress perpendicular to the slip direction is
considered. Clearly, the two distributions in \reffig{fig_sgtria_tcasym} are very different.
Therefore, the tension-compression yield stress asymmetry is \emph{not} solely due to the
twinning-antitwinning asymmetry. This discrepancy was identified already by \citet{seeger:00} and
attributed to ``a modification of the Peierls potential by stress components other than the resolved
shear stress''. In the language of the non-associated plastic flow model, we identify this
contribution as an effect of the shear stress perpendicular to the slip direction. It can be easily
shown that the two non-glide stresses of the $\tau^*$ criterion that are proportional to $\tau$ do
not vanish even for $\chi=0$, which thus gives rise to nonzero $\Delta\sigma_{t,c}$ for this
orientation of the MRSSP.

In summary, the variation of the tension-compression yield stress asymmetry predicted by the
$\tau^*$ criterion and plotted in \reffig{fig_sgtria_tcasym}a agrees qualitatively with the
experimental observations of \citet{seeger:00}. The comparison of the magnitudes of the stress
differential is given in \reftab{tab_tcasym_cmp}, where the values predicted from the $\tau^*$
criterion are taken from \reffig{fig_sgtria_tcasym}a and b. The reason why the agreement is only
qualitative is that the experimental data correspond to 150~K, while the calculations using the
$\tau^*$ criterion correspond to 0~K.

\begin{table}[!htb]
  \centering
  \parbox{13cm}{\caption{Magnitudes of the stress differential calculated from the $\tau^*$
  criterion and obtained from the experimental data of \citet{seeger:00} at the temperature 150~K. The
  theoretical data are obtained from the restricted and full forms of the $\tau^*$ criterion at 0~K,
  i.e. without and with shear stresses perpendicular to the slip direction.}
  \label{tab_tcasym_cmp}} \\[1em]
  \begin{tabular}{lccc}
    \hline
    & \multicolumn{3}{c}{$\Delta\sigma_{t,c}$} \\ \cline{2-4}
    & from $\tau^*$ (restricted) & from $\tau^*$ (full) & experiment \\
    \hline
    $[\bar{1}49], \chi=0$   &  0    & -0.28 & -0.06 \\
    $[\bar{5}79], \chi=+21\deg$ & +0.16 & -0.21 & -0.04 \\
    $[\bar{1}22], \chi=+29\deg$ & +0.21 &  0    & +0.07 \\
    \hline
  \end{tabular}
\end{table}

\begin{figure}[!htb]
  \centering
  \includegraphics[width=12cm]{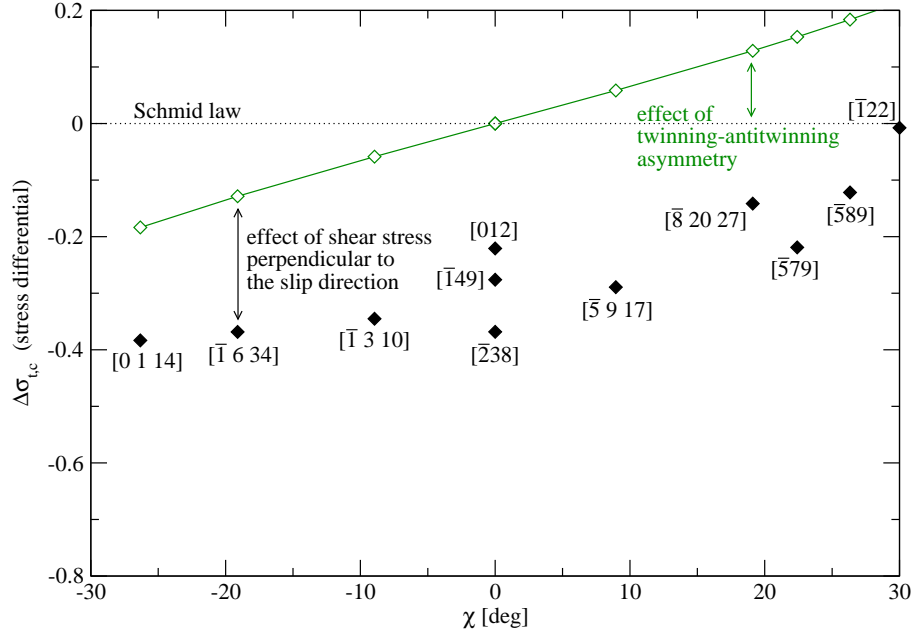}
  \parbox{14cm}{\caption{Variation of the stress differential (\ref{eq_stressdiff}) in molybdenum
      with the orientation of the MRSSP. The empty symbols are obtained from the restricted $\tau^*$
      criterion (\ref{eq_tstar_restr}) and the filled symbols from the full $\tau^*$ criterion
      (\ref{eq_tstar_full}).}
  \label{fig_tc_asym_MoBOP}}
\end{figure}

To compare the contributions of the shear stresses parallel and perpendicular to the slip direction
to the stress differential, we plot in \reffig{fig_tc_asym_MoBOP} a variation of
$\Delta\sigma_{t,c}$ with $\chi$. We include all the angles $\chi$ that correspond to the uniaxial
loading axes in \reffig{fig_CRSS_chi_MoBOP} as well as the three orientations in
\reftab{tab_tcasym_cmp}. For each of the loading axes, we first use the restricted form of the
$\tau^*$ criterion to calculate the yield stresses $\sigma_t$, $\sigma_c$ which, in turn, determine
the stress differential (\ref{eq_stressdiff}). This yields the data plotted in
\reffig{fig_tc_asym_MoBOP} with empty symbols that are further connected to highlight the
antisymmetric character of $\Delta\sigma_{t,c}$ (see also \reffig{fig_sgtria_tcasym}b). It is
important to emphasize, that the strength of the twinning-antitwinning asymmetry is proportional to
the angle that this line makes with the prediction of the Schmid law. Subsequently, the same process
is repeated using the full form of the $\tau^*$ criterion, which gives the data plotted with the filled
symbols. The difference between the data obtained from the full and restricted models is obviously
related to the effect of the shear stress perpendicular to the slip direction, which thus provides
an important measure of the strength of these non-glide stresses. Clearly, the complex
tension-compression asymmetry in molybdenum, shown in \reffig{fig_sgtria_tcasym}a, is a consequence
of the combined effect of \emph{both} the twinning-antitwinning asymmetry and the shear stress
perpendicular to the slip direction.

%----------------------------------------------------------------------------------------------------

\subsection{Six degrees of freedom microstrain experiments}

In recent years, the group of D.~Lassila at Lawrence Livermore National Laboratory proposed and
successfully executed a novel experimental technique in which five components of the elastic-plastic
strain tensor, $\eps_{ij}$, have been measured \emph{independently}. The distinguishing feature of
this technique is that the sample is essentially unconstrained with its bottom part able to slide
freely on the substrate. This unconstrained motion prevents the rotation of the slip planes during
loading. Knowing the total plastic strain tensor, obtained by eliminating the elastic contribution
in $\eps_{ij}$, the activities of 5 slip systems have been analyzed. Besides the identification of
the dominant systems, this method also quantifies the degree of prominence of the slip on each
of these systems, which is the key factor that can be used to correlate the predictions of the
$\tau^*$ criterion with experiments.

\begin{table}[!htb]
  \centering

  \parbox{14cm}{\caption{List of the slip systems whose activity was observed in the experiments 
      of \citet{lassila:03}. The systems are sorted in order of descending Schmid factors. The
      values of $\tau^*$ are calculated for unit applied loading. The anomalous system is written in
      bold.}

  \label{tab_dof6_lassila}} \\[5mm]
  \begin{tabular}{cccc}
    \hline
    $\alpha$ (see \reftab{tab_bcc24sys}) & slip system & Schmid factor & $\tau^*$ \\
    \hline
    14 & $(\bar{1}01)[\bar{1}\bar{1}\bar{1}]$ & 0.50 & 0.48 \\ 
     7 & $(0\bar{1}\bar{1})[\bar{1}\bar{1}1]$ & 0.32 & 0.24 \\
    17 & $(0\bar{1}1)[1\bar{1}\bar{1}]$ & 0.29 & 0.29 \\
     1 & $\bf (0\bar{1}1)[\bar{1}\bar{1}\bar{1}]$ & 0.25 & 0.27 \\
    23 & $(011)[\bar{1}1\bar{1}]$ & 0.22 & 0.16 \\
    10 & $(10\bar{1})[1\bar{1}1]$ & 0.17 & 0.23 \\
    \hline
  \end{tabular}
\end{table}

In the following, we concentrate on the experiment of \citet{lassila:02, lassila:03} done on
molybdenum under compression at room temperature, where the orientation of the loading axis is
$[2\,\bar{9}\,\overline{20}]$. This orientation is characterized by angles $\lambda=45\deg$, between
the loading axis and the $[\bar{1}\bar{1}\bar{1}]$ slip direction, and $\chi=0$, between the MRSSP
and the $(10\bar{1})$ reference plane. Because the loading in compression reverses the sense of
shearing on the $(10\bar{1})$ plane, the most highly stressed system with positive Schmid factor is
not $(10\bar{1})[\bar{1}\bar{1}\bar{1}]$ but, instead, its conjugate $(10\bar{1})[111]$. The list of
the most prominent reference systems identified by \citet{lassila:03} is given in
\reftab{tab_dof6_lassila}, where the Miller indices are always written in such a way that each
system can be found in \reftab{tab_bcc24sys}. For example, instead of $(10\bar{1})[111]$, which does
not appear in \reftab{tab_bcc24sys}, we use an equivalent notation
$(\bar{1}01)[\bar{1}\bar{1}\bar{1}]$, which corresponds to system 14.

\begin{table}[!htb]
  \centering
  \parbox{14cm}{\caption{List of potentially active slip systems under loading in compression along
      $[2\,\bar{9}\,\overline{20}]$, orientation of the MRSSP of each system, and the
      stress ratio $\eta$ corresponding to positive shear stress
      $\sigma$ parallel to the slip direction. The anomalous system is written in bold.}
  \label{tab_dof6_tstar}}
  \vskip5mm
  \begin{tabular}{ccccc}
    \hline
    ref. system & $\chi~[\deg]$ & $\tau/\sigma~(=\eta)$ & predicted slip system &
      $\alpha$ in \reftab{tab_dof6_lassila} \\
    \hline
    $(\bar{1}0\bar{1})[\bar{1}11]$ & 14.0 & -0.38 & 
      $(\bar{1}0\bar{1})[\bar{1}11]$, $(0\bar{1}1)[1\bar{1}\bar{1}]$ & --, 17 \\
    $(0\bar{1}\bar{1})[\bar{1}\bar{1}1]$ & 12.3 & -0.88 & 
      $(\bar{1}0\bar{1})[\bar{1}11]$, $(0\bar{1}1)[1\bar{1}\bar{1}]$ & --, 17 \\
    $(10\bar{1})[1\bar{1}1]$ & 25.2 & -2.45 & 
      $(0\bar{1}\bar{1})[\bar{1}\bar{1}1]$ & 7 \\
    $(\bar{1}01)[\bar{1}\bar{1}\bar{1}]$ & 0.0 & -0.50 & 
      $(\bar{1}01)[\bar{1}\bar{1}\bar{1}]$, $\bf (0\bar{1}1)[\bar{1}\bar{1}\bar{1}]$ & 14, 1 \\
    \hline
  \end{tabular} \\[5mm]
\end{table}

For comparison, we list in \reftab{tab_dof6_tstar} the four reference systems that can become
operative because the orientation of their MRSSPs lies between $\pm30\deg$ from their reference
planes, and the shear stress parallel to the slip direction is positive. The orientation of the
MRSSP in each system is given by the angle $\chi$, and the ratio of the shear stress perpendicular to
the shear stress parallel to the slip direction, resolved in the MRSSP, by the parameter $\eta$. Note,
that all values of $\eta$ are negative, and, therefore, all loading paths in the $\CRSS-\tau$ projection
of the corresponding MRSSP pass to negative $\tau$. The last two columns of the table give the slip
systems predicted by the $\tau^*$ criterion and, if observed experimentally, the designation of this
system referring to \reftab{tab_dof6_lassila}.

\begin{figure}[!htb]
  \centering
  \includegraphics[width=12cm]{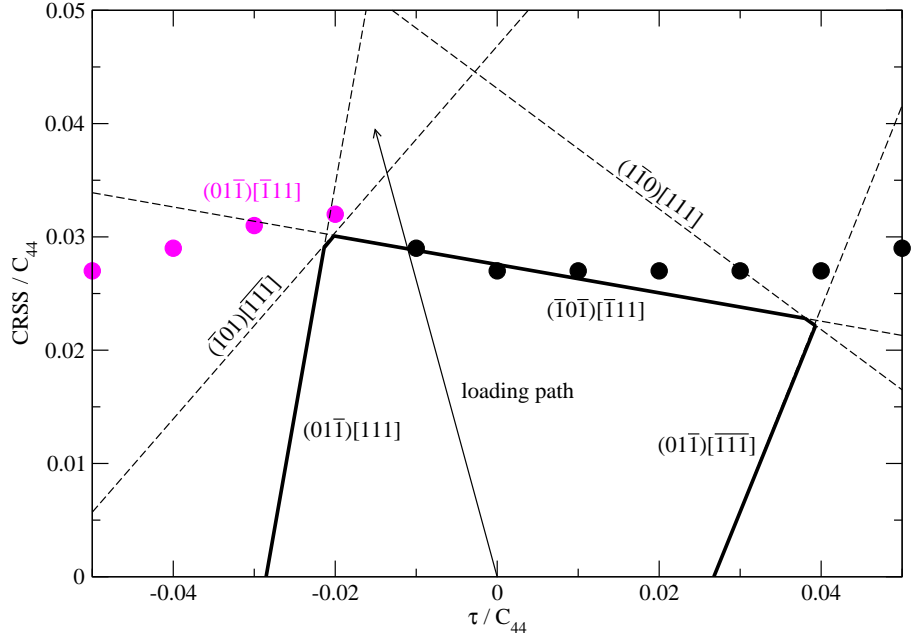}
  \parbox{14cm}{\caption{Projection of the critical lines (dashed lines) and the yield surface
  (solid polygon) in the $\CRSS-\tau$ graph corresponding to $\chi\approx+19\deg$, which is close to
  the MRSSP of the reference system $(\bar{1}0\bar{1})[\bar{1}11]$ (see the first line
  in \reftab{tab_dof6_tstar}). The slope of the loading path is $\eta=-0.38$.}
    \label{fig_chi+19_Lassila_MoBOP}}
\end{figure}

For example, the MRSSP of the $(\bar{1}0\bar{1})[\bar{1}11]$ system lies at $\chi=14\deg$ and the
stress ratio $\eta=-0.38$. This angle $\chi$ is close to $\chi\approx19\deg$ for which we have
obtained earlier the $\CRSS-\tau$ dependence by means of atomistic simulations. These results were
further generalized to real single crystals to arrive at \reffig{fig_CRSS_tau_fit_MoBOP}c. However,
since the reference system is now $(\bar{1}0\bar{1})[\bar{1}11]$ and not $(\bar{1}01)[111]$ as
considered earlier, we need to transform the indices of the slip systems in
\reffig{fig_CRSS_tau_fit_MoBOP}c into this new basis. These transformations are explained in detail
in Appendix~\ref{sec_eulerang} and provide the indices of the slip planes for the case when the
$(\bar{1}0\bar{1})[\bar{1}11]$ is the reference system; see \reffig{fig_chi+19_Lassila_MoBOP}. Note,
that the loading path plotted in this figure, corresponding to $\eta=-0.38$, first reaches the
critical line for the $(\bar{1}0\bar{1})[\bar{1}11]$ system, which is thus predicted by the $\tau^*$
criterion to be the primary slip system. However, the loading path reaches the \emph{atomistic data}
between the points corresponding to slip on $(\bar{1}0\bar{1})$ and $(01\bar{1})$%
\footnote{Unfortunately, the slip on this plane cannot be directly predicted by the $\tau^*$
  criterion, since its linear character does not accurately resolve the subtle nonlinearity in the
  $\CRSS-\tau$ dependence at negative $\tau$. This was explained in
  Section~\ref{sec_expt_jeffcoat} and shown for a different orientation in compression in
  \reffig{fig_chi0_0-11slip_MoBOP}. Nevertheless, these detailed predictions can always be done
  using directly the atomistic data.} 
planes. Consequently, two of the most operative slip systems predicted by the $\tau^*$ criterion
(and atomistic results) are $(\bar{1}0\bar{1})[\bar{1}11]$ and $(01\bar{1})[\bar{1}11]$. The latter
system can be equivalently written as $(0\bar{1}1)[1\bar{1}\bar{1}]$, which is identical to the
experimentally observed slip system 17 with the Schmid factor 0.29 (see \reftab{tab_dof6_lassila}).

\begin{figure}[!htb]
  \centering
  \includegraphics[width=12cm]{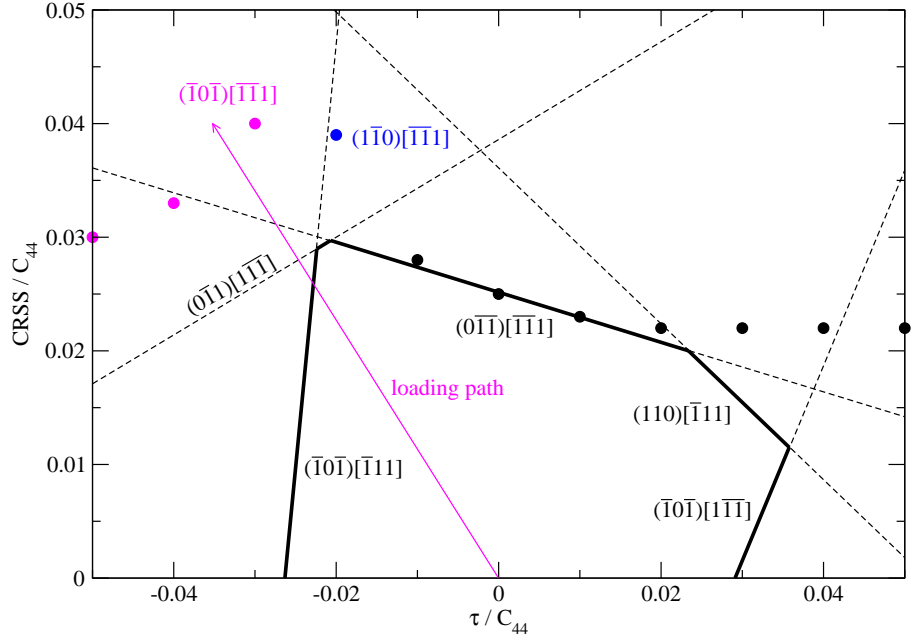}
  \parbox{14cm}{\caption{Projection of the critical lines (dashed lines) and the yield surface
  (solid polygon) in the $\CRSS-\tau$ graph corresponding to $\chi\approx+9\deg$, which is close to
  the MRSSP of the reference system $(0\bar{1}\bar{1})[\bar{1}\bar{1}1]$ (see the second line
  in \reftab{tab_dof6_tstar}). The slope of the loading path is $\eta=-0.88$.}
    \label{fig_chi+9_Lassila_MoBOP}}
\end{figure}

The next potentially operative slip system listed in \reftab{tab_dof6_tstar} is
$(0\bar{1}\bar{1})[\bar{1}\bar{1}1]$ with MRSSP at $\chi=12.3\deg$ and $\eta=-0.88$. In
\reffig{fig_chi+9_Lassila_MoBOP}, we plot the $\CRSS-\tau$ data for $\chi\approx9\deg$ calculated
from the atomistic simulations together with the critical lines obtained from the $\tau^*$ criterion
for molybdenum.  The loading path plotted in this figure, corresponding to $\eta=-0.88$, first
reaches the $(\bar{1}0\bar{1})[\bar{1}11]$ system, which thus becomes the predicted primary slip
system. Furthermore, since the CRSS for slip on the secondary system, $(0\bar{1}1)[1\bar{1}\bar{1}]$,
is virtually identical, one may expect that the two systems will operate simultaneously. This
secondary system coincides with that obtained by cross-slip from the $(\bar{1}0\bar{1})[\bar{1}11]$
system above and is indeed one of the operative systems in the experiments of \citet{lassila:03}.

The same analysis for the remaining two slip systems in \reftab{tab_dof6_tstar} and their comparison
with experimental observations of \citet{lassila:03} is summarized in the last two columns of this
table. For the last system listed, $(\bar{1}01)[\bar{1}\bar{1}\bar{1}]$, the loading path given by
$\eta=-0.5$ is close to that for compression along $[\bar{2}38]$ (\reffig{fig_CRSS_tau_MoBOP}a), and,
therefore, the slip on the same slip plane, at $\psi=+60\deg$, should be anticipated. The
corresponding index of the slip plane is found from \reffig{fig_8slipdir} to be
$(0\bar{1}1)$. Because the Schmid factor for the system $(0\bar{1}1)[\bar{1}\bar{1}\bar{1}]$ is only
the fourth highest one (see \reftab{tab_dof6_lassila}), this gives rise to the anomalous slip whose
activity is also observed in the experiments of \citet{lassila:03}.

From the analysis given above, one can observe that the $\tau^*$ criterion predicts the activation
of four out of six slip systems whose operation is observed experimentally at the room
temperature. Although the six degrees of freedom experiments were originally designed to validate the
results of large-scale dislocation dynamics experiments \citep{lassila:02}, the agreement with the
$\tau^*$ criterion is a strong indication that macroscopic plastic flow of bcc molybdenum can be
accurately obtained using simple linear yield criteria based on single-dislocation studies.

%----------------------------------------------------------------------------------------------------
%----------------------------------------------------------------------------------------------------

\section{Origin of the $(0\bar{1}1)$ slip in tension}

Under tensile loading, the shear stress perpendicular to the slip direction is always positive
(i.e. $\eta>0$), and the atomistic data predict that a $1/2[111]$ screw dislocation will move on
the $(\bar{1}01)$ plane. This is in agreement with many experimental studies, some of which have
been cited earlier in this chapter \citep{lau:70, aono:83}. In principle, the operation of the
$(0\bar{1}1)[111]$ system is possible even in tension and arises naturally from the proposed
$\tau^*$ criterion (\ref{eq_tstar_tensor}).

\begin{figure}[!htb]
  \centering
  \includegraphics[width=12cm]{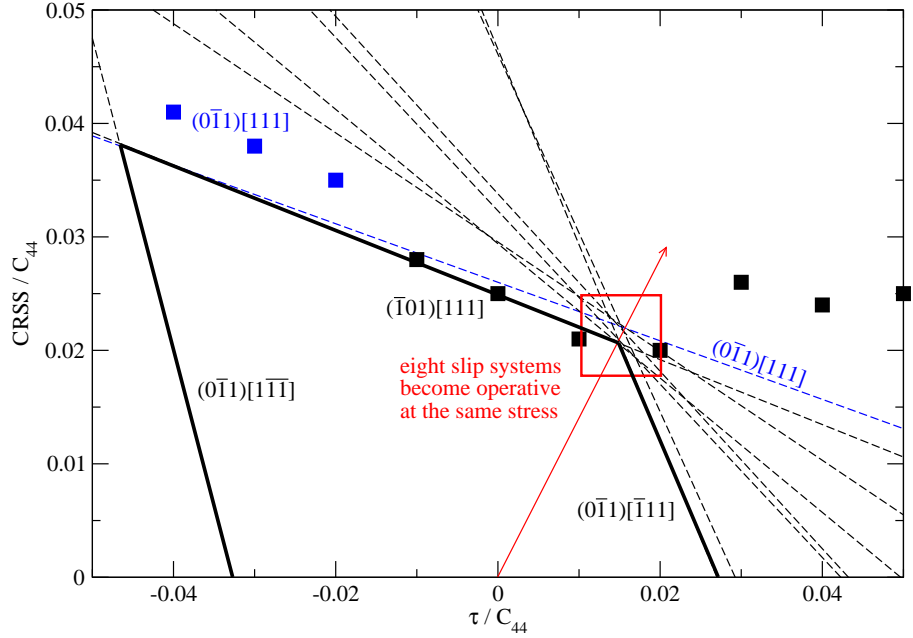}
  \parbox{14cm}{\caption{Onset of slip on the $(0\bar{1}1)[111]$ system under tension. For the
  loading path depicted, many slip systems become operative simultaneously. The calculated Schmid
  factors and the values of $\tau^*$ are listed in \reftab{tab_tstar_t0_1_14}.}
    \label{fig_chi-26_fit_MoBOP_multislip}}
\end{figure}

Let us look at \reffig{fig_chi-26_fit_MoBOP_multislip} showing a set of critical lines calculated
from the $\tau^*$ criterion and projected in the $\CRSS-\tau$ graph of the MRSSP $(\bar{5}\bar{4}9)$
corresponding to $\chi=-26\deg$. Consider uniaxial loading in tension for which the loading path
emanates from the origin and passes to the region of positive $\tau$. When this loading path reaches
the inner envelope of the critical lines at its corner, many slip systems become spontaneously
operative. One of these systems is $(0\bar{1}1)[111]$, whose critical line should ideally
interpolate the atomistic data corresponding to slip on the $(0\bar{1}1)$ plane at negative
$\tau$. It is important to emphasize that this must not be confused with the anomalous slip, because
$(0\bar{1}1)[111]$ is now one of the dominant slip systems with a large Schmid factor. Consequently,
the observation of the $(0\bar{1}1)$ slip in tension at $\chi\rightarrow-30\deg$ and in compression
corresponds to two different modes of slip.

To support the above conclusions, we show in \reftab{tab_tstar_t0_1_14} the complete list of all
slip systems that can be activated under loading in tension along $[0\,1\,14]$, corresponding to
$\chi=-26\deg$. The loading path for this orientation agrees closely with that plotted in
\reffig{fig_chi-26_fit_MoBOP_multislip}. The observation that the eight systems will be actived
simultaneously is equivalent to saying that all these systems have similarly high values of
$\tau^*$. This is confirmed by looking at the actual values of both Schmid stresses and $\tau^*$
given in \reftab{tab_tstar_t0_1_14}. Since $\tau^*$ for all eight systems is within 91\% of that
for the most-highly stressed $(\bar{1}01)[111]$ system, they will in reality act concomitantly,
which will result in a burst of slip activity at the stress for which the loading reaches the corner
of the inner polygon in \reffig{fig_chi-26_fit_MoBOP_multislip}.

\begin{table}[!htb]
  \centering
  \parbox{14cm}{\caption{The list of all slip systems that can become operative under loading in
      tension along $[0\,1\,14]$ for which the corresponding MRSSP is at $\chi\approx-26\deg$ and
      $\eta=0.64$. The Schmid stresses and the values of $\tau^*$ are
      normalized by the values obtained for the primary slip system, $(\bar{1}01)[111]$.}
    \label{tab_tstar_t0_1_14}} \\[1em]
  \begin{tabular}{ccc}
    \hline
    slip system & Schmid stress & $\tau^*$ \\
    \hline
    $(\bar{1}01)[111]$ & 1.00 & 1.00 \\
    $(101)[\bar{1}11]$ & 1.00 & 1.00 \\
    $(011)[1\bar{1}1]$ & 0.93 & 0.96 \\
    $(0\bar{1}\bar{1})[11\bar{1}]$ & 0.93 & 0.96 \\
    $(0\bar{1}1)[\bar{1}11]$ & 0.93 & 0.94 \\
    \color{blue} $(0\bar{1}1)[111]$ & \color{blue} 0.93 & \color{blue} 0.94 \\
    $(101)[\bar{1}\bar{1}1]$ & 0.87 & 0.91 \\
    $(10\bar{1})[\bar{1}1\bar{1}]$ & 0.87 & 0.91 \\
    \hline
  \end{tabular}
\end{table}

The fact that the $(0\bar{1}1)[111]$ system increases its prominence as $\chi\rightarrow-30\deg$
implies that, for these orientations of tensile loading, $1/2[111]$ dislocations can equally well
glide on both $(\bar{1}01)$ and $(0\bar{1}1)$ planes. This gives rise to the zig-zag slip on these
two planes that manifests itself as a macroscopic slip on one of the intermediate planes. In
particular, macroscopic slip on the $(\bar{1}\bar{1}2)$ plane emerges if the elementary steps on
these two planes are of equal size. This $\gplane{112}$ slip was indeed observed in tensile
experiments of \citet{lau:70} for orientation $[\bar{1}17]$, corresponding to $\chi=-16\deg$, denoted
thereafter as A. For all other orientations at $\chi>-16\deg$, the glide of the dislocation proceeded
entirely on the most highly stressed $(\bar{1}01)$ plane. Interestingly, \citet{aono:89} report the
``anomalous'' slip in \emph{tension} for orientations $\chi<+16\deg$ and temperature 77~K. However,
slip on the $(0\bar{1}1)$ plane was dominant only to about 1\% of plastic strain and became
completely suppressed at higher strains, where the $(\bar{1}01)[111]$ system governed the plastic
flow. According to \citet{aono:83}, the $(0\bar{1}1)$ slip in tension completely vanishes at 4.2~K,
and the dislocation moves on the $(\bar{1}01)$ plane for any orientation of the MRSSP.

The main result of the analysis presented above is that the glide of a $1/2[111]$ dislocation on
the $(0\bar{1}1)$ plane is indeed possible and is correctly predicted by the $\tau^*$ criterion for
large negative angles $\chi$. However, this is not to be confused with the anomalous slip because
the Schmid stress for the $(0\bar{1}1)[111]$ system is comparable to that of the primary slip
system, $(\bar{1}01)[111]$; see \reftab{tab_tstar_t0_1_14}.

  \chapter{Thermally activated plastic flow in molybdenum}
\label{chap_thermalact}

\begin{flushright}
  Once you eliminate the impossible, whatever remains, \\
  no matter how improbable, must be the truth.\\
  \emph{Sherlock Holmes}
\end{flushright}

We will now proceed to formulate a mesoscopic phenomenological theory of thermally activated motion
of screw dislocations that is based on the results of atomistic calculations. The link between the
microscopic behavior of screw dislocations and the mesoscale will be conveniently traversed by the
effective yield criterion that we constructed in Chapter~\ref{chap_tstarcrit} and that was shown to
reproduce correctly experimental measurements.

%----------------------------------------------------------------------------------------------------
%----------------------------------------------------------------------------------------------------

\section{Thermodynamics of dislocation glide}
\label{sec_thermoglide}

In this development we consider that the motion of screw dislocations in bcc crystals is driven by
the applied stress and opposed by the lattice friction. At 0~K a screw dislocation can be regarded
as a straight line that assumes a position at which the Peach-Koehler force exerted on the
dislocation by the applied stress and the restoring force due to the lattice friction balance each
other. In order to move the dislocation, the applied stress must be large enough to transform the
dislocation into a glissile configuration. At finite temperatures a saddle-point configuration is
attained at a stress lower than that needed for the dislocation glide at 0~K, because a part of the
energy needed to reach this configuration is supplied by thermal fluctuations. This is the reason
why the stress needed to sustain a certain rate of dislocation motion decreases with increasing
temperature. Such process leads to thermally activated dislocation glide.

The Gibbs free energy of activation for the dislocation glide depends crucially on the choice of the
thermodynamic variables that remain constant during the activation. Following \citet{schoeck:65} and
\citet{hirth:69}, we will consider that temperature $T$, pressure $p$, applied stress $\sigma$, and
surface energy $\gamma$ remain constant. The elementary change of the Gibbs free energy during the
activation can then be written as
\begin{equation}
  \d{G} = \d{U} - T\d{S} + p\d{V} - \sigma\d{\eps} - \gamma \d{A} \ ,
  \label{eq_dG_act}
\end{equation}
where $U, S, V, \sigma, \eps, \gamma, A$ are the internal energy, entropy, volume, applied stress,
strain, surface energy, and surface area of the system, respectively. The change of volume and
surface area during the activation is usually small and can thus be safely neglected. Defining the
change of enthalpy as $\d{H}=\d{U}-\sigma\d{\eps}$, one directly arrives at an equality $\d{G} =
\d{H} - T\d{S}$.  The total change of the Gibbs free energy during the activation can then be
obtained directly by integrating the last equation between the configuration of the dislocation and
the saddle-point, which yields the well-known formula
\begin{equation}
  \Delta{G} = \Delta{H} - T\Delta{S} \ .
\end{equation}
According to the transition state theory \citep{caillard:03, kocks:75}, the rate of arriving at
the saddle-point follows an Arrhenius law
\begin{equation}
  \nu = \nu_0 \exp\left( -\frac{\Delta{G}}{kT} \right) \equiv
  \nu_0^* \exp\left( -\frac{\Delta{H}}{kT} \right)
  \ ,
\end{equation}
where $\nu_0$ is the frequency of unconstrained vibrations, and $\nu_0^*=\nu_0\exp(\Delta S/k)$ is an
effective vibrational frequency that accounts for the change in entropy during the
activation. Within the framework of the rate theory of \citet{vineyard:57}, the effective frequency
$\nu_0^*$, involving the entropy $\Delta{S}$, is the ratio of the product of $N$ normal frequencies,
$\nu_j$, of the system at the starting point of the transition to the product of $N-1$ normal
frequencies of the system at the saddle-point configuration, $\nu_j'$, so that
\begin{equation}
  \nu_0^* = \prod_{j=1}^N \nu_j \left/\ \prod_{j=1}^{N-1} \nu_j' \right. \ .
  \label{eq_nu0star}
\end{equation}
\citet{celli:63} suggested the separation of these vibrational modes into two mutually
noninteracting classes: dislocation modes and crystal modes. The crystal modes in the starting and
the saddle-point configurations are assumed to be identical and to cancel each other in
\refeq{eq_nu0star}. Then, in order to apply Vineyard's formula, it is sufficient to calculate
the ratio of the product of the \emph{dislocation modes} in the starting configuration to that in
the saddle-point activated state. This product was explicitly evaluated for square-shaped activation
barriers by \citet{celli:63}. In general, the usual approach is to consider in the starting
configuration only the frequency of vibration in the direction of motion, i.e. the so-called attack
frequency. This frequency can be approximated as $\nu_D(l/b)$, where $\nu_D$ is the Debye frequency,
$l$ the dislocation length involved in the activation process, and $b$ the magnitude of the Burgers
vector. Since this frequency is missing in the activated state, $\nu_0^*$ is written as
$\nu_D(l/b)$. The velocity of the dislocation can then be written as
$v_D=a_0\nu_0^*\exp(-\Delta{H}(\sigma)/kT)$, where $a_0$ is the distance traversed during the
activation event. Assuming that only one slip system operates, the strain rate can be written as
$\dot\gamma=\rho_m b v_D$ or, equivalently, as
\begin{equation}
  \dot\gamma = \dot\gamma_0 \exp\left( -\frac{\Delta H(\sigma)}{kT} \right) \ ,
  \label{eq_gammadot}
\end{equation}
with the pre-exponential factor $\dot\gamma_0=\rho_m a_0 b \nu_0^*$ \citep{kocks:75,caillard:03}. Here,
$\rho_m$ is the density of mobile dislocations actively participating in the gliding process, and $b$
the magnitude of their Burgers vector. The stress dependence of $\dot\gamma_0$ is usually neglected,
since it is much weaker than the exponential dependence on stress via $\Delta{H}(\sigma)$, and, for
similar reasons, one usually makes the assumption that the mobile dislocation density, $\rho_m$, is a
constant. The determination of the stress dependence of the activation enthalpy,
$\Delta{H}(\sigma)$, based on the results of atomistic studies, is the main topic of this
chapter. For the sake of brevity, we will further drop the symbol $\Delta$ in the designation of the
activation enthalpy.

\begin{figure}[!htb]
  \centering
  \includegraphics[width=10cm]{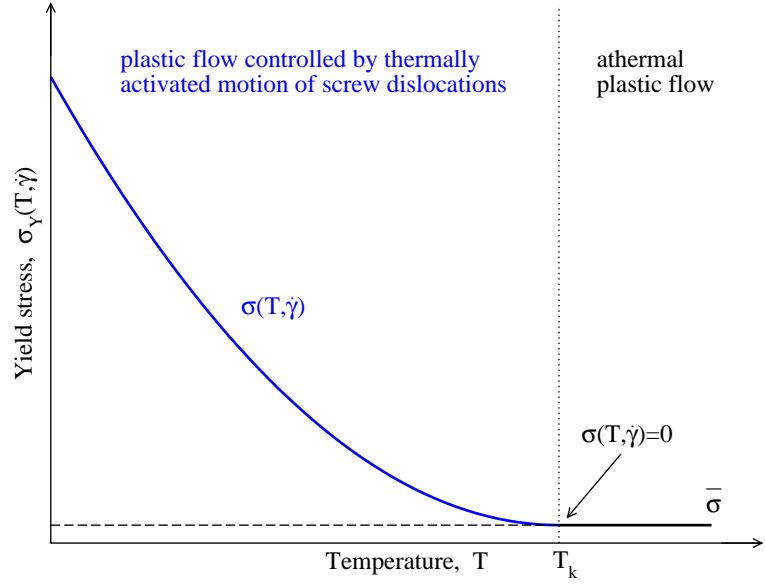}
  \parbox{13cm}{\caption{Schematic illustration of the temperature dependence of the yield stress in
      bcc metals.}
    \label{fig_ystress_T}}
\end{figure}

At high temperatures, the yield stress of bcc metals is only weakly dependent on temperature via the
temperature dependence of elastic moduli and can be approximated by a constant athermal stress
$\bar\sigma$. On the other hand, at low temperatures, the plastic flow is controlled by the
thermally activated motion of screw dislocations, and the yield stress is a strong function of both
temperature and plastic strain rate. One can, therefore, write the actual yield stress as a sum of two
terms,
\begin{equation}
  \sigma_Y(T,\dot\gamma) = \sigma(T,\dot\gamma) + \bar\sigma \ ,
\end{equation}
where $\sigma(T,\dot\gamma)$ is the thermal component of the yield stress that depends on both
temperature $T$ and plastic strain rate $\dot\gamma$. $\sigma(T,\dot\gamma)$ vanishes at the temperature
$T_k$, where the yield stress approaches its athermal value, $\bar\sigma$. Below $T_k$, the thermal
component $\sigma(T,\dot\gamma)$ gradually increases and reaches its maximum at $0~\K$ (see
\reffig{fig_ystress_T}).

In the following, we will be concerned with the thermally activated motion of screw dislocations at
temperatures below $T_k$. This will provide us with the variation of the \emph{thermal component} of
the yield stress, $\sigma(T,\dot\gamma)$, which can be compared with the thermal component of the
yield stress obtained from experiments by subtracting from the measured yield stress the constant
athermal stress $\bar\sigma$.

%----------------------------------------------------------------------------------------------------
%----------------------------------------------------------------------------------------------------

\section{Activation enthalpy of formation of pairs of kinks}

It is generally accepted that at finite temperatures screw dislocations in bcc metals overcome the
Peierls barriers by nucleating pairs of kinks that subsequently migrate, which leads to the glide of
the dislocation in the direction of the kink formation. As discussed in the preceding chapters, the
dislocation core changes under the influence of the applied stress, and this is considered
phenomenologically as shifting the dislocation as a straight line toward the top of the barrier.  In
this work, the shape of the Peierls barrier is considered to be a function of non-glide stresses,
while the work on shifting the dislocation toward the top of this barrier is done purely by the
applied Schmid stress.

\begin{figure}[!htb]
  \centering
  \includegraphics[width=10cm]{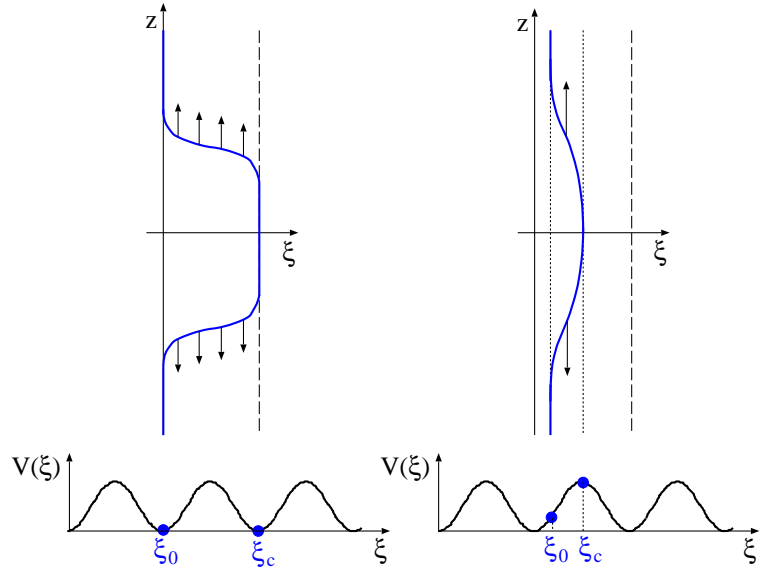}
  \parbox{13cm}{\caption{Schematic illustration of the nucleation of a pair of kinks at low stresses
      and high temperatures (left panel) and of the critical bow-out at high stresses and low
      temperatures (right panel). $\xi$ designates the activation path and $V(\xi)$ the Peierls
      barrier along this path.}
    \label{fig_kinkpair}}
\end{figure}

At low temperatures and high stresses the dislocation core is almost transformed to the glissile
form and, phenomenologically, the dislocation has been moved by stress almost to the top of the
barrier.  Hence, a relatively small thermal energy is needed to overcome the barrier. In this
regime, the activated state consists of a continuous ``bow-out'' developed on otherwise straight
screw dislocation \citep{dorn:64, duesbery:89}, as shown in the right panel of
\reffig{fig_kinkpair}. On the other hand, at high temperatures (left panel in
\reffig{fig_kinkpair}), thermal fluctuations are large and only a small stress is needed to move the
dislocation through the crystal. In this case, the dislocation core is only slightly distorted,
which in the phenomenological analysis means that the dislocation moves as a straight line only a
short distance. Thermal fluctuations then help to nucleate a pair of kinks that attract each other
elastically via their long-range strain fields \citep{seeger:56}, and the saddle-point configuration
corresponds to two well-separated kinks at the distance for which the attraction between the kinks
is compensated by the force exerted on them by the applied Schmid stress.

Dislocations in crystalline materials experience lattice friction that varies periodically with the
position of the dislocation in the crystal. At finite temperatures, the Peierls barrier is overcome
partly due to the work done with the aid of thermal fluctuations and the work done by the applied
Schmid stress during the activation process. In contrast, the \emph{activation barrier} is the
fraction of the Peierls barrier that is left after subtracting the work done by the Schmid stress
when shifting the dislocation up on the barrier. The construction of the Peierls barrier will be
explained in detail in Section \ref{sec_EffPeierlsPot}.

%----------------------------------------------------------------------------------------------------

\subsection{Energy of an isolated kink}

The energy of a single isolated kink was first derived by \citet{seeger:56} in an attempt to explain
the low-temperature internal friction peak in polycrystals. In his work, the saddle-point energy
corresponds to a pair of two fully developed kinks, interacting with each other, at the separation
for which the mutual attraction of the kinks is compensated by the applied stress. These
calculations were later revisited by \citet{dorn:64} who successfully demonstrated that the
kink-pair formation model can be utilized to explain the low-temperature increase of the yield
stress in bcc metals. In contrast to the model of \citet{seeger:56}, the saddle-point configuration
is represented by a continuous bow-out on an otherwise straight screw dislocation, which is attained
before a pair of two fully developed interacting kinks is formed.

In the original theory \citep{seeger:56,dorn:64}, the Peierls potential was written as a periodic
function of the position of the dislocation in a well-defined slip plane. The only component of the
stress tensor considered was the shear stress in the slip plane acting parallel to the Burgers
vector. Although widely observed in experiments on molybdenum and other bcc metals, no
twinning-antitwinning asymmetry was predicted from these early models. More recently,
\citet{edagawa:97} suggested that the Peierls potential should be a two-dimensional function of the
position of the intersection of the dislocation with the $\gplane{111}$ plane that is perpendicular
to the dislocation line. The Peierls barrier was identified with the profile of the two-dimensional
Peierls potential along a curved transition path between two adjacent low-energy sites in the
$\gplane{111}$ plane, and was considered to be a function of the shear stress parallel to the slip
direction. This model captured not only the twinning-antitwinning asymmetry in molybdenum but also
the stress dependence of the activation enthalpy and thus the trend of the temperature dependence of
the yield stress.

If the functional form of the Peierls potential is given as a function $V(x,y)$ defined in the
$\gplane{111}$ plane, the energy of a single kink can be determined as
\begin{equation}
  H_k = \int_{-\infty}^{\infty} \left[ V(x,y)\, \d{s} - V_0\, \d{z} \right] \ ,
  \label{eq_Hk_start}
\end{equation}
where $\d{s}$ is an elementary segment along the kinked dislocation, $V_0$ the magnitude of the
potential at the bottom of the Peierls valley, and $\d{z}$ an element along the $z$ axis that is
parallel to the straight dislocation. In this formulation, $V$ can be regarded as the energy per
unit length of the dislocation found at a particular position in the Peierls potential valley. The
kink energy $H_k$, given by equation (\ref{eq_Hk_start}), is defined as the difference between the
line energy of the kinked dislocation and the line energy of a straight dislocation at the bottom of
the Peierls valley. Because the dislocation is now a three-dimensional curve,
$\d{s}=\sqrt{(\d{x})^2+(\d{y})^2+(\d{z})^2}$ and (\ref{eq_Hk_start}) then reads
\begin{equation}
  H_k = \int_{-\infty}^{\infty} \left[ V(x,y) \sqrt{ 1 + \left(\frac{\d{x}}{\d{z}}\right)^2 +
      \left(\frac{\d{y}}{\d{z}}\right)^2 } - V_0  \right] \d{z} \ .
  \label{eq_Hk_tang}
\end{equation}

From the 0~K atomistic simulations, we know that screw dislocations in molybdenum move on
$\gplane{110}$ planes by discrete jumps between two adjacent minima of the Peierls potential
$V(x,y)$. The transition path between these minima is always a continuous curve and corresponds to
the minimum energy path (MEP). When forming the kink, the dislocation bows-out along the MEP
specified by a curve $\xi(z)$. The elements $\d{x}$ and $\d{y}$, defined in the $\gplane{111}$ plane
perpendicular to the dislocation line, can be written as $\d{x}=\d{\xi}\cos\theta$ and
$\d{y}=\d{\xi}\sin\theta$, where $\theta$ is at any point along the MEP the angle between the
tangent of the MEP and the $x$ axis. Substituting these expressions into (\ref{eq_Hk_tang}) and
using the identity $\sin^2\theta+\cos^2\theta=1$ reduces the expression for the kink energy to:
\begin{equation}
  H_k = \int_{-\infty}^{\infty} \left[ V(\xi) \sqrt{ 1 + \left(\xi'(z)\right)^2 } - 
    V_0  \right] \d{z} \ ,
  \label{eq_Hk_VR}
\end{equation}
where we replaced $V(x,y)$ by $V(\xi)$, profile of the Peierls potential along the MEP, and
wrote $\xi'(z)=\d{\xi}/\d{z}$. Following \citet{dorn:64}, the stationary solution of
\refeq{eq_Hk_VR}, corresponding to the maximum of $H_k$, is given by the Euler-Lagrange equation
\begin{equation}
  \left( \frac{\partial}{\partial{\xi}} - \frac{\d}{\d{z}} \frac{\partial}{\partial \xi'(z)} \right)
  \left[ V(\xi)\sqrt{1+\left(\xi'(z)\right)^2} \right] = 0 \ ,
\end{equation}
which reads, after differentiation,
\begin{equation}
  \frac{\d}{\d{\xi}} \left( \frac{V(\xi)}{\sqrt{1+\left(\xi'(z)\right)^2}} \right) = 0 \ .
\end{equation}
This equation can be integrated once to get the slope of the dislocation line with a kink,
\begin{equation}
  \xi'(z) = \pm \sqrt{ \left( \frac{V(\xi)}{C} \right)^2 - 1 } \ ,
  \label{eq_dR_dz_kink}
\end{equation}
where $C$ is an integration constant. Assume that one end of the dislocation lies in the Peierls
valley determined by $\xi=0$ and the other in the neighboring valley at $\xi_{max}$ (measured along
the curved MEP). The slope of the dislocation at $\xi=0$ and $\xi_{max}$ then vanishes and this
suggests two boundary conditions: (i) $\xi=0 \Rightarrow z\rightarrow-\infty$, $\xi'(-\infty)=0$ and
$V(0)=V_0$, and (ii) $\xi=\xi_{max} \Rightarrow z\rightarrow+\infty$, $\xi'(+\infty)=0$ and
$V(\xi_{max})=V_0$. In the case of an opposite kink, one has to interchange the plus and minus
signs. Because the slope $\xi'(z)$ of the dislocation is zero at $\xi=0$ and $\xi_{max}$, where the
corresponding height of the Peierls potential is $V_0$, \refeq{eq_dR_dz_kink} gives the magnitude of
the integration constant, $C=V_0$. Finally, substituting \refeq{eq_dR_dz_kink} together with the
value of $C$ into \refeq{eq_Hk_VR} gives the energy of an isolated kink,
\begin{equation}
  H_k = V_0 \int_0^{\xi_{max}} \sqrt{ \left(\frac{V(\xi)}{V_0}\right)^2 - 1 }\ \d{\xi} \ ,
  \label{eq_Hk_0stress}
\end{equation}
where we changed the integration variable by substituting $\d{z}$ as a function of $\d{\xi}$ from
(\ref{eq_dR_dz_kink}). In \refeq{eq_Hk_0stress}, the kink energy is determined as a path integral
between two potential minima calculated along the curved MEP.

%----------------------------------------------------------------------------------------------------

\subsection{High stresses: dislocation bow-out}
\label{sec_bowout}

\begin{figure}[!htb]
  \centering
  \includegraphics{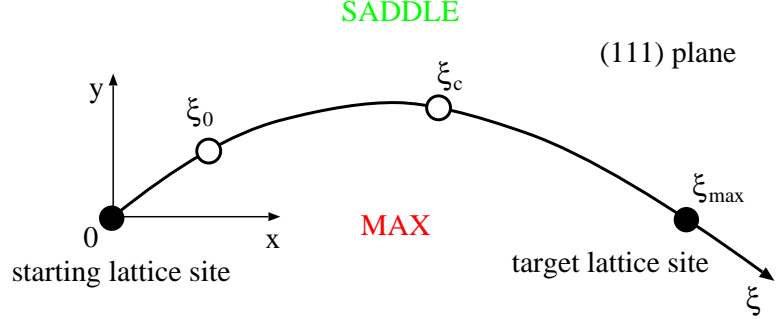}
  \parbox{13cm}{\caption{Schematic depiction of a curved transition coordinate $\xi$ for a
      dislocation moving between two minimum energy sites (filled circles). SADDLE and MAX
      designate a saddle-point and a maximum of the Peierls potential, respectively. At zero applied
      stress the kink develops from $0$ to $\xi_{max}$, whereas at finite stresses the kink only
      extends from $\xi_0$ to $\xi_c$.}
    \label{fig_reaction_path}}
\end{figure}

At a nonzero stress, the dislocation is moved away from the bottom of the Peierls valley as a
straight line by the action of the applied Schmid stress $\sigma$. The position at which the force
$\d{V}/\d{\xi}$, originating from the Peierls potential, is equal to the Peach-Koehler force $\sigma
b$, induced by the applied stress, will be denoted as $\xi_0$. At high stresses, the dislocation
bows-out toward the top of the Peierls barrier and the activated state corresponds to the maximum of
the enthalpy. Starting with \refeq{eq_Hk_VR}, the enthalpy associated with the dislocation
bowing-out along the MEP is
\begin{equation}
  H_b = \int_{-\infty}^{\infty} \left[ V(\xi) \sqrt{ 1 + \left(\xi'(z)\right)^2 } - 
    V(\xi_0) - \sigma b (\xi-\xi_0) \right] \d{z} \ .
  \label{eq_Eb_VR_stress}
\end{equation}
Here, the first term in the integrand is the line energy of the curved dislocation bowed-out along
the MEP, and the last two terms are the line energy of a straight dislocation at $\xi_0$ and the work
done by the stress $\sigma$ when displacing the dislocation from $\xi_0$ to $\xi$. As in the case of
zero stress, we are looking for a stationary shape of the dislocation that is given by the
Euler-Lagrange equation
\begin{equation}
  \left( \frac{\partial}{\partial{\xi}} - \frac{\d}{\d{z}} \frac{\partial}{\partial \xi'(z)} \right)
  \left[ V(\xi)\sqrt{1+\left(\xi'(z)\right)^2} - V(\xi_0) - \sigma b (\xi-\xi_0) \right] = 0 \ .
\end{equation}
Performing the indicated differentiation, this immediately simplifies to
\begin{equation}
  \sigma b = \frac{\d}{\d{\xi}} \left( \frac{V(\xi)}{\sqrt{1+\left( \xi'(z) \right)^2}}
  \right) \ .
  \label{eq_sigmab_bulge}
\end{equation}
Integrating and solving for the slope $\xi'(z)$ gives
\begin{equation}
  \xi'(z) = \pm \sqrt{ \left( \frac{V(\xi)}{\sigma b \xi + C} \right)^2 - 1 } \ ,
  \label{eq_dR_dz_bulgeC}
\end{equation}
where $C$ is an integration constant that can be determined by demanding that the slope of the
dislocation at $z \rightarrow \pm \infty$ and $\xi \rightarrow \xi_0$ remains zero. These boundary
conditions yield $C=V(\xi_0)-\sigma b \xi_0$, which can be substituted back into
(\ref{eq_dR_dz_bulgeC}) to obtain the slope of the dislocation in the activated state:
\begin{equation}
  \xi'(z) = \pm \sqrt{ \left( \frac{V(\xi)}{\sigma b (\xi-\xi_0) + V(\xi_0)} \right)^2 - 1 } \ .
  \label{eq_dR_dz_bulge}
\end{equation}
Each sign in this equation defines the slope in one-half of the bowed dislocation that meet at
$z=0$. The stationary magnitude of the bow-out, denoted hereafter as $\xi_m$, is determined from the
condition $\xi'=0$ for $z=0$. It then follows from (\ref{eq_dR_dz_bulge}) that $\xi_m$ is
determined by the relation
\begin{equation}
  V(\xi_m) - V(\xi_0) = \sigma b (\xi_m - \xi_0) \ .
\end{equation}
One solution is $\xi_m=\xi_0$ (\reffig{fig_reaction_path}), and this corresponds to the straight
dislocation at $\xi_0$ when $H_b=0$, which is the minimum value of the enthalpy given by
\refeq{eq_Eb_VR_stress}. The second solution, $\xi_m=\xi_c$ (\reffig{fig_reaction_path}), defines
the segment of the dislocation in the critical configuration for which the activation enthalpy
reaches a maximum and the bow-out represents a saddle-point configuration. No stationary solution
for the bow-out exists for $\xi_0<\xi_m<\xi_c$, since in this case the two kinks of the bow-out
annihilate each other. Similarly, no stationary solution exists for $\xi_m>\xi_c$, because in this
case the two kinks are driven apart by the applied stress.

\begin{figure}[!htb]
  \centering
  \includegraphics[width=8cm]{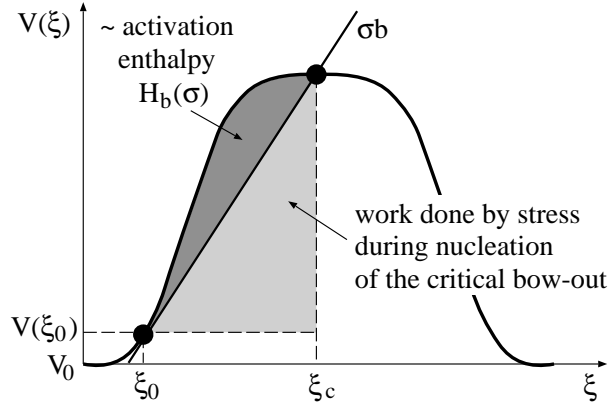}
  \parbox{13cm}{\caption{Graphical interpretation of \refeq{eq_Hb_stress} in which the activation
  enthalpy is calculated as a function of the applied stress $\sigma$.}
    \label{fig_Hb_area}}
\end{figure}

The enthalpy of nucleation of the critical bow-out on the dislocation line at high stresses is
obtained by substituting (\ref{eq_dR_dz_bulge}) into (\ref{eq_Eb_VR_stress}) that gives
\begin{equation}
  H_b = 2 \int_{\xi_0}^{\xi_c} \sqrt{ \left[ V(\xi) \right]^2 - \left[ \sigma b (\xi-\xi_0) + V(\xi_0) \right]^2 }
  \ \d{\xi} \ ,
  \label{eq_Hb_stress}
\end{equation}
where we again changed the integration variable by realizing that $\d{z}$ can be written as a
function of $\d{\xi}$ by inverting the slope (\ref{eq_dR_dz_bulge}). The factor of 2 ahead of the
integral arises because the integrand is a double-valued function of $\xi$ that is even about $z=0$
and, therefore, the integration gives only one-half of the activation enthalpy. The activation
enthalpy (\ref{eq_Hb_stress}) is illustrated in \reffig{fig_Hb_area} as an area bounded by the
Peierls barrier, $V(\xi)$, and the line with the slope $\sigma b$. It is important to realize that
$V(\xi)$ is the line energy of the dislocation at position $\xi$ along the MEP, which can never be
lower than $V_0$, the line energy of the dislocation lying at the bottom of the Peierls valley. This
requires that the zero of energy, i.e. the minimum of $V(x,y)$, coincides with $V_0$.  In the
following, we write $V_0=p \mu b^2$, where $\mu$ is the $\gdir{111}$ shear modulus, and $p$ a
constant whose magnitude will be determined in Section~\ref{sec_kinkpar}.

The activation enthalpy (\ref{eq_Hb_stress}) ceases to be applicable at low stresses (high
temperatures), where an alternative saddle-point configuration, consisting of two fully developed
kinks spanning from one minimum of the Peierls potential to the next, becomes energetically more
favorable. The saddle-point configuration then corresponds to the separation of these kinks at which
their attraction is compensated by the force exerted on them by the Schmid stress $\sigma$ that
pushes them apart. The derivation of the activation enthalpy in this low-stress regime will concern
us in the following section. In the limit of zero stress, one can easily see that $H_b \rightarrow
2H_k$, and the activation enthalpy coincides with the energy of two isolated (non-interacting)
kinks.

Based on the stress dependence of the activation enthalpy given by \refeq{eq_Hb_stress}, one can
calculate the so-called activation volume, $V_b=-\partial{H_b}/\partial{\sigma}$, which can be
directly measured in experiments \citep{conrad:63}. The explicit forms of the activation volumes for
a limited class of simple one-dimensional Peierls potentials can be found in \citet{caillard:03}.

%---------------------------------------------------------------------------------------------------

\subsection{Low stresses: elastic interaction of fully developed kinks}
\label{sec_low_stresses}

At low applied stresses, i.e. high temperatures, the shape of the dislocation in the activated
configuration is represented by a pair of kinks that interact elastically via the attractive Eshelby
potential $\mu a_0^2 b^2/8\pi\Delta{z}$ \citep{caillard:03}, where $a_0$ is the height of the kinks
or, equivalently, the distance between two neighboring Peierls valleys in the slip plane, and
$\Delta{z}$ their separation. This attractive interaction is opposed by the applied shear stress
$\sigma$ during the nucleation which does work equal to $\sigma a_0 b \Delta{z}$. The enthalpy of
nucleation of a pair of interacting kinks in the low-stress regime, $H_{kp}$, is then determined
by competition between the attractive elastic interaction and the work done on displacing the two
kinks a distance $\Delta{z}$ apart,
\begin{equation}
  H_{kp} = 2H_k - \frac{\mu a_0^2 b^2}{8\pi\Delta{z}} - \sigma a_0 b \Delta{z} \ ,
  \label{eq_Hkp_def}
\end{equation}
where $2H_k$ is the energy of two isolated kinks. At short separations $\Delta{z}$, the elastic
attraction dominates the enthalpy $H_{kp}$, and the two kinks tend to annihilate each other. This
attraction weakens with increasing separation between the kinks, and, at large separation
$\Delta{z}$, the activation enthalpy $H_{kp}$ is dominated by the work $\sigma a_0 b \Delta{z}$ done
while separating the two kinks. In this case, the pair of kinks is unstable and the two kinks
migrate apart, which moves the screw dislocation into the neighboring Peierls valley. Apparently,
there exists a critical separation $\Delta{z_c}$ corresponding to a saddle-point configuration for
which the enthalpy $H_{kp}$ attains a maximum. This critical separation is determined by the
condition that $\partial{H_{kp}}/\partial{\Delta{z}}$ vanishes at the saddle-point, which implies
\begin{equation}
  \Delta{z_c} = \sqrt{\frac{a_0b}{8\pi} \frac{\mu}{\sigma}} \ .
  \label{eq_Deltax}
\end{equation}
Substituting $\Delta{z_c}$ for $\Delta{z}$ in \refeq{eq_Hkp_def} gives the activation enthalpy to
nucleate the pair of kinks of critical separation $\Delta{z_c}$ at applied stress $\sigma$,
\begin{equation}
  H_{kp} = 2H_k - (a_0 b)^{3/2} \sqrt{\frac{\mu \sigma}{2\pi}} \ .
  \label{eq_Hkp_final}
\end{equation}
At zero stress, $H_{kp}$ reduces to the energy of two isolated kinks, $2H_k$.

The activation volume at low stresses can be calculated in a straightforward manner as
$V_{kp}=-\partial{H_{kp}}/\partial{\sigma}$, which gives
\begin{equation}
  V_{kp} = \frac{(a_0 b)^{3/2}}{2} \sqrt{\frac{\mu}{2\pi}} \sigma^{-1/2} \ .
  \label{eq_Vkp}
\end{equation}
It is important to realize that $V_{kp} \propto \sigma^{-1/2}$, and, therefore, the activation volume
diverges as the applied stress approaches zero. The origin of this divergence is the increase of the
separation between the two kinks as the applied stress decreases, see \refeq{eq_Deltax}.

%---------------------------------------------------------------------------------------------------

\subsection{Magnitudes of kink parameters}
\label{sec_kinkpar}

A necessary prerequisite to the calculation of both the kink energy (\ref{eq_Hk_0stress}) and the
activation enthalpy (\ref{eq_Hb_stress}) is knowledge of the line energy of a straight screw
dislocation, $V_0$, that is often written as $V_0=p\mu b^2$. Here, $p$ is a constant that can be
calculated theoretically by demanding that the enthalpy (\ref{eq_Hb_stress}) approaches in the limit
of zero stress the energy of two isolated kinks, $2H_k$. The magnitude of $2H_k$ can, in turn, be
determined by extrapolating the experimentally measured stress dependence of the activation enthalpy
to the limit of zero stress, e.g. by virtue of \refeq{eq_Hkp_final}. This procedure was used by
\citet{conrad:63}, who showed that the magnitude of $2H_k$ for any bcc metal can be closely
approximated as $2H_k=0.1\mu b^3$, which, in the case of molybdenum, yields $2H_k \approx
1.4\ \eV$. Although this value is in reasonable agreement with the results of LMTO calculations of
\citet{xu:98}, which predict $2H_k$ between 1 and 2 eV, the range of values considered in the
literature and summarized in \reftab{tab_2Hk_lit} calls for a deeper analysis.

\begin{table}[!htb]
  \centering
  \parbox{14cm}{\caption{Summary of the values of $2H_k$, considered by different authors, for
    nucleation of two individual kinks on $\gplane{110}$ planes.}
  \label{tab_2Hk_lit}} \\[1em]
  \begin{tabular}{lcl}
    \hline
    reference & $2H_k$~[eV] & comment \\
    \hline
    \citet{conrad:63}   & 1.40 & extrapolating $H(\sigma)$ to $\sigma=0$ \\
    \citet{dorn:64}     & 0.70 & -- \\
    \citet{suzuki:95}   & 1.08 & $V_0\approx \mu b^2$ considered \\
    \citet{edagawa:97}  & 0.55 & $V_0=0.5 \mu b^2$ considered \\
    \citet{hollang:97}  & 1.27 & fitting the experiment (see their Fig. 10)\\
    \hline
  \end{tabular}
\end{table}

\begin{figure}[!htb]
  \centering
  \includegraphics[width=12cm]{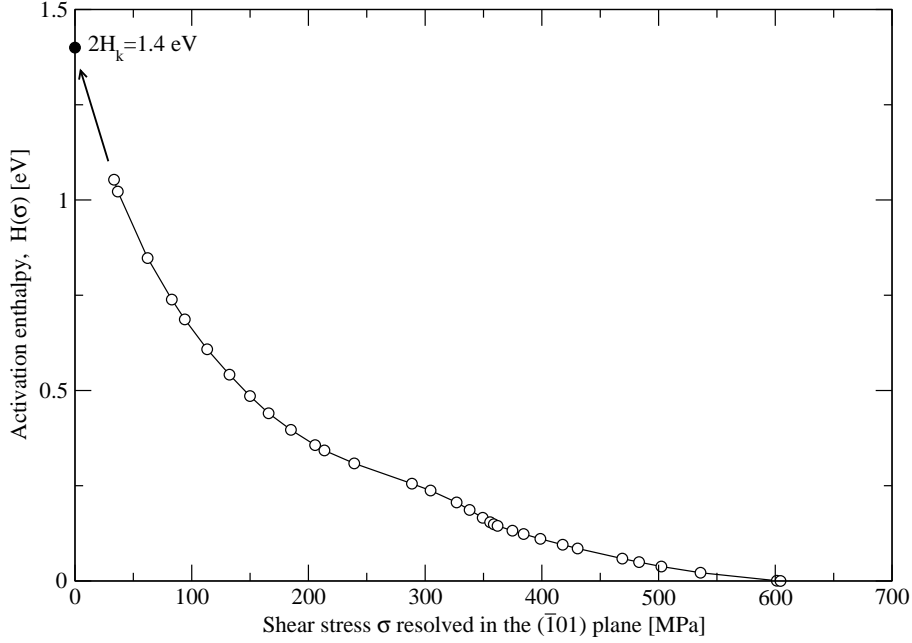}
  \parbox{14cm}{\caption{Stress dependence of the activation enthalpy calculated by integrating the
      activation volume as a function of stress \citep{aono:83} according to \refeq{eq_H_sigma}.}
  \label{fig_actene_from_actvol}}
\end{figure}

Alternatively, one can determine $2H_k$ by integrating the stress dependence of the activation volume
that can be measured in strain rate sensitivity experiments (for details, see \citet{caillard:03}
and references therein). This measurement is based on the relation
\begin{equation}
  V(\sigma) = kT  \left.\frac{\partial \ln \dot{\gamma}}{\partial \sigma} \right|_T \ ,
\end{equation}
where $\dot{\gamma}$ is the actual plastic strain rate. Assuming that the pre-exponential factor
$\dot{\gamma}_0$ in the Arrhenius law (\ref{eq_gammadot}) is independent of stress, one obtains the
classical prescription for the calculation of the activation volume, $V(\sigma)=-\d
H(\sigma)/\d\sigma$, that we have used earlier. Integrating both sides with respect to stress and
identifying the integration constant as $2H_k$ gives a formula for the calculation of the stress
dependence of the activation enthalpy:
\begin{equation}
  H(\sigma) = 2H_k - \int_0^\sigma V(\sigma^*) \d{\sigma^*} \ .
  \label{eq_H_sigma}
\end{equation}
Substituting in \refeq{eq_H_sigma} the experimentally measured stress dependence of the activation
volume, $V(\sigma)$, from the work of \citet{aono:83} on molybdenum and carrying out the
integration, one arrives at the stress dependence of the activation enthalpy shown in
\reffig{fig_actene_from_actvol}. As the applied stress $\sigma$ increases, the activation enthalpy
decreases and eventually vanishes. On the other hand, as the stress approaches zero, the activation
enthalpy approaches the energy of two isolated kinks, $2H_k$. Extrapolating this value from the
low-stress regime yields for molybdenum $2H_k=1.4~\eV$, which is identical to the value obtained by
\citet{conrad:63}.

More recently, \citet{hollang:97} devised a method that allows one to calculate $2H_k$ directly from
measurements of the dependence of $\ln \dot\gamma$ on $1/T_k$, an inverse temperature at which the
thermal component of the yield stress vanishes. All experimental measurements of the temperature
dependence of the yield stress were performed on \emph{one} sample using the cyclic deformation
technique of \citet{ackermann:83}, which removes the ambiguity of interpretation of data obtained
from different samples. The rate equation (\ref{eq_gammadot}) implies that the logarithm of the
plastic strain rate, $\ln \dot\gamma$, is linearly proportional to $1/T_k$, since
\begin{equation}
  \ln \dot\gamma = \ln \dot\gamma_0 - \left(\frac{2H_k}{k}\right)\frac{1}{T_k} \ ,
  \label{eq_fit2Hk}
\end{equation}
where $k$ is the Boltzmann constant. Fitting the experimentally measured dependence of $\ln
\dot\gamma$ on $1/T_k$ using \refeq{eq_fit2Hk} yields $2H_k=1.27~\eV$, which is in good agreement
with the values obtained by extrapolation \citep{conrad:63} and direct integration of the activation
volume, \refeq{eq_H_sigma}. Because the experiments of \citet{hollang:97} represent currently the
most accurate studies of the plastic deformation of molybdenum, we will take in the following
$2H_k=1.27~\eV$ for molybdenum. This value can now be used together with the experimentally measured
temperature $T_k$ at which the thermal component of the yield stress vanishes \citep{ackermann:83}
to obtain an estimate for $\dot\gamma_0$. As $T \rightarrow T_k$, the activation enthalpy reaches
its maximum, $2H_k$, and the Arrhenius law suggests that
\begin{equation}
  q \equiv \ln(\dot\gamma_0/\dot\gamma) = \frac{2H_k}{kT_k} \ .
  \label{eq_qcalc}
\end{equation}
For a given plastic strain rate $\dot\gamma$ and measured temperature $T_k$, \refeq{eq_qcalc}
provides an estimate for $\dot\gamma_0$. As we mentioned earlier, the mobile dislocation density,
included in $\dot\gamma_0$, is typically a function of temperature and thus $\dot\gamma_0$ estimated
this way is likely to vary as the temperature is lowered. For theoretical calculations, one usually
requires an effective value of $\dot\gamma_0$ that is independent of temperature. This value may be
slightly different from that calculated above and can be obtained from the requirement that the
theoretically calculated temperature dependence of the yield stress matches the experimental data at
high temperatures. In the case of the experiments of \citet{hollang:97} on molybdenum, this yields
$\dot\gamma_0=3\times10^{10}~\s^{-1}$ and then $q=31.2$.

For an assumed shape of the Peierls potential, the magnitude of the coefficient $p$ in the line
energy of a straight screw dislocation, $V_0=p\mu b^2$, can be calculated numerically from
\refeq{eq_Hk_0stress} by requiring that $2H_k=1.27\ \eV$. The magnitude of $p$ depends on the shape
of the Peierls potential and will be given explicitly below.

%----------------------------------------------------------------------------------------------------
%----------------------------------------------------------------------------------------------------

\section{The Nudged Elastic Band method}
\label{sec_NEB}

Eqs.\,\ref{eq_Hk_0stress} and \ref{eq_Hb_stress}, derived in the last chapter, determine the energy
of an isolated kink and the activation enthalpy of the critical bow-out representing the
saddle-point configuration at high stresses, respectively. We have mentioned that integrations in
these equations are performed along the minimum energy path (MEP) that defines the ``reaction
coordinate'' along which the transformation of the dislocation core takes place. In most cases, the
direct calculation of these pathways using, for example, molecular dynamics is not realistic, because
these transitions occur typically at time scales many orders of magnitude longer than vibrations of
atoms. On the other hand, a number of approximate methods have been developed that predict these
transition paths from only a few fundamental parameters of the system, such as the shape of the
energy surface and the positions of the two low-energy configurations between which the transition takes
place. These methods have been very successful in providing the transition paths in a wide variety
of problems, such as chemical reactions, conformational changes of molecules, and diffusion in solids
(see \citet{mckee:93} for review). The maximum of the potential energy along the MEP is the
saddle-point for motion of the straight dislocation, defining the activation barrier that plays a key
role in the harmonic transition state theory \citep{vineyard:57}.

The method that has recently received a considerable attention \citep{maragakis:02} and that we employ
to determine the activation paths and calculate the activation enthalpies for motion of dislocations
is the Nudged Elastic Band (NEB) method, put forward by \citet{jonsson:98} and
\citet{henkelman:00}. Within this method, one works with replicas of the system that are connected
together with springs to obtain a discrete representation of the reaction coordinate. In the
terminology of the NEB method, the individual replicas of the system are called ``images''. The
string of $N+2$ images is represented by a chain of states $[\mat{\xi}_0, \mat{\xi}_1, \mat{\xi}_2,
\ldots, \mat{\xi}_{N+1}]$. Two of these images, namely $\mat{\xi}_0$ and $\mat{\xi}_{N+1}$, are
fixed in the two low-energy configurations that represent the initial and target state of the system
and the positions of the remaining $N$ images, forming the so-called elastic band, are adjusted by
the optimization algorithm. In our case, $\mat{\xi}_0$ is the position of the dislocation in the
original and $\mat{\xi}_{N+1}$ in the target site, and the energy surface is the Peierls potential
$V(\mat{\xi})$ that will be constructed in Section~\ref{sec_EffPeierlsPot}. The most straightforward
way of obtaining the coordinates of the $N$ intermediate images is by connecting the nearest
neighbors by identical linear springs and subsequently minimizing the total potential energy,
represented by the objective function
\begin{equation}
  {\cal S}(\mat{\xi}_1, \ldots, \mat{\xi}_N) = \sum_{i=1}^N V(\mat{\xi}_i) + 
  \sum_{i=1}^{N+1} \frac{1}{2} k (\mat{\xi}_i-\mat{\xi}_{i-1})^2 
\end{equation}
with respect to the positions of images $\mat{\xi}_1, \ldots, \mat{\xi}_N$. This formulation
comprises an elastic band of $N$ images connected by nearest-neighbor springs with stiffness $k$.

\begin{figure}[!htb]
  \centering
  \includegraphics{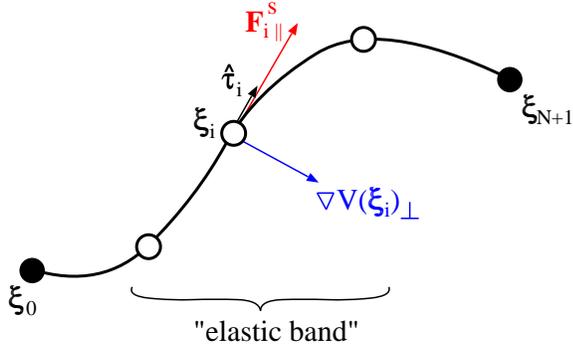}
  \parbox{13cm}{\caption{Elastic band strung between two fixed images
  $\mat{\xi}_0$ and $\mat{\xi}_{N+1}$.}
    \label{fig_elastic_band}}
\end{figure}

As pointed out by \citet{henkelman:00}, the method as formulated is not always well-behaved in that
the elastic band tends to straighten in the regions where the MEP is curved, and the images tend to
slide down toward the fixed endpoints $\mat{\xi}_0$ and $\mat{\xi}_{N+1}$, which gives a poor
resolution of the path close to the saddle-point. It was soon realized that the straightening, or
corner-cutting, is caused by the component of the spring force perpendicular to the elastic band,
while the down-sliding is caused by the parallel component of the so-called ``true force'' arising
from the potential $V(\mat{\xi})$. The original idea to avoid these problems is referred to as
``nudging'' in which each image is subjected only to the component of the spring force parallel to
the elastic band and the perpendicular component of the true force. If we denote the unit vector
tangent to the elastic band at image $i$ as $\mat{\hat{\tau}}_i$, the force on each image is
\begin{equation}
  \mat{F}_i = \mat{F}_{i\,||}^s - \nabla V(\mat{\xi}_i)_\perp \ ,
  \label{eq_force_NEB}
\end{equation}
where $\mat{F}_{i\,||}^s=(\mat{F}_i^s \mat{\hat{\tau}}_i) \mat{\hat{\tau}}_i$ denotes the parallel
component of the spring force at image $i$, and $-\nabla V(\mat{\xi}_i)_\perp$ is the positive
perpendicular component of the true force; see \reffig{fig_elastic_band}. During the minimization,
the parallel component of the spring force does not interfere with the perpendicular component of
the true force and $\nabla V(\mat{\xi}_i)_\perp \stackrel{{\rm all}~i}{\longrightarrow} 0$ as the
shape of the elastic band approaches the MEP. In the relaxed configuration, the force on each image
coincides with the parallel component of the spring force which determines the spacing between
individual images along the path. In the simplest version of the NEB method, the parallel
component of the spring force is determined by subtracting the forces calculated using the
neighboring images:
\begin{equation}
  \mat{F}_{i\,||}^s = k \left\{ \left[ (\mat{\xi}_{i+1}-\mat{\xi}_i) - 
    (\mat{\xi}_i-\mat{\xi}_{i-1}) \right] \mat{\hat{\tau}}_i \right\} \mat{\hat{\tau}}_i \ .
\end{equation}
The perpendicular component of the true force can be easily obtained by subtracting the
component of the true force projected along the elastic band from the total true force vector, i.e.
\begin{equation}
  \nabla V(\mat{\xi}_i)_\perp = \nabla V(\mat{\xi}_i) - 
  [\nabla V(\mat{\xi}_i) \mat{\hat{\tau}}_i] \mat{\hat{\tau}}_i \ .
\end{equation}
A convenient way to calculate the unit vector tangent to the elastic band at image $i$ is to bisect
the tangents calculated with the help of the two neighboring images,
\begin{equation}
  \mat{\tau}_i = \frac{\mat{\xi}_i-\mat{\xi}_{i-1}}{|\mat{\xi}_i-\mat{\xi}_{i-1}|} +
  \frac{\mat{\xi}_{i+1}-\mat{\xi}_i}{|\mat{\xi}_{i+1}-\mat{\xi}_i|} \ ,
\end{equation}
and then normalize, $\mat{\hat{\tau}}_i=\mat{\tau}_i/|\mat{\tau}_i|$. If the spring constant is
identical for each image, defining the tangent this way ensures that the images are equispaced, even
in regions of a large curvature. In certain cases, mainly when the energy of the system changes
rapidly along the path, this method may cause the elastic band to get ``kinky'' which often results
in a slow convergence of the optimization algorithm. To remedy this problem, \citet{henkelman:00b}
devised a new way of defining the local tangent at image $i$. Instead of considering the two
neighboring images, $i-1$ and $i+1$, only the image with the \emph{higher} energy and $i$ are
retained in the estimate. Within this improved scheme, the tangent is calculated as
\begin{equation}
  \mat\tau_i = \left\{ 
    \begin{array}{ll}
      \mat\tau_i^+ & {\rm if} \quad V(\mat{\xi}_{i+1}) > V(\mat{\xi}_i) \geq V(\mat{\xi}_{i-1}) \\
      \mat\tau_i^- & {\rm if} \quad V(\mat{\xi}_{i+1}) \leq V(\mat{\xi}_i) < V(\mat{\xi}_{i-1}) 
    \end{array} \right. \ .
  \label{eq_tangent1}
\end{equation}
If the two neighboring images have both higher or lower energy than that of image $i$, i.e. if
$V(\mat{\xi}_{i+1}) > V(\mat{\xi}_i) < V(\mat{\xi}_{i-1})$ or $V(\mat{\xi}_{i+1}) < V(\mat{\xi}_i) >
V(\mat{\xi}_{i-1})$, the tangent is determined by a weighted average of the tangents calculated from
the two neighboring images:
\begin{equation}
  \mat\tau_i = \left\{ 
    \begin{array}{ll}
      \mat\tau_i^+ \Delta V^{\max}(\mat{\xi}_i) + \mat\tau_i^- \Delta V^{\min}(\mat{\xi}_i) & 
	  {\rm if} \quad V(\mat{\xi}_{i+1}) > V(\mat{\xi}_{i-1}) \\
      \mat\tau_i^+ \Delta V^{\min}(\mat{\xi}_i) + \mat\tau_i^- \Delta V^{\max}(\mat{\xi}_i) & 
	  {\rm if} \quad V(\mat{\xi}_{i+1}) < V(\mat{\xi}_{i-1}) 
    \end{array} \right.  \ ,
  \label{eq_tangent2}
\end{equation}
where
\begin{eqnarray}
  \Delta V^{\min}(\mat{\xi}_i) &=& \min( |V(\mat{\xi}_{i+1})-V(\mat{\xi}_i)|,
  |V(\mat{\xi}_{i-1})-V(\mat{\xi}_i)| ) \\
  \Delta V^{\max}(\mat{\xi}_i) &=& \max( |V(\mat{\xi}_{i+1})-V(\mat{\xi}_i)|,
  |V(\mat{\xi}_{i-1})-V(\mat{\xi}_i)| )
\end{eqnarray}
and the tangent vectors obtained from (\ref{eq_tangent1}) and (\ref{eq_tangent2}) have to be
normalized. This formulation guarantees that the unit tangent $\mat{\hat{\tau}}_i$ varies smoothly as
the energies of the two neighboring images get either higher or lower relative to the energy of
image $i$, and, therefore, the convergence of the optimization process is not affected by any abrupt
changes.

When the restoring forces on the images perpendicular to the path are weak compared to the rapidly
changing parallel forces, the images caught in the regions of large parallel forces tend to slide
down. However, equal spacing between the images is guaranteed if the springs are identical, and,
therefore, the only way the objective function ${\cal S}(\mat{\xi}_1,\ldots,\mat{\xi}_N)$ can be
lowered is by buckling the chain of images into another, nearly force-free region. To avoid this,
\citet{jonsson:98} suggested the use of a smooth switching function that gradually turns on the
perpendicular component of the spring force whenever the path becomes kinky. The force on image $i$
is then calculated as
\begin{equation}
  \mat{F}_i = \mat{F}_{i\,||}^s - \nabla V(\mat{\xi}_i)_\perp + f(\phi_i) \mat{F}^s_{_i\,\perp} \ ,
  \label{eq_force_NEBswitch}
\end{equation}
where $f(\phi_i)=1/2[1+\cos(\pi \cos\phi_i)]$ is a switching function that varies between 0 for a
straight path and 1 if the adjacent segments joined at image $i$ are at a right angle. In this
expression, the angle $\phi_i$ is defined as
\begin{equation}
  \cos \phi_i = \frac{(\mat{\xi}_{i+1}-\mat{\xi}_i)(\mat{\xi}_{i}-\mat{\xi}_{i-1})}
       {|\mat{\xi}_{i+1}-\mat{\xi}_i| \cdot |\mat{\xi}_{i}-\mat{\xi}_{i-1}|} \ .
\end{equation}
Note, that if there are no kinks along the path, $f(\phi_i)=0$ for each image $i$, and
\refeq{eq_force_NEBswitch} then becomes identical to \refeq{eq_force_NEB}.

The most recent advancement of the NEB method \citep{henkelman:00c}, called the Climbing Image
NEB method (CI-NEB), constitutes a modification of the scheme by introducing variable spring
constants and an alternative way of calculating the force on an image, $\mat{F}_i$, which gives a
better resolution of the saddle-point configuration. After a few iterations with the conventional
NEB method, the image $i_{\max}$ with the highest energy is identified. In further calculations, the
force on this image is not given by \refeq{eq_force_NEB}, but rather by
\begin{equation}
  \mat{F}_{i_{\max}} = \nabla V(\mat{\xi}_{i_{\max}})_{||} \ ,
\end{equation}
which corresponds to zeroing the spring force on this image completely and including only the
inverted parallel component of the true force. The image $i_{\max}$ is thus dragged uphill towards
the saddle-point, while the direction of the drag is determined by the position of the two
neighboring images.

The NEB method has been used extensively in the present work to calculate the transition paths of
the dislocation between two minimum energy configurations. All searches utilize the method of
variable spring constants in combination with the switching function defined above. The method is
rather insensitive to the chosen range of magnitudes of the spring constants and performed
efficiently for all shapes of the Peierls potential. The velocity Verlet algorithm \citep{allen:87}
was used to update the positions of images $\mat{\xi}_1$ to $\mat{\xi}_N$ in each iteration
step.

%----------------------------------------------------------------------------------------------------
%----------------------------------------------------------------------------------------------------

\section{Construction of the Peierls potential}
\label{sec_EffPeierlsPot}

%----------------------------------------------------------------------------------------------------

\subsection{Peierls stress and Peierls potential}

Because $1/2\gdir{111}$ screw dislocations in bcc metals do not have well-defined slip planes and
can thus easily cross-slip between different planes of the zone of the slip direction, we will
construct the Peierls potential, $V$, as a function of two variables, $x$ and $y$, representing
the position of the intersection of the dislocation line with the $\gplane{111}$ plane perpendicular
to the corresponding $\gdir{111}$ slip direction. At 0~K, an isolated screw dislocation moves
between two low-energy sites by transforming its core along a minimum energy path, described
by a curvilinear coordinate $\xi$ that connects the two potential minima. During this transition, the
dislocation surmounts the Peierls barrier $V(\xi)$ that is obtained as a cross-section of the
Peierls potential, $V(x,y)$, along $\xi$. Atomistic simulations do not provide the overall form of
$V(\xi)$ but merely its maximum slope that is proportional to the Peierls stress, $\sigma_P$, at
which the dislocation moves. The relation between the Peierls stress and the Peierls barrier is
\begin{equation}
  \sigma_P b = \max \left( \frac{\d{V}(\xi)}{\d{\xi}} \right) \ .
  \label{eq_sigmaP}
\end{equation}

For a given orientation of the MRSSP, characterized by the angle $\chi$, and the angle the slip
plane makes with the most highly stressed $\gplane{110}$ plane, $\psi$, the Peierls stress can be
easily calculated as a projection of the CRSS applied in the MRSSP to the slip plane,
i.e. $\sigma_P=\CRSS\cos(\chi-\psi)$. Unlike the Peierls stress that is readily obtained from the
CRSS found in atomistic studies, there is no direct way of extracting the Peierls barrier, $V(\xi)$,
either from experimental data or atomic-level calculations. We will construct the Peierls potential
by assuming a certain functional form obeying the symmetry and the periodicity of the lattice, whose
height is then parametrized to satisfy \refeq{eq_sigmaP}.

In bcc metals, the Schmid stress is not the only stress component that affects the dislocation
glide. As we have seen in Chapter \ref{chap_MoBOP}, the glide of an isolated screw dislocation in
molybdenum displays a significant twinning-antitwinning asymmetry for pure shear and asymmetries
related to the effect of the shear stress perpendicular to the slip direction. These features must
be reflected in the Peierls potential as well as in the corresponding Peierls barrier $V(\xi)$. One
can imagine that non-glide stresses that are responsible for the above-mentioned asymmetries
``distort'' the Peierls potential in such a way that the glide of the dislocation occurs at either
lower or higher stresses than those needed for the glide if the Peierls potential were undistorted.

%----------------------------------------------------------------------------------------------------

\subsection{Symmetry-mapping function}

The shape of the Peierls potential will be based on the so-called mapping function that merely
captures the three-fold symmetric character of $\gplane{111}$ planes that are perpendicular to the
lines of screw dislocations. This function can be written as a product of three basis functions
defined along the three characteristic directions in the $(111)$ plane (for complete derivation, see
Appendix \ref{sec_mapterm}). For the sake of simplicity, we will use a particularly simple
sinusoidal forms of the basis functions for which the mapping term reads
\begin{equation}
  m(x,y) = \frac{1}{2} + \frac{4}{3\sqrt{3}} \sin{ \frac{\pi}{3a_0} \left( 2y\sqrt{3}+a_0 \right)  } \,
  \sin{ \frac{\pi}{a_0} \left( \frac{y}{\sqrt{3}}-x-\frac{a_0}{3} \right) } \,
  \sin{ \frac{\pi}{a_0} \left( \frac{y}{\sqrt{3}}+x+\frac{2a_0}{3} \right) }
  \label{eq_mxy}
\end{equation}
and is identical to the undeformed Peierls potential of \citet{edagawa:97}. In \refeq{eq_mxy},
$(x,y)$ defines the position of the intersection of the dislocation line with the $(111)$ plane,
where $x$ coincides with the $[\bar{1}2\bar{1}]$ direction that is a trace of the $(\bar{1}01)$
plane on the $(111)$ plane, and $y=[10\bar{1}]$ is perpendicular to $x$ and to the $[111]$
direction. It can be shown that $m(x,y)$ is a three-fold symmetric function bounded such that $0
\leq m \leq 1$. The minima and maxima of $m$ form a triangular lattice with the lattice parameter
$a_0=a\sqrt{2/3}$ where $a$ is the $\gdir{100}$ lattice parameter.

\begin{figure}[!htb]
  \centering
  \includegraphics[width=7cm]{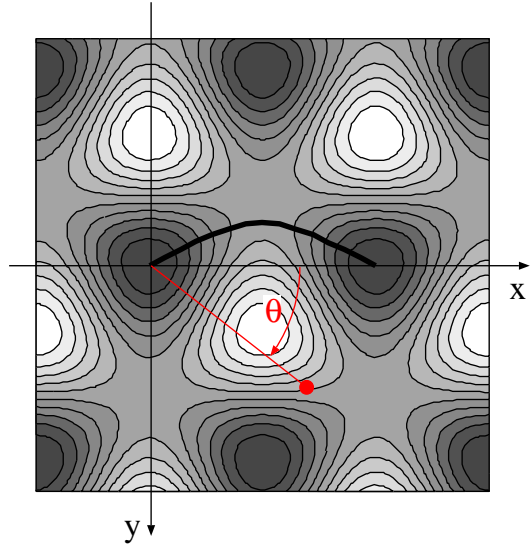}\\
  \parbox{10cm}{\caption{Mapping function $m(x,y)$ of the Peierls potential. Dark domains correspond
      to potential minima and bright domains to potential maxima.}
    \label{fig_V_chi0_tau0}}
\end{figure}

The procedure for constructing the Peierls potential is not unique, and the outlined approach is one
possible way. We start by considering first that: (i) the loading is applied as a pure shear acting
in the $(\bar{1}01)$ slip plane parallel to the $[111]$ direction, and (ii) the glide commences when
this stress reaches the Peierls stress. A consequence of this loading is that the Peierls stress is
determined directly by the Schmid law. This provides a good starting point for the construction of
the potential that will eventually be modified to take the non-Schmid effects into account. In this
first approximation, the Peierls potential is simply
\begin{equation}
  V(x,y) = \Delta{V} \, m(x,y) \ ,
  \label{eq_Vm}
\end{equation}
where $\Delta{V}$ is an unknown height of the potential and $m(x,y)$ is the mapping function
(\ref{eq_mxy}). Considering the case that $\chi=\psi=0$, i.e. loading by pure shear parallel to the
$[111]$ direction in the $(\bar{1}01)$ plane, the Peierls stress in \refeq{eq_sigmaP} is directly
equal to the CRSS for $\chi=0$ obtained from atomistic simulations in Chapter
\ref{chap_MoBOP}. Hence, \refeq{eq_sigmaP} represents a condition that allows one to determine the
height of the Peierls potential, $\Delta V$, as follows. For a trial value of $\Delta{V}$, use the
NEB method described in Section~\ref{sec_NEB} to calculate the MEP between two adjacent potential
minima on the $(\bar{1}01)$ plane. Sufficient accuracy has been obtained using a total of fifteen
images, where thirteen of them formed the elastic band and the last two were fixed at the beginning
and end of the transition path. Since the elastic band determines the contour $\xi$ of the
transition path, it is easy to obtain the Peierls barrier, $V(\xi)$, by interpolating the potential
values corresponding to the images along the elastic band. One can then differentiate the Peierls
barrier and find the maximum. Finally, we evaluate the difference between the left and right side of
\refeq{eq_sigmaP}. If the error is large, we adjust $\Delta{V}$ and repeat the whole process. Once the
error falls into the specified tolerance, which in our case is $10^{-4}~\eV/\A^2$, we terminate the
optimization process and $\Delta{V}$ is then the height of the Peierls potential. For molybdenum,
this approach yields 
\begin{equation}
  \Delta{V}=0.0757~\eV/\A \ .
\end{equation}

%----------------------------------------------------------------------------------------------------

\subsection{Effect of shear stress parallel to the slip direction}

If the Peierls potential were stress-independent and had the form defined in the preceding section,
the orientation dependence of the CRSS would follow exactly the Schmid law in that $\CRSS \propto
1/\cos\chi$. As we can see from \reffig{fig_CRSS_chi_MoBOP}, this is not the case in
molybdenum. Provided that only the shear stress parallel to the slip direction is applied, the CRSS
varies with the orientation of the MRSSP in such a way that it is higher for the antitwinning shear
($\chi>0$) and lower for the twinning shear ($\chi<0$), relative to the value for $\chi=0$
when the MRSSP coincides with the $(\bar{1}01)$ plane. This orientation dependence of the CRSS
implies that the activation barrier for motion of the dislocation is higher for $\chi>0$ and
lower for $\chi<0$.

In order to account for the twinning-antitwinning asymmetry, we augment the Peierls potential by an
angular-dependent function $V_\sigma(\chi,\theta)$ that represents a distortion of the three-fold
symmetric basis by a non-glide shear stress parallel to the slip direction. Here, $\theta$ is the
angle between the $x$ axis and the line connecting the origin with the point $(x,y)$, as shown in
\reffig{fig_V_chi0_tau0}. The Peierls potential then reads
\begin{equation}
  V(x,y) = \left[ \Delta{V} + V_\sigma(\chi,\theta) \right] m(x,y) \ .
  \label{eq_V1}
\end{equation}
We assume that $V_\sigma$ is proportional to the shear stress parallel to the slip direction
resolved in the plane defined by the angle $\theta$, i.e. $V_\sigma \propto
\sigma\cos(\chi-\theta)$. In the case of molybdenum, loading by pure shear parallel to the slip
direction always leads to slip on the $(\bar{1}01)$ plane. Hence, the points along the MEP
are represented by small angles $\theta$, and this implies that $\cos(\chi-\theta)\propto
\cos\theta$. The dependence of $V_\sigma$ on the angle $\chi$ will be determined by a dimensionless
function $K_\sigma(\chi)$. One of the simplest forms of $V_\sigma$ is then
\begin{equation}
  V_\sigma(\chi,\theta) = K_\sigma(\chi) \sigma b^2 \cos\theta \ ,
  \label{eq_Vsigma}
\end{equation}
where the factor $b^2$ is included to arrive at correct dimension of $V_\sigma$. Since $\Delta{V}$
in \refeq{eq_V1} is already known, we can proceed to determine the $K_\sigma(\chi)$ that will
recover the twinning-antitwinning asymmetry. For any orientation of the MRSSP, the dislocation
subjected to pure shear parallel to the slip direction glides on the $(\bar{1}01)$ plane and thus
$\psi=0$. The Peierls stress in \refeq{eq_sigmaP} can then be written as $\sigma_P=\CRSS\cos\chi$,
where the CRSS is obtained from the restricted form of the $\tau^*$ criterion
(\ref{eq_tstar_restr_angle}) as
\begin{equation}
  \CRSS = \frac{\tau_{cr}^*}{\cos\chi + a_1\cos(\chi+\pi/3)} \ .
  \label{eq_CRSS@V1}
\end{equation}

For each value of $\chi$ and the CRSS obtained from \refeq{eq_CRSS@V1}, the corresponding $K_\sigma$
can be determined as follows. Starting with an initial guess for $K_\sigma$ for the given $\chi$, we
find the MEP for transition of the dislocation into the adjacent equivalent lattice site on the
$(\bar{1}01)$ plane in the potential valley defined by \refeq{eq_V1}. We then extract the Peierls
barrier $V(\xi)$ from the set of discrete images along the elastic band, differentiate the Peierls
barrier, find its maximum and evaluate the difference between the two sides of \refeq{eq_sigmaP}. If
the error is greater than $10^{-4}~\eV/\A^2$, we adjust $K_\sigma$ and repeat the procedure
above. Once the error falls into the specified tolerance, we have found one point of the dependence
of $K_\sigma$ on $\chi$.

\begin{figure}[!htb]
  \centering
  \includegraphics[width=12cm]{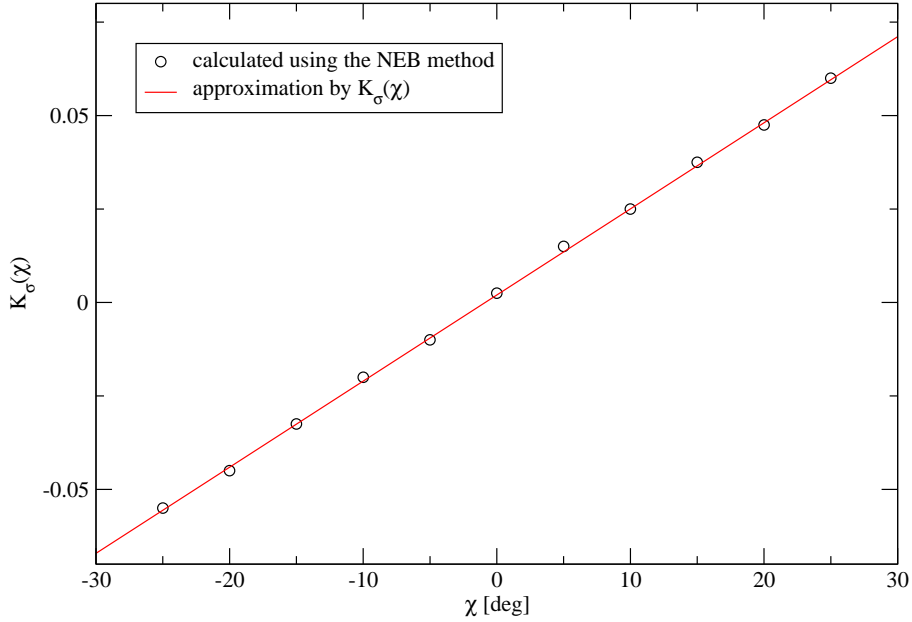}
  \parbox{14cm}{\caption{Fitting of the $K_\sigma(\chi)$ dependence to the values calculated for a
  discrete set of orientations of the MRSSP.}
    \label{fig_Ka1}}
\end{figure}

In \reffig{fig_Ka1}, we show by symbols the data calculated for a set of orientations of the
MRSSP. Apparently, $K_\sigma(\chi)$ can be described by the linear function
\begin{equation}
  K_\sigma(\chi) = 0.132\chi \ ,
  \label{eq_Ka1}
\end{equation}
in which the coefficient has been obtained by fitting the numerical data, and the angle $\chi$ is in
radians. When $\chi=0$, no non-glide stresses are present and the function $K_\sigma$ becomes
zero. In this case, the dislocation glide is governed by the Schmid law, and the Peierls potential
(\ref{eq_V1}) reduces to that given by \refeq{eq_Vm}. For positive $\chi$, i.e. shearing in the
antitwinning sense, $V_\sigma$ is positive, and both the Peierls barrier and the Peierls stress for
the $(\bar{1}01)$ slip increase relative to $\chi=0$. In contrast, for negative $\chi$,
i.e. twinning shear, $V_\sigma$ is negative, and both the Peierls barrier and the Peierls stress
decrease relative to $\chi=0$. Therefore, \refeq{eq_V1} represents the Peierls potential that
reproduces the twinning-antitwinning asymmetry of glide for loading by the shear stress parallel to
the slip direction.

Upon reversing the sense of shearing, \refeq{eq_Vsigma} becomes
$V_\sigma(\chi,\theta)=K_\sigma(\chi) (-\sigma) b^2 \cos\theta$. Since $K_\sigma(\chi)$ is odd,
$-K_\sigma(\chi)\sigma = K_\sigma(-\chi)\sigma$, and this implies the well-known result that the
reversal of the sense of shearing is identical to keeping the stress positive and reversing the sign
of the angle $\chi$. In this case, however, the planes of the $[111]$ zone that were sheared in the
twinning sense become sheared in the antitwinning sense and vice versa. Another important point to
note is that $V_\sigma$ does not obey the translational symmetry of $m(x,y)$, and, when introduced
into \refeq{eq_V1}, it causes a breakdown of the translational symmetry of the potential. As a
consequence, one has to make sure that the origin of $V(x,y)$ always coincides with the lattice site
\emph{from} which the dislocation moves into the adjacent lattice site. After this elementary step
is completed, the origin of $V(x,y)$ must be shifted to this new position, and the search for the
minimum energy path can be repeated. This loss of the translational symmetry of the potential is the
price that we pay for a particularly simple form of the added term $V_\sigma(\chi,\theta)$.

%----------------------------------------------------------------------------------------------------

\subsection{Effect of shear stress perpendicular to the slip direction}

To incorporate the effect of the shear stress perpendicular to the slip direction, we will
complement $V(x,y)$ in \refeq{eq_V1} by a term $V_\tau(\chi,\theta)$ that represents an angular
distortion of the Peierls potential by the shear stress $\tau$ perpendicular to the slip
direction. The full form of the Peierls potential involving effects of both non-glide stresses,
i.e. shear stress parallel to the slip direction in a plane other than the slip plane and the shear
stress perpendicular to the slip direction, then reads
\begin{equation}
  V(x,y) = \left[ \Delta{V} + V_\sigma(\chi,\theta) + V_\tau(\chi,\theta) \right] m(x,y) \ .
  \label{eq_V3}
\end{equation}
For simplicity, we require that $V_\tau$ is a linear function of $\tau$. Furthermore, because the
stress tensor (\ref{eq_tensor_tau}) is invariant with respect to rotations by integral multiples of
$\pi$ about the $z$ axis, and since $V_\tau$ is a linear function of $\tau$, it follows that
$V_\tau$ also has to obey the same symmetry. One of the simplest ways to reproduce this symmetry is
to write $V_\tau$ proportional to $\cos(2\theta+\theta_0)$, where $\theta_0$ is a constant that has
to be determined such that $V_\tau$ is a \emph{linear} function of $\tau$. It can be shown that this
is true provided $\theta_0=\pi/3$, and, therefore, the simplest form that satisfies the
above-mentioned requirements is
\begin{equation}
  V_\tau(\chi,\theta) = K_\tau(\chi) \tau b^2 \cos(2\theta+\pi/3) \ ,
  \label{eq_Vtau}
\end{equation}
where $K_\tau(\chi)$ is yet an unknown function. If $\theta_0\not=\pi/3$, $K_\tau$ would also be an
explicit function of $\tau$, and this would make $V_\tau$ a nonlinear function of $\tau$. Because
$K_\tau(\chi)$ is dimensioneless, we have again included in \refeq{eq_Vtau} the factor $b^2$ that
ensures the correct dimension of $V_\tau$. In order to keep the calculation of $K_\tau$ relatively
simple, we considered only two different shear stresses perpendicular to the slip direction, namely
$\tau=\pm 0.01C_{44}$. For these stresses, the dislocation glides on the $(\bar{1}01)$ plane, and thus
$\psi=0$ for any orientation of the MRSSP. The Peierls stress in \refeq{eq_sigmaP} can then be
written as $\sigma_P=\CRSS\cos\chi$, where the CRSS is obtained directly from the full form of the
$\tau^*$ criterion (\ref{eq_tstar_full_angle}) as
\begin{equation}
  \CRSS = \frac{\tau_{cr}^*-\tau [ a_2\sin2\chi + a_3\cos(2\chi+\pi/6) ]} {\cos\chi +
	a_1\cos(\chi+\pi/3)} \ .
  \label{eq_CRSS@V3}
\end{equation}
The advantage of this approach is that we do not assume \apriori{} that the dislocation can
cross-slip into different $\gplane{110}$ planes as a result of the transformation of its core
induced by $\tau$. If the Peierls potential is constructed correctly, this feature would be
predicted automatically, which is an important test of the predictive power of the potential.

To determine $K_\tau$ corresponding to a given $\chi$, we first fix both $\Delta V$ and $K_\sigma$
to the values determined earlier. For a trial value of $K_\tau$, we then find the MEP for an
elementary $(\bar{1}01)$ jump of the dislocation in the potential valley given by \refeq{eq_V3}. As
before, we extract the Peierls barrier $V(\xi)$ along the MEP by interpolating the values calculated
from the set of discrete points along the elastic band, differentiate $V(\xi)$, find its maximum
slope, and calculate the difference between the two sides of \refeq{eq_sigmaP} for the CRSS from
\refeq{eq_CRSS@V3}. If the error is large, we adjust $K_\tau$ and repeat the process until
\refeq{eq_sigmaP} becomes valid within the desired accuracy.

\begin{figure}[!htb]
  \centering
  \includegraphics[width=12cm]{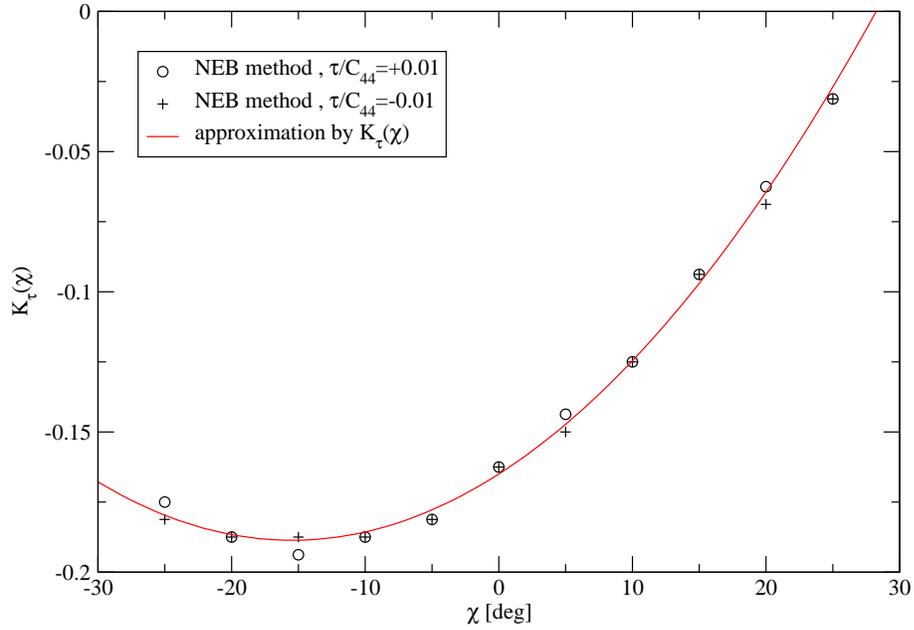}
  \parbox{14cm}{\caption{Fitting of the $K_\tau(\chi)$ dependence from values calculated for a
  discrete set of orientations of the MRSSPs and two values of $\tau$.}
    \label{fig_Ka3}}
\end{figure}

Note, that only two optimization calculations have been performed for each $\chi$, yielding
$K_\tau(\chi)$ for $\tau=-0.01C_{44}$ and $\tau=+0.01C_{44}$. The variation of $K_\tau$ with $\chi$
for these two values of $\tau$ is shown by symbols in \reffig{fig_Ka3} and proves that $K_\tau$ is
independent of $\tau$. As required, $V_\tau$ is then a linear function of $\tau$. The curve plotted
in this figure represents an approximation of the discrete data and takes on a particularly
convenient polynomial form
\begin{equation}
  K_\tau(\chi) = -0.165 + 0.173\chi + 0.316\chi^2 \ ,
  \label{eq_Ka3}
\end{equation}
where the angle $\chi$ is in radians. If the angle $\chi$ of the MRSSP is close to $+30\deg$,
\reffig{fig_Ka3} shows that $K_\tau$ virtually vanishes and thus, for this $\chi$, $V(x,y)$ is not
distorted by the shear stress perpendicular to the slip direction. Provided that $V(x,y)$ is
independent of $\tau$, \refeq{eq_sigmaP} dictates that the Peierls stress also cannot be a function
of $\tau$. This conclusion is consistent with the atomistic data for $\chi=+26\deg$ plotted in
\reffig{fig_CRSS_tau_MoBOP}d showing that the CRSS is almost independent of $\tau$.

\refeq{eq_V3} represents the final form of the Peierls potential that was constructed to reproduce
the 0~K atomistic results from Chapter \ref{chap_MoBOP} with the help of the functional form of the
effective yield criterion (\ref{eq_tstar_full}). In order to keep the fitting procedure reasonably
transparent, we considered only two points in each $\CRSS-\tau$ plot, namely those corresponding to
$\tau=\pm 0.01C_{44}$. No fitting has been done for the large $\tau$ regime and, therefore, we do
not include \apriori{} any information about the slip of the dislocation on planes other than the
$(\bar{1}01)$ plane. Because the $\CRSS-\tau$ atomistic data vary almost linearly at small $\tau$,
the dependence of the Peierls potential on the shear stress perpendicular to the slip direction has
also been chosen to be linear. Thus the Peierls potential is likely to reproduce poorly the
large $\tau$ regime in which the CRSS values are predicted by linear extrapolation from the regime
of small $\tau$. However, as we have shown in \reffig{fig_CRSS_tau_fit_MoBOP}, this region is
virtually inaccessible in real single crystals where another $\gplane{110}\gdir{111}$ system becomes
dominant.

\begin{figure}[!htb]
  \centering
  \includegraphics[width=7cm]{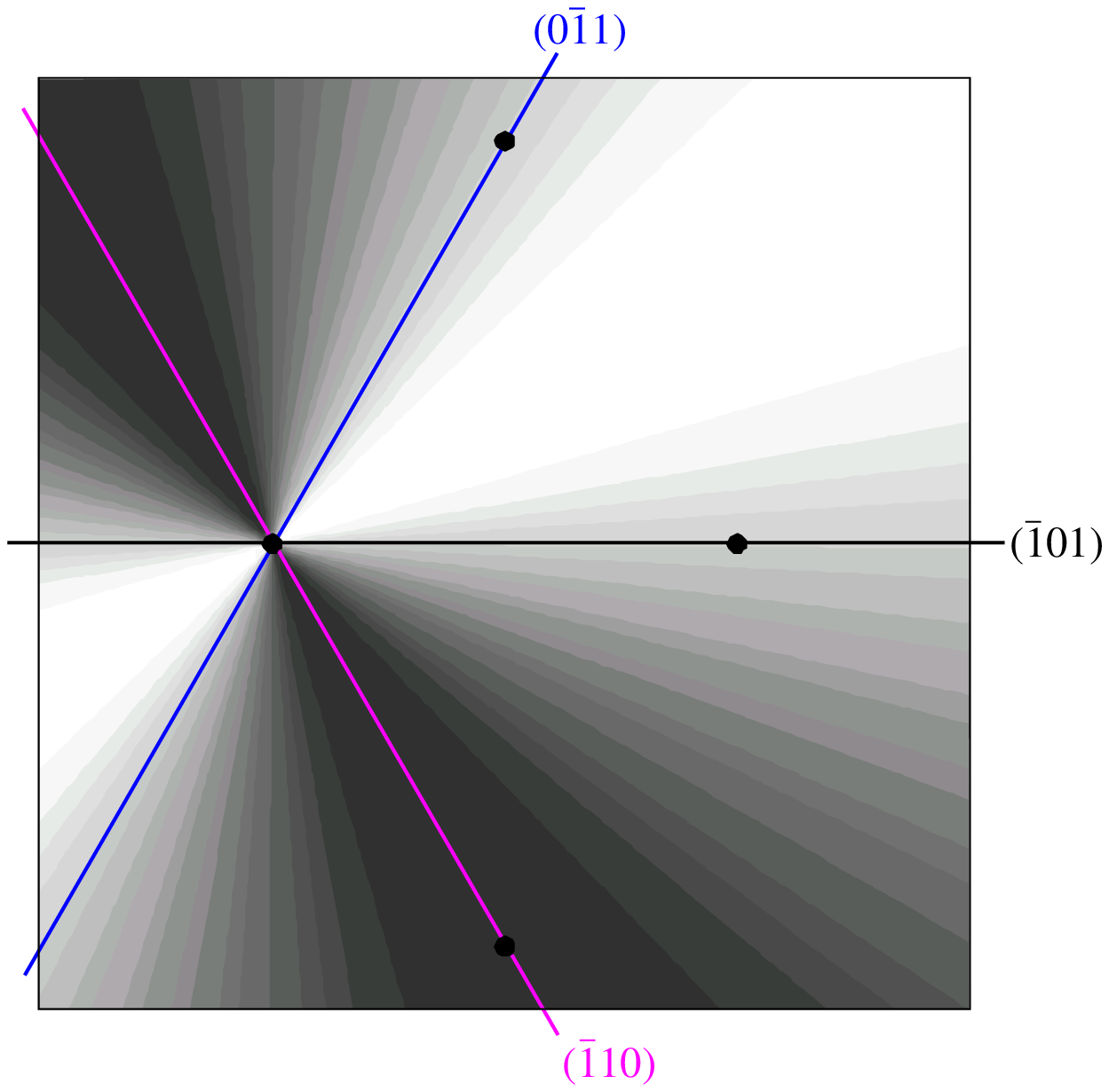} \hskip1cm
  \includegraphics[width=7cm]{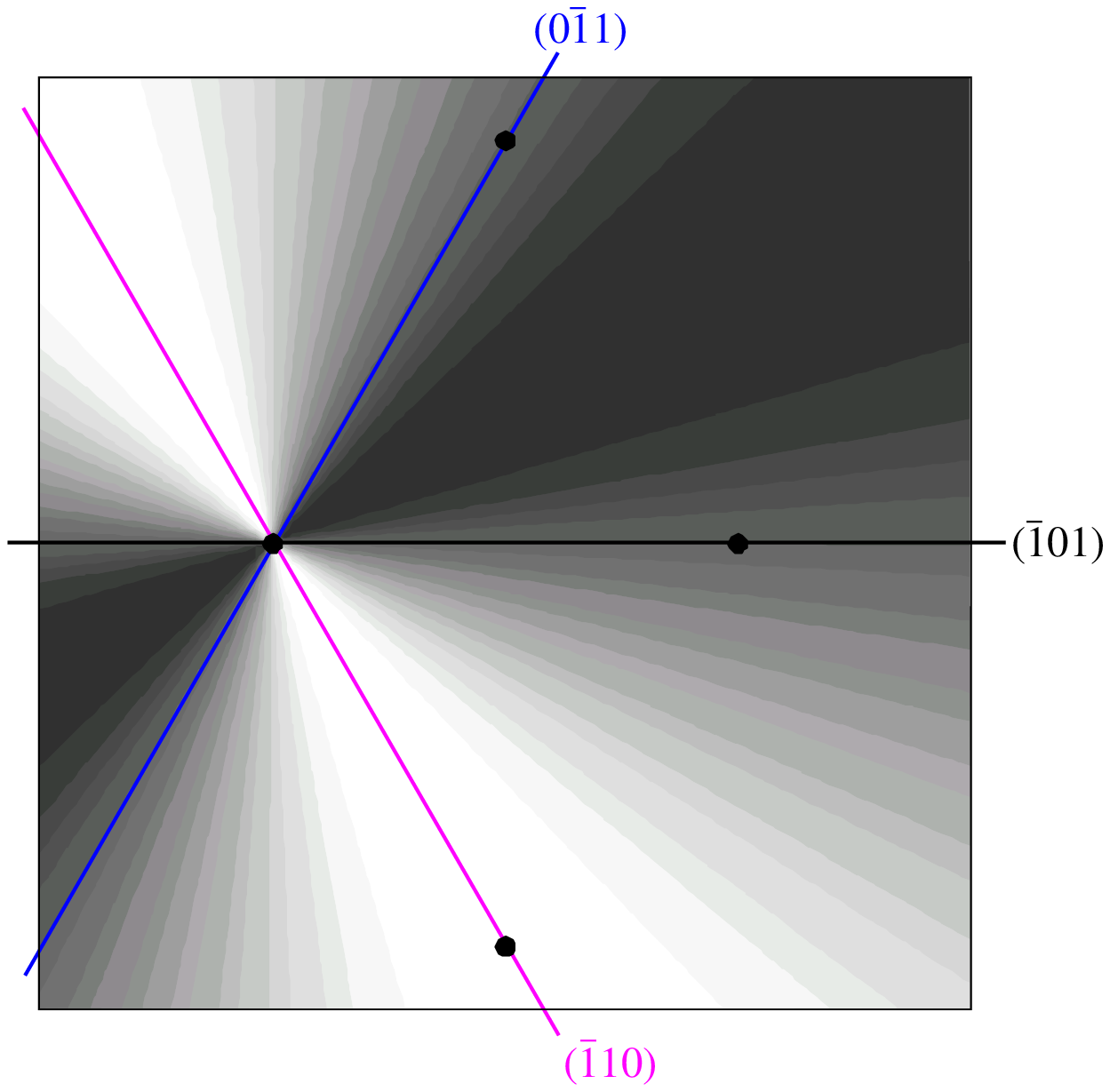} \\
  a) $\tau/C_{44}=-0.05$ \hskip5cm b) $\tau/C_{44}=+0.05$ \\
  \parbox{15cm}{\caption{Effect of the shear stress perpendicular to the slip direction, $\tau$, on
      the shape of $V_{\tau}$. The contours correspond to $V_\tau$ given by \refeq{eq_Vtau} for
      $\chi=0$ for which $K_\tau$ is negative. Dark regions are negative and bright regions positive
      values.}
  \label{fig_potsurf_V3}}
\end{figure}

The effect of the shear stress perpendicular to the slip direction on the shape of the
Peierls potential is shown graphically in \reffig{fig_potsurf_V3}. The two contour maps show
$V_{\tau}$ for positive and negative $\tau$ of the same magnitude, respectively. The dark regions
contribute a negative value to the Peierls potential and cause its flattening. On the other hand,
bright regions are domains of positive values and result in increase of the height of the
potential. It is evident that positive shear stress perpendicular to the slip direction causes
lowering of the potential barrier for the slip on the $(\bar{1}01)$ plane and, therefore, makes the
slip on this plane easier. This slip mode is suppressed at negative $\tau$, which causes a flattening
of the potential along $(\bar{1}10)$ and close to the $(0\bar{1}1)$ plane, while the Peierls barrier
along the $(\bar{1}01)$ plane increases. In this case, the slip is more likely to proceed on the
$(0\bar{1}1)$ or $(\bar{1}10)$ plane. The slip planes at positive and negative $\tau$ inferred from
\reffig{fig_potsurf_V3} are identical to those observed in atomistic simulations (see
\reffig{fig_CRSS_tau_MoBOP}) which proves that the predictions made directly from the Peierls
potential (\ref{eq_V3}) are consistent with the results of 0~K atomistic simulations.

%----------------------------------------------------------------------------------------------------
%----------------------------------------------------------------------------------------------------

\section{Correlations of the Peierls potential with results of 0~K atomistic simulations}
\label{sec_testpot}

Because the Peierls potential (\ref{eq_V3}) is a function of the orientation of the MRSSP
(i.e. angle $\chi$) and both shear stresses parallel and perpendicular to the slip direction, it is
instructive to investigate its shape under various loadings. Besides these qualitative tests, the
Peierls potential can be used to \emph{predict} the $\CRSS-\chi$ and $\CRSS-\tau$ dependencies that
should be in agreement with those determined directly from the $\tau^*$ criterion and shown
in \reffig{fig_CRSS_tau_fit_MoBOP}.

In the first test, we will consider the loading by pure shear stress parallel to the $[111]$
direction applied in the MRSSPs determined by angles $\chi=-20\deg$, $\chi=0$, and
$\chi=+20\deg$. For this loading, the Peierls potential is given by \refeq{eq_V1} and the Peierls
stress in \refeq{eq_sigmaP} reads $\sigma_P=\CRSS\cos\chi$. Note, that $V(x,y)$ depends explicitly
on the actual magnitude of the shear stress parallel to the slip direction, and the shape of the
Peierls potential thus \emph{evolves} with the applied loading.

\begin{figure}[!htb]
  \centering
  \includegraphics[width=12cm]{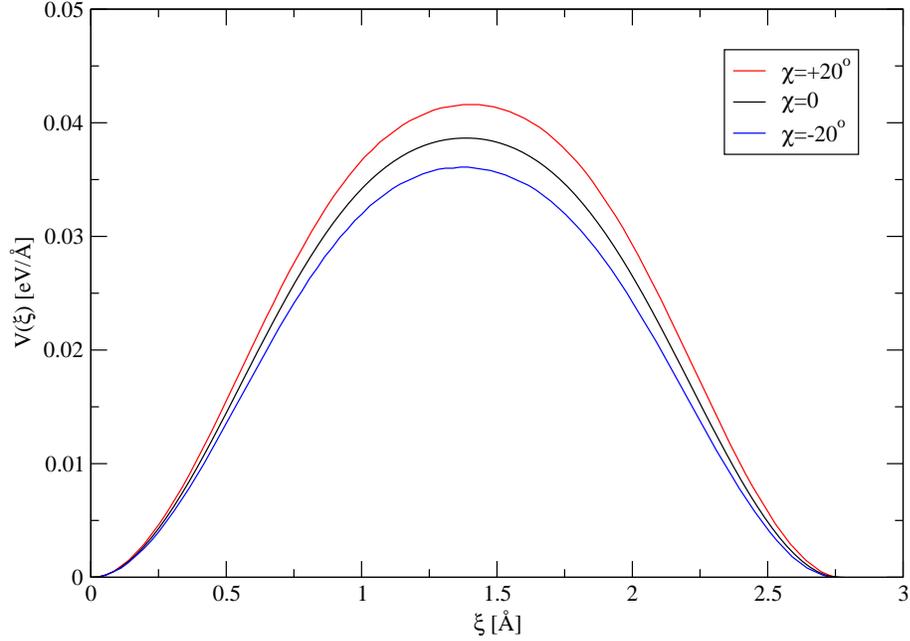}
  \parbox{14cm}{\caption{The Peierls barrier $V(\xi)$ along the MEP for different angles $\chi$ of
      the MRSSP of the applied shear stress. The Peierls stress is proportional to the maximum slope
      of $V(\xi)$.}
    \label{fig_Vbar_tau0}}
\end{figure}

\begin{figure}[!htb]
  \centering
  \includegraphics[width=12cm]{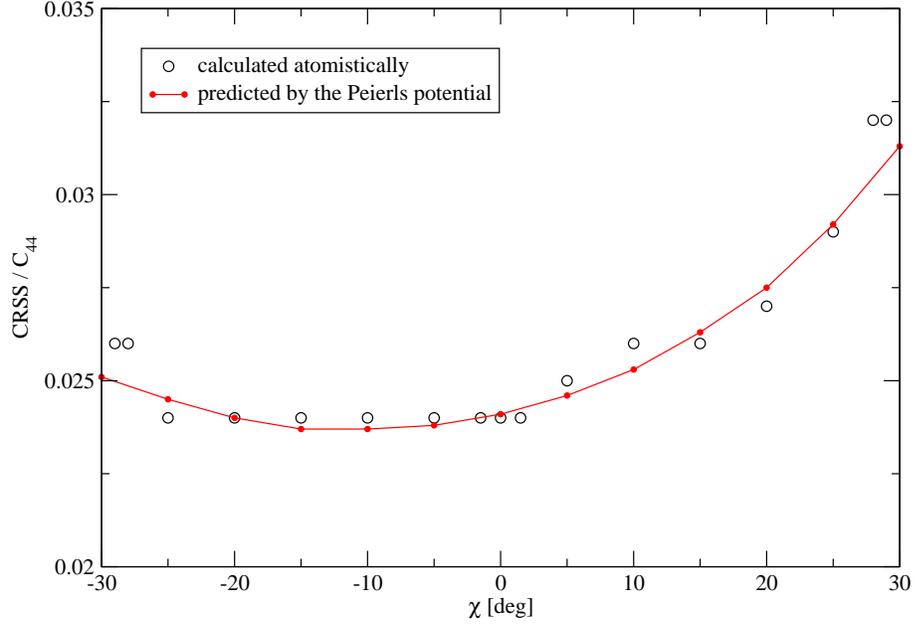}
  \parbox{14cm}{\caption{Comparison of the orientation dependence of the CRSS calculated
      atomistically (circles) and predicted from \refeq{eq_sigmaP} with $\sigma_P = \CRSS\cos\chi$.}
    \label{fig_CRSS_tau_MoBOP_fromV}}
\end{figure}

The Peierls barriers for the three orientations of the MRSSP, calculated from the constructed
Peierls potential using the NEB method, are plotted in \reffig{fig_Vbar_tau0}. If the shear stress
is applied in the antitwinning sense, i.e. $\chi>0$, the Peierls barrier increases relative to the
case when $\chi=0$. The opposite is true for negative $\chi$, i.e. twinning shear, in which case the
activation barrier decreases compared to that for $\chi=0$. Because the Peierls stress is
proportional to the maximum slope of this barrier, which clearly increases with increasing $\chi$,
the potential predicts higher Peierls stress for positive $\chi$ and lower for negative $\chi$. This
qualitative trend is consistent with the atomistic simulations in which the CRSS at positive $\chi$
is always higher than that at negative $\chi$. The Peierls potential not only correctly reproduces
the twinning-antitwinning asymmetry of shear stress parallel to the slip direction, but also
quantitatively agrees with the $\CRSS-\chi$ dependence calculated from atomistic simulations. This
agreement is shown in \reffig{fig_CRSS_tau_MoBOP_fromV}.

The second test concerns the effect of a pure shear stress perpendicular to the slip direction on the
shape of the Peierls potential and the MEP determined by the NEB method for $(\bar{1}01)$,
$(0\bar{1}1)$, and $(\bar{1}10)$ slip. From \reffig{fig_bl_tau0.05_MoBOP}, we know that the shear
stress perpendicular to the slip direction changes the structure of the dislocation core such that
further glide of the dislocation, induced by the shear stress parallel to the slip direction, occurs
at either lower or higher stresses as compared to the case when $\tau=0$. Within the Peierls
potential, the transformation of the dislocation core under stress is reproduced via its dependence
on both applied shear stresses perpendicular and parallel to the slip direction.

\begin{figure}[!htb]
  \centering
  \includegraphics[width=5cm]{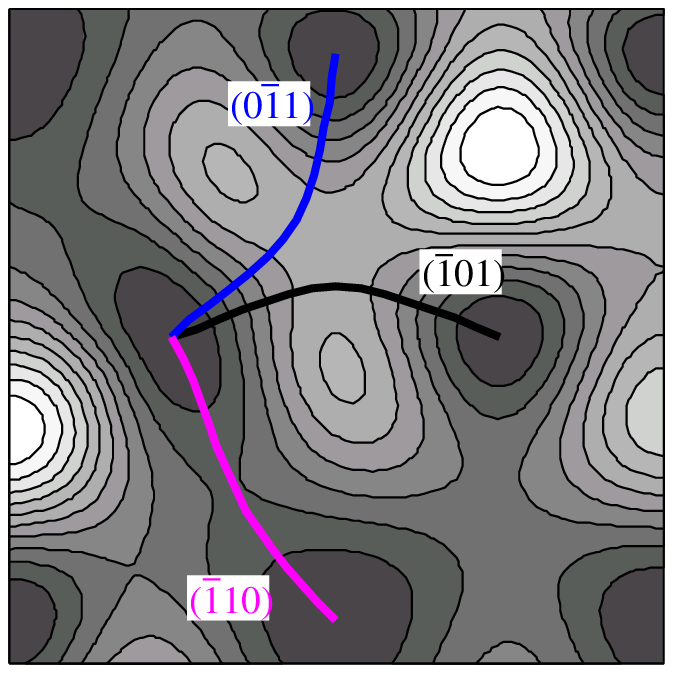}
  \includegraphics[width=5cm]{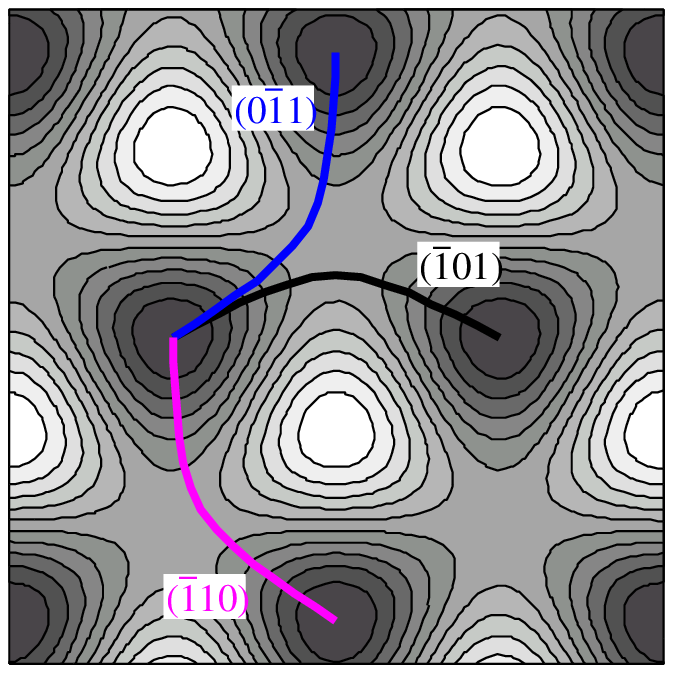}
  \includegraphics[width=5cm]{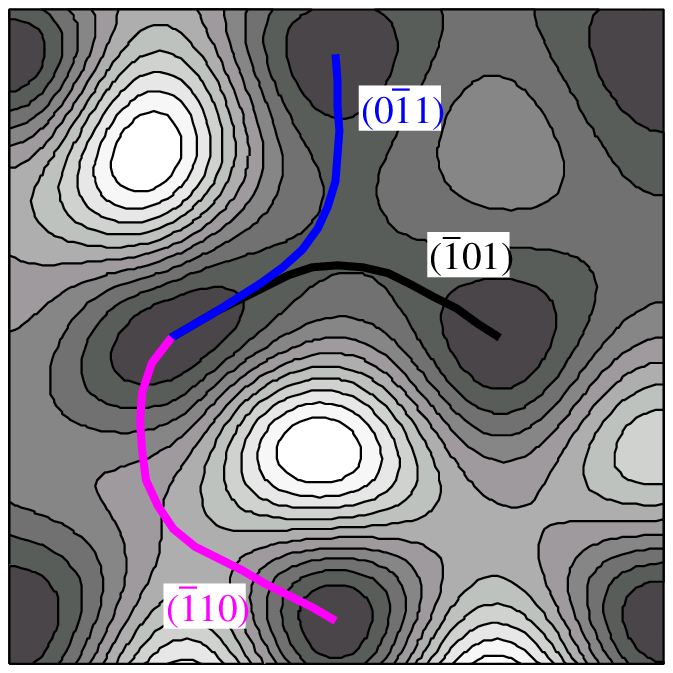} \\
  a) $\tau/C_{44}=-0.05$ \hskip2.2cm b) $\tau/C_{44}=0$ \hskip2.2cm c) $\tau/C_{44}=+0.05$
  \caption{Shape of the Peierls potential under applied shear stress perpendicular to the slip
    direction.}
    \label{fig_V_tau}
\end{figure}

In an unstressed case, the barriers for the motion of the dislocation on the three $\gplane{110}$
planes are identical, and, therefore, there is no \apriori{} tendency towards glide on any particular
plane (see the middle panel in \reffig{fig_V_tau}). In contrast, positive stress $\tau=+0.05C_{44}$
deforms the Peierls potential such that it develops a low-energy path along the trace of the
$(\bar{1}01)$ plane. This modification of the Peierls potential reflects the transformation of the
dislocation core by positive $\tau$, which makes the dislocation more glissile in the $(\bar{1}01)$
plane. In contrast, for $\tau=-0.05C_{44}$, the Peierls potential is modified such that the lowest
energy path is along the trace of the $(\bar{1}10)$ plane, while the path along the trace of the
$(\bar{1}01)$ plane passes through a large Peierls barrier. This is again consistent with the
atomistic results, which show that the slip plane for this magnitude of the shear stress
perpendicular to the slip direction is the $(\bar{1}10)$ plane. This result is remarkable, since the
Peierls potential was constructed by fitting the $\CRSS-\tau$ data only for $\tau=\pm0.01C_{44}$,
and, therefore, the possibility of glide of the dislocation on any other plane but $(\bar{1}01)$ was
not assumed. Hence, the change of the slip plane at negative $\tau$ can be understood as a natural
consequence of the asymmetry of the Peierls potential induced by the shear stress perpendicular to
the slip direction.

Finally, we use the Peierls potential (\ref{eq_V3}) to: (i) identify the most operative slip systems
in real single crystals of bcc molybdenum in which 12 independent slip systems are present, and (ii)
find the critical loadings that mark the onset of slip on each individual system. As in
\reffig{fig_chi0_real_MoBOP}, we will consider a set of uniaxial loadings that can be represented in
the $\CRSS-\tau$ plot for MRSSP $(\bar{1}01)$ as a set of straight lines emanating from the origin,
with slopes $\eta=\tau/\sigma$. Small values of $\eta$ correspond to uniaxial loadings in the
stereographic triangle of the $(\bar{1}01)[111]$ system for which the MRSSP is $(\bar{1}01)$. For
larger $|\eta|$, the loading axis rotates into the stereographic triangle of another
$\gplane{110}\gdir{111}$ system, and, therefore, a system other than $(\bar{1}01)[111]$ becomes
dominant. For a fixed slope $\eta$ of the loading path in the MRSSP of the $(\bar{1}01)[111]$
system, we can always find the orientations of the MRSSPs corresponding to other $\gdir{111}$ slip
directions. As was discussed in Section~\ref{sec_4slipsys}, only four such MRSSPs need to be
considered in which the resolved shear stress parallel to the slip direction is positive and
$-30\deg<\chi<+30\deg$. Since all reference systems are equivalent, the Peierls potential
(\ref{eq_V3}) applies to any of them. At zero applied stress, the Peierls potential is the same for
all four systems, as dictated by symmetry. However, it differs from system to system when the
crystal is loaded, since each system is stressed differently. The Peierls potential in each system
$\alpha$ is determined by the orientation of the MRSSP, $\chi_\alpha$, and the resolved shear
stresses perpendicular and parallel to the slip direction, $\tau_\alpha$ and $\sigma_\alpha$. The
$\CRSS_\alpha$ at which a particular $1/2\gdir{111}$ dislocation moves can be predicted from the
corresponding Peierls potential by seeking the solution of
\begin{equation}
  \CRSS_\alpha\, b \cos(\chi_\alpha-\psi_\alpha) = \max \left( \frac{\d{V(\xi)}}{\d{\xi}} \right)_\alpha \ ,
  \label{eq_CRSS_from_V}
\end{equation}
where $\psi_\alpha$ is the orientation of the slip plane for system $\alpha$. For example, if
$(\bar{1}01)[111]$ is the reference system with the highest Schmid stress, $\psi_\alpha=0$ for
$(\bar{1}01)$, $\psi_\alpha=-60\deg$ for $(0\bar{1}1)$, and $\psi_\alpha=+60\deg$ for $(\bar{1}10)$
plane. For other reference systems, the orientations of the three possible $\gplane{110}$ slip
planes can be easily deduced from \reffig{fig_8slipdir}.

\begin{figure}[!htb]
  \centering
  \includegraphics[width=13cm]{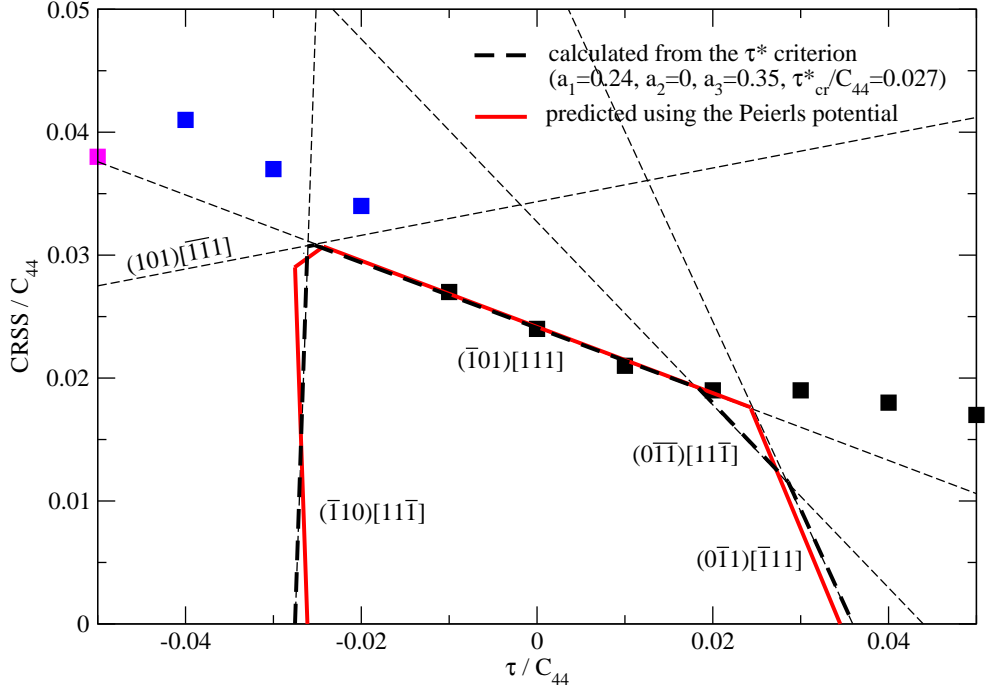}
  \parbox{14cm}{\caption{Projection of the yield surface calculated from the $\tau^*$ criterion
  (dashed lines) and predicted using the Peierls potential
  (solid lines) for MRSSP $(\bar{1}01)$.}
    \label{fig_chi0_ysurf_MoBOP}}
\end{figure}

The four stresses $\CRSS_\alpha$ calculated from \refeq{eq_CRSS_from_V} and the corresponding shear
stresses perpendicular to the slip direction $\tau_\alpha=\eta_\alpha\CRSS_\alpha$, in the MRSSP
given by the angle $\chi_\alpha$, can now be mapped back onto the loading path in the MRSSP of the
$(\bar{1}01)[111]$ system. This gives four critical points that mark the onset of slip on the four
slip systems $\alpha$. If the same calculation is performed for all possible orientations of the
loading path, i.e. for $-\infty<\eta<\infty$, the points corresponding to activation of individual
slip systems trace the critical lines for these systems. The inner envelope of these lines
represents a projection of the yield surface that is plotted in \reffig{fig_chi0_ysurf_MoBOP} as the
solid polygon. Superimposed in \reffig{fig_chi0_ysurf_MoBOP} are also the critical lines calculated
earlier from the $\tau^*$ criterion (\ref{eq_tstar_tensor}) and their inner envelope, plotted as the
dashed polygon. One can see that the yield surface obtained from the Peierls potential by means of
\refeq{eq_CRSS_from_V} is in full agreement with that calculated from the $\tau^*$ criterion. This
proves that the predictions made on the basis of the constructed Peierls potential (\ref{eq_V3}) are
fully consistent with the $\tau^*$ criterion, which is itself a close approximation of the results of
0~K atomistic simulations.

%----------------------------------------------------------------------------------------------------
%----------------------------------------------------------------------------------------------------

\section{Macroscopic yield behavior predicted by the constructed Peierls potential}
\label{sec_macroyield_sine}

The Peierls potential (\ref{eq_V3}) will now be used in conjunction with the Dorn-Rajnak expression
for the activation enthalpy (\ref{eq_Hb_stress}), or the model of elastic interaction of kinks
(\ref{eq_Hkp_final}), to predict for single crystals of molybdenum the dependence of the activation
enthalpy and the activation volume on the applied stress, and the temperature and strain rate
dependence of the yield stress. These calculated dependencies can be readily compared with
experimental measurements.

\begin{figure}[!htb]
  \centering
  \includegraphics[width=12cm]{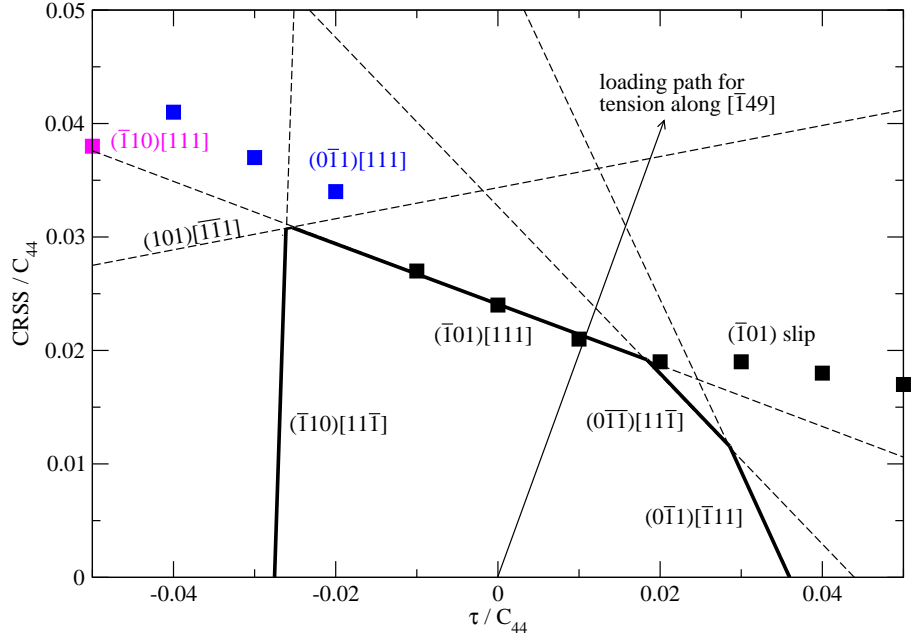}
  \parbox{14cm}{\caption{Orientation of the loading path corresponding to the tensile loading along
      $[\bar{1}49]$ used in the experiments of \citet{hollang:97}. The $\CRSS-\tau$ atomistic data
      (symbols) and the lines calculated from the $\tau^*$ criterion correspond to the MRSSP
      $(\bar{1}01)$.}
    \label{fig_chi0_fit_MoBOP_t-149}}
\end{figure}

The following calculations will be made for tensile loading along the $[\bar{1}49]$ axis that allows
subsequent comparison with tensile experiments by \citet{hollang:97}. For this loading axis, the
MRSSP is $(\bar{1}01)$ and $\chi=0$, and the ratio of the two resolved shear stresses is
$\eta=\tau/\sigma=0.51$. In the calculation, the loading is expressed as a combination of the shear
stresses parallel and perpendicular to the slip direction, where the ratio $\eta$ determines the
orientation of the loading path plotted in \reffig{fig_chi0_fit_MoBOP_t-149}. It is important to
emphasize that the loading path first reaches the critical line for the $(\bar{1}01)[111]$ system,
which thus becomes the most prominent, and the dislocations will proceed by normal slip on the
$(\bar{1}01)$ plane. Moreover, since no other system is activated for slip at the comparatively low
CRSS, the macroscopic plastic deformation will correspond to single slip on the $(\bar{1}01)[111]$
system. In other words, the activation enthalpy for the $(\bar{1}01)[111]$ system is markedly lower
than those for other slip systems, which will greatly simplify the calculation of the temperature
dependence of the yield stress.

For each stress state along the loading path, we can determine the orientation of the MRSSP in the
zone of every $\gdir{111}$ direction and thus identify the four $\gplane{110}\gdir{111}$ reference
systems for which the MRSSP lies within the angular region $\chi=\pm30\deg$ and the shear stress
parallel to the slip direction acting in this MRSSP is positive. For each of these systems, the
shear stresses perpendicular and parallel to the slip direction are substituted in \refeq{eq_V3} to
determine the Peierls potential in the $\gplane{111}$ plane of this system. In each case, we find
the MEP that connects the original Peierls valley with that of the neighboring minimum energy
lattice site and calculate the profile along this MEP, $V(\xi)$. Note, that the dislocation can move
into three adjacent lattice sites on planes at $\psi=0$, $\psi=-60\deg$, and $\psi=+60\deg$, marked
in \reffig{fig_V_tau}. For each such contour, one can obtain the activation enthalpy either from the
model of the dislocation bow-out (\ref{eq_Hb_stress}) or from the model of elastic interaction of
fully developed kinks (\ref{eq_Hkp_final}).

At this point, it is important to emphasize that the yield stresses predicted by the model are
uniformly about a factor of 3 higher than experiments. This is a well-known discrepancy that appears
in all 0~K atomistic simulations of an isolated screw dislocation and that is routinely removed by
scaling the theoretical data to reproduce the experimental yield stress extrapolated to 0~K. The
physical origin of this discrepancy is explained in detail in Appendix~\ref{chap_interact}. This
scaling can be easily accomplished by renormalizing the shear stresses in \refeqs{eq_Hb_stress},
\ref{eq_Hkp_final}, and \ref{eq_V3}, as $\tau=m\tilde{\tau}$ and $\sigma=m\tilde{\sigma}$, where
$\tilde{\tau}$ and $\tilde{\sigma}$ are the renormalized shear stresses perpendicular and parallel
to the slip direction, respectively. The constant $m=2.9$ was determined by requiring that the yield
stress corresponding to zero activation enthalpy (and thus also zero temperature) equals to the
experimental yield stress extrapolated to 0~K, i.e. $870\ \MPa$. For comparisons with experiments
given in the following text, we will always use the renormalized stress $\tilde\sigma$ given in
$\MPa$ and drop the tilde for brevity.

\begin{figure}[!htb]
  \centering
  \includegraphics[width=12cm]{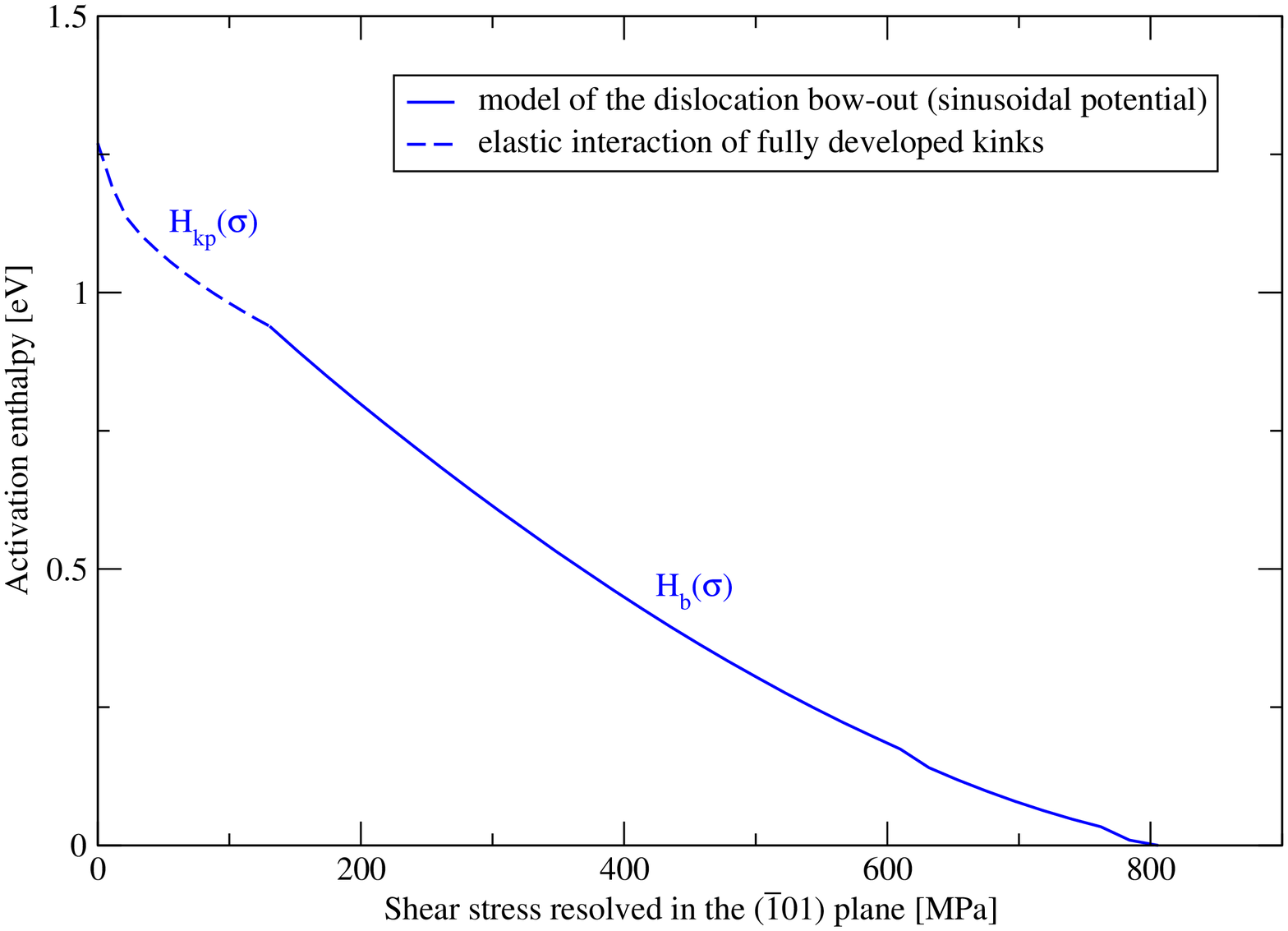}
  \parbox{14cm}{\caption{Stress dependence of the activation enthalpy for the $(\bar{1}01)[111]$ primary
  slip system under loading in tension along $[\bar{1}49]$.}
    \label{fig_actene_hollang-149}}
\end{figure}

The stress dependence of the activation enthalpy for the most prominent slip system,
$(\bar{1}01)[111]$, is shown in \reffig{fig_actene_hollang-149}, where the horizontal axis
represents the renormalized shear stress $\tilde\sigma$ parallel to the $[111]$ slip direction
resolved in the $(\bar{1}01)$ plane. At low stresses, the dislocation moves by nucleating a pair of
interacting kinks, and the activation enthalpy is given by $H_{kp}$ obtained from
\refeq{eq_Hkp_final}. In the high-stress regime, the dislocation moves by bowing-out in the
direction of the applied stress, and the activation enthalpy, $H_b$, is given by
\refeq{eq_Hb_stress}. In the limit of zero stress, the activation enthalpy approaches the energy of
two isolated kinks, $2H_k=1.27\ \eV$.

\begin{figure}[!htb]
  \centering
  \includegraphics[width=12cm]{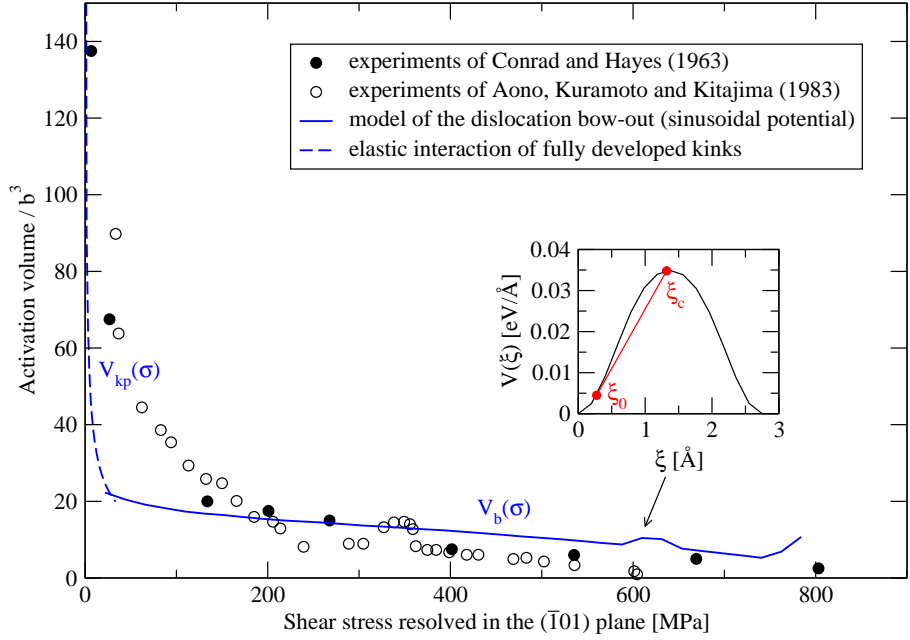}
  \parbox{14cm}{\caption{Stress dependence of the activation volume for loading in tension along
  $[\bar{1}49]$. Two different sources of experimental data are plotted for comparison.}
    \label{fig_actvol_hollang-149}}
\end{figure}

The stress dependence of the activation volume calculated by differentiating \refeq{eq_Hb_stress}
(high stresses) with respect to the thermal component of the yield stress, $\sigma$, and from
\refeq{eq_Vkp} (low stresses), is shown in \reffig{fig_actvol_hollang-149}, where the symbols
correspond to the values measured by \citet{conrad:63}. At intermediate stresses, the calculated
activation volume agrees reasonably well with the experiment. The origin of the local maximum close
to $600\ \MPa$, corresponding to a change in the slope of the activation enthalpy at this stress,
can be traced back to the shape of the activation barrier. From \reffig{fig_Hb_area}, we can see
that the activation enthalpy is proportional to the area bounded between the Peierls barrier and the
line with slope $\sigma b$ that is a tangent of the Peierls barrier at $\xi_0$. At large stresses,
i.e. when $\xi_0$ is close to $\xi_c$, this area, and thus also the activation enthalpy, increases
proportionally to the decrease of the applied stress. At lower stresses, when the position of the
critical point $\xi_c$ reaches the top of the Peierls barrier (see the inset of
\reffig{fig_actvol_hollang-149}), the bounded area increases temporarily more rapidly with the
decrease of the stress, which gives rise to a change of slope of the activation enthalpy at this
stress. Hence, this sudden increase is revealed in the stress dependence of the activation volume as
an intermediate maximum, shown in \reffig{fig_actvol_hollang-149} at about $600~\MPa$.

Finally, the temperature dependence of the yield stress can be calculated from the stress dependence
of the activation enthalpy plotted in \reffig{fig_actene_hollang-149}. Following
Section~\ref{sec_thermoglide}, the total plastic strain rate is determined from the Arrhenius law
\begin{equation}
  \dot{\gamma} = \dot{\gamma}_0 \sum_\alpha \sum_\psi 
    \exp \left( -\frac{H_b^{(\alpha,\psi)}(\sigma)}{kT} \right) \ ,
  \label{eq_gdot24}
\end{equation}
where $\dot{\gamma}$ is the total plastic strain rate and $k$ the Boltzmann constant. In general,
the summation is carried out over all possible slip planes $\psi=\{0,\pm 60\deg\}$ of those
$\gplane{110}\gdir{111}$ slip systems $\alpha$ that can become activated for slip at the given
loading. However, in many cases, one slip system has a markedly smaller activation enthalpy than the
others, and, since its activation enthalpy appears in the exponential of (\ref{eq_gdot24}), the
contribution of all other slip systems can be neglected. Hence, one arrives at a simple expression
of the total plastic strain rate dominated by the most prominent slip system,
\begin{equation}
  \dot{\gamma} = \dot{\gamma}_0 \exp \left( -\frac{H_b(\sigma)}{kT} \right) \ ,
  \label{eq_gdot1}
\end{equation}
which gives the activation enthalpy as a function of temperature and plastic strain rate:
\begin{equation}
  H_b(\sigma) = kT \ln\left( \frac{\dot{\gamma}_0}{\dot{\gamma}} \right) \ .
  \label{eq_Eb_log}
\end{equation}
It is worthwhile noting that the simple expression (\ref{eq_Eb_log}) can be used even when two slip
systems have almost equal prominence, i.e. two slip systems have equally small activation enthalpy
$H_b^{(\alpha,\psi)}(\sigma)$. In this case, the double sum in \refeq{eq_gdot24} reduces effectively
to a factor of 2, and the activation enthalpy then becomes
$H_b(\sigma)=kT\ln(2\dot\gamma_0/\dot\gamma)=kT\ln 2 + kT\ln(\dot\gamma_0/\dot\gamma)$. Since
$\ln(\dot\gamma_0/\dot\gamma)$ is typically 20--40, which is much larger than $\ln 2\ (\approx0.7)$,
the activation enthalpy is still closely approximated by the relation (\ref{eq_Eb_log}).

\begin{figure}[!hb]
  \centering
  \includegraphics[width=12cm]{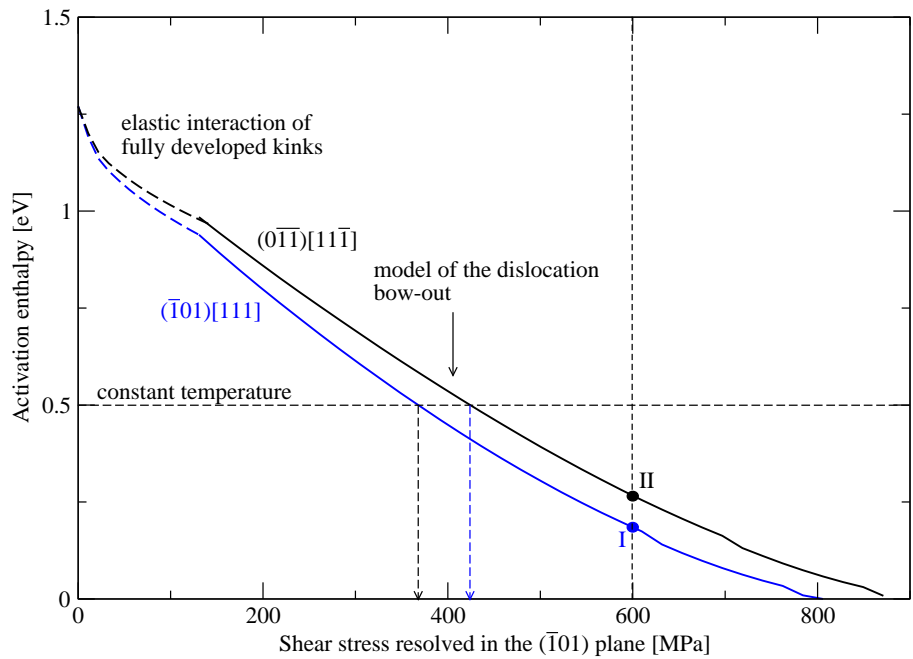}
  \parbox{14cm}{\caption{Stress dependence of the activation enthalpy for the two most operative
      slip systems under loading in tension along $[\bar{1}49]$. The labels I and II mark the
      activation enthalpies calculated for the two systems at $600~\MPa$.}
  \label{fig_actene_hollang_-149_allsys}}
\end{figure}

For loading in tension along $[\bar{1}49]$, \reffig{fig_actene_hollang_-149_allsys} shows that the
calculated activation enthalpy for the $(\bar{1}01)[111]$ system is at any stress $\sigma$
significantly lower than that for the next most operative system,
$(0\bar{1}\bar{1})[11\bar{1}]$. This means that, at constant temperature $T$, the $(\bar{1}01)[111]$
system will be activated for slip at a \emph{lower} applied stress than that needed for operation of
the secondary system, $(0\bar{1}\bar{1})[11\bar{1}]$. The order of activation of these two systems
is clearly consistent with the $\tau^*$ criterion, particularly with
\reffig{fig_chi0_fit_MoBOP_t-149}, where the loading path first reaches the critical line for the
$(\bar{1}01)[111]$ primary system, then that for the $(0\bar{1}\bar{1})[11\bar{1}]$ secondary system,
etc. Hence, the rate equation (\ref{eq_gdot24}) is clearly dominated by the term involving the
activation enthalpy for the $(\bar{1}01)[111]$ system and thus effectively reduces to the single
slip rate equation (\ref{eq_gdot1}), where $H_b(\sigma)$ is the stress dependence of the activation
enthalpy for the $(\bar{1}01)[111]$ primary system.

\begin{figure}[!htb]
  \centering \includegraphics[width=12cm]{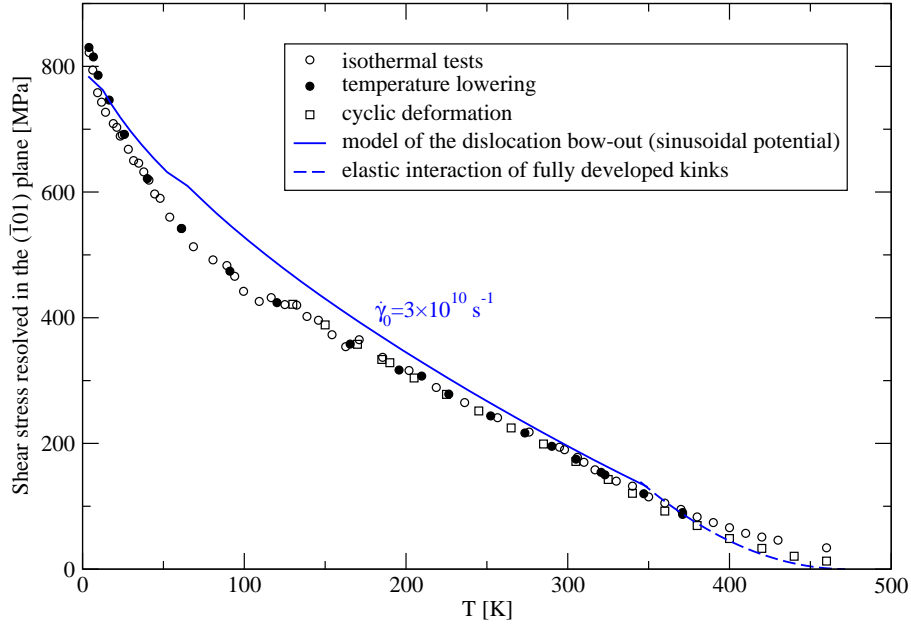}
  \parbox{14cm}{\caption{Temperature dependence of the yield stress for the $(\bar{1}01)[111]$
      system under loading in tension along $[\bar{1}49]$. The solid curve is calculated from the
      model of the dislocation bow-out using the constructed Peierls potential. The experimental data
      are from \citet{hollang:01}.}
    \label{fig_yieldt_hollang_-149}}
\end{figure}

The temperature dependence of the yield stress can now be calculated directly from
\refeq{eq_Eb_log}. We will consider the same plastic strain rate as in the experiments of
\citet{hollang:97}, $\dot{\gamma}=8.6\times 10^{-4}\ \s^{-1}$ and, following the estimates in
Section~{\ref{sec_kinkpar}}, we set $\dot\gamma_0=3\times10^{10}~\s^{-1}$. The activation enthalpy
can then be written as $H_b(\sigma)=qkT$, where $q=\ln(\dot\gamma_0/\dot\gamma)=31.2$. For a given
temperature $T$, this equation gives $H_b(\sigma)$ for which one can find the corresponding yield
stress, $\sigma$, from \reffig{fig_actene_hollang-149}. By repeating this calculation for various
temperatures, one obtains the temperature dependence of the yield stress that is plotted in
\reffig{fig_yieldt_hollang_-149}, where the data measured by \citet{hollang:97} using three
different methods are shown by symbols. The overall trend of $\sigma(T)$ follows the trend of the
experimental data. At high temperatures, the predicted yield stress agrees well with the experiment,
but, at low temperatures, the theoretical calculations overestimate the experiment. The cause of
this deviation can be traced back to the formulation of the mapping function of the Peierls
potential, $m(x,y)$, given by \refeq{eq_mxy}. While this function has the correct periodicity, its
sinusoidal form imposes a specific shape of the Peierls barrier with a sharp maximum (see
\reffig{fig_Vbar_tau0}). The effect of the shape of the Peierls barrier on the temperature
dependence of the yield stress was studied before by \citet{suzuki:95}, who found that the
temperature dependence of the yield stress agrees with experiments if the Peierls potential exhibits
a flat maximum.

%----------------------------------------------------------------------------------------------------
%----------------------------------------------------------------------------------------------------

\section{Adjustment of the shape of the symmetry-mapping function}

When constructing the Peierls potential, we have assumed that the mapping term $m(x,y)$ can be
written as a product of three \emph{sinusoidal} functions. This naturally leads to the Peierls
barrier with sharp maximum. In order to arrive at different shapes of the Peierls barriers, consider
multiplication of $m$ in \refeq{eq_mxy} by a function $f$ that affects only the immediate
neighborhood of each saddle-point of $m$. The final form of the adjusted mapping function will
be written as $\hat{f} m$, where we use an operator $\hat{f}$ to emphasize that the function $f$ is
applied to \emph{every} saddle-point of $m$. Since the MEP passes in the neighborhood of a
saddle-point of $m$, this operation merely adjusts the maximum of the Peierls barrier. The important
feature of this method is that the positions and heights of minima and maxima of $m$ remain
unaffected. A simple way to accomplish this perturbation is to define $f$ in terms of the
Fermi-Dirac function, as
\begin{equation}
  f(r) = 1 - \frac{\beta}{1+\exp(\frac{r-r_0}{\alpha})} \ ,
  \label{eq_fFD}
\end{equation}
which ranges between $1-\beta/[1+\exp(-r_0/\alpha)]$ (for $r=0$) and $1$ (for $r \rightarrow
\infty$), and the three adjustable parameters are $\alpha \geq 0$, $\beta \geq 0$ and $r_0 \in
\langle 0, r_0^{max} \rangle$, where $r_0^{max}$ denotes the maximum acceptable radius for which the
adjacent potential extrema are unaffected by $f$. The magnitude of $\beta$ determines the
\emph{height} of $f$; the mapping function is unaffected if $\beta=0$. The parameter $\alpha$
determines the \emph{shape} of $f$ in that it becomes a sharper function as $\alpha$ gets
smaller. Finally, the radius $r_0$ defines the position of the inflection point of $f$.

As we mentioned before, $f$ has to be applied at each saddle-point of the mapping function,
$m$. These saddle-points are arranged in a triangular lattice with primitive translation vectors
$\mat{t}_1=(1,0)a_0$ and $\mat{t}_2=(1/2,\sqrt{3}/2)a_0$, whose origin is identified with any
saddle-point, e.g. $\mat{t}_0=(1/2,-\sqrt{3}/6)a_0$. The positions of the saddle-points are then
defined by a set of crystallographic vectors $\mat{t}=\mat{t}_0+k\mat{t}_1+l\mat{t}_2$, where $k,l$
are integers. For the saddle-points located at $\mat{t}$, the perturbation is a set of
radially-symmetric functions $f(|\mat{r}-\mat{t}|)$, where the argument simply means the distance
from a particular saddle-point determined by integers $k,l$.

\begin{figure}[!htb]
  \centering
  \includegraphics[width=8cm]{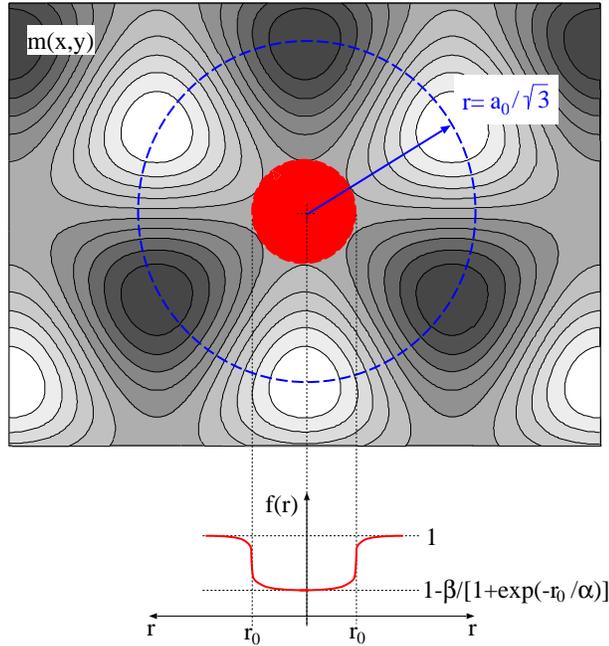}
  \parbox{12cm}{\caption{Region in which the adjustment of $m$ (contour plot) by the function $f(r)$
      takes place. Virtually no change of $m$ takes place beyond $r_0$ measured from the
      saddle-point. The inset below shows the shape of the perturbing function $f(r)$.}
    \label{fig_fpert_region}}
\end{figure}

\begin{figure}[!htb]
  \centering
  \includegraphics[width=7cm]{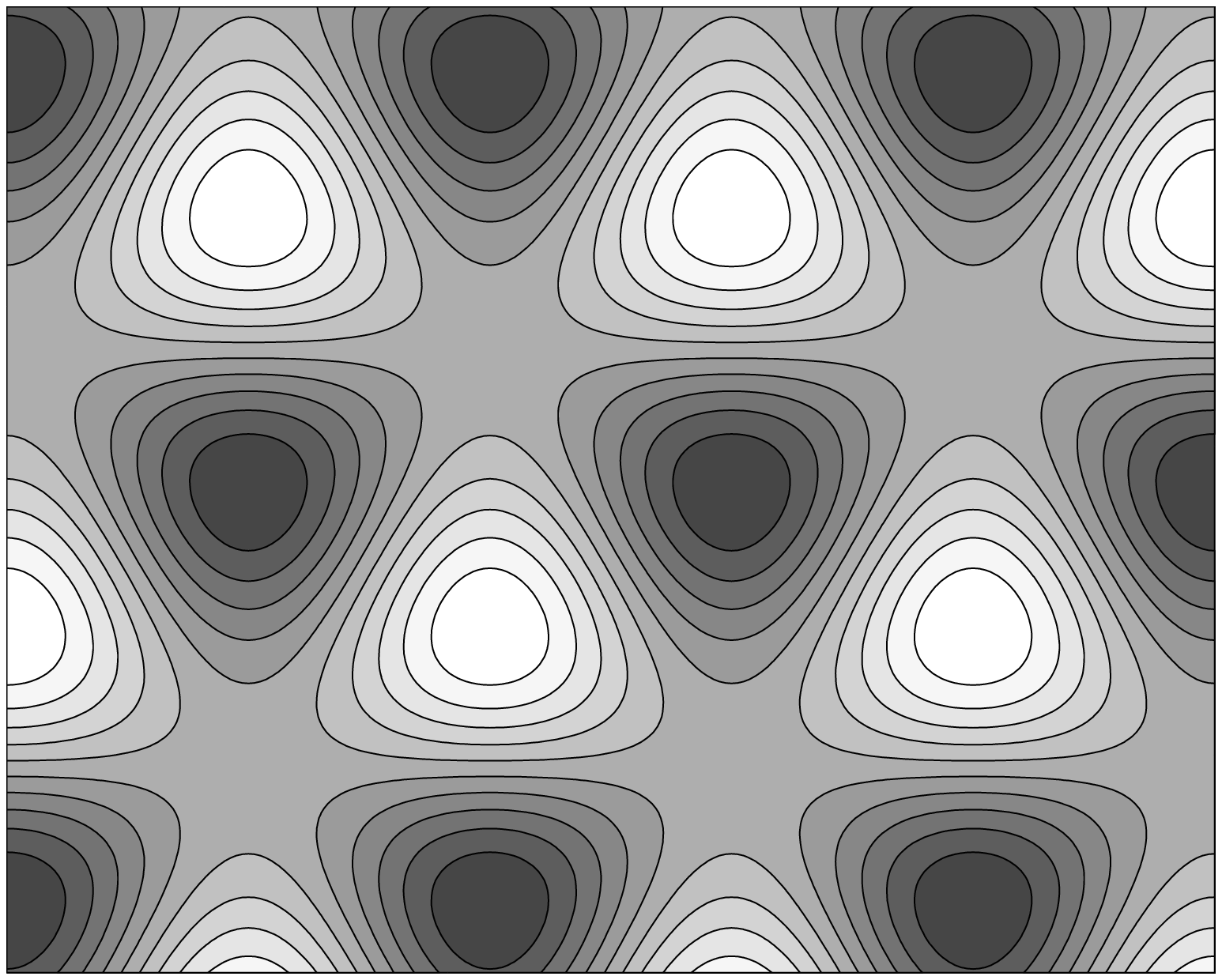} \hfill
  \includegraphics[width=7cm]{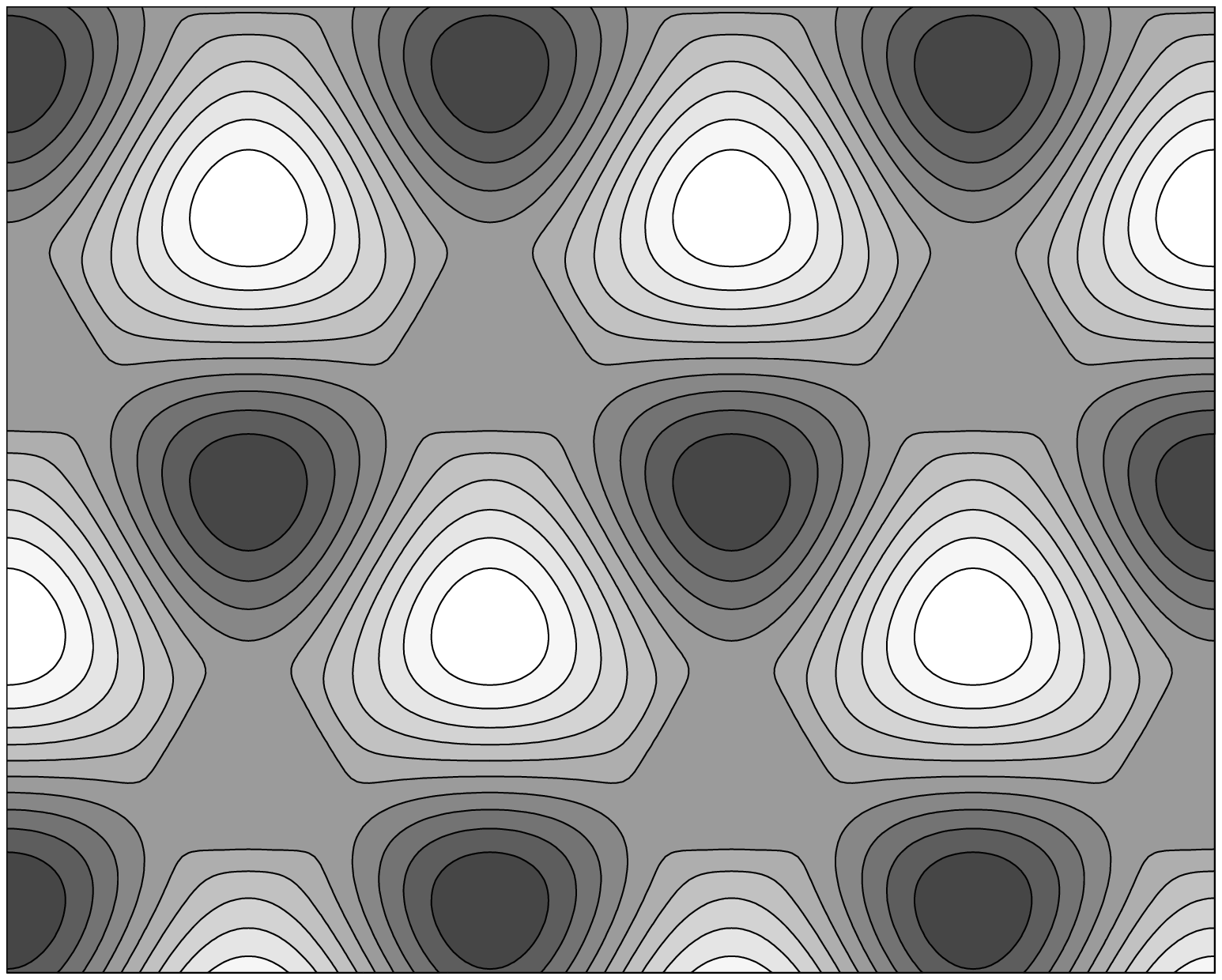} \\
  a) $\beta=0$ (no perturbation) \hskip4cm b) $\alpha=0.05, \beta=0.2$
  \caption{Contour plots for: (a) unperturbed mapping function $m(x,y)$, and (b) modified mapping
    function obtained after the perturbation. Different shades in these plots mean different heights
    of the potential; white regions are potential maxima and black domains potential minima. The
    values of $\alpha$ and $\beta$ in (b) are chosen such that the change of the mapping function is
    visible.}
  \label{fig_fm_alpha_beta}
\end{figure}

In the following, we consider that the parameter $r_0$, which controls the radial extent of the
saddle-point perturbation, is $r_0=a_0/(3\sqrt{3})$, which is one-third of the distance from a
saddle-point to the nearest maximum/minimum of $m$. This ensures that the perturbation is applied
only locally to the given saddle-point and does not affect the adjacent extrema of the
potential. The choice of the parameters $\alpha$ and $\beta$ is not unique, but the rule of thumb
for their selection is that the mapping function should not lead to sudden potential drops (if
$\alpha$ is too small) or intermediate local minima (if $\beta$ is too large). The reasonable values
of $\alpha$ are between 0.05 and 0.2, and $\beta$ should be between 0.1 and 0.4. For comparison,
\reffig{fig_fm_alpha_beta} shows the contour plots of the product $\hat{f}m$ calculated for: a)
$\beta=0$, i.e. without the saddle-point perturbation, and b) $\alpha=0.05$, $\beta=0.2$. The most
important feature to notice is that the modified mapping function $\hat{f} m$
(\reffig{fig_fm_alpha_beta}b) still possesses the three-fold rotational and the long-range
translational symmetry, as required. The proposed perturbation flattens all saddle-points which
implies that the activation barrier calculated along the MEP that always passes in the neighborhood
of a saddle-point will also exhibit a flat maximum.

%----------------------------------------------------------------------------------------------------
%----------------------------------------------------------------------------------------------------

\section{Modified Peierls potential}

We stated above that the saddle-point adjustment by the function $f$, \refeq{eq_fFD}, modifies only
the mapping function $m(x,y)$. The functional forms of all the other terms enclosed in the square
bracket of \refeq{eq_V3}, representing the effects of the shear stresses parallel and perpendicular
to the slip direction, remain unaffected. In analogy with \refeq{eq_V3}, the modified Peierls
potential with adjusted saddle-points of $m$ can now be written as
\begin{equation}
  V(x,y) = \left[ \Delta{V} + V_\sigma(\chi,\theta) + V_\tau(\chi,\theta) \right] \hat{f} m(x,y)
  \ ,
    \label{eq_V3_flattop}
\end{equation}
where the operator $\hat{f}$ acts on every saddle-point of $m$. It is important to realize that even
though the saddle-point perturbation acts only on $m$, the potential height $\Delta V$ and the two
adjustable functions $K_\sigma(\chi)$ and $K_\tau(\chi)$ in \refeqs{eq_Vsigma} and \ref{eq_Vtau}
have to be re-fitted.

As already mentioned above, the parameters $\alpha$, $\beta$, $r_0$ of the function $f$
(\ref{eq_fFD}) determine the shape and the extent of the perturbation and must be chosen such that
the neighboring potential extrema remain unaffected. In the following, we use $\alpha=0.12$,
$\beta=0.2$, $r_0=a_0/3\sqrt{3}$ for which the saddle-point perturbation leads to the Peierls
barrier with a flat maximum. Considering $\chi=0$ and $\tau=0$, and fitting the potential height such
that the dislocation moves at the Peierls stress $\sigma_P=\CRSS(\chi=0)=0.024C_{44}$, calculated by 0~K
atomistic simulations, one arrives at
\begin{equation}
  \Delta{V}=0.0787~\eV/\A \ .
\end{equation}
Similarly, $K_\sigma(\chi)$ can be fitted by requiring that the orientation dependence of the shear
stress parallel to the slip direction, calculated from \refeq{eq_sigmaP}, reproduces the
$\CRSS-\chi$ dependence in \reffig{fig_CRSS_chi_MoBOP}. As previously, one again arrives at a linear
dependence of $K_\sigma$ on $\chi$,
\begin{equation}
  K_\sigma(\chi) = 0.139\chi \ .
\end{equation}
Finally, the last adjustable function, $K_\tau(\chi)$, is determined such that the dependence of the
CRSS on the shear stress perpendicular to the slip direction, $\tau$, predicted by
\refeq{eq_sigmaP}, reproduces the $\CRSS-\tau$ data for $\tau=\pm0.01C_{44}$ and a number of
orientations of the MRSSP. As before, $K_\tau(\chi)$ can be expressed as a polynomial which now has
the form
\begin{equation}
  K_\tau(\chi) =  -0.171 + 0.182\chi + 0.319\chi^2 \ .
\end{equation}

\begin{figure}[!htb]
  \centering
  \includegraphics[width=12cm]{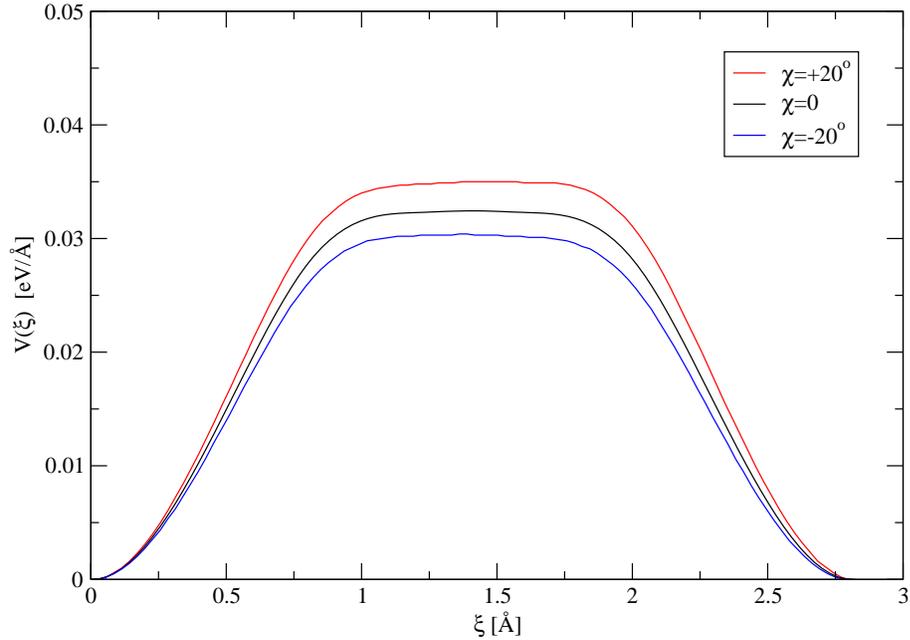}
  \parbox{14cm}{\caption{The Peierls barrier $V(\xi)$ for slip along the minimum energy path under
      loading by pure shear stress parallel to the slip direction. The curves are calculated from
      \refeq{eq_sigmaP} for the Peierls potential with the saddle-point perturbation,
      \refeq{eq_V3_flattop}.}
    \label{fig_Vbar_tau0_flattop}}
\end{figure}

We can now perform the same tests for the Peierls potential with flat saddle-points as we have
done for the sinusoidal potential earlier in Section \ref{sec_testpot}. The Peierls barriers for
loading by pure shear parallel to the slip direction acting in three different MRSSPs are shown in
\reffig{fig_Vbar_tau0_flattop}. In contrast to \reffig{fig_Vbar_tau0}, the maxima of the barriers
are now flattened, which results from flat saddle-points of the modified mapping function,
$\hat{f}m$. Since the stress to move the dislocation over the barrier is proportional to
$\max(\d{V}/\d{\xi})$, \reffig{fig_Vbar_tau0_flattop} again implies that the CRSS for shearing in
the antitwinning sense ($\chi>0$) is higher than the CRSS for shearing in the twinning sense
($\chi<0$), which is consistent with the results of our 0~K atomistic simulations, particularly with
\reffig{fig_CRSS_chi_MoBOP}.

\begin{figure}[!htb]
  \centering
  \includegraphics[width=4.5cm]{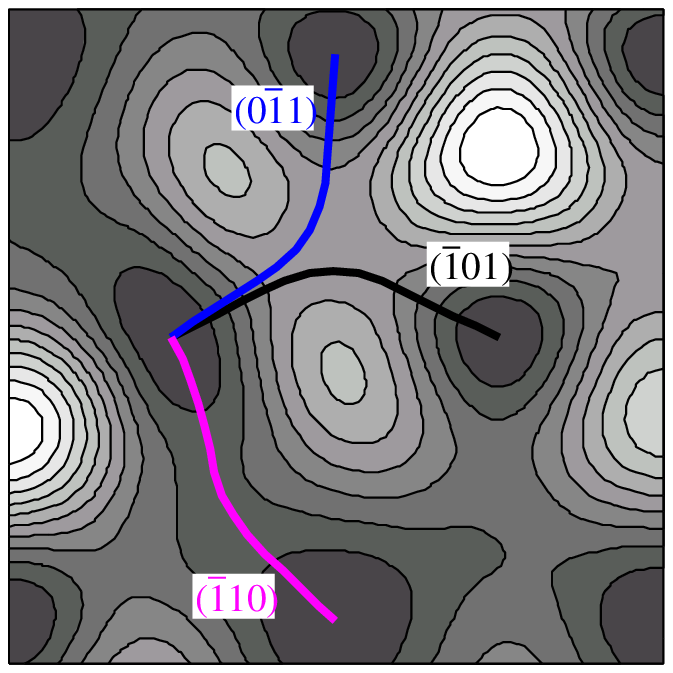} \quad
  \includegraphics[width=4.5cm]{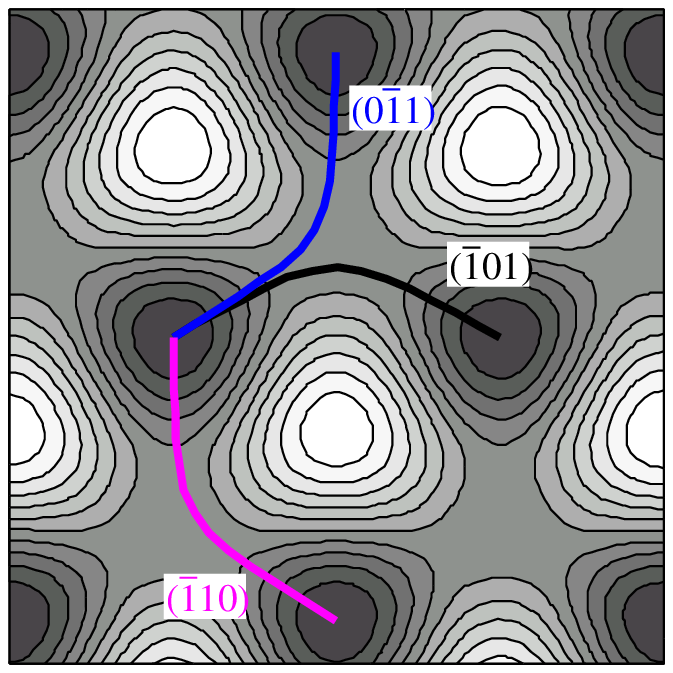} \quad
  \includegraphics[width=4.5cm]{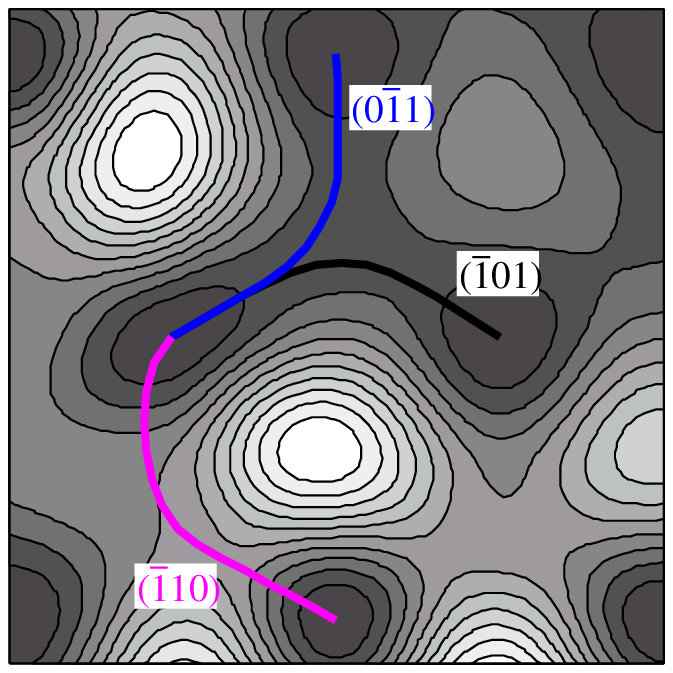} \\
  a) $\tau/C_{44}=-0.05$ \hskip2.2cm b) $\tau/C_{44}=0$ \hskip2.2cm c) $\tau/C_{44}=+0.05$
  \caption{Shape of the Peierls potential (\ref{eq_V3_flattop}) with adjusted saddle-points under
    applied shear stress perpendicular to the slip direction.}
    \label{fig_V_tau_flattop}
\end{figure}

The effect of the shear stress perpendicular to the slip direction on the shape of the activation
barrier for the motion of the dislocation along the $\gplane{110}$ planes is qualitatively similar
to that discussed for the sinusoidal Peierls potential. For comparison, we show the three minimum
energy paths calculated for the Peierls potential (\ref{eq_V3_flattop}) in
\reffig{fig_V_tau_flattop}. The shape of the potential unaffected by $\tau$, shown in
\reffig{fig_V_tau_flattop}b, reveals the flattened saddle-points controlled by the parameters
$\alpha$, $\beta$, and $r_0$. Negative $\tau$ causes the Peierls barrier to flatten along
$(\bar{1}10)$ and close to the $(0\bar{1}1)$ plane, see \reffig{fig_V_tau_flattop}a, which lowers
the activation barrier for slip on these planes. On the other hand, positive $\tau$ causes
flattening of the Peierls barrier along the $(\bar{1}01)$ plane, see \reffig{fig_V_tau_flattop}c,
which thus promotes the slip on this plane. These conclusions are again consistent with the results
of our 0~K atomistic studies that show a change of the slip plane to $(0\bar{1}1)$ or $(\bar{1}10)$
at negative applied shear stress perpendicular to the slip direction.

%----------------------------------------------------------------------------------------------------
%----------------------------------------------------------------------------------------------------

\section{Macroscopic yield behavior predicted by the modified Peierls potential}
\label{sec_expt_hollang}

The macroscopic predictions shown earlier for the sinusoidal potential will now be recalculated
using the modified Peierls potential (\ref{eq_V3_flattop}) with adjusted saddle-points. Since the
details of these calculations have been presented already in Section \ref{sec_macroyield_sine}, we
concentrate only on the differences in the results obtained using the two Peierls potentials.

\begin{figure}[!htb]
  \centering 
  \includegraphics[width=12cm]{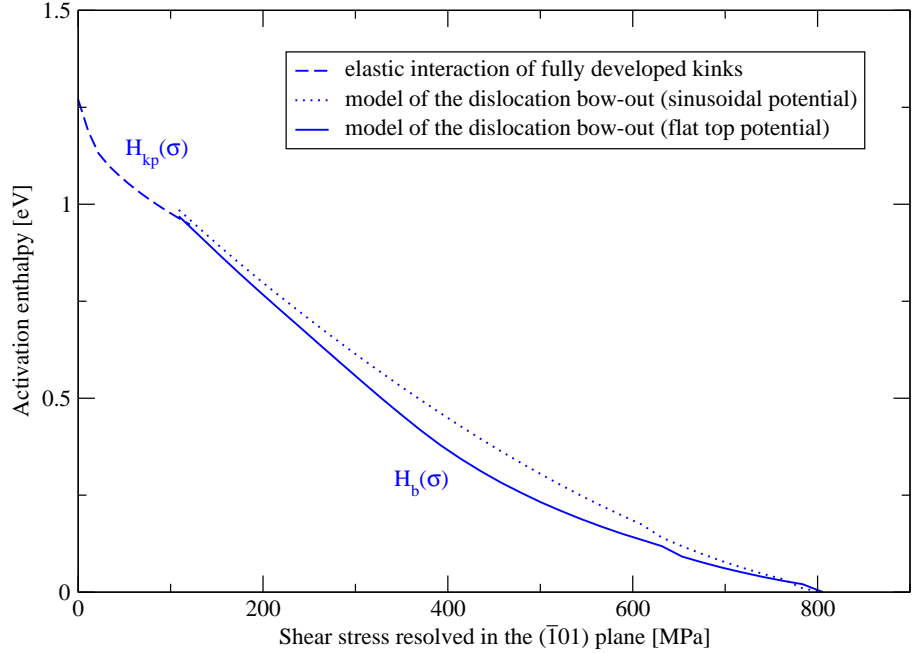}
  \parbox{14cm}{\caption{Stress dependence of the activation enthalpy for the $(\bar{1}01)[111]$
      system under loading in tension along $[\bar{1}49]$.}
    \label{fig_actene_hollang_-149_flattop}}
\end{figure}

The stress dependence of the activation enthalpy for tensile loading along $[\bar{1}49]$ is shown in
\reffig{fig_actene_hollang_-149_flattop}, where the solid curve corresponds to the modified Peierls
potential and the dotted curve is the dependence obtained previously
(\reffig{fig_actene_hollang-149}) for the sinusoidal potential. The curves plotted in this figure
again correspond to the primary slip system $(\bar{1}01)[111]$. All other systems have markedly
larger activation enthalpies, and their operation can thus be neglected in the following
treatment. One can clearly see that the adjustment of the saddle-points affects the gradient of this
dependence mainly at intermediate and high stresses. This implies that the temperature dependence of
the yield stress will also change significantly which will be apparent mainly at low and
intermediate temperatures.

\begin{figure}[!b]
  \centering 
  \includegraphics[width=12cm]{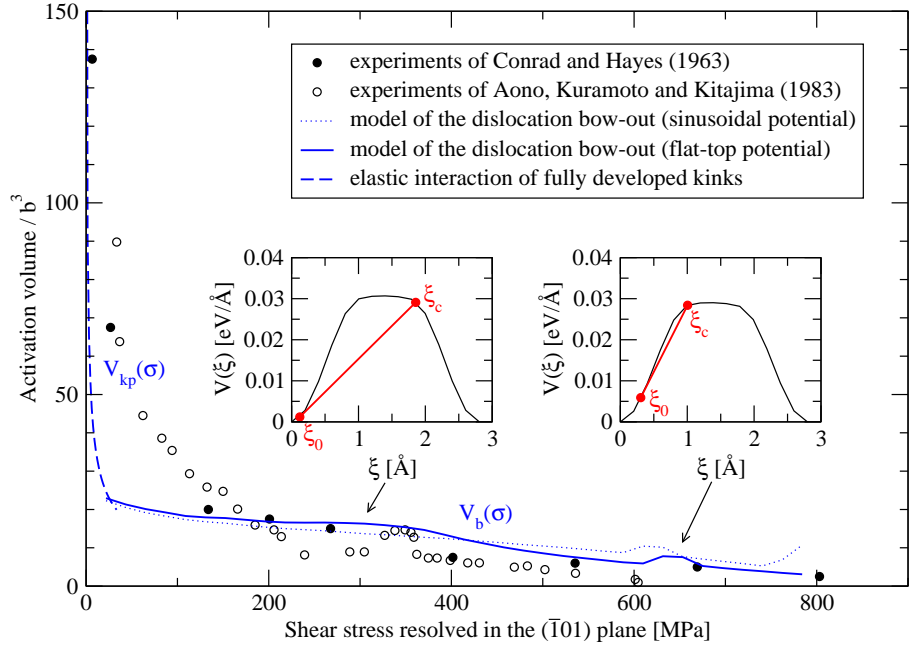}
  \parbox{14cm}{\caption{Stress dependence of the activation volume for loading in tension along
  $[\bar{1}49]$. Two different sources of experimental data are plotted for comparison.}
    \label{fig_actvol_hollang_-149_flattop}}
\end{figure}

The stress dependence of the activation volume derived from \reffig{fig_actene_hollang_-149_flattop}
is plotted in \reffig{fig_actvol_hollang_-149_flattop} by a solid line. This new dependence exhibits
a broad plateau close to $300\ \MPa$ whose origin can be understood with reference to the shape of
the activation barrier. As explained earlier, if the position of the activated segment, $\xi_c$, at
a given stress $\sigma$ meets a point on the activation barrier where its curvature changes rapidly,
the activation enthalpy $H_b(\sigma)$ suddenly increases with decreasing the applied stress. In our
case, the activation barrier has a flat top and so two such points exist, both corresponding to the
positions $\xi$ at which $V(\xi)$ drops from its maximum (see the insets of
\reffig{fig_actvol_hollang_-149_flattop}). By calculating the position $\xi_c$ of the activated
segment of the dislocation as a function of $\sigma$, one can see that the plateau and the local
maximum in \reffig{fig_actvol_hollang_-149_flattop} correspond exactly to the two cases where
$\xi_c$ approaches or leaves the flat top of $V(\xi)$.

\begin{figure}[!htb]
  \centering 
  \includegraphics[width=12cm]{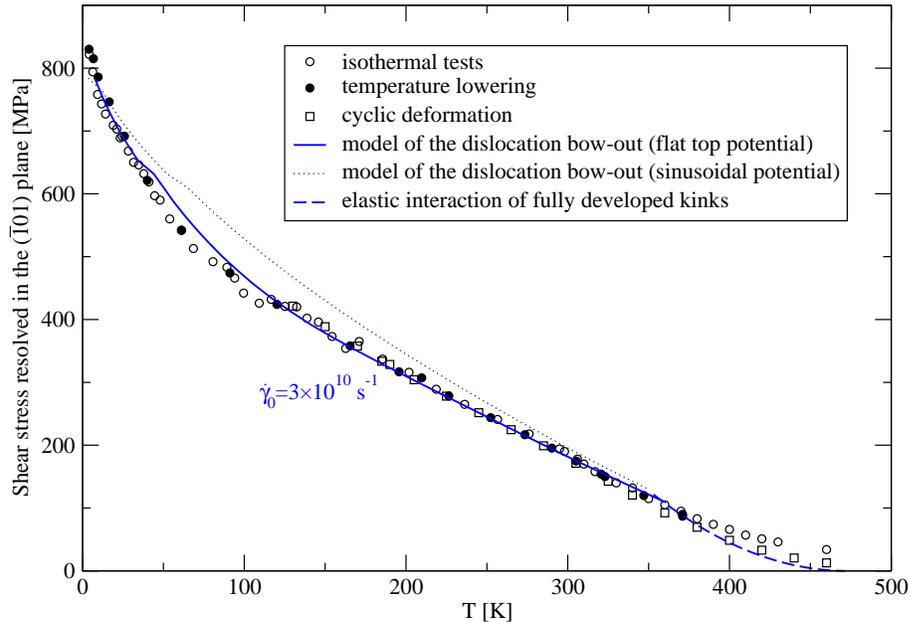}
  \parbox{14cm}{\caption{Temperature dependence of the yield stress for the $(\bar{1}01)[111]$
      system under loading in tension along $[\bar{1}49]$. The experimental data are from
      \citet{hollang:01}.}
    \label{fig_yieldt_hollang_-149_flattop}}
\end{figure}

Finally, the temperature dependence of the yield stress can be calculated similarly as for the
sinusoidal Peierls potential. We have shown earlier that the value of $\dot\gamma_0$ entering the
rate equation (\ref{eq_Eb_log}) cannot be calculated by extrapolation from experimental
data. Instead, an effective value of this parameter can be conveniently determined by requiring that
the temperature dependence of the yield stress, calculated from \refeq{eq_sigmaP}, approaches the
experimental data at high temperatures, which gave us $\dot\gamma_0=3\times10^{10}~\s^{-1}$. The
temperature dependence of the yield stress obtained using the modified Peierls potential
(\ref{eq_V3_flattop}), as well as the previously calculated curve for the sinusoidal potential, are
plotted in \reffig{fig_yieldt_hollang_-149_flattop}. One can clearly see that the saddle-point
perturbation significantly improves the agreement with experiment at low and intermediate
temperatures, where the sinusoidal potential led to a substantial disagreement. In other words, if
the Peierls potential exhibits flat saddle-points, the top of the activation barrier also flattens,
and this then gives rise to a strongly nonlinear temperature dependence of the yield stress that is
in good agreement with experiments.

%----------------------------------------------------------------------------------------------------
%----------------------------------------------------------------------------------------------------

\section{Dislocation glide on higher-index planes}

An isolated screw dislocation at 0~K moves by elementary steps on one of the three $\gplane{110}$
planes in the zone of the $\gdir{111}$ slip direction. This motion is now widely accepted as the
mechanism of slip at low temperatures. Nevertheless, it is often argued that the observation of the
$\gplane{110}$ slip in 0~K atomistic simulations does not automatically guarantee that this
mechanism of slip holds also at higher temperatures. For example, \citet{seeger:00} assert that the
dislocation core capable of gliding on $\gplane{110}$ planes at low temperatures undergoes, at
approximately 100~K, a transformation to another configuration that can move by elementary steps
directly on $\gplane{112}$ planes. However, this core transformation, as well as the hypothetical
elementary steps of the dislocation on $\gplane{112}$ planes, have never been observed in atomistic
simulations, and, therefore, the plane of elementary steps of the dislocation at temperatures above
100~K (for molybdenum) still remains a matter of debate.

\begin{figure}[!b]
  \centering
  \begin{minipage}{0.35\textwidth}
    \centering
    \includegraphics[width=5cm]{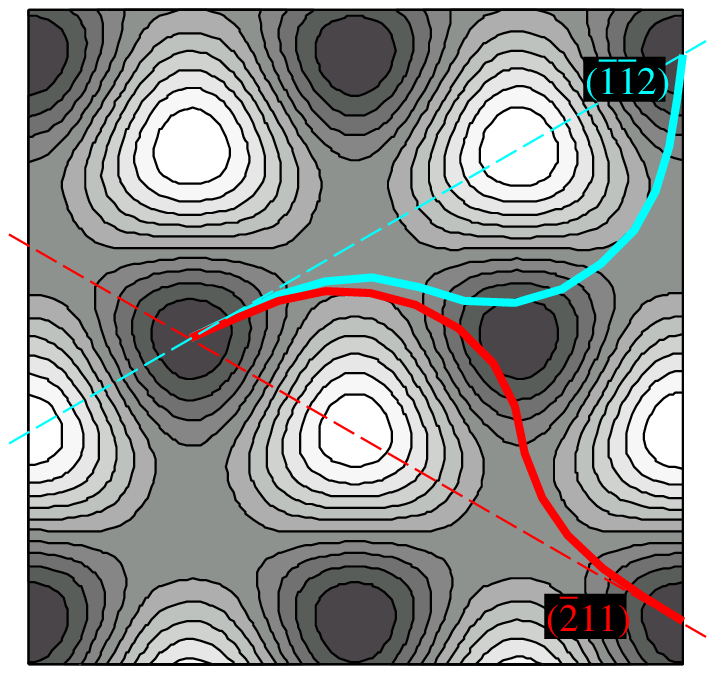} \\
    a) minimum energy paths for the two $\gplane{112}$ jumps
  \end{minipage}
  \hfill
  \begin{minipage}{0.64\textwidth} 
    \centering
    \includegraphics[width=9.5cm]{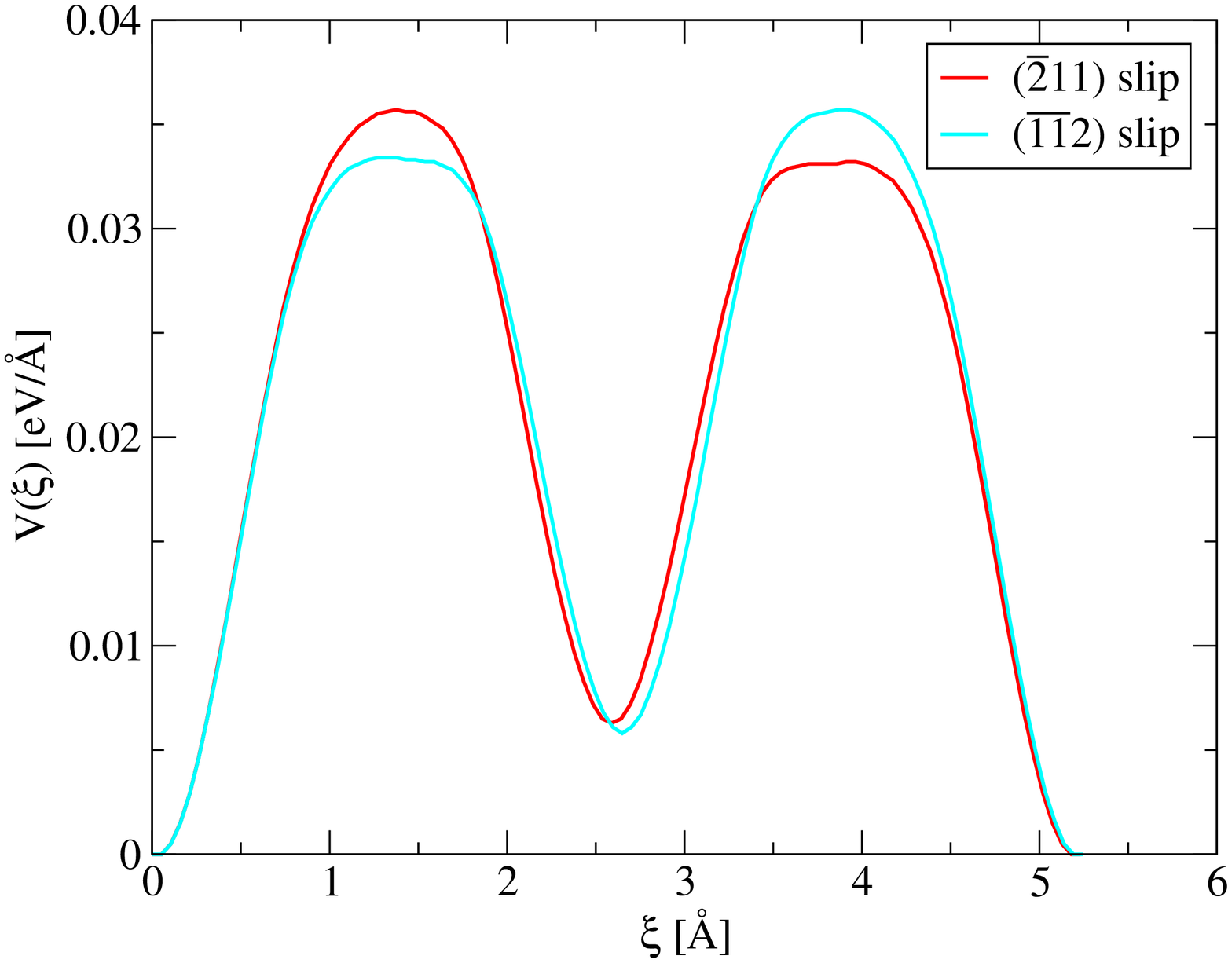} \\
    \hskip2em b) Peierls barriers along the MEPs shown in a)
  \end{minipage} \\[1em]
  \caption{Minimum energy paths calculated for two $\gplane{112}$ jumps using the NEB method (a) and
    the corresponding Peierls barriers measured along these paths (b).}
  \label{fig_V_chi0_tau0_flattop_211}
\end{figure}

In order to investigate the glide of the dislocation on $\gplane{112}$ planes, we calculated the
MEPs for slip on the two $\gplane{112}$ planes in the zone of the $[111]$ slip direction by
following the approach introduced earlier for the $\gplane{110}$ slip. For simplicity, we will
discuss here only the case when the MRSSP coincides with the $(\bar{1}01)$ plane and the shear
stress perpendicular to the slip direction is zero. Within the NEB method, the minimum energy path
for an elementary step of the dislocation on a $\gplane{112}$ plane can be obtained by providing the
original and the target position of the dislocation. In our case, these positions coincide with two
adjacent low-energy lattice sites on: (i) the $(\bar{2}11)$ plane at $\psi=+30\deg$, and (ii)
the $(\bar{1}\bar{1}2)$ plane at $\psi=-30\deg$. The calculated MEPs for these two transitions are
plotted in \reffig{fig_V_chi0_tau0_flattop_211}a. Due to the high potential hill on the
$(\bar{1}01)$ plane, both MEPs are initially identical and pass around the potential maximum on the
$(\bar{1}01)$ plane. After leaving the first (common) saddle-point, the two paths divert and
continue to the second saddle-point on the corresponding $\gplane{112}$ plane, close to the target
lattice site.

It is important to realize that the two MEPs are effectively composed of two steps, the first on the
$(\bar{1}01)$ plane and the second either on $(0\bar{1}1)$, if the target lattice site lies on the
$(\bar{1}\bar{1}2)$ plane, or on $(\bar{1}10)$ if it lies on the $(\bar{2}11)$ plane. Because the
Peierls potential is not a function of temperature, one may conclude that the glide of the
dislocation proceeds by elementary steps on the two $\gplane{110}$ planes at any temperature where
the plastic flow is induced by thermally activated motion of screw dislocations.

  \chapter{Temperature and strain rate dependent yield criterion for molybdenum}
\label{chap_yieldtemp}

\begin{flushright}
  Problems worthy of attack prove their worth by fighting back. \\
  \emph{Paul Erd\"os}
\end{flushright}

In the previous chapter, we have shown that the Peierls potential developed on the basis of the 0~K
effective yield criterion reproduces quite well the experimental measurements of the temperature
dependence of the yield stress. For a given orientation of the loading axis, the calculation of the
stress dependence of the activation enthalpy requires the following steps: (i) identification of
potentially operative slip systems, (ii) calculation of shear stresses parallel and perpendicular to
the slip direction for each potentially operative system, (iii) searching the MEP for slip on the
three $\gplane{110}$ planes of each system, and (iv) numerical integration of the activation
barriers along these MEPs. For practical calculations of macroscopic flow characteristics, it is
highly desirable to carry out these numerical calculations only once and approximate the obtained
dependencies using sufficiently accurate analytical relations. This idea is not new and was
originally used for formulations of the phenomenological laws of plastic deformation
\citep{kocks:75}. Using the approximate expression for the activation enthalpy, one can obtain the
temperature dependence of the yield stress by virtue of the formula $H(\sigma)=qkT$, where $q=\ln(
\dot{\gamma}_0/\dot{\gamma} )$ is a constant and $H(\sigma)$ is the obtained analytical expression
for the stress dependence of the activation enthalpy. If we now write the yield stress $\sigma$ as a
function of $\tau^*_{cr}$ and substitute this into the relation $H(\sigma)=qkT$, we arrive at the
temperature and strain rate dependent effective yield stress, $\tau^*_{cr}(T,\dot\gamma)$.

%----------------------------------------------------------------------------------------------------
%----------------------------------------------------------------------------------------------------

\section{Restricted model for slip at low temperatures}
\label{sec_yieldtemp_restr}

\begin{figure}[!htb]
  \centering
  \includegraphics[width=12cm]{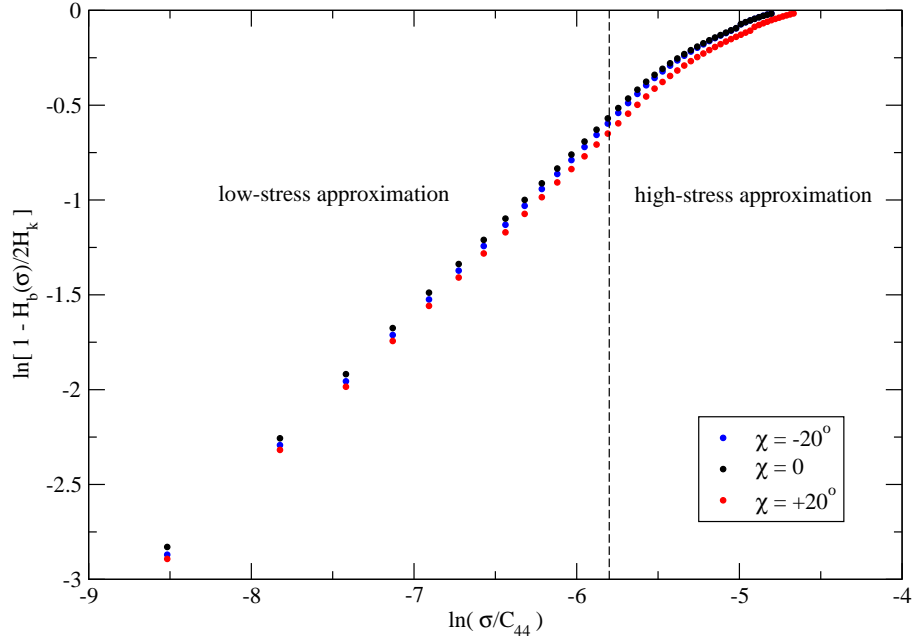}
  \parbox{14cm}{\caption{Stress dependence of the activation enthalpy calculated from the model of
      the dislocation bow-out, developed in Section~\ref{sec_bowout}. The yield stress, $\sigma$, is
      expressed as the CRSS applied in the MRSSP determined by the angle $\chi$.}
  \label{fig_eallcases_restr.data}}
\end{figure}

The key step that will allow us later to obtain the closed-form expression for the temperature and
strain rate dependence of the yield stress is to approximate the stress dependence of the activation
enthalpy using a simple analytical formula. An important requirement imposed on this equation is its
invertibility, i.e. the possibility to express the yield stress $\sigma$ as an explicit analytical
function of other parameters. For the sake of simplicity, we will initially consider loading by
pure shear stress parallel to the $[111]$ slip direction acting in the MRSSP whose orientation is
given by the angle $\chi$.  We will consider a minimal set of three discrete angles $\chi$, namely
$\chi=0$ and $\chi=\pm20\deg$, that essentially cover the region of orientations
$-30\deg<\chi<+30\deg$. Recall, that for any orientation of the applied shear stress, the
dislocation always moves on the $(\bar{1}01)$ plane. For a given angle $\chi$, we calculate the
stress dependence of the Dorn-Rajnak activation enthalpy, $H_b(\sigma)$, for an elementary jump of
the dislocation on the $(\bar{1}01)$ plane according to \refeq{eq_Hb_stress}. This yields three sets
of data points corresponding to $\chi=\{-20\deg,0\deg,+20\deg\}$ that are plotted in
\reffig{fig_eallcases_restr.data} in non-dimensional logarithmic representation. Here,
$2H_k=1.27~\eV$ is the energy of two isolated kinks. As the applied stress $\sigma$ approaches zero,
i.e. $\ln(\sigma/C_{44}) \rightarrow -\infty$, $H_b(\sigma) \rightarrow 2H_k$ and
$\ln[1-H_b(\sigma)/2H_k] \rightarrow -\infty$. At high stresses, as $\sigma$ approaches its maximum,
$H_b(\sigma) \rightarrow 0$ and $\ln[1-H_b(\sigma)/2H_k] \rightarrow 0$. The fact that
$\ln[1-H_b(\sigma)/2H_k]$ increases monotonically with $\ln(\sigma/C_{44})$ suggests that a simple
analytical expression for the stress dependence of the activation enthalpy can be constructed.

The goal is now to approximate the dependencies shown in \reffig{fig_eallcases_restr.data} using
primitive functions that can be inverted to obtain a simple expression for the yield stress
$\sigma$. Due to the change in the slope of the curve, it is necessary to consider two functional
forms. At low stresses, the dependence is almost linear which allows us to write
\begin{equation}
  \ln \left[ 1-\frac{H_b(\sigma)}{2H_k} \right] = \ln{a} + b\ln \left( \frac{\sigma}{C_{44}} \right) \ ,
  \label{eq_eallcases_restr_low}
\end{equation}
where $a$ and $b$ are two adjustable parameteres. At high stresses, one could in principle use the
same functional form with different parameters $a$ and $b$. However, this regime is more closely
approximated by an exponential
\begin{equation}
  \ln \left[ 1-\frac{H_b(\sigma)}{2H_k} \right] = \ln{a'} - b'{\rm e}^{-\sigma/C_{44}} \ ,
  \label{eq_eallcases_restr_high}
\end{equation}
where we explicitly use the minus sign in front of the second term in the right-hand side to
emphasize the concave character of the function. For each of the three angles $\chi$, the
coefficients $a,b$ are determined by fitting the low-stress data plotted in
\reffig{fig_eallcases_restr.data}. Similarly, the other two coefficients, $a'$ and $b'$, are
determined by fitting the high-stress data. The cross-over between the low and the high stress
regime is approximately at $\ln(\sigma/C_{44})\approx -5.8$ which corresponds to
$\sigma/C_{44}\approx0.003$. The coefficients $a,b,a',b'$ for the two approximations are given
numerically in Appendix~\ref{apx_thermalpar}.

\begin{figure}[!htb]
  \centering
  \includegraphics[width=12cm]{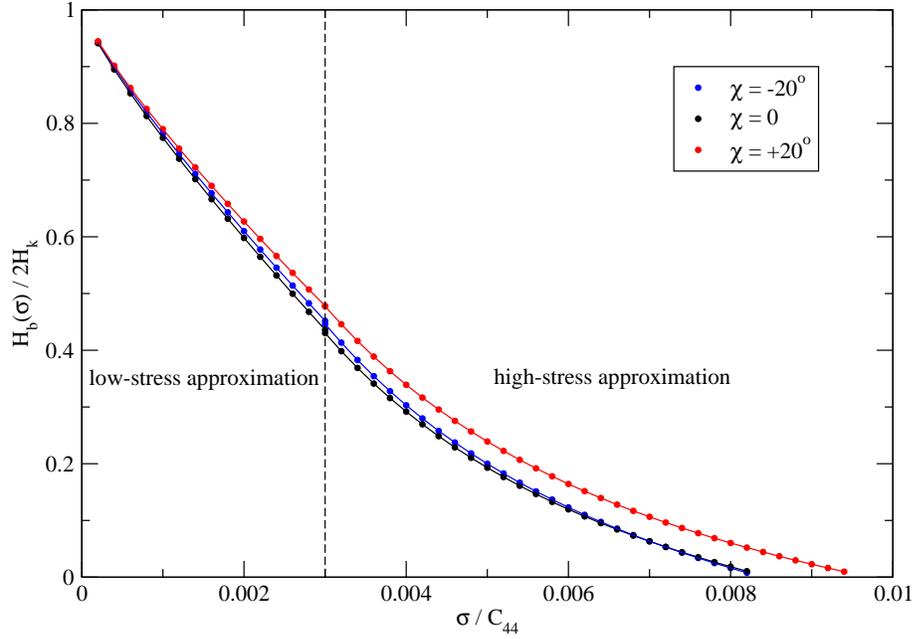}
  \parbox{14cm}{\caption{Stress dependence of the activation enthalpy calculated from the model of
      the dislocation bow-out (dots). The curves are calculated from
      \refeq{eq_eallcases_restr_both}.}
  \label{fig_eallcases_restr_interp}}
\end{figure}

The difference in the data for the three angles $\chi$, plotted in
\reffig{fig_eallcases_restr.data}, implies that the stress dependence of the activation enthalpy
depends on the orientation of the MRSSP and, therefore, the four coefficients $a,b,a',b'$, obtained
by fitting, are clearly functions of the angle $\chi$. Since we now have for each coefficient three
values, corresponding to the three angles $\chi=\{-20\deg,0,+20\deg\}$, we can approximate the
dependence of each coefficient on $\chi$ by a cubic polynomial. For example, $a(\chi) = a_0 +
a_1\chi + a_2\chi^2$, where the constants $a_0,a_1,a_2$ are determined by fitting. Finally, the
approximate expression for the stress dependence of the activation enthalpy can be obtained from
\refeqs{eq_eallcases_restr_low} and \ref{eq_eallcases_restr_high}, and reads
\begin{equation}
  \frac{H_b(\sigma)}{2H_k} = \left\{
  \begin{array}{ll}
    1 - a(\chi) \left( \sigma/C_{44} \right)^{b(\chi)} & 
    ,\quad {\rm for}\ \sigma/C_{44}\leq0.003 \\
    1 - a'(\chi) \exp\left[ -b'(\chi)(\sigma/C_{44})^{-1} \right] & 
    ,\quad {\rm for}\ \sigma/C_{44}\geq0.003 \ .
  \end{array} \right.
  \label{eq_eallcases_restr_both}
\end{equation}

The stress dependencies of the activation enthalpy, $H_b(\sigma)/2H_k$, predicted from
\refeq{eq_eallcases_restr_both}, are shown in \reffig{fig_eallcases_restr_interp} as solid lines;
the dots correspond to the data plotted in \reffig{fig_eallcases_restr.data}. The excellent
agreement with the data calculated directly from the model of the dislocation bow-out
(Section~\ref{sec_bowout}) proves that \refeq{eq_eallcases_restr_both} is indeed a good
approximation of the stress dependence of the activation enthalpy for a wide range of orientations
of the MRSSP. As the applied stress approaches zero, the activation enthalpy becomes equal to the
energy of two isolated kinks, $2H_k$. At large stresses approaching the Peierls stress, the
activation enthalpy vanishes and the dislocation is moved purely mechanically by the action of the
applied stress.

\begin{figure}[!htb]
  \centering
  \includegraphics[width=12cm]{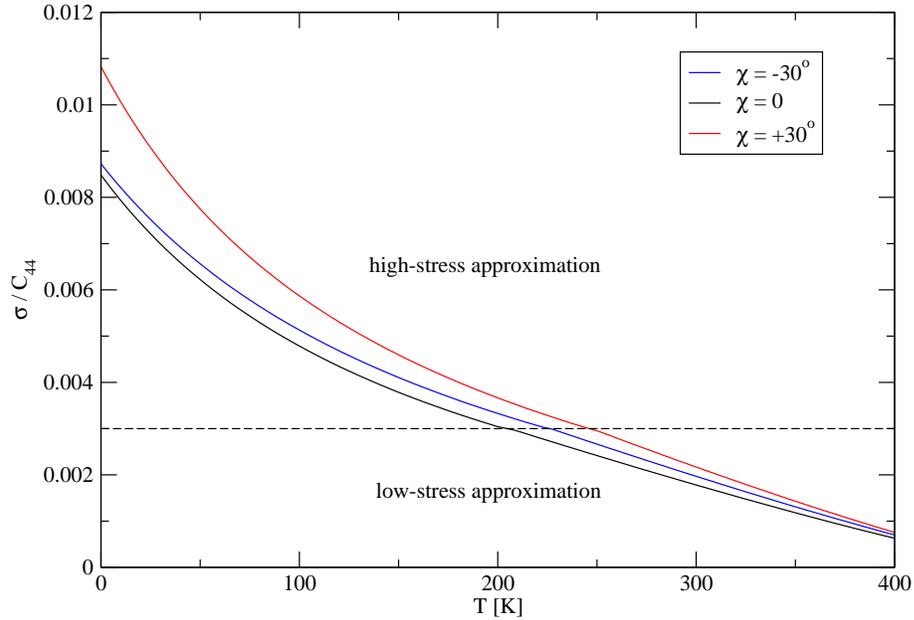}
  \parbox{14cm}{\caption{Temperature dependence of the shear stress $\sigma$ parallel to the slip
      direction and acting in the MRSSP, as predicted from \refeq{eq_eallcases_restr_sigma-T}.}
  \label{fig_eallcases_restr_sigma-T}}
\end{figure}

The temperature dependence of the yield stress $\sigma$ can now be obtained by recalling the
Arrhenius law for the total plastic strain rate, \refeq{eq_gammadot}. For a given temperature $T$,
the activation enthalpy can be written as $H_b(\sigma)=qkT$. Substituting $H_b(\sigma)$ from
\refeq{eq_eallcases_restr_both} and recovering $\sigma$ then provides the temperature and strain
rate dependence of the yield stress:
\begin{equation} 
  \sigma/C_{44} = \left\{
  \begin{array}{ll}
    \left[ \frac{1-qkT/2H_k}{a(\chi)} \right]^{1/b(\chi)} & 
    ,\quad {\rm for}\ \sigma/C_{44}\leq0.003 \\
    -b'(\chi) \left/ \ln \left[ \frac{1-qkT/2H_k}{a'(\chi)} \right] \right. & 
    ,\quad {\rm for}\ \sigma/C_{44}\geq0.003 \ .
  \end{array} \right.
  \label{eq_eallcases_restr_sigma-T}
\end{equation}
This dependence is shown graphically in \reffig{fig_eallcases_restr_sigma-T} for $\chi=0$ and
$\chi=\pm30\deg$ that represent the boundaries of the angular region of all independent orientations
of the MRSSP. The most important feature to note is that non-glide stresses, although so far only
those parallel to the slip direction, play an important role at low temperatures and gradually
weaken as the temperature is raised. This can be clearly seen on the magnitudes of $\sigma$ for the
three MRSSPs which are very different at low temperatures but virtually identical as $T \rightarrow
400~\K$. At $T=0~\K$, larger shear stress is needed to move the dislocation on the $(\bar{1}01)$
plane if the crystal is sheared in the antitwinning sense ($\chi>0$) than if the shear is applied in
the twinning sense ($\chi<0$). This observation is fully consistent with the results of our 0~K
atomistic simulations, particularly with \reffig{fig_CRSS_chi_MoBOP}.

\begin{figure}[!b]
  \centering
  \includegraphics[width=12cm]{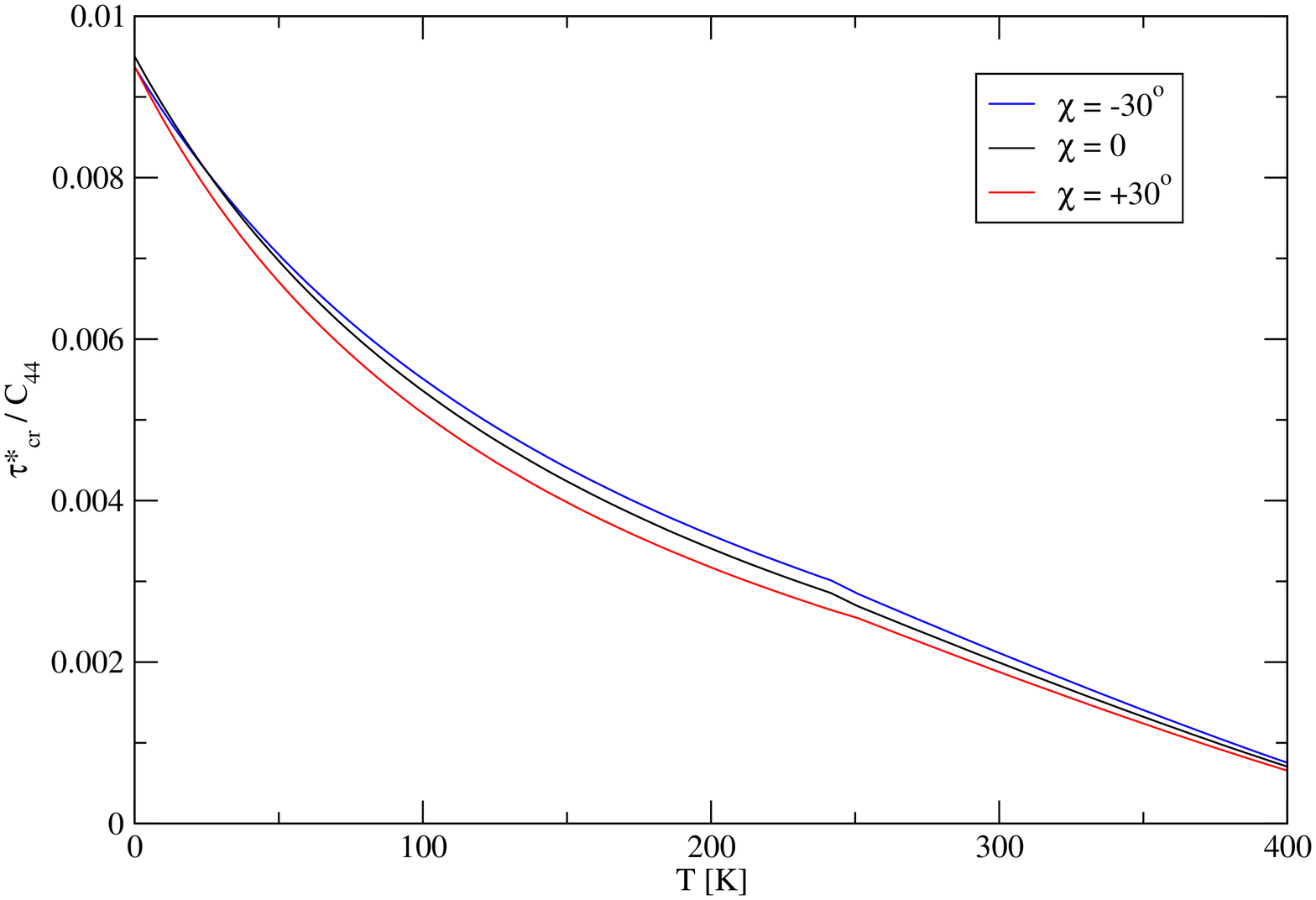}
  \parbox{14cm}{\caption{Temperature dependence of $\tau^*_{cr}$ predicted from
  \refeq{eq_eallcases_restr_tstarC-T}.}
  \label{fig_eallcases_restr_tstarC-T}}
\end{figure}

The expression for the temperature and strain rate dependence of the effective yield stress,
$\tau^*_{cr}(T,\dot{\gamma})$ can now be obtained by noting that, within the restricted model,
$\tau^*_{cr}$ can be written as a product $\tau^*_{cr}=\sigma\, t(\chi)$, where
$t(\chi)=\cos\chi+a_1\cos(\chi+\pi/3)$ and $\sigma$ is the yield stress. Substituting $\sigma$ from
\refeq{eq_eallcases_restr_sigma-T} yields
\begin{equation} 
  \tau^*_{cr}/C_{44} = \left\{
  \begin{array}{ll}
    t(\chi) \left[ \frac{1-qkT/2H_k}{a(\chi)} \right]^{1/b(\chi)} & 
    ,\quad {\rm for}\ \sigma/C_{44}\leq0.003 \\
    -b'(\chi)t(\chi) \left/ \ln \left[ \frac{1-qkT/2H_k}{a'(\chi)} \right] \right. & 
    ,\quad {\rm for}\ \sigma/C_{44}\geq0.003 \ .
  \end{array} \right.
  \label{eq_eallcases_restr_tstarC-T}
\end{equation}
The temperature dependence of $\tau^*_{cr}$ for the three angles $\chi$ is shown in
\reffig{fig_eallcases_restr_tstarC-T}. In contrast to $\sigma(T)$, plotted in
\reffig{fig_eallcases_restr_sigma-T}, $\tau^*_{cr}$ is almost independent of the orientation of the
MRSSP, and, therefore, \refeq{eq_eallcases_restr_tstarC-T} can be closely approximated as a function
of temperature and strain rate only, without any dependence on $\chi$.

%----------------------------------------------------------------------------------------------------
%----------------------------------------------------------------------------------------------------

\section{Full model for slip at low temperatures}
\label{sec_yieldtemp_full}

The extension of the restricted model presented in the last section to account for the effect of the
shear stress perpendicular to the slip direction can now be developed in a straightforward
manner. For the sake of simplicity, we will consider three different combinations of the shear
stresses perpendicular and parallel to the slip direction applied in the three MRSSPs at
$\chi=\{-20\deg,0,+20\deg\}$, namely $\eta=\{-0.5,0,+0.5\}$, where $\eta=\tau/\sigma$ and $\tau$
is the shear stress perpendicular to the slip direction. It is important to realize that for any
combination of $\chi$ and $\eta$ from these sets, the dislocation always moves on the $(\bar{1}01)$
plane, and, therefore, one does not have to consider the possibility of the glide of the dislocation
on the other two $\gplane{110}$ planes of the $[111]$ zone. The activation enthalpy for an
elementary step of the dislocation on the $(\bar{1}01)$ plane can be calculated directly from
\refeq{eq_Hb_stress}, the only difference from the loading by pure shear parallel to the slip
direction being that now the Peierls potential changes shape due to both shear stresses perpendicular
and parallel to the slip direction.

\begin{figure}[!htb]
  \centering
  \includegraphics[width=15cm]{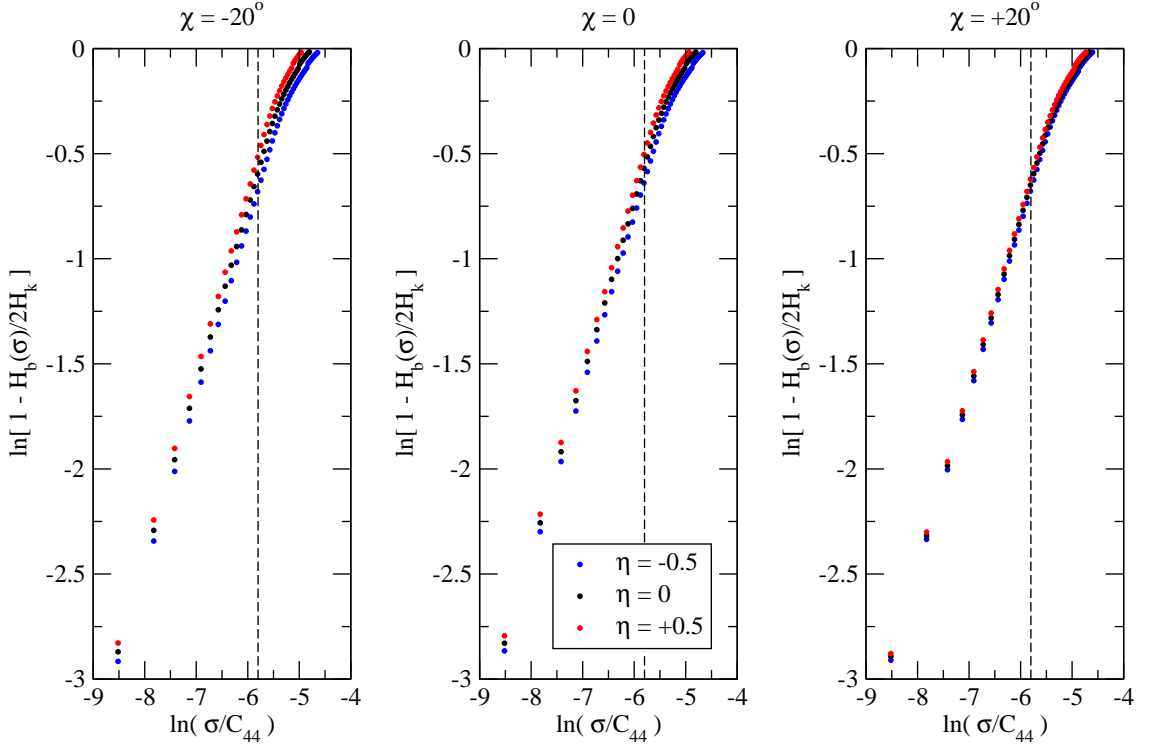}
  \caption{Stress dependence of the activation enthalpy calculated using the model of the
    dislocation bow-out for $\chi=\{-20\deg,0,+20\deg\}$ and $\eta=\{-0.5,0,0.5\}$.}
  \label{fig_eallcases_full.data}
\end{figure}

The stress dependence of the activation enthalpy calculated from the model of the dislocation
bow-out (Section~\ref{sec_bowout}) for all nine combinations of $\chi$ and $\eta$ from the sets
above is shown in \reffig{fig_eallcases_full.data}. One can again see that, if the dependence is
plotted in logarithmic representation, the low stress behavior can be accurately approximated by a
straight line, whereas the high-stress regime has to be described using an inverse
exponential. Moreover, the effect of the shear stress perpendicular to the slip direction is
significant for $\chi=-20\deg$ but almost vanishes for $\chi=+20\deg$. This observation is
consistent with the results of 0~K atomistic simulations plotted in \reffig{fig_CRSS_tau_MoBOP}a-d,
where the slope $|\partial{\CRSS}/\partial{\tau}|_{\tau=0}$ is largest for negative $\chi$ and
almost vanishes at positive $\chi$.  Furthermore, the fact that the shear stress perpendicular to
the slip direction, i.e. nonzero value of $\eta$, does not lead to the change of the shape of the
dependence in \reffig{fig_eallcases_full.data} implies the same approximate forms for the activation
enthalpy as used in the restricted model in Section~\ref{sec_yieldtemp_restr}. The only additional
complication in this case is that the coefficients $a,b$ in the low-stress approximation, and $a',b'$
in the high-stress approximation, are generally functions of both $\chi$ and $\eta$, i.e.
\begin{equation}
  \frac{H_b(\sigma)}{2H_k} = \left\{
  \begin{array}{ll}
    1 - a(\chi,\eta) \left( \sigma/C_{44} \right)^{b(\chi,\eta)} & 
    ,\quad {\rm for}\ \sigma/C_{44}\leq0.003 \\
    1 - a'(\chi,\eta) \exp \left[ - b'(\chi,\eta) (\sigma/C_{44})^{-1} \right] & 
    ,\quad {\rm for}\ \sigma/C_{44}\geq0.003 \ .
  \end{array} \right.
  \label{eq_eallcases_full_both}
\end{equation}
Each of the four functions $a,b,a',b'$ can again be expressed as a second-order polynomial in $\chi$
with coefficients defined by another second-order polynomial in $\eta$. For example,
\begin{equation}
  a(\chi,\eta) = a_0(\eta) + a_1(\eta)\chi + a_2(\eta)\chi^2 \ ,
\end{equation}
where the coefficient functions $a_0(\eta)$, $a_1(\eta)$ and $a_2(\eta)$ can be written as
\begin{equation}
  \left[
    \begin{array}{c}
      a_0 \\ a_1 \\ a_2
    \end{array}
  \right] = \left[
    \begin{array}{ccc}
      a_{00} & a_{01} & a_{02} \\
      a_{10} & a_{11} & a_{12} \\
      a_{20} & a_{21} & a_{22}
    \end{array}
  \right] \left[
    \begin{array}{c}
      1 \\ \eta \\ \eta^2
    \end{array} \right] \ .
  \label{eq_a0a1a2_eta}
\end{equation}
The remaining three functions $b,a',b'$ can be expressed in a similar fashion. Note, that $a_{ij}$
in \refeq{eq_a0a1a2_eta} are independent adjustable constants that have to be determined by fitting
the stress dependence of the activation enthalpy shown in \reffig{fig_eallcases_full.data}. Since we
have calculated $H_b(\sigma)$ for three different angles $\chi$ and three ratios $\eta$, we have
exactly nine conditions required for each low and high stress approximation that are needed to
determine the constants $a_{ij}$, $b_{ij}$, $a'_{ij}$, $b'_{ij}$ unambiguously. For convenience, we
list the numerical values of these coefficients in Appendix~\ref{apx_thermalpar}.

\begin{figure}[!htb]
  \centering
  \includegraphics[width=15cm]{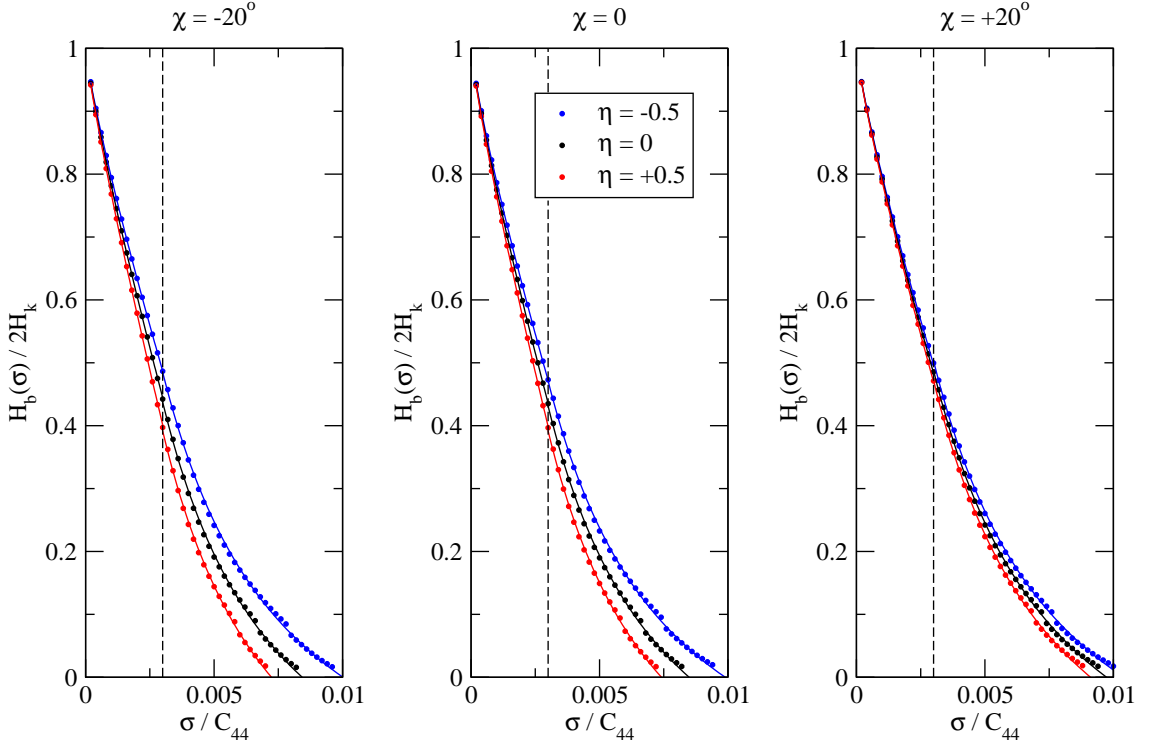}
  \caption{Stress dependence of the activation enthalpy calculated using the model of the
    dislocation bow-out (dots) and from the approximation (\ref{eq_eallcases_full_both})
    (curves).}
  \label{fig_eallcases_full.interp}
\end{figure}

The accuracy of fitting the stress dependence of the activation enthalpy by
\refeq{eq_eallcases_full_both} can be easily checked by calculating the dependence $H_b(\sigma)$
directly from this equation and comparing it with the data shown in
\reffig{fig_eallcases_full.data}. This comparison is presented in
\reffig{fig_eallcases_full.interp}, where the dots represent the original data calculated from the
model of the dislocation bow-out (Section~\ref{sec_bowout}) and the lines are the approximations
calculated from \refeq{eq_eallcases_full_both}. The magnitudes of the yield stress $\sigma$,
corresponding to zero activation enthalpy, agree well with the values of the CRSS obtained from
atomistic studies on a single isolated screw dislocation at 0~K. For example, if the MRSSP lies at
$\chi=0$ and $H_b(\sigma)=0$, one can clearly see that
$\sigma(\eta=-0.5)>\sigma(\eta=0)>\sigma(\eta=+0.5)$, and the yield stress at negative $\tau$ is thus
larger than that at positive $\tau$. This conclusion is consistent with the $\CRSS-\tau$ data
plotted in \reffig{fig_CRSS_tau_MoBOP}a. The same is true for the remaining two angles $\chi$.

\begin{figure}[!htb]
  \centering
  \includegraphics[width=12cm]{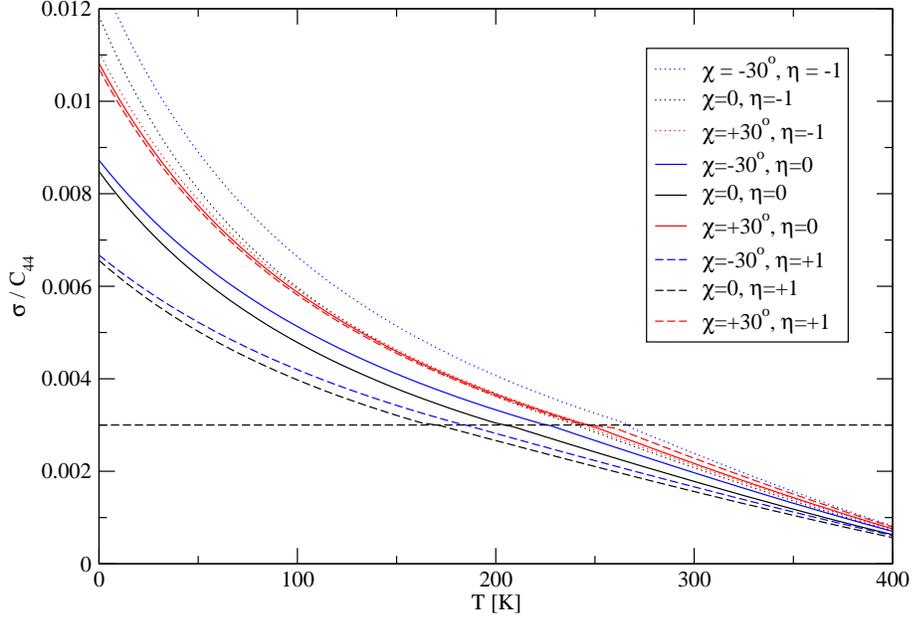}
  \parbox{14cm}{\caption{Temperature dependence of the yield stress $\sigma$, acting in the MRSSP,
      as calculated from \refeq{eq_eallcases_full_sigma-T} for three different angles $\chi$ and three
      ratios $\eta$.}
  \label{fig_eallcases_full_sigma-T}}
\end{figure}

The temperature dependence of the yield stress, $\sigma$, can be obtained similarly as in the
restricted model presented in the previous section. Substitution of $H_b(\sigma)$ from
\refeq{eq_eallcases_full_both} into the relation $H_b(\sigma)=qkT$ yields
\begin{equation} 
  \sigma/C_{44} = \left\{
  \begin{array}{ll}
    \left[ \frac{1-qkT/2H_k}{a(\chi,\eta)} \right]^{1/b(\chi,\eta)} & 
    ,\quad {\rm for}\ \sigma/C_{44}\leq0.003 \\
    -b'(\chi,\eta) \left/ \ln \left[ \frac{1-qkT/2H_k}{a'(\chi,\eta)} \right] \right. & 
    ,\quad {\rm for}\ \sigma/C_{44}\geq0.003 \ .
  \end{array} \right.
  \label{eq_eallcases_full_sigma-T}
\end{equation}
For convenience, we plot in \reffig{fig_eallcases_full_sigma-T} the temperature dependence of
$\sigma$ calculated for nine different combinations of $\chi$ and $\eta$ from the sets
$\chi={-30\deg,0,+30\deg}$ and $\eta={-1,0,1}$. At low temperatures, the non-glide stresses play 
dominant role in that the magnitude of $\sigma$ strongly depends on both $\chi$ and $\eta$. The
range of stresses at 0~K is now much wider than that calculated from the restricted model and
plotted in \reffig{fig_eallcases_restr_sigma-T}, which is due to nonzero shear stress perpendicular
to the slip direction. As the temperature increases, all stresses gradually decay and thus the
importance of the non-glide stresses diminishes.

At this point, it is important to emphasize that \reffig{fig_eallcases_full_sigma-T} shows the
temperature dependence of the yield stress, $\sigma$, corresponding to a set of limiting
combinations of $\chi$ and $\eta$. Due to the crystal symmetry, it is sufficient to consider the
MRSSPs in the angular region $-30\deg\leq\chi\leq+30\deg$. Similarly, $-1\leq\eta\leq1$ defines the
bounds of the loading paths plotted in the $\CRSS-\tau$ projection for which $(\bar{1}01)[111]$ is
the most operative $\gplane{110}\gdir{111}$ system. If $|\eta|>1$, which means that the magnitude of
shear stress perpendicular to the slip direction is larger than the corresponding shear stress
parallel to the slip direction, \reffig{fig_CRSS_tau_fit_MoBOP}a-c imply that another
$\gplane{110}\gdir{111}$ system becomes most operative. Importantly, the ratio $\eta$ calculated for
the MRSSP of this latter system always falls into the bounds $\langle -1; 1 \rangle$ unless the
loading axis is very close to the $[\bar{1}11]$ corner of the stereographic triangle (see
\reffig{fig_sgtria_eta}). Since all $\gplane{110}\gdir{111}$ systems are mutually equivalent and
provided that we know both the angle $\chi$ and the corresponding ratio $\eta$ for each of these
systems, \reffig{fig_eallcases_full_sigma-T} can be used to find the magnitude of the shear stress
parallel to the slip direction for which each individual $\gplane{110}\gdir{111}$ system becomes
operative at temperature $T$ and strain rate $\dot\gamma$.

\begin{figure}[!b]
  \centering
  \includegraphics[width=12cm]{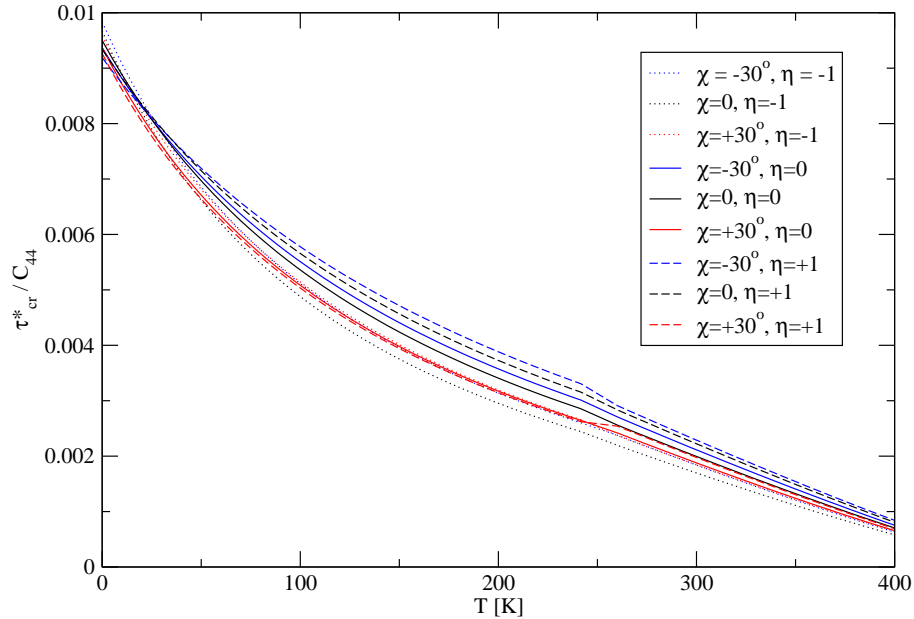}
  \parbox{14cm}{\caption{Temperature dependence of the effective yield stress, $\tau^*_{cr}$,
  calculated for a fixed strain rate (corresponding to $q=31.2$) and various combinations of $\chi$
  and $\eta$.}
  \label{fig_eallcases_full_tstarC-T}}
\end{figure}

Finally, \refeq{eq_eallcases_full_sigma-T} can be used to derive the temperature and strain rate
dependent effective yield stress, $\tau^*_{cr}(T,\dot\gamma)$. In this case, $\tau^*_{cr}=\sigma
t(\chi,\eta)$, where $t(\chi,\eta) = \cos\chi + a_1\cos(\chi+\pi/3) + \eta[ a_2\sin2\chi +
a_3\cos(2\chi+\pi/6) ]$ is now a function of both the angle of the MRSSP, $\chi$, and the ratio
$\eta=\tau/\sigma$. Substituting in this equation the shear stress $\sigma$ from
\refeq{eq_eallcases_full_sigma-T} yields
\begin{equation} 
  \tau^*_{cr}/C_{44} = \left\{
  \begin{array}{ll}
    t(\chi,\eta) \left[ \frac{1-qkT/2H_k}{a(\chi,\eta)} \right]^{1/b(\chi,\eta)} & 
    ,\quad {\rm for}\ \sigma/C_{44}\leq0.003 \\
    -b'(\chi,\eta)t(\chi,\eta) \left/ \ln \left[ \frac{1-qkT/2H_k}{a'(\chi,\eta)} \right] \right. & 
    ,\quad {\rm for}\ \sigma/C_{44}\geq0.003 \ .
  \end{array} \right.
  \label{eq_eallcases_full_tstarC-T}
\end{equation} 
This is the sought expression for the effective yield stress $\tau^*_{cr}(T,\dot{\gamma})$ that is
plotted for $q=31.2$ in \reffig{fig_eallcases_full_tstarC-T}. In contrast to $\sigma-T$, shown in
\reffig{fig_eallcases_full_sigma-T}, that was strongly dependent on non-glide stresses within a
broad range of temperatures, $\tau^*_{cr}$ depends on both $\chi$ and $\eta$ at intermediate
temperatures but this dependence almost vanishes close to 0~K and 400~K.

If the shear stress perpendicular to the slip direction is very different from zero, $\tau^*_{cr}$
is not only a function of temperature and strain rate but, at intermediate temperatures, also quite
strong function of both $\chi$ and $\eta$. This is best illustrated by realizing that the range of
values of $\tau^*_{cr}$ close to 200~K is $\Delta\tau^*_{cr}=0.001C_{44}$ which, for molybdenum, is
approximately 100~\MPa. This value represents the maximum error that may occur if one neglects the
dependence of $\tau^*_{cr}$ on $\eta$.

%----------------------------------------------------------------------------------------------------
%----------------------------------------------------------------------------------------------------

\section{High temperature, low stress regime}
\label{sec_highT_approx}

At high temperatures, the configuration corresponding to the saddle-point energy consists of a pair
of fully developed interacting kinks at a finite distance $\Delta{z}$, whose attraction is opposed
by the applied Schmid stress. The activation enthalpy to nucleate this critical configuration is
given by \refeq{eq_Hkp_final}, and, when substituted into $H_{kp}(\sigma)=qkT$, provides the
temperature and strain rate dependence of the yield stress $\sigma$, i.e. the shear stress parallel
to the slip direction applied in the MRSSP with angle $\chi$:
\begin{equation}
   \sigma = \frac{2\pi}{a_0^3 \cos(\chi-\psi)} \frac{(2H_k)^2}{\mu b^3} \left( 1 - \frac{q k T}{2H_k}
   \right)^2 \ .
   \label{eq_lowstress_sigma-T}
\end{equation}
Here, $\psi$ is the angle of the slip plane, and $a_0$ the distance between two adjacent sites in
the slip plane. For example, if we consider that the dislocation at elevated temperatures moves by
elementary jumps on the $\gplane{110}$ plane, similarly as at low temperatures, $a_0=a\sqrt{2/3}$,
where $a$ is the $\gdir{100}$ lattice parameter.  The temperature and strain rate dependence of the
effective yield stress, $\tau^*_{cr}(T,\dot\gamma)$, can then be obtained the same way as in the
low-temperature approximations presented in Sections~\ref{sec_yieldtemp_restr} and
\ref{sec_yieldtemp_full}. Noting that $\tau^*_{cr}=\sigma\, t(\chi,\eta)$, where $t(\chi,\eta)$ has
been given in the last section, and substituting $\sigma$ from \refeq{eq_lowstress_sigma-T}, yields
\begin{equation}
    \tau^*_{cr} = \frac{2\pi t(\chi,\eta)}{a_0^3 \cos(\chi-\psi)} 
    \frac{(2H_k)^2}{\mu b^3} \left( 1 - \frac{q k T}{2H_k}\right)^2 \ .
   \label{eq_lowstress_tstarC-T} 
\end{equation}
The thermal component of the yield stress, $\sigma$, as well as $\tau^*_{cr}$, vanish at the
temperature $T_k=2H_k/qk$. Because \refeq{eq_lowstress_tstarC-T} is applicable only at high
temperatures where non-glide stresses play only negligible role and thus the ratio
$t(\chi,\eta)/\cos(\chi-\psi)$ is of the order of one, the last equation reduces to
\begin{equation}
    \tau^*_{cr} \approx \frac{2\pi}{a_0^3} 
    \frac{(2H_k)^2}{\mu b^3} \left( 1 - \frac{q k T}{2H_k}\right)^2 \ ,
   \label{eq_lowstress_tstarC-T_eff} 
\end{equation}
which is only a function of temperature and strain rate but \emph{not} of the orientation of the
MRSSP or of the magnitude of the shear stress perpendicular to the slip direction.

Theoretically, the high-temperature approximation (\ref{eq_lowstress_tstarC-T_eff}) should be used
in molybdenum only at temperatures $T\approx\langle 360;460\rangle~\K$. However, at these
temperatures, \refeq{eq_lowstress_tstarC-T_eff} gives values of $\tau^*_{cr}$ that are only within
$5~\MPa$ from those predicted by \refeq{eq_eallcases_full_tstarC-T}. For practical purposes, one may,
therefore, safely use the approximations of the data obtained from the model of the dislocation
bow-out that were derived in Section~\ref{sec_yieldtemp_full}.

%----------------------------------------------------------------------------------------------------
%----------------------------------------------------------------------------------------------------

\section{Activity of individual slip systems derived from the $\tau^*$ criterion}

The temperature and strain rate dependent effective yield criterion in which $\tau^*_{cr}$ is given
by \refeq{eq_eallcases_full_tstarC-T} can now be used to predict the activity of individual slip
systems. For a given orientation and character of applied loading, this can be achieved by
calculating the positions of the critical lines in the $\CRSS-\tau$ projection that determine the
onset of slip on each individual slip system. For a given orientation of the MRSSP, we will consider
a number of straight loading paths in the $\CRSS-\tau$ plot that correspond to a set of uniaxial
loadings. For each loading path, we will find the orientation of the MRSSP in the zone of each
$\langle{111}\rangle$ slip direction and eliminate those with $|\chi|>30\deg$ or negative resolved
shear stress parallel to the slip direction. For finite values of $\eta$, i.e. if the shear stress
parallel to the slip direction applied in the MRSSP of the $(\bar{1}01)[111]$ system is not zero,
this process yields four potentially operative systems $\alpha$. For each system $\alpha$, one can
easily calculate how large the CRSS in the MRSSP of this system has to be in order to activate the
slip on this system at temperature $T$ and plastic strain rate $\dot\gamma$, i.e. to satisfy
$\tau^*=\tau^*_{cr}(T,\dot\gamma)$. By repeating this process for all straight loading paths in the
chosen MRSSP, one obtains a set of points defining the critical lines for activation of individual
slip systems. Recall that the inner envelope of these lines coincides with the yield surface.

\begin{figure}[!htb]
  \centering \includegraphics[width=12cm]{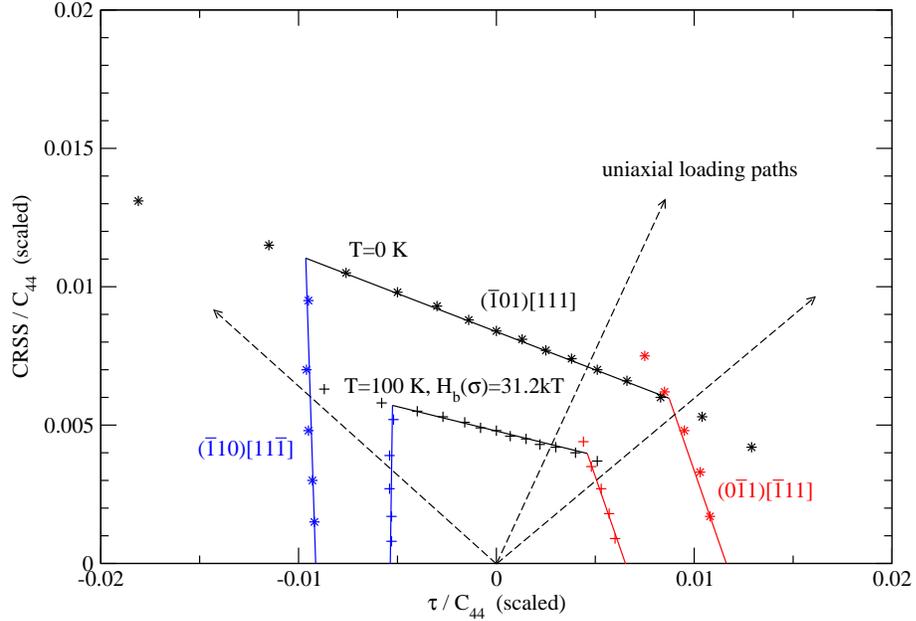}
  \parbox{14cm}{\caption{Combinations of shear stresses perpendicular and parallel to the slip
  direction that induce yielding of real single crystals of bcc molybdenum. The calculation is done
  solely using the temperature and strain rate dependent effective yield criterion in which
  $\tau^*_{cr}(T,\dot\gamma)$ is obtained from \refeq{eq_eallcases_full_tstarC-T}.}
  \label{fig_chi0_MoBOP_temp}}
\end{figure}

The critical lines calculated using the above-mentioned procedure for $\chi=0$ are shown in
\reffig{fig_chi0_MoBOP_temp} for $T=0~\K$ and $T=100~\K$; in the latter case, the activation
enthalpy is chosen as $H_b(\sigma)=31.2kT$ which corresponds to the strain rate
$\dot\gamma=8.6\times10^{-4}~\s^{-1}$ used in the experiments of \citet{hollang:97}. In the limit of
zero temperature, the critical lines correspond to those obtained earlier and plotted in
\reffig{fig_chi0_ysurf_MoBOP}. Due to the scaling of stresses that we introduced earlier in the
calculation of the activation enthalpy, the stresses $\tau$ and $\sigma$ are now smaller than those
in \reffig{fig_chi0_ysurf_MoBOP} and correspond to the magnitudes of stresses encountered in
experiments. As one expects, at higher temperatures and/or lower strain rates $\dot\gamma$
(i.e. larger $q$), the stress to move the dislocation decreases and, therefore, the yield surface
shrinks. This can be seen clearly in \reffig{fig_chi0_MoBOP_temp}, where we plot the critical lines
for the primary slip systems only.

It is important to note that more than two slip systems can be activated if the loading path
intersects the yield polygon in \reffig{fig_chi0_MoBOP_temp} at one of its corners. This can be seen
also in the $\pi$-plane projection of the yield surface in \reffig{fig_yieldsurf_allsys_MoBOP}.

%----------------------------------------------------------------------------------------------------
%----------------------------------------------------------------------------------------------------

\section{Comparisons with experiments}

We have shown in Section \ref{sec_expt_hollang} that the temperature dependence of the yield stress
can be obtained theoretically by first calculating numerically the stress dependence of the
activation enthalpy and then inverting it to recover the yield stress. This tedious process can now
be replaced by using the derived formulas for $\sigma(T,\dot\gamma)$ or $\tau^*_{cr}(T,\dot\gamma)$
which, for uniaxial loading, only require a substitution of temperature, strain rate, angle $\chi$
of the MRSSP, and ratio $\eta=\tau/\sigma$.

\begin{figure}[!htb]
  \centering 
  \includegraphics[width=12cm]{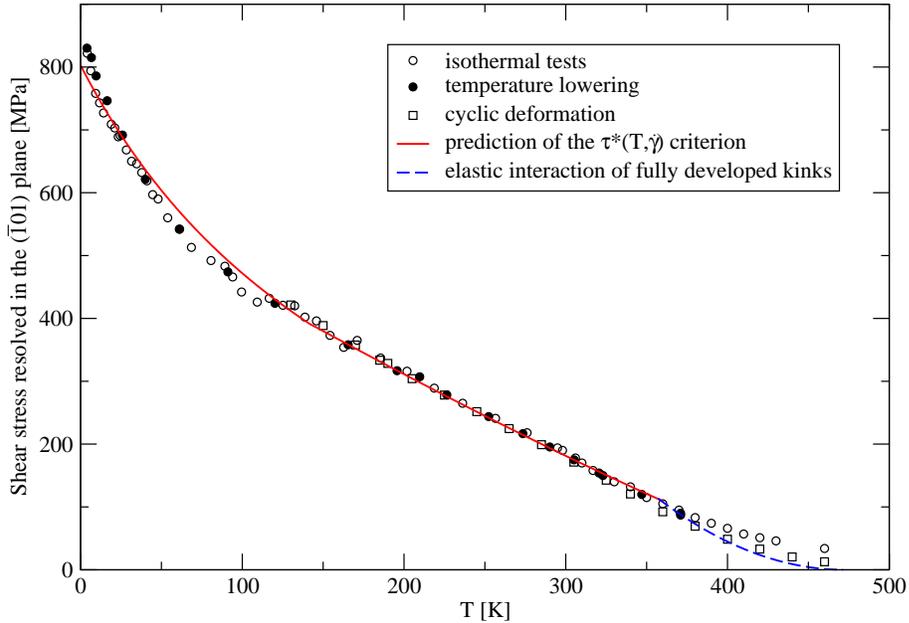}
  \parbox{14cm}{\caption{Temperature dependence of the yield stress, resolved as shear stress
      parallel to the $[111]$ slip direction and acting in the $(\bar{1}01)$ plane, for uniaxial
      tension along $[\bar{1}49]$ (curves). The symbols represent the experimental data of
      \citet{hollang:01} obtained using three different techniques.}
  \label{fig_yieldt-tstarC.hollang_-149_flattop}}
\end{figure}

The tensile loading along the $[\bar{1}49]$ axis, used in the experiments of \citet{hollang:97},
corresponds to $\chi=0$, and the ratio of the two shear stresses to $\eta=0.51$. We use the experimental
strain rate $\dot\gamma=8.6\times10^{-4}~\s^{-1}$ \citep{hollang:97} and the previously estimated
value $\dot\gamma_0=3\times10^{10}~\s^{-1}$ to arrive at
$q=\ln(\dot\gamma_0/\dot\gamma)=31.2$. Substituting these parameters in
\refeq{eq_eallcases_full_sigma-T} yields directly the temperature dependence of the yield stress
$\sigma$ acting in the $(\bar{1}01)$ plane that is plotted in
\reffig{fig_yieldt-tstarC.hollang_-149_flattop}. One can observe a close agreement with experiment
within a wide range of temperatures. At low temperatures, the theoretical prediction slightly
departs from the experiment with the maximum deviation less than $20\ \MPa$. The same conclusion was
reached previously by comparing the $\sigma-T$ dependence calculated numerically from the model of
the dislocation bow-out (Section~\ref{sec_bowout}) and from the experiment, see
\reffig{fig_yieldt_hollang_-149_flattop}.

\begin{figure}[!b]
  \centering
  \includegraphics[width=12cm]{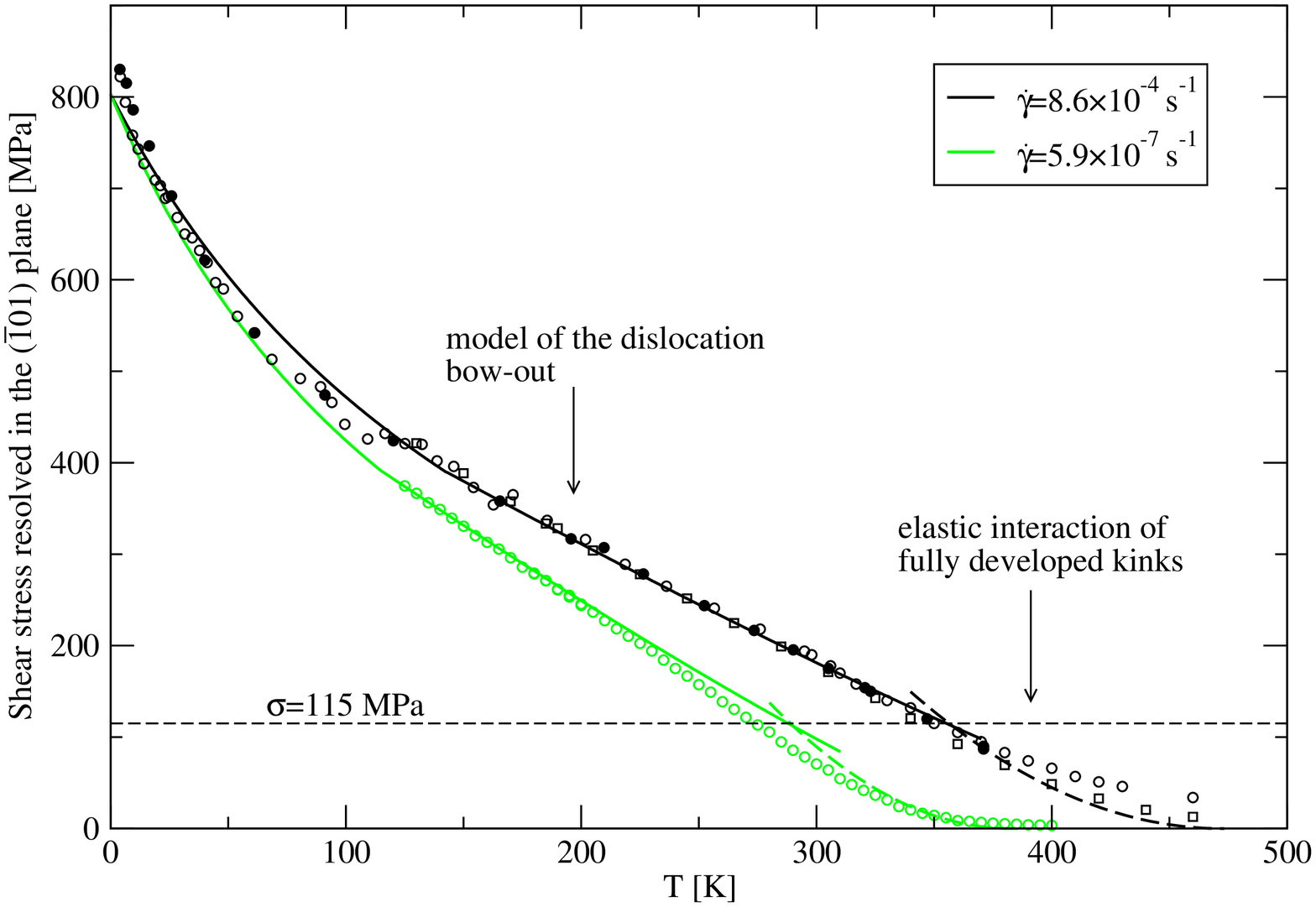}
  \parbox{14cm}{\caption{Temperature dependence of the yield stress for two different strain rates
      $\dot\gamma$. The curves are calculated from \refeq{eq_eallcases_full_sigma-T} for low
      temperatures and from \refeq{eq_lowstress_sigma-T} for high temperatures. The symbols are the
      experimental data for two different strain rates from \citet{hollang:97} and
      \citet{hollang:01a}.}
  \label{fig_yieldt_gdot_hollang}}
\end{figure}

An important feature of $\tau^*_{cr}(T,\dot\gamma)$, given by \refeq{eq_eallcases_full_tstarC-T},
and also the temperature dependence of the yield stress in \refeq{eq_eallcases_full_sigma-T} is
their validity at different strain rates $\dot\gamma$. Since the parameter $q$ in the rate equation
is inversely proportional to the plastic strain rate $\dot\gamma$, the temperature dependence of the
yield stress decays to zero the more quickly the lower the value of $\dot\gamma$. In
\reffig{fig_yieldt_gdot_hollang}, we show the temperature dependence of the yield stress calculated
for two different plastic strain rates, and their comparison with experimental data. At high
temperatures, the yield stress is calculated from the approximation of the data obtained from the
model of elastic interaction of fully developed kinks (\ref{eq_lowstress_sigma-T}), whereas the low
and intermediate temperatures are treated using \refeq{eq_eallcases_full_sigma-T}. For any strain
rate, the cross-over between the low-temperature and high-temperature approximation occurs at the
yield stress equal to $115~\MPa$. This value is in good agreement with the \emph{experimental data}
plotted in Fig.~11 of \citet{hollang:97}, obtained for a wide range of plastic strain rates
$\dot\gamma$, while the kink-pair formation theory presented thereafter predicts that this
cross-over occurs at 180~\MPa.

At high strain rates, both the theory and the experiments exhibit a marked change of slope at
$\sigma=115~\MPa$ that can be attributed to the change of the shape of the dislocation in the
activated state, particularly to the nucleation of two interacting kinks at high temperatures and to
a continuous bow-out at low temperatures. In experiments, this change of slope completely vanishes
at low strain rates and the linear regime at intermediate temperatures extends almost to the
temperature at which the thermal component of the yield stress vanishes. The slight deviation of the
theoretical calculations from experiments close to $115~\MPa$ does not constitute any major problem
because the maximum error in $\sigma$ is only about $20~\MPa$.

%----------------------------------------------------------------------------------------------------
%----------------------------------------------------------------------------------------------------

\section{Simplified ``engineering'' expressions for plastic flow}
\label{sec_gdot_engng}

For practical calculations, it is always convenient to write the plastic strain rate $\dot\gamma$ as
a simple function of temperature, stress, and other parameters. Due to the lack of detailed
understanding of the microscopic nature of plastic deformation, these expressions were originally
constructed purely phenomenologically and adjusted such that they correspond to experimental
measurements. This approach has been adopted already by \citet{kocks:75} who showed that the plastic
strain rate can be approximated by the power law expression in which the Schmid stress is raised to
a constant power that is obtained by fitting the experiment. Since none of these approximations
involve the effect of non-glide stresses, which have been shown in preceding chapters to play an
important role in bcc molybdenum, they are not applicable at low temperatures and high applied
stresses.

\begin{figure}[!htb]
  \centering
  \includegraphics[width=15cm]{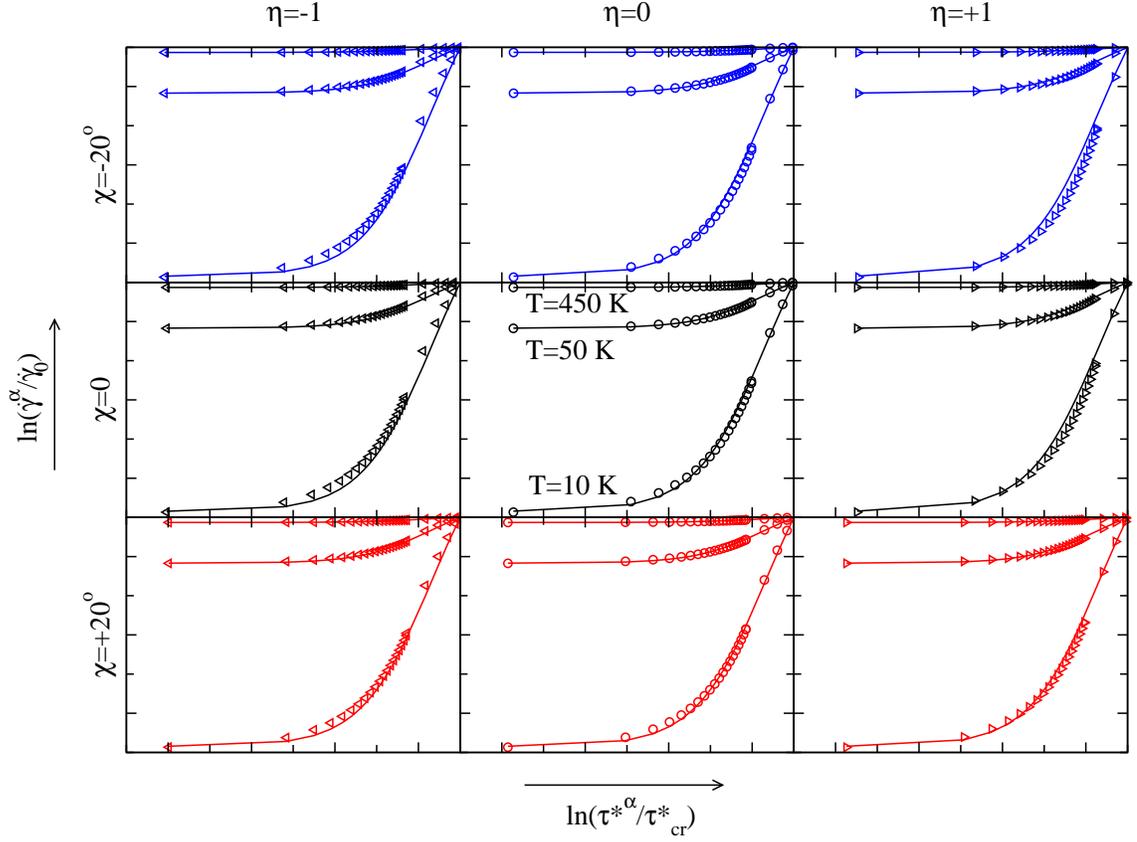}
  \caption{Stress dependence of the plastic strain rate obtained from the simplified formula
  (\ref{eq_gdot_sigma_eng}) (curves) and its comparison with the data calculated from the
  original expressions (\ref{eq_gdot_sigma1}) and (\ref{eq_gdot_sigma2}) (symbols). Here,
  $\alpha$ corresponds to the most operative $(\bar{1}01)[111]$ system. Nine characteristic
  combinations of $\chi$ and $\eta$ are shown here; the ranges of values along the two axes are
  $\ln(\tau^{*\alpha}/\tau^*_{cr}) \in \langle -8, 0 \rangle$, $\ln(\dot\gamma^\alpha/\dot\gamma_0)
  \in \langle -1500, 0 \rangle$.}
  \label{fig_gdot-tstar}
\end{figure}

Following our derivation of the temperature and strain rate dependent effective yield criterion, we
can now derive the rate equation which will explicitly contain the effect of non-glide stresses. In
principle, this can be done by considering the Arrhenius law $\dot\gamma = \dot\gamma_0
\exp(-H_b(\sigma)/kT)$ in which the activation enthalpy, $H_b(\sigma)$, takes the form given by
\refeq{eq_eallcases_full_both}. For simplicity, consider a particular slip system $\alpha$ whose
MRSSP is determined by the angle $\chi$, and $\eta$ is the ratio of the shear stress perpendicular
to the shear stress parallel to the slip direction resolved in this MRSSP. Hence, the plastic strain
rate in this system can be written as
\begin{equation}
  \dot\gamma^\alpha = \dot\gamma_0 \exp\left\{ -\frac{2H_k}{kT}\left[ 1-a(\chi,\eta)
    \left(\frac{\sigma^\alpha}{C_{44}}\right)^{b(\chi,\eta)} \right] \right\}
  \label{eq_gdot_sigma1}
\end{equation}
if $\sigma^\alpha/C_{44}\leq0.003$ and as
\begin{equation}
  \dot\gamma^\alpha = \dot\gamma_0 \exp\left\{ -\frac{2H_k}{kT}\left[
    1-a'(\chi,\eta) \exp\left(-\frac{b'(\chi,\eta)}{\sigma^\alpha/C_{44}}\right) \right] \right\}
  \label{eq_gdot_sigma2}
\end{equation}
if $\sigma^\alpha/C_{44}\geq0.003$. In these equations, $a,b,a',b'$ are quadratic functions of
$\chi$ and $\eta$, and their coefficients obtained by fitting are summarized in
Appendix~\ref{apx_thermalpar}. The plastic strain rates have been calculated for nine characteristic
combinations of $\chi$ and $\eta$ according to \refeq{eq_gdot_sigma1} and \ref{eq_gdot_sigma2} and
are plotted as symbols in \reffig{fig_gdot-tstar}. The slip system $\alpha$ considered for the
calculations corresponds to $(\bar{1}01)[111]$ that is the dominant system for all chosen
combination of $\chi$ and $\eta$. For convenience, we plot $\ln(\dot\gamma^\alpha/\dot\gamma_0)$ as
a function of $\ln(\tau^{*\alpha}/\tau^*_{cr})$, where the latter can be easily evaluated by noting
that $\tau^{*\alpha}=\sigma^\alpha t(\chi,\eta)$; $t(\chi,\eta)$ was given earlier. Each cell
in \reffig{fig_gdot-tstar} corresponds to a fixed straight loading path, whose angle is determined by
the stress ratio $\eta$ in the MRSSP characterized by the angle $\chi$, and contains three sets of data,
for temperatures $T=10, 50, 450~\K$.

For engineering calculations, it is necessary to replace \refeqs{eq_gdot_sigma1} and
\ref{eq_gdot_sigma2} by \emph{one} simple expression of $\dot\gamma^\alpha$ that could be written
using primitive functions and still correctly accounts for the effect of non-glide stresses. In
order to find this functional form, it is first helpful to look at the shape of the dependencies
plotted in \reffig{fig_gdot-tstar}. It is clear that the approximate form of
$\ln(\dot\gamma^\alpha/\dot\gamma_0)$ for a given temperature $T$ must approach a constant at large
negative $\ln(\tau^{*\alpha}/\tau^*_{cr})$ and vanish as $\ln(\tau^{\alpha*}/\tau^*_{cr})$
vanishes. The asymptotic character of $\ln(\dot\gamma^\alpha/\dot\gamma_0)$ at negative
$\ln(\tau^{*\alpha}/\tau^*_{cr})$ eliminates most of the primitive functions down to either $\tanh$
or $\arctan$ \citep{jeffrey:00}. Note, that both functions can approximate the data in
\reffig{fig_gdot-tstar} equally well and, therefore, there is no advantage of using one functional
form over the other. In the following, we will utilize $\tanh$ only because it can be evaluated
easily using exponentials. Hence, the data plotted in \reffig{fig_gdot-tstar} can be closely
approximated using the following functional form:
\begin{equation}
  \dot\gamma^\alpha = \dot\gamma_0 \exp\left\{ -\frac{A}{kT} \left[ B - 
    \tanh\left( C \left(\ln\frac{\tau^{*\alpha}}{\tau^*_{cr}}+D\right) \right) \right] \right\} \ ,
  \label{eq_gdot_sigma_eng}
\end{equation}
where $k$ is the Boltzmann constant, $\dot\gamma_0=3\times10^{10}~\s^{-1}$ is the pre-exponential
term of the Arrhenius law in which the mobile dislocation density is a constant, and $A,B,C,D$
parameters that have to be adjusted such that the dependencies
$\ln(\dot\gamma^\alpha/\dot\gamma_0)-\ln(\tau^{*\alpha}/\tau^*_{cr})$, calculated from
\refeq{eq_gdot_sigma_eng}, match the data obtained from the two unsimplified expressions and plotted
in \reffig{fig_gdot-tstar} by symbols. At zero stress, as $\tau^{*\alpha} \rightarrow 0$, the
product of $\tanh$ in \refeq{eq_gdot_sigma_eng} becomes -1 and the rate equation reduces to
$\dot\gamma^\alpha = \dot\gamma_0\exp(-2H_k/kT)$, where $2H_k = A(B+1)$ is the energy of two
isolated kinks. At large stresses, as $\tau^{*\alpha} \rightarrow \tau^*_{cr}$, the activation
enthalpy vanishes and the rate equation has to read $\dot\gamma^\alpha=\dot\gamma_0$, which implies that
the argument of the exponential of \refeq{eq_gdot_sigma_eng} is $A[B-\tanh(CD)] = 0$. The four
unknown parameters $A,B,C,D$ can now be determined by specifying $A$ and $C$ that control the height
and slope of the $\tanh$ function in \refeq{eq_gdot_sigma_eng}, after which $B$ and $D$ can be
calculated from the two conditions above.  For molybdenum, $\tau^*_{cr}=1014~\MPa$ and the best
agreement with the data obtained from \refeq{eq_gdot_sigma1} and \ref{eq_gdot_sigma2} is for the
parameters listed in \reftab{tab_parABCD_Mo}.

\begin{table}[!htb]
  \centering
  \parbox{10cm}{\caption{Parameters $A,B,C,D$ for molybdenum used in 
      \refeqs{eq_gdot_sigma_eng}, \ref{eq_sigma_T_eng} and \ref{eq_tstarC_engng}.}
    \label{tab_parABCD_Mo}}\\[1em]
  \begin{tabular}{cccc}
    \hline
    $A$ & $B$ & $C$ & $D$ \\
    \hline
    0.90 & 0.41 & 0.60 & 0.73 \\
    \hline
  \end{tabular} \ .
\end{table}

The dependencies calculated from \refeq{eq_gdot_sigma_eng} with $A,B,C,D$ from the table above are
plotted in \reffig{fig_gdot-tstar} by solid curves. The total plastic strain rate can now be easily
calculated by summing over the strain rates generated by each slip system $\alpha$, i.e.
\begin{equation}
  \dot\gamma = \sum_\alpha \dot\gamma^\alpha \ .
\end{equation}
Note, that the sum includes only the slip systems that contribute to the plastic flow, i.e. those
for which $\tau^{*\alpha}>0$.

\begin{figure}[!htb]
  \centering 
  \includegraphics[width=12cm]{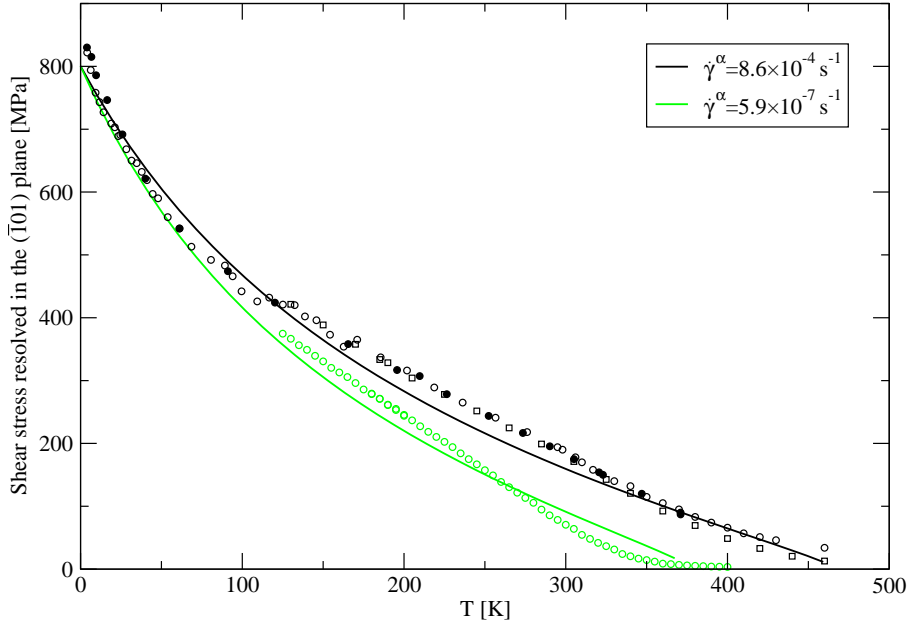}
  \parbox{14cm}{\caption{Temperature dependence of the yield stress calculated from the simplified
  expression (\ref{eq_sigma_T_eng}) for loading in tension along $[\bar{1}49]$. Here, $\alpha$ is
  the most highly stressed $(\bar{1}01)[111]$ system. The symbols are the experimental data for two
  different strain rates from \citet{hollang:97} and \citet{hollang:01a}.}
  \label{fig_yieldt_gdot_engng_hollang_-149}}
\end{figure}

\refeq{eq_gdot_sigma_eng} also provides an elegant formula for the temperature and strain rate
dependence of the yield stress that can be used instead of the original approximations
(\ref{eq_eallcases_full_sigma-T}) in which one would have to determine first the values of
$a,b,a',b'$. Writing $\tau^{*\alpha}=\sigma^\alpha t(\chi,\eta)$, where $t(\chi,\eta)$ is the
angular factor given earlier in Section~\ref{sec_yieldtemp_full}, one obtains an expression for the
yield stress, i.e. the shear stress $\sigma^\alpha$ parallel to the slip direction and applied in the
MRSSP of the system $\alpha$:
\begin{equation}
  \sigma^\alpha = \frac{\tau^*_{cr}}{t(\chi,\eta)} \exp\left\{ -D+\frac{1}{C} \tanh^{-1} \left[ B +
    \frac{kT}{A} \ln\frac{\dot\gamma^\alpha}{\dot\gamma_0} \right] \right\} \ ,
  \label{eq_sigma_T_eng}
\end{equation}
where the parameters $A,B,C,D$ are listed in \reftab{tab_parABCD_Mo}. Since \refeq{eq_sigma_T_eng}
is a simplified ``engineering'' expression of the temperature dependence of the yield stress, one
should not expect that this formula will be capable of reproducing the experimental data as closely
as its unsimplified version (\ref{eq_eallcases_full_sigma-T}). To demonstrate this, we plot in
\reffig{fig_yieldt_gdot_engng_hollang_-149} the temperature dependence of the yield stress
calculated from \refeq{eq_sigma_T_eng} for loading in tension along the $[\bar{1}49]$ axis and its
comparison with experimental data. This graph should be compared with
\reffig{fig_yieldt_gdot_hollang} that shows the same dependence but calculated from the unsimplified
\refeq{eq_eallcases_full_sigma-T} at low temperatures (high stresses) and from
\refeq{eq_lowstress_sigma-T} at high temperatures (low stresses).

As expected, \refeq{eq_sigma_T_eng} correctly reproduces the overall trend of the experimental data
but cannot resolve their details. This is the price that we pay for the demand to simplify the
expressions for the plastic strain rate (\ref{eq_gdot_sigma1}) and (\ref{eq_gdot_sigma2}), and
describe them by one analytical formula that is valid at all temperatures in which the plastic flow
is controlled by thermally activated motion of screw dislocations. Nevertheless, the maximum
deviation of the dependence predicted by \refeq{eq_sigma_T_eng} from the experiment is less than
$30~\MPa$ and, therefore, \refeqs{eq_gdot_sigma_eng} and \ref{eq_sigma_T_eng} represent reasonably
accurate macroscopic representations of the plastic flow of single crystals of molybdenum. It is
important to note that these expressions explicitly involve the effect of non-glide stresses that
are included through $\tau^*$. Hence, a question arises naturally as to how different would the
predictions of \refeqs{eq_gdot_sigma_eng} and \ref{eq_sigma_T_eng} be if one considered the Schmid
law instead of the full $\tau^*$ criterion. For a given $\chi$ and stress ratio $\eta$, this can be
easily checked by setting $\tau^*$ equal to the Schmid stress, $\sigma\cos\chi$.

\begin{figure}[p]
  \centering
  \includegraphics[width=12cm]{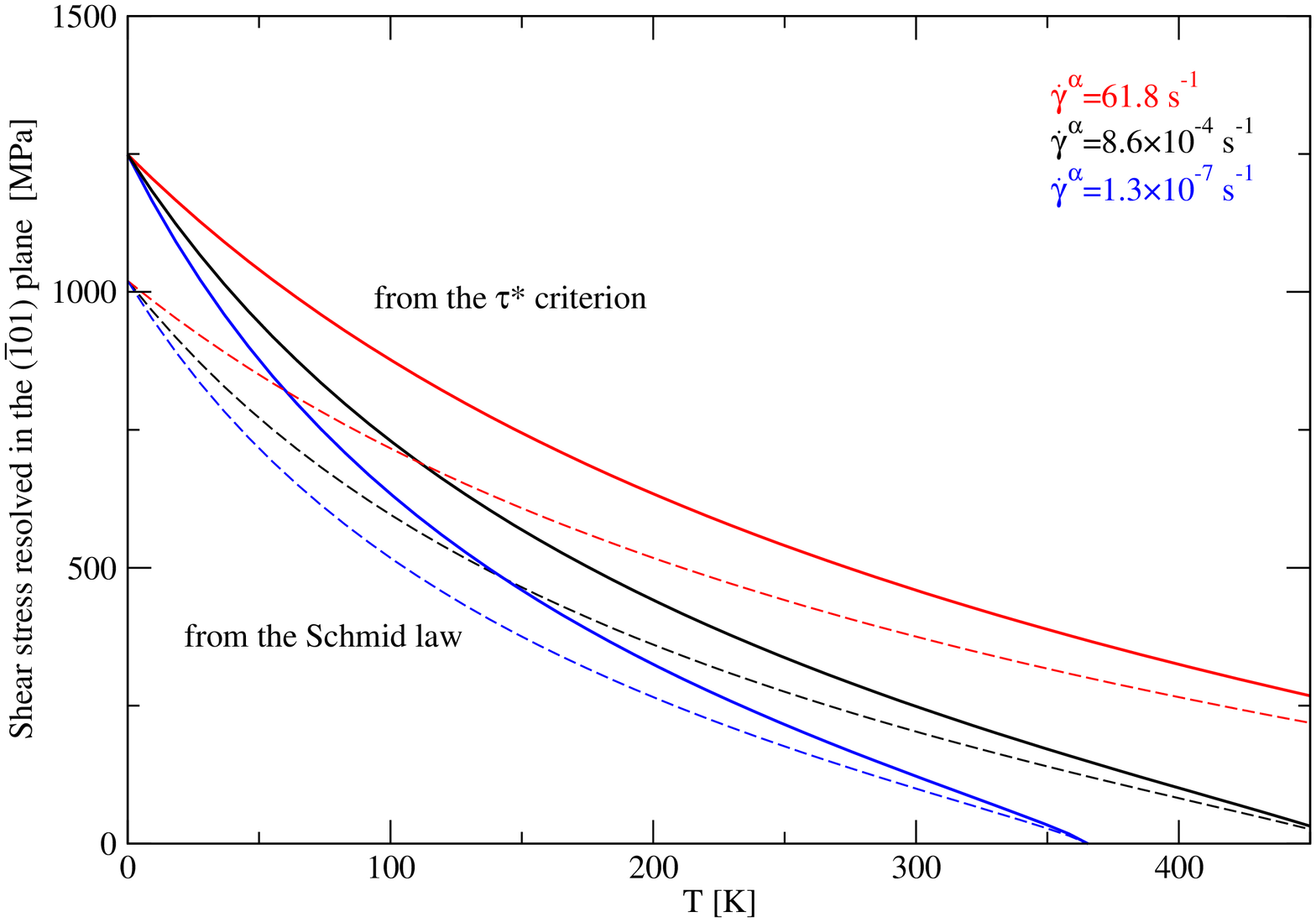} \\
  a) $\chi=0$, $\eta=\tau/\sigma=-1$ \\[1em]
  \includegraphics[width=12cm]{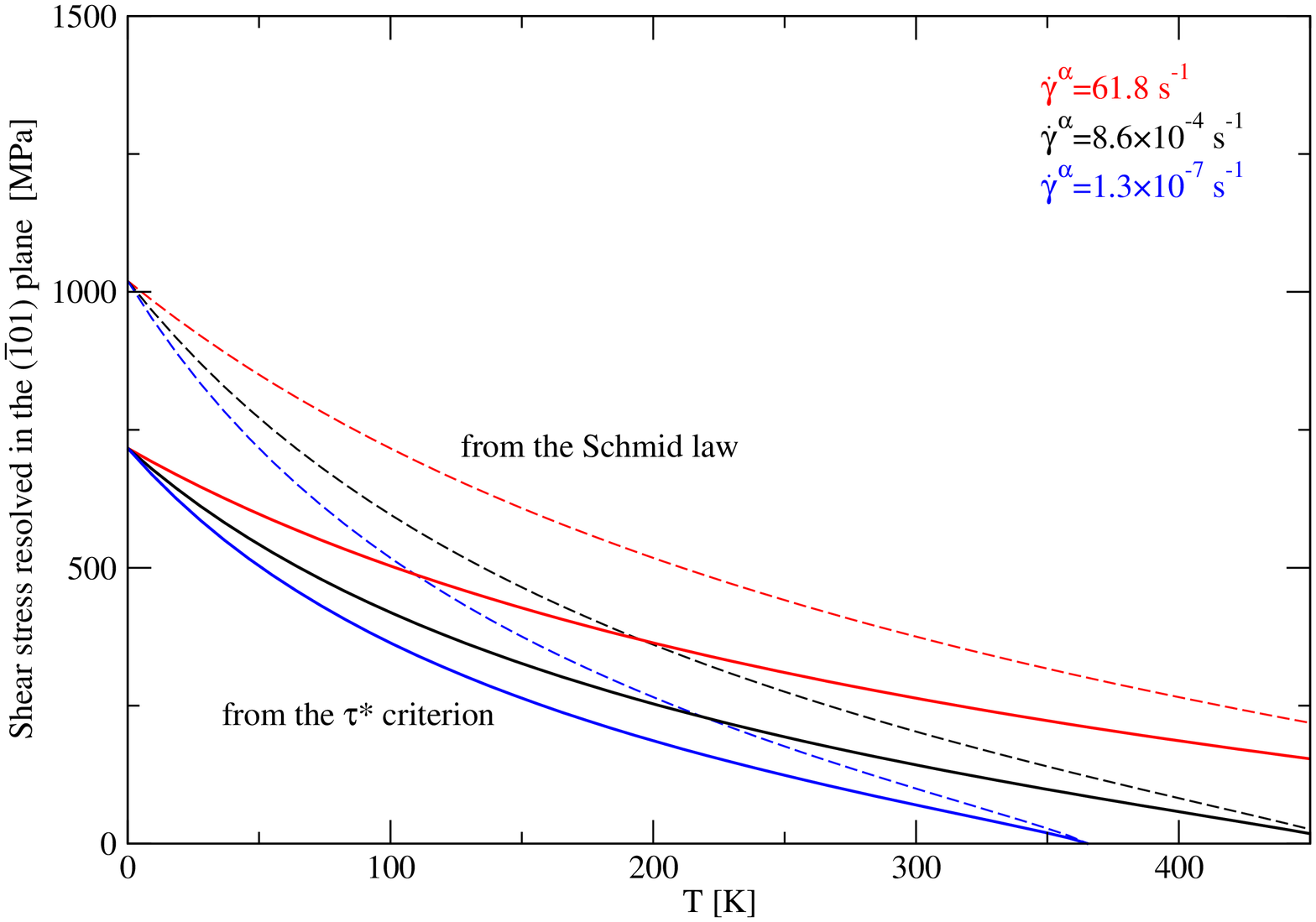} \\
  b) $\chi=0$, $\eta=\tau/\sigma=+1$ \\
  \caption{Temperature dependence of the yield stress calculated for three different strain rates
    $\dot\gamma^\alpha$ from \refeq{eq_sigma_T_eng}. The solid lines are obtained by considering the
    $\tau^*$ criterion and the dashed lines are calculated from the Schmid law.}
  \label{fig_yieldt_engng_chi0}
\end{figure}

For illustration, we will consider two loading paths, specified by $\chi=0$ and $\eta=\{-1,+1\}$,
and three different strain rates given by $q=\ln(\dot\gamma_0/\dot\gamma^\alpha)=\{20,31.2,40\}$,
where the middle value corresponds to the plastic strain rate $\dot\gamma=8.6\times10^{-4}~\s^{-1}$
of the tensile experiments of \citet{hollang:97}. For this loading, we have already shown in
\reffig{fig_actene_hollang_-149_allsys} that the plastic flow is dominated by operation of the
$(\bar{1}01)[111]$ system that we will denote as $\alpha$. The comparison of the dependencies
calculated using the Schmid law, $\tau^{*\alpha}=\sigma^\alpha\cos\chi$, with those obtained from
the full effective yield criterion, $\tau^{*\alpha}=\sigma^\alpha t(\chi,\eta)$, that involves all
non-glide stresses is shown in \reffig{fig_yieldt_engng_chi0} by the dashed and the solid curves,
respectively. Obviously, the predictions of the Schmid law are independent of $\eta$ (and also of
$\tau$), while the curves calculated from the $\tau^*$ criterion correctly reproduce the yield
stress asymmetry for positive and negative shear stress perpendicular to the slip direction (see
also \reffig{fig_CRSS_tau_MoBOP}a). At low temperatures and any plastic strain rate, the predictions
made from the Schmid law deviate significantly from those calculated from the $\tau^*$ criterion,
which is a clear manifestation of importance of non-glide stresses at low temperatures. This
deviation gradually vanishes as the temperature is raised, which implies that the importance of
non-glide stresses also diminishes with increasing temperature.

\begin{figure}[!htb]
  \centering 
  \includegraphics[width=12cm]{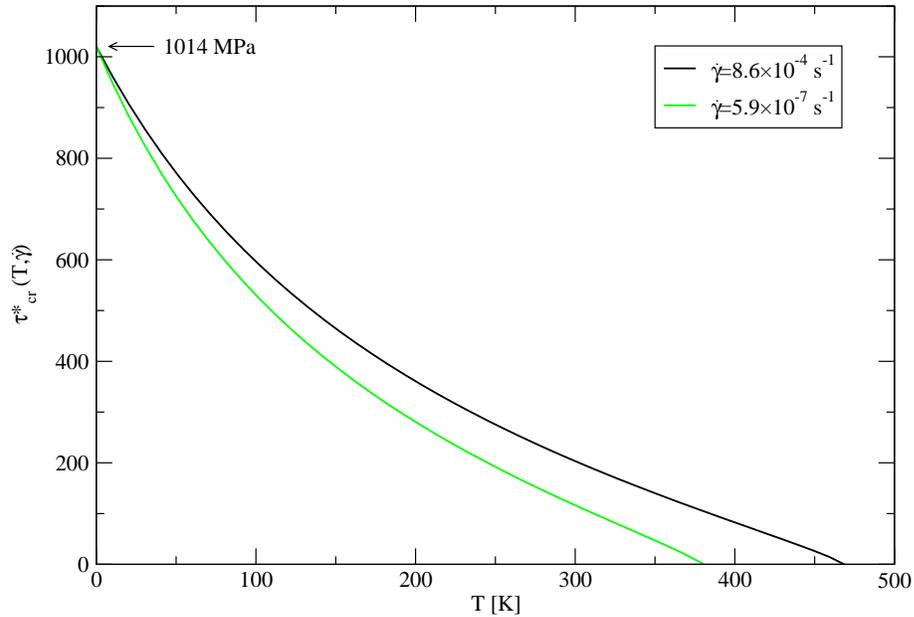}
  \parbox{14cm}{\caption{Variation of the effective yield stress with temperature, calculated from
  the simplified formula (\ref{eq_tstarC_engng}).}
  \label{fig_tstarC_engng}}
\end{figure}

\refeq{eq_sigma_T_eng} also implies a convenient form of the temperature and strain rate dependent
effective yield stress that is a simplified version of \refeq{eq_eallcases_full_tstarC-T}. At yield,
$\tau_{cr}^*=\sigma^\alpha t(\chi,\eta)$, and, upon substituting $\sigma^\alpha$ from
\refeq{eq_sigma_T_eng}, this leads to
\begin{equation}
  \tau^*_{cr}(T,\dot\gamma) = \tau^*_{cr}(0) \exp\left\{ -D+\frac{1}{C} \tanh^{-1} \left[ B +
    \frac{kT}{A} \ln\frac{\dot\gamma}{\dot\gamma_0} \right] \right\} \ ,
  \label{eq_tstarC_engng}
\end{equation}
where $\tau^*_{cr}(0)=1014~\MPa$ is the magnitude of the effective yield stress at zero temperature. The
variation of $\tau^*_{cr}$ with temperature is shown in \reffig{fig_tstarC_engng} for the two strain
rates considered earlier. At a given temperature and strain rate, $\tau^*_{cr}(T,\dot\gamma)$
represents the critical value of $\tau^{*\alpha}$ at which the slip system $\alpha$ becomes
activated.

\begin{figure}[p]
  \centering
  \includegraphics[width=12cm]{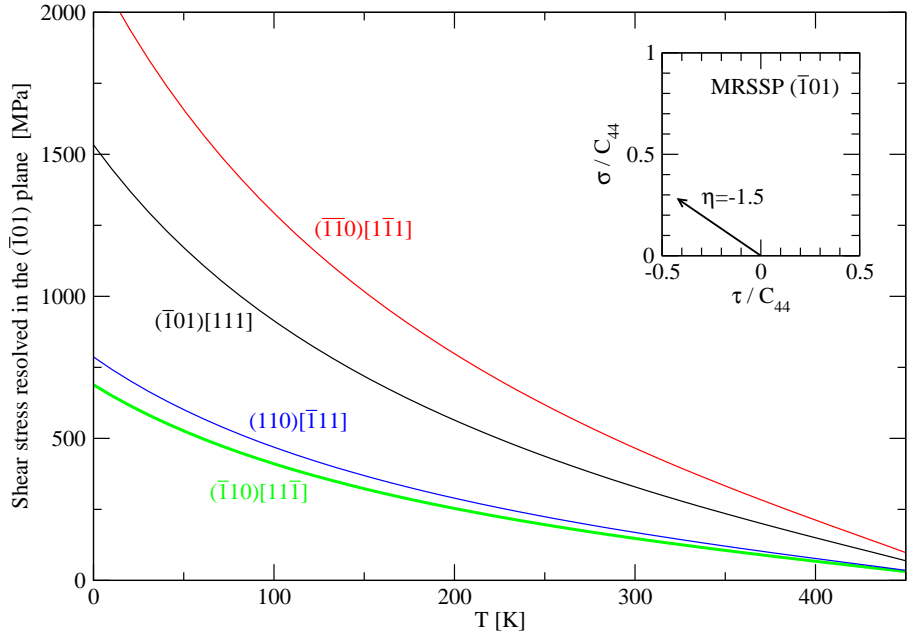} \\
  a) $\chi=0$, $\eta=\tau/\sigma=-1.5$ in the reference system $(\bar{1}01)[111]$ \\[1em]
  \includegraphics[width=12cm]{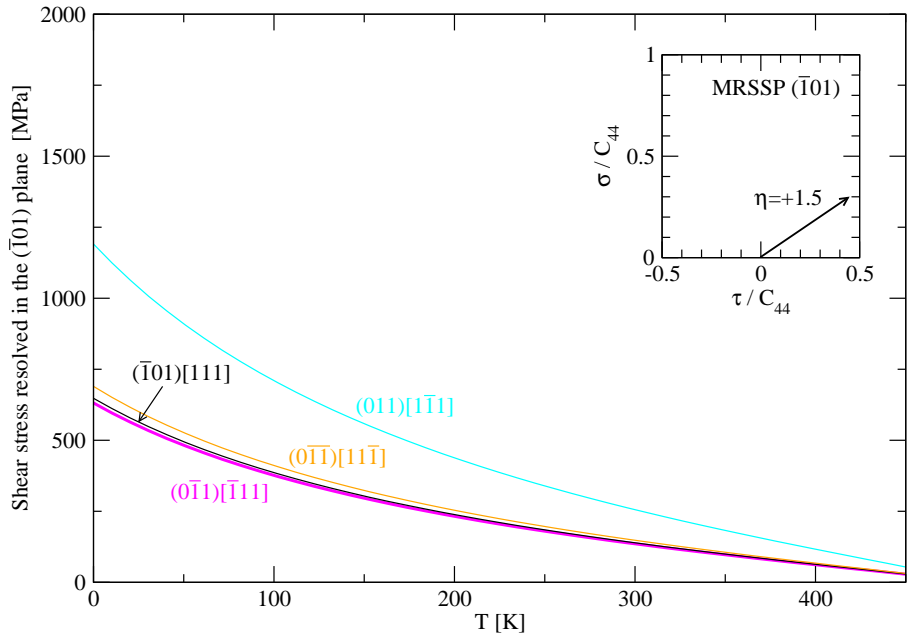} \\
  b) $\chi=0$, $\eta=\tau/\sigma=+1.5$ in the reference system $(\bar{1}01)[111]$ \\
  \caption{Temperature dependence of the yield stress for a fixed strain rate ($q=30$) and two
    loading paths for which the shear stress perpendicular to the slip direction is negative (a) and
    positive (b). The lowest curve corresponds to the primary slip system.}
  \label{fig_yieldt_engng_chi0_allsys}
\end{figure}

Finally, it is instructive to check the order of activity of individual slip systems at larger
applied shear stress perpendicular to the slip direction. Without the loss of generality, we will
consider loading by shear stress perpendicular ($\tau$) and parallel ($\sigma$) to the $[111]$ slip
direction and choose the $(\bar{1}01)$ plane to be the MRSSP. We will consider two loading paths in
the $\CRSS-\tau$ projection, given by the stress ratios $\eta=\pm1.5$. For each of these loadings,
we will identify the four slip systems $\alpha$ that can become operative, calculate the
orientations of the MRSSPs, $\chi^\alpha$, in the zones of their slip directions, and determine the
stress ratios $\eta^\alpha$ from the two shear stresses resolved in these
MRSSPs. \refeq{eq_sigma_T_eng} then provides a critical shear stress $\sigma^\alpha$ that induces
slip on each of the four slip systems $\alpha$. By resolving this stress back into the MRSSP of the
$(\bar{1}01)[111]$ system, one arrives at the dependencies plotted in
\reffig{fig_yieldt_engng_chi0_allsys}, where the loading paths are shown in the insets. At negative
$\tau$, the plastic flow occurs by virtually simultaneous operation of the primary
$(\bar{1}10)[11\bar{1}]$ and the secondary $(110)[\bar{1}11]$ system. These systems are identical to
those predicted for $\eta=-1.5$ directly from the temperature and strain rate dependent effective
yield criterion; the critical line for the former system is also shown in
\reffig{fig_chi0_MoBOP_temp}. At positive $\tau$, three slip systems are almost equally operative
and the plastic flow occurs by simultaneous operation of $(0\bar{1}1)[\bar{1}11]$,
$(\bar{1}01)[111]$ and $(0\bar{1}\bar{1})[11\bar{1}]$ systems. The operation of the two former
systems is directly predicted for $\eta=+1.5$ from \reffig{fig_chi0_MoBOP_temp}; the third system is
not plotted in this figure because its critical line passes beyond the corner of the yield polygon.

Based on the observed agreement between the theoretical prediction of the temperature dependence of
the yield stress and experimental measurements, one may conclude that the ``engineering''
expressions (\ref{eq_gdot_sigma_eng}), (\ref{eq_sigma_T_eng}) and (\ref{eq_tstarC_engng}) correctly
capture: (i) the effect of non-glide stresses, particularly twinning-antitwinning asymmetry and the
effect of the shear stress perpendicular to the slip direction, (ii) decreasing role of non-glide
stresses with increasing temperature, and (iii) the order of activation of individual slip systems.
The approximate relations given in this section thus represent physically-based rules for plastic
flow of single crystals of molybdenum in which the microscopic details of the motion of dislocations
in each slip system $\alpha$ are condensed into the form of the effective stress
$\tau^{*\alpha}$. Due to its relative simplicity, \refeq{eq_gdot_sigma_eng} provides for the first
time a rate equation which explicitly accounts for the effect of non-glide stresses through
physically based understanding of their effect on dislocation glide, and can thus be used in
continuum finite-element calculations.

  \chapter{Plastic flow of bcc tungsten}
\label{chap_WBOP}

\begin{flushright}
  The significant problems we face cannot be solved at \\
  the same level of thinking we were at when we created them.\\
  \emph{Albert Einstein}
\end{flushright}

In previous chapters we have shown in detail how the results of atomistic simulations of an isolated
screw dislocation can be used to construct the effective yield criterion for molybdenum that
correctly reproduces the effects of non-glide stresses in this material. Subsequently, this yield
criterion was used to construct the Peierls potential and to calculate the stress dependence of
the activation enthalpy, activation volume, and the temperature dependence of the yield stress. These
dependencies were finally used to determine the temperature and strain rate dependent effective
yield criterion that was shown to correctly reproduce the experimentally observed asymmetries of the
yield stress.

In order to demonstrate that this approach can be used also for other bcc metals, we will now repeat
the whole process and construct the plastic flow rules for tungsten. Because the details of this
development have been explained thoroughly for molybdenum, we will focus only on the features that
distinguish these two metals. The atomistic simulations presented in this chapter were performed
using the screened BOP constructed specifically for tungsten \citep{mrovec:07}. This potential
correctly reproduces the elastic moduli, lattice parameter and the cohesive energy.

The crystallographic data and the elastic moduli for tungsten used in the forthcoming calculations
are summarized below.

\begin{table}[!htb]
  \begin{minipage}{0.5\textwidth}
    \centering
    \begin{tabular}{lcl}
      \hline
      $\gdir{100}$ lattice parameter & $a$   & 3.165 \\
      Shortest periodicity in $\gplane{110}\gdir{112}$ & $a_0$ & 2.584 \\
      Magnitude of the Burgers vector & $b$   & 2.741 \\
      \hline
    \end{tabular} \\
    Dimensions:  [\A]
  \end{minipage}
  \hfill
  \begin{minipage}{0.4\textwidth}
    \centering
    \begin{tabular}{lcl}
      \hline
                     & $C_{11}$ & 5.224 \\
      Elastic moduli & $C_{12}$ & 2.044 \\
                     & $C_{44}$ & 1.606 \\
      $\gdir{111}$ shear modulus & $\mu$ & 1.595 \\
      Anisotropy factor & & 1.01 \\
      \hline
    \end{tabular} \\
    Dimensions:  $C_{ij},\mu~[10^{5}~\MPa]$
  \end{minipage}
\end{table}

%----------------------------------------------------------------------------------------------------
%----------------------------------------------------------------------------------------------------

\section{$\gamma$-surface and the structure of the dislocation core}
\label{sec_Wgsurf}

\begin{figure}[!htb]
  \centering
  \includegraphics[width=12cm]{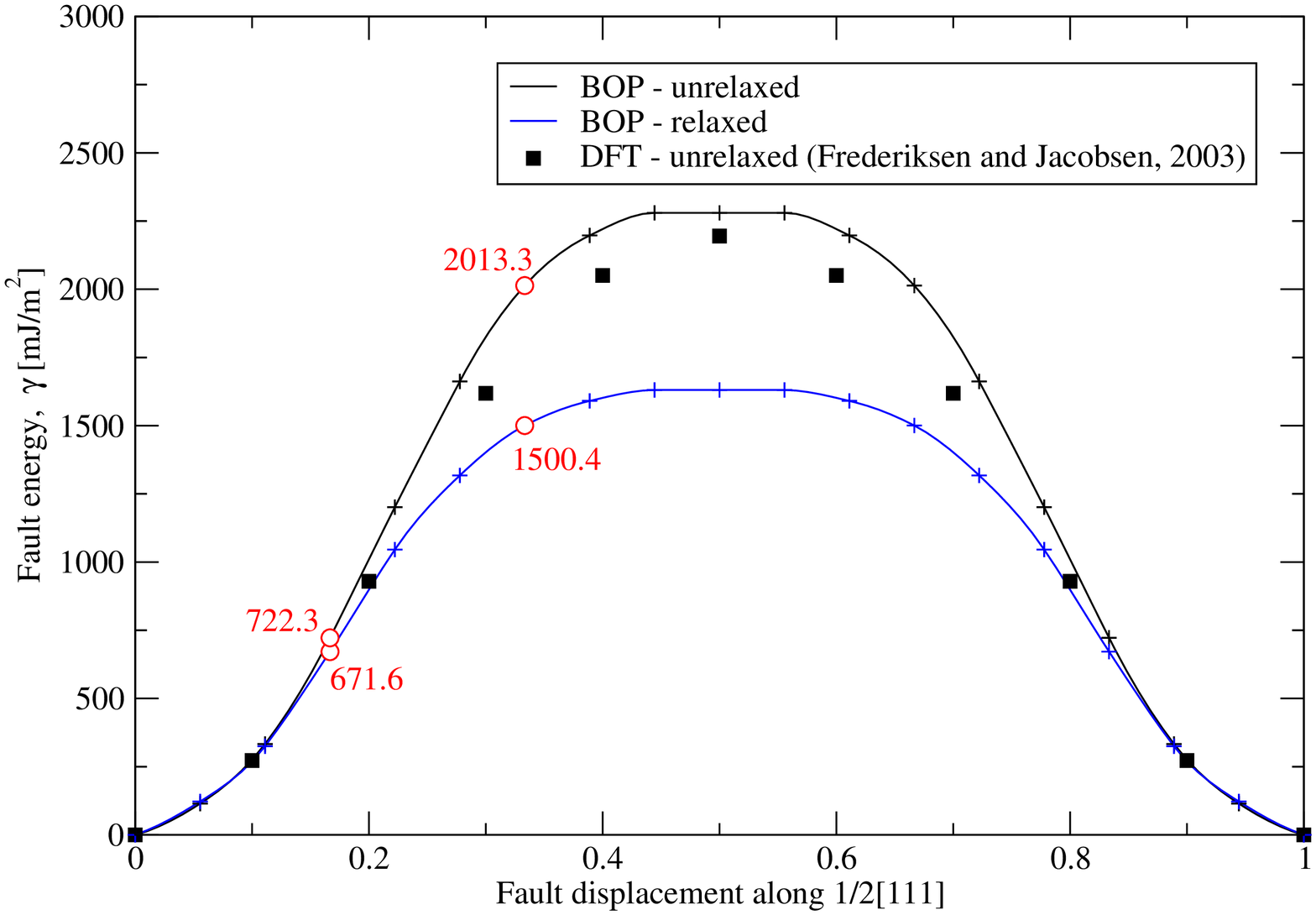}
  \parbox{14cm}{\caption{$\gdir{111}$ cross-section of the $\gplane{110}$ $\gamma$-surface in
  tungsten calculated using the screened BOP and DFT.}
  \label{fig_scrW_110gsurf}}
\end{figure}

In \reffig{fig_scrW_110gsurf}, we show the unrelaxed and relaxed $\gamma$-surface calculated for bcc
tungsten using the screened BOP. One can clearly see that there are no intermediate minima, and,
therefore, no metastable stacking faults exist in tungsten. The unrelaxed $\gamma$-surface is in
good agreement with the data obtained from full electron DFT simulations of \citet{frederiksen:03},
which demonstrates the accuracy of the constructed potential.

Comparison of the fault energies corresponding to the displacements $b/3$ and $b/6$ (see
\reffig{fig_scrW_110gsurf}) implies that $3\gamma(b/3)>6\gamma(b/6)$. Therefore, according to the
criterion discussed in Section~\ref{sec_gsurf_MoBOP} screw dislocations in tungsten favor
non-degenerate cores. Because the same type of dislocation core has been found in
Section~\ref{sec_gsurf_MoBOP} for molybdenum, one may expect that the behavior of these two metals
under stress will display common characteristics.

In subsequent 0~K atomistic studies of the effect of the applied stress on the screw dislocation, we
adopted the same geometry of the simulation cell as used previously for molybdenum. The block was
oriented such that the $z$-axis coincided with the dislocation line and thus the slip direction
(i.e. the Burgers vector of the dislocation), $x$ axis coincided with the $[\bar{1}2\bar{1}]$
direction that is a trace of the $(\bar{1}01)$ plane on the $(111)$ plane, and $y$ was perpendicular
to the $(\bar{1}01)$ plane. The block was effectively infinite along the dislocation line, which was
achieved using periodic boundary conditions. In the $xy$ plane, perpendicular to the dislocation
line, the block consisted of (see \reffig{fig_simul_block}): (i) \emph{active region} with atoms that
move during the relaxation, and (ii) \emph{inert region} that represents an infinitely large block in
which the active region is embedded.

\begin{figure}[!htb]
  \centering
  \includegraphics[width=9cm]{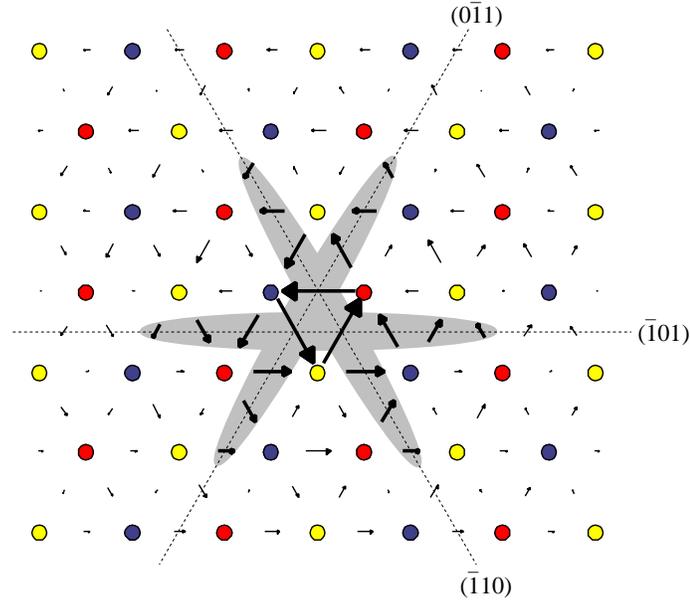}
  \parbox{10cm}{\caption{Structure of the $1/2[111]$ screw dislocation core calculated
  using the BOP for tungsten.}
  \label{fig_scrW_core}}
\end{figure}

Starting with a perfect single crystal, we inserted a $1/2[111]$ screw dislocation by displacing all
atoms in the block according to the long-range strain field of the dislocation \citep{hirth:82}.
This configuration was subsequently relaxed using the BOP for tungsten. The differential
displacement map showing the relaxed dislocation core is plotted in \reffig{fig_scrW_core}. In this
projection, the circles again depict atoms in three successive $(111)$ layers separated by the
distance $a/2\sqrt{3}$, where $a$ is the $\gdir{100}$ lattice parameter.  Each arrow corresponds to
the mutual displacement of two atoms in the direction of the Burgers vector (i.e. perpendicular to
the plane of the figure) and calculated relative to their distance in the perfect lattice.

%----------------------------------------------------------------------------------------------------
%----------------------------------------------------------------------------------------------------

\section{Behavior of dislocations under stress}
\label{sec_Wstress}

In order to investigate the behavior of an isolated screw dislocation in tungsten under stress, we
carried out a series of atomistic simulations at $0~\K$ using the BOP. From the original work of
\citet{ito:01} and the Chapter~\ref{chap_MoBOP} of this Thesis, we know that the motion of screw
dislocations in molybdenum is affected only by shear stresses parallel and perpendicular to the slip
direction. The latter exerts zero Peach-Koehler force on the dislocation and affects its motion only
indirectly via modification of the structure of the dislocation core. On the other hand, the former
stress exerts a nonzero Peach-Koehler force on the dislocation and can thus directly move it through
the crystal when the applied stress reaches the Peierls stress. Owing to the fact that the
dislocation core in tungsten is of the same type as that of molybdenum, one may assume that the
plastic flow of tungsten is also affected only by the shear stresses parallel and perpendicular to the
slip direction.

%----------------------------------------------------------------------------------------------------

\subsection{Loading by shear stress parallel to the slip direction}

In this calculation we started with a completely relaxed block with a $1/2[111]$ screw dislocation
in the middle, as shown in \reffig{fig_scrW_core}. The shear stress parallel to the slip direction,
$\sigma$, was imposed on the block by applying to all atoms the anisotropic displacements
corresponding to the stress tensor (\ref{eq_tensor_sigma}). To ensure a smooth convergence, the
shear stress $\sigma$ was built up incrementally in steps of $0.001C_{44}$, where $C_{44}$ is the
elastic modulus. In each loading step, the simulated block was fully relaxed before increasing
$\sigma$. At stresses lower than the CRSS, the dislocation core transforms from its initially
three-fold symmetric non-degenerate structure, to a less symmetric form. This transformation is
purely elastic in that the block returns back into its original configuration when the stress is
removed. Once the applied shear, $\sigma$, attains its critical value, the critical resolved shear
stress (CRSS), the transformation is complete and the dislocation moves through the crystal.

\begin{figure}[!htb]
  \centering
  \includegraphics[width=12cm]{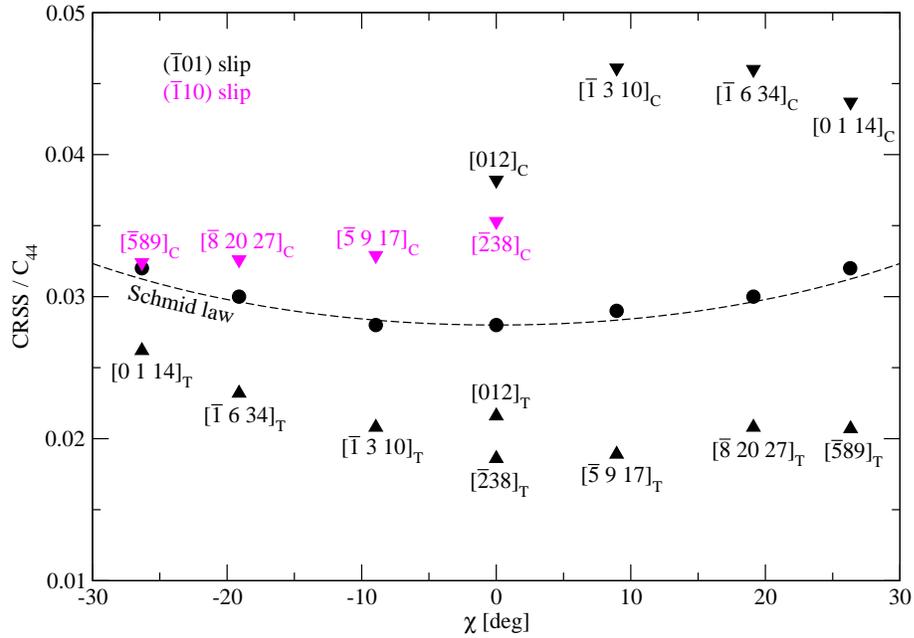}
  \parbox{13cm}{\caption{Orientation dependence of the CRSS for loading by pure shear stress
      parallel to the slip direction (circles) and for uniaxial loading (triangles). The subscripts
      $_{\rm T}$ and $_{\rm C}$ correspond to the loading in tension and compression, respectively.}
  \label{fig_CRSS_chi_WBOP}}
\end{figure}

In order to obtain the dependence of the CRSS on the orientation of the MRSSP, i.e. the angle
$\chi$, we considered seven different angles $\chi$ spanning effectively the entire angular region
$-30\deg<\chi<+30\deg$. The $\CRSS-\chi$ dependence obtained for loading by shear stress parallel to
the slip direction is plotted in \reffig{fig_CRSS_chi_WBOP} with circles. For all orientations of
the MRSSP, the dislocation moved by single slip on the $(\bar{1}01)$ plane. Importantly, the
$\CRSS-\chi$ data follow the Schmid law, which means that no twinning-antitwinning asymmetry exists
in tungsten. This is in contrast to molybdenum for which \reffig{fig_CRSS_chi_MoBOP} reveals a
strong orientation dependence of the CRSS for loading by pure shear stress parallel to the slip
direction.

%----------------------------------------------------------------------------------------------------

\subsection{Loading in tension and compression}

Superimposed in \reffig{fig_CRSS_chi_WBOP} are also the values of the CRSS calculated for loading in
tension and compression (triangles). The loading axes were distributed uniformly throughout the
entire interior of the stereographic triangle (\reffig{fig_sgtria_tc}) to cover a broad variety of
possible orientations. For any loading axis, one can always find the orientation of the MRSSP in the
zone of the $[111]$ slip direction that lies within $\pm30\deg$ from the $(\bar{1}01)$ plane. The
shear stress parallel to the slip direction resolved in the MRSSP, characterized by the angle
$\chi$, that moves the dislocation in tension/compression, is the CRSS that can be directly
compared with the value calculated for the same angle $\chi$ when applying the pure shear stress
parallel to the slip direction.

If, for a given $\chi$, the CRSS for tension/compression (triangles) were identical to those
obtained for pure shear stress parallel to the slip direction (circles), we would conclude that only
the shear stress parallel to the slip direction controls the plastic flow of tungsten. However,
\reffig{fig_CRSS_chi_WBOP} clearly shows that this is not the case, and, therefore, the CRSS for slip
must depend also on another stress component that is identified as the shear stress perpendicular to
the slip direction.

%----------------------------------------------------------------------------------------------------

\subsection{Loading by shear stress perpendicular to the slip direction}

Because the shear stress perpendicular to the slip direction cannot move the dislocation, the
deformation induced by this stress in the crystal is purely elastic. This loading is applied by
imposing the stress tensor (\ref{eq_tensor_tau}) in steps of $0.005C_{44}$, as explained in
Section~\ref{sec_sperp_section}. 

\begin{figure}[!b]
  \centering
  \includegraphics[width=7.3cm]{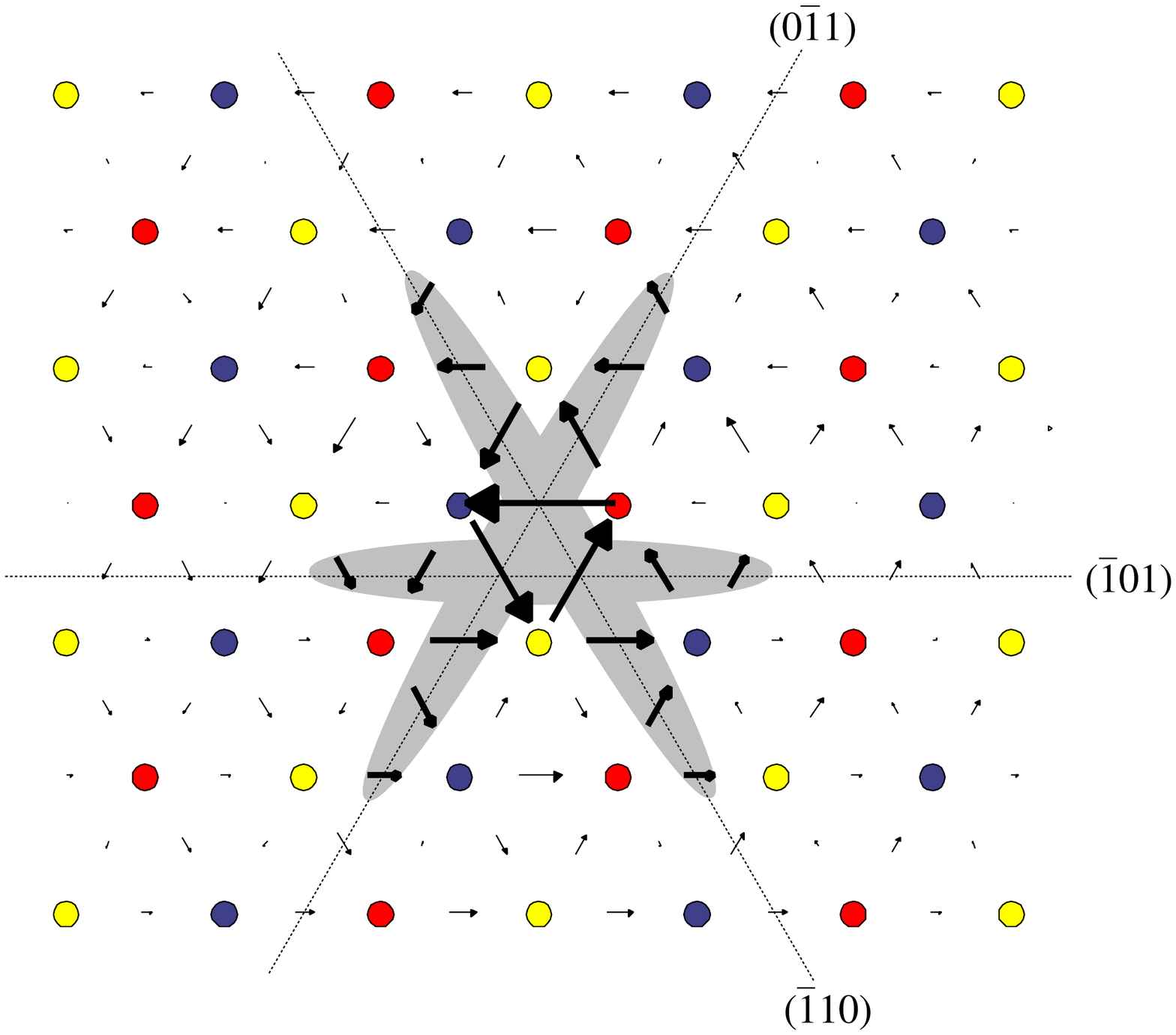} \hfill
    \includegraphics[width=7.3cm]{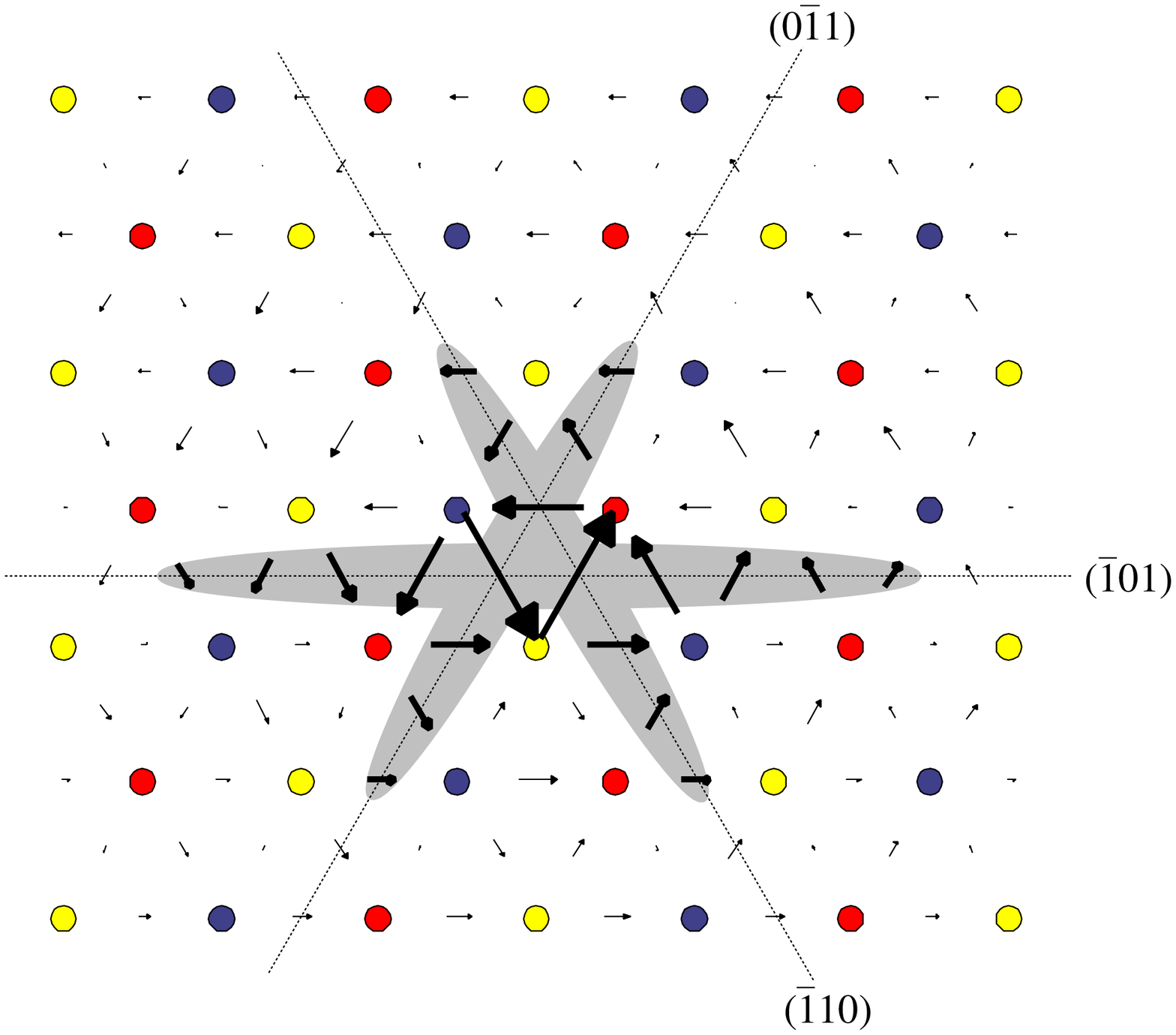}\\
  a) $\tau/C_{44}=-0.05$  \hskip5cm b) $\tau/C_{44}=+0.05$ \\
  \caption{Structure of the relaxed dislocation core upon applying: (a) negative, and (b)
      positive shear stress perpendicular to the slip direction. The stress tensor
      (\ref{eq_tensor_tau}) is applied in the coordinate system where the $y$-axis coincides with
      the normal of the $(\bar{1}01)$ plane.}
  \label{fig_scrW_tau0.05}
\end{figure}

For simplicity, we will discuss here only the case when the stress tensor is applied in the
orientation for which the $y$-axis coincides with the normal of the $(\bar{1}01)$ plane, and the
$z$-axis is parallel to the $[111]$ slip direction. The final structures of the dislocation core
obtained by relaxing the simulated block at $\tau=\pm 0.05C_{44}$ are shown in
\reffig{fig_scrW_tau0.05}. At negative $\tau$, the dislocation core constricts on the $(\bar{1}01)$
plane and extends on both $(0\bar{1}1)$ and $(\bar{1}10)$ planes. Due to the larger spreading of the
core on the two low-stressed $\gplane{110}$ planes, the dislocation can move more easily in
$(0\bar{1}1)$ and $(\bar{1}10)$ planes than in the most-highly stressed $(\bar{1}01)$ plane. On the
other hand, at positive $\tau$, the dislocation core extends on the $(\bar{1}01)$ plane and
constricts significantly on both $(0\bar{1}1)$ and $(\bar{1}10)$ planes. Since the core is now
extended in the $(\bar{1}01)$ plane, one may expect that the subsequent loading by the shear stress
parallel to the slip direction will move the dislocation in this plane.

In order to quantify the effect of the shear stress perpendicular to the slip direction on the
magnitude of the CRSS for slip, we simulated a combined loading by these two stresses. The MRSSPs of
applied loading considered in these simulations were identical to those involved in the studies of
molybdenum, i.e. $(\bar{1}01)$ at $\chi=0$, $(\bar{6}15)$ and $(\bar{5}\bar{1}6)$ at
$\chi=\pm9\deg$, $(\bar{3}12)$ and $(\bar{2}\bar{1}3)$ at $\chi=\pm19\deg$, and $(\bar{9}45)$ and
$(\bar{5}\bar{4}9)$ at $\chi=\pm26\deg$. In each simulation, we first applied the shear stress
perpendicular to the slip direction, $\tau$, according to the stress tensor
(\ref{eq_tensor_tau}). For a completely relaxed block at a given $\tau$, we superimposed the shear
stress parallel to the slip direction, $\sigma$, by incrementally building up the stress tensor
(\ref{eq_tensor_sigma}). When $\sigma$ reaches its critical value, identified as the CRSS for slip,
the total applied stress tensor is given by \refeq{eq_tensor_full} and the dislocation moves through
the crystal. These simulations were carried out for several discrete values of $\tau$ that cover the
usual magnitudes of stresses accessible in tension/compression tests. The dependence of the CRSS on
$\tau$ for the above-mentioned orientations of the MRSSP is shown in
\reffig{fig_CRSS_tau_WBOP}. Recall that, if the Schmid law were valid in tungsten, the CRSS would be
independent of $\tau$, and its magnitude would be equal to the Schmid stress,
i.e. $\CRSS(\chi=0)=0.028C_{44}$. However, the obtained data clearly prove that the shear stress
perpendicular to the slip direction strongly affects the magnitude of the CRSS for slip of screw
dislocations in tungsten.

\begin{figure}[p]
  \centering
  \includegraphics[width=12cm]{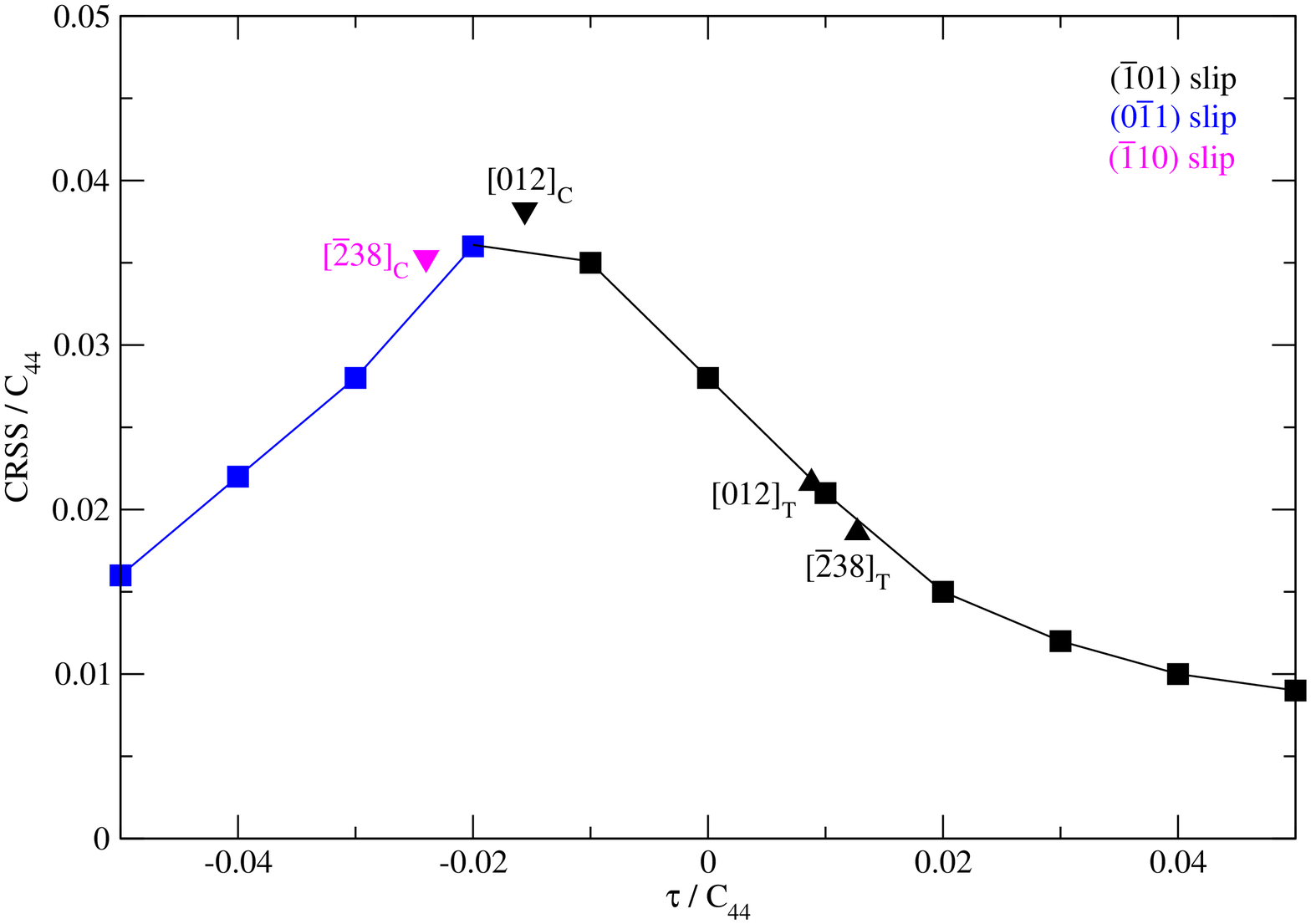} \\
  a) MRSSP $(\bar{1}01)$, $\chi=0$
  \vskip2em
  \includegraphics[width=12cm]{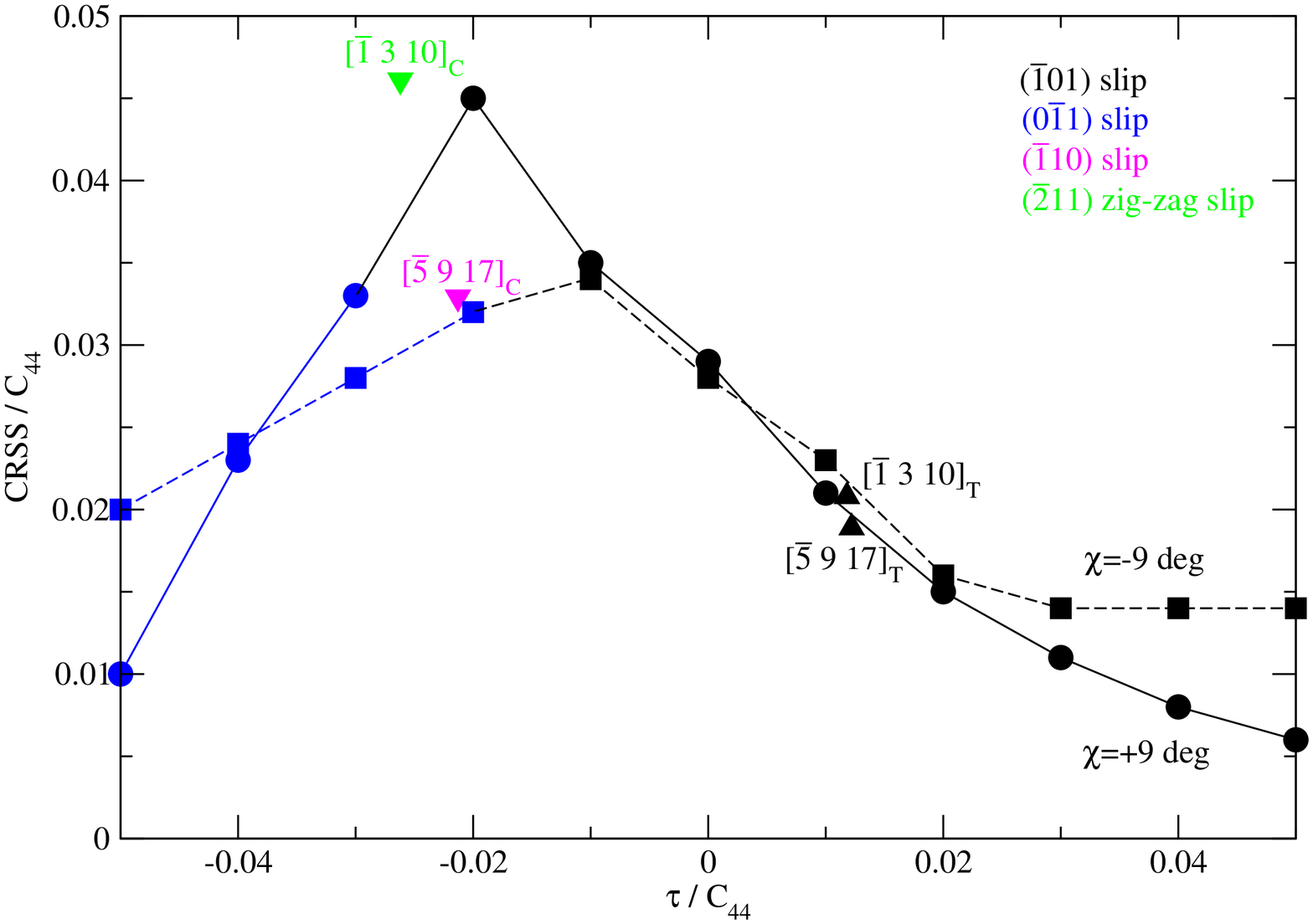} \\
  b) MRSSP $(\bar{6}15)$, $\chi=+9\deg$ and 
    $(\bar{5}\bar{1}6)$, $\chi=-9\deg$ 
\end{figure}
\begin{figure}[p]
  \centering
  \includegraphics[width=12cm]{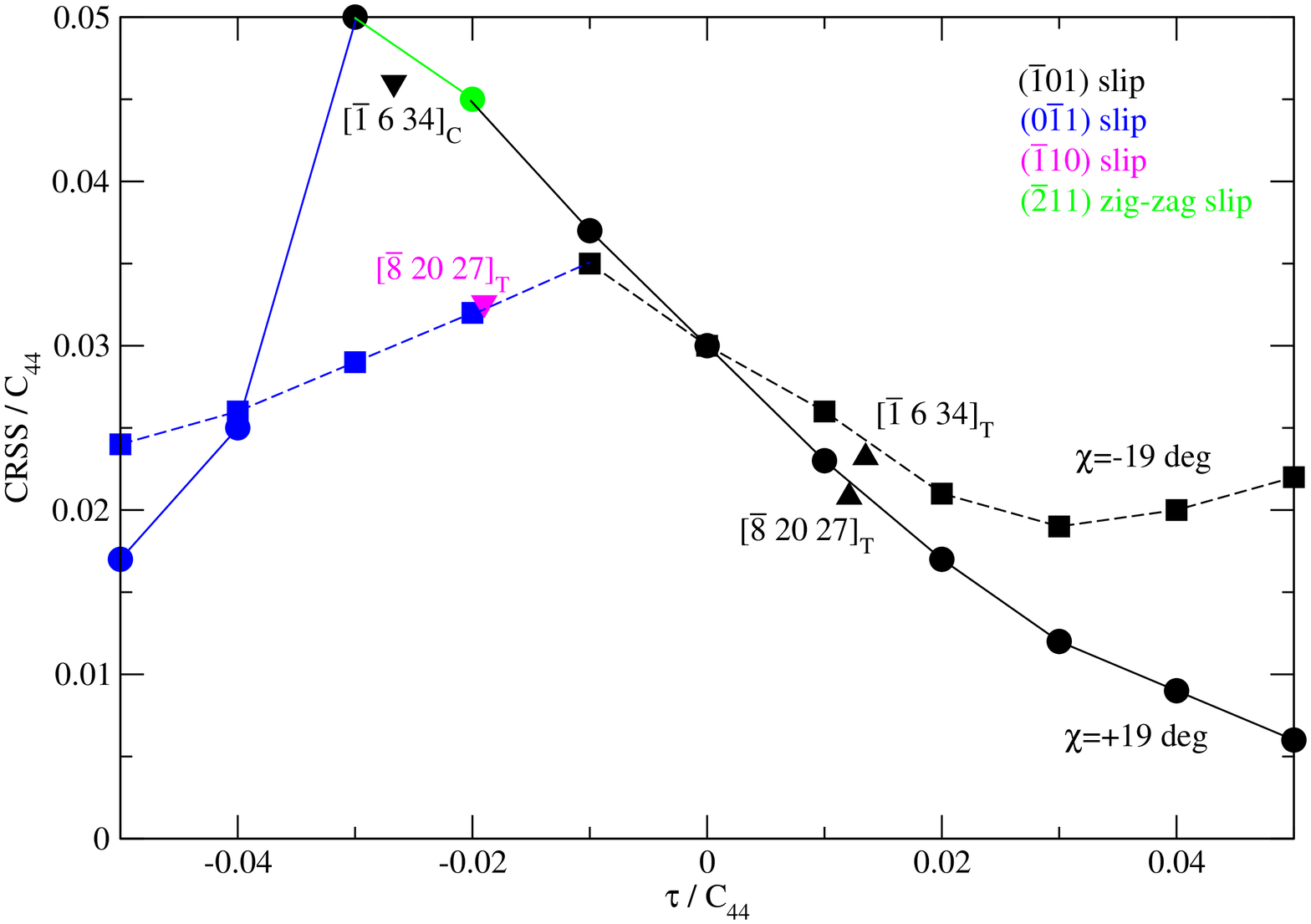} \\
  c) MRSSP $(\bar{3}12)$, $\chi=+19\deg$ and 
    $(\bar{2}\bar{1}3)$, $\chi=-19\deg$ 
  \vskip2em
  \includegraphics[width=12cm]{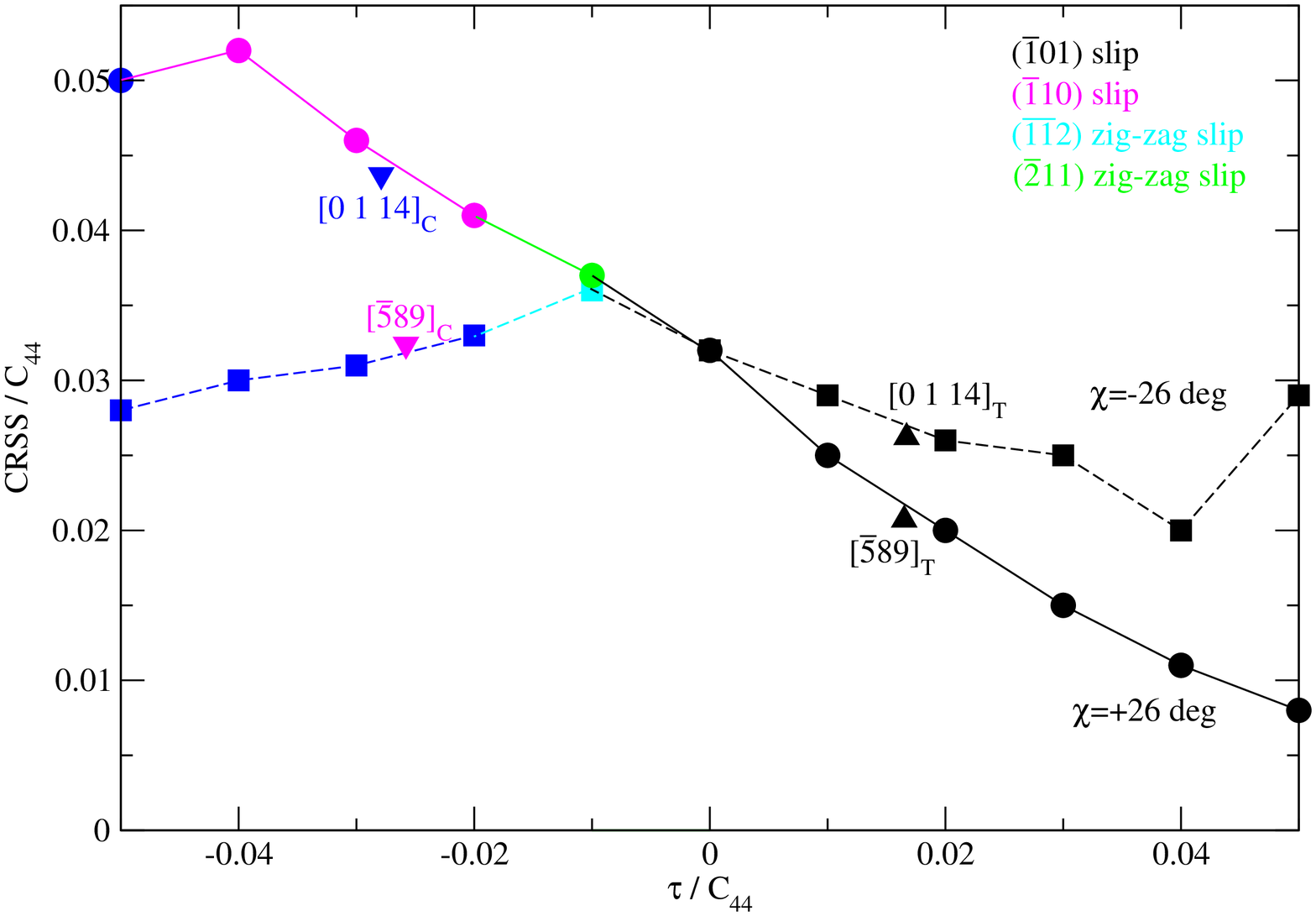} \\
  d) MRSSP $(\bar{9}45)$, $\chi=+26\deg$ and 
    $(\bar{5}\bar{4}9)$, $\chi=-26\deg$ \\[1em]
  \parbox{14cm}{\caption{Dependence of the CRSS on the shear stress perpendicular to the slip
  direction, $\tau$, for various orientations of the MRSSP.}
  \label{fig_CRSS_tau_WBOP}}
\end{figure}

At positive $\tau$, the CRSS decreases relative to $\tau=0$ and the slip always occurs on the most
highly stressed $(\bar{1}01)$ plane. At small negative values of $\tau$, the dislocation glides on
the most highly stressed $(\bar{1}01)$ plane. However, larger negative values of $\tau$ induce a
significant core distortion, and the slip then occurs preferentially on either the $(0\bar{1}1)$ or
the $(\bar{1}10)$ plane. The zig-zag slip on $\gplane{112}$ planes for $\chi=\pm26\deg$ and negative
$\tau$ is caused by virtually equal shearing of the two adjacent $\gplane{110}$ planes. If we
consider all possible slip systems, the Schmid factors corresponding to the $(0\bar{1}1)[111]$ and
the $(\bar{1}10)[111]$ systems are typically only the fourth or the fifth highest among available
slip systems, much lower than the most highly stressed $(\bar{1}01)[111]$ system. Similarly as in
molybdenum, this gives rise to the anomalous slip which occurs as a consequence of the effect of the
shear stress perpendicular to the slip direction.

It should be noted that, for positive $\tau$, the CRSS for antitwinning shear ($\chi>0$) is
\emph{lower} than the corresponding CRSS for the twinning shear ($\chi<0$). This is completely
opposite to what we observed for molybdenum in \reffig{fig_CRSS_tau_MoBOP}a-d where, for positive
$\tau$, the CRSS for antitwinning shear is always \emph{higher} than that for the twinning shear.

Superimposed in \reffig{fig_CRSS_tau_WBOP}a-d are also the results for several different
orientations of uniaxial loadings. It is important to recognize that for such loadings the resolved
$\tau$ is always positive for tension and negative for compression. One can observe a close
agreement of the $\CRSS-\tau$ data calculated for tension/compression with the data for loading by a
combination of the shear stresses perpendicular and parallel to the slip direction. This suggests
that we have identified unambiguously all stress components that affect the plastic flow of single
crystals of bcc tungsten. These stresses are: (i) shear stress perpendicular to the slip direction
that merely changes the structure of the dislocation core, and (ii) shear stress parallel to the
slip direction whose critical value (CRSS) determines the onset of the dislocation glide. The
concomitant effect of these stress components on the onset of glide of an isolated $1/2[111]$ screw
dislocation in tungsten is then given by the calculated $\CRSS-\tau$ dependencies. Unlike in
molybdenum, however, in tungsten the CRSS for loading by pure shear stress parallel to the slip
direction follows the Schmid law. Since there is no twinning-antitwinning asymmetry in tungsten, the
only non-glide stress is the shear stress perpendicular to the slip direction.

%----------------------------------------------------------------------------------------------------
%----------------------------------------------------------------------------------------------------

\section{Construction of the 0~K effective yield criterion}

\subsection{Fitting the atomistic results}

The results of the atomistic studies, presented in the last section, can now be used to construct
the effective yield criterion for tungsten. Similarly as in Section~\ref{sec_yieldcrit_MoBOP} for
molybdenum, we will first determine the parameters $a_1$ and $\tau^*_{cr}$ in the restricted form of
the $\tau^*$ criterion by fitting the $\CRSS-\chi$ dependence for loading by pure shear stress
parallel to the slip direction, see \reffig{fig_CRSS_chi_WBOP}. Keeping $a_1$ and $\tau^*_{cr}$
fixed, we subsequently obtain the coefficients $a_2$ and $a_3$ by fitting the $\CRSS-\tau$
dependencies for $\tau/C_{44}=\pm0.01$. Again, we consider \apriori{} only those atomistic data
for which the slip occurs on the most highly stressed $(\bar{1}01)$ plane. If the criterion is
constructed correctly, the change of the slip plane at larger negative $\tau$, shown in
\reffig{fig_CRSS_tau_WBOP}a-d, will be reproduced automatically.

\begin{table}[!htb]
  \centering
  \parbox{10cm}{\caption{Coefficients of the effective yield criterion for tungsten determined by
  fitting the 0~K atomistic data.}
  \label{tab_tstar_params_W}}\\[1em]

  \begin{tabular}{cccc}
    \hline
    $a_1$ & $a_2$ & $a_3$ & $\tau^*_{cr}/C_{44}$ \\
    \hline
    0 & 0.56 & 0.75 & 0.028 \\
    \hline
  \end{tabular}
\end{table}

The coefficients of the $\tau^*$ criterion for tungsten, determined by fitting the atomistic results,
are given in \reftab{tab_tstar_params_W}. Because the $\CRSS-\chi$ dependence obtained from loading
by pure shear stress parallel to the slip direction follows the Schmid law, the parameter $a_1$ is
obviously zero. This means that there is no twinning-antitwinning asymmetry in tungsten, which is
corroborated already in \reffig{fig_CRSS_chi_WBOP}. Both parameters $a_2$ and $a_3$ are now large,
and this implies a strong dependence of $\tau^*$ on the shear stress perpendicular to the slip
direction. The full form of the $\tau^*$ criterion is given by \refeq{eq_tstar_full} and, in the
case of tungsten, involves only three terms:

\begin{equation}
  \tau^* = \sigma^{(\bar{1}01)} + a_2 \tau^{(\bar{1}01)} + a_3\tau^{(0\bar{1}1)} \leq \tau_{cr}^* \ .
  \label{eq_tstar_full_WBOP}
\end{equation}

\begin{figure}[!p]
  \centering 
  \includegraphics[width=12cm]{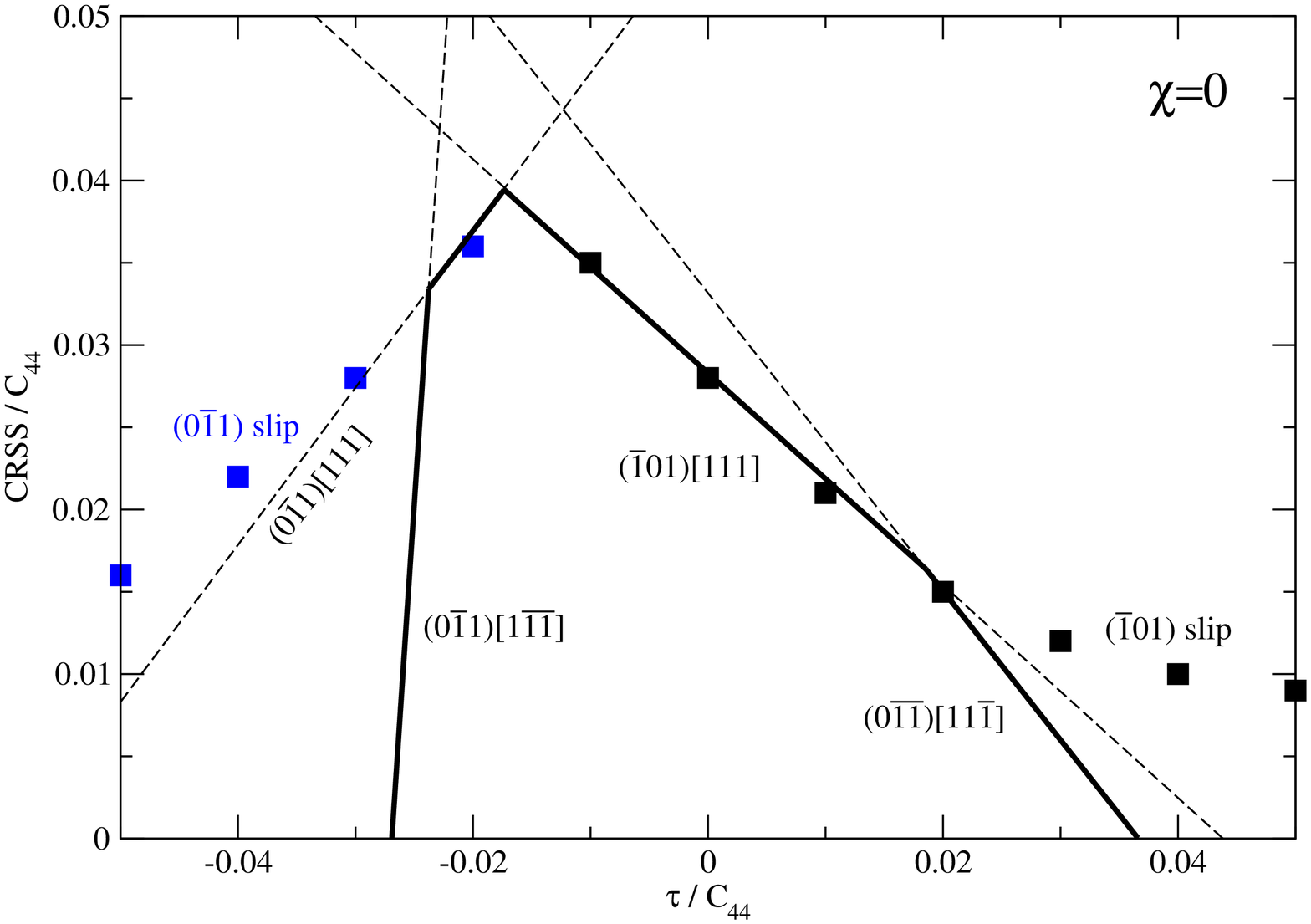} \\
  a) MRSSP $(\bar{1}01)$ at $\chi=0$ \\[3em]
  \includegraphics[width=12cm]{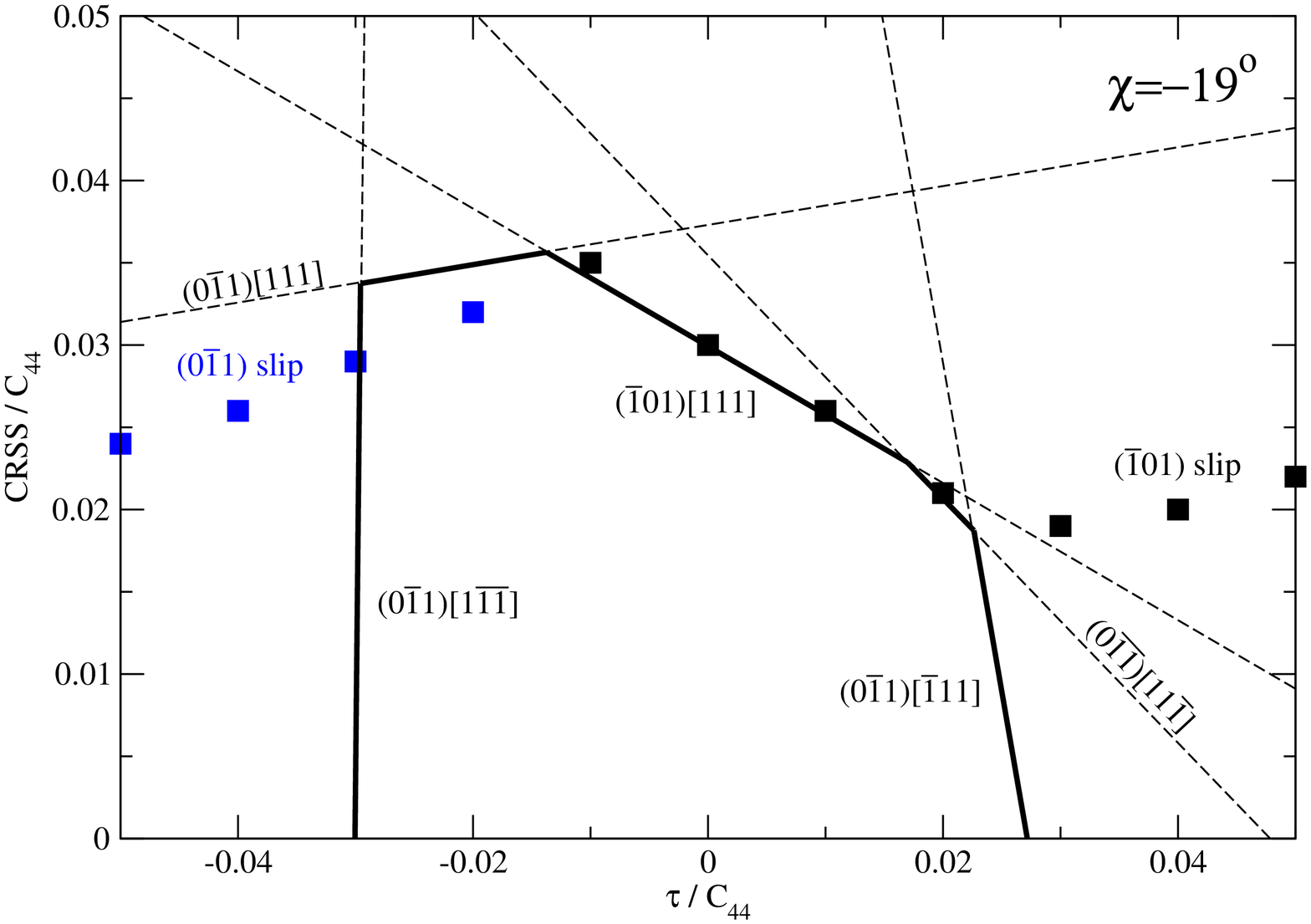} \\
  b) MRSSP $(\bar{2}\bar{1}3)$ at $\chi=-19\deg$
\end{figure}
\begin{figure}[!htb]
  \centering
  \includegraphics[width=12cm]{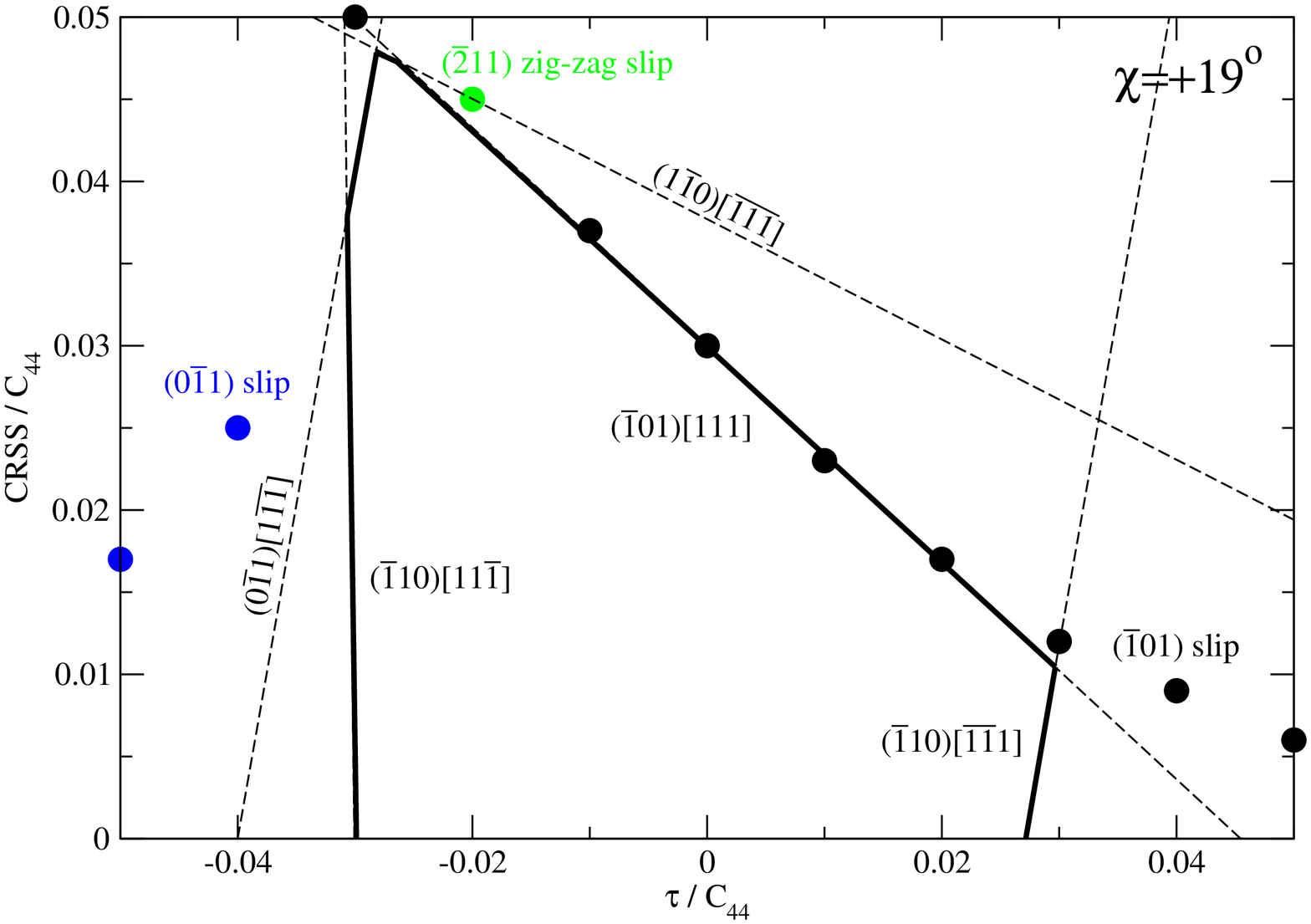} \\
  c) MRSSP $(\bar{3}12)$ at $\chi=+19\deg$ \\[1em]
  \parbox{14cm}{\caption{Critical lines from the $\tau^*$ criterion for tungsten calculated for
  three characteristic orientations of the MRSSP. Compare with
  \reffig{fig_CRSS_tau_fit_MoBOP} for molybdenum.}
    \label{fig_CRSS_tau_fit_WBOP}}
\end{figure}

Similarly as in molybdenum, the $\tau^*$ criterion (\ref{eq_tstar_full_WBOP}) with the coefficients
from \reftab{tab_tstar_params_W} can now be used to generalize the $\CRSS-\tau$ dependencies in
\reffig{fig_CRSS_tau_WBOP} to real single crystals in which any $\gplane{110}\gdir{111}$ system from
\reftab{tab_bcc24sys} can be activated for slip by the applied loading. The details of these
calculations are given in Section~\ref{sec_4slipsys} and will not be repeated here. The results of
these calculations are shown in \reffig{fig_CRSS_tau_fit_WBOP}, where the critical lines represent
the loci of the loading states that cause the activation of individual slip systems. The inner
envelope of these critical lines represents the projection of the yield surface.

For any loading, one can easily resolve the loading path in the $\CRSS-\tau$ plots in
\reffig{fig_CRSS_tau_fit_WBOP} corresponding to the actual MRSSP. If this path reaches the yield
polygon at its edge, the plastic flow occurs by single slip on the system labeled in the figure. On
the other hand, if it meets the yield polygon at its corner, two or more slip systems are activated
for slip at once, and the plastic flow is then generated by simultaneous operation of these
systems. These conclusions are very general and hold for any orientation of the MRSSP, as has been
shown already in the case of molybdenum.

%----------------------------------------------------------------------------------------------------

\subsection{The 0~K yield surface}

\begin{figure}[!b]
  \centering \includegraphics[width=10cm]{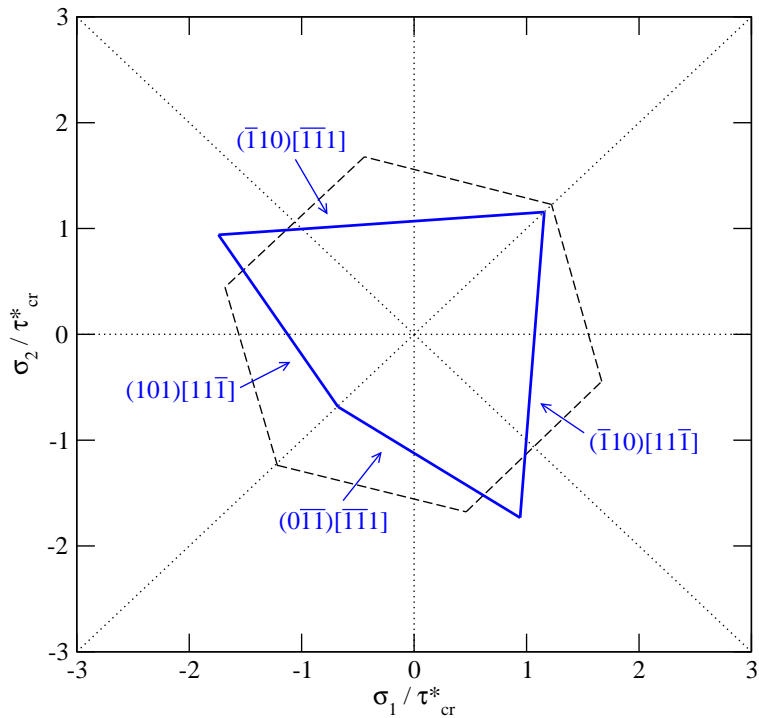}
  \parbox{14cm}{\caption{$\pi$-plane projection of the yield surface for tungsten calculated from the
  $\tau^*$ criterion (solid lines). The dashed Tresca hexagon corresponds to the purely Schmid
  behavior for which $a_1=a_2=a_3=0$.}
    \label{fig_yieldsurf_WBOP}}
\end{figure}

For simplicity, we will again discuss the $\pi$-plane projection of the yield surface that
represents its cut by a plane whose normal is defined in the principal space as
$\sigma_1=\sigma_2=\sigma_3$. If the plastic flow of tungsten were governed by the Schmid law, the
$\pi$-plane projection of the yield surface would coincide with the Tresca's regular hexagon. The
actual shape of the yield surface is plotted in \reffig{fig_yieldsurf_WBOP} and compared to the
prediction of the Schmid law. Each edge of the polygon corresponds to operation of a particular slip
system, while the corners mark a simultaneous slip on more than one system. 

Recall that, in the case of molybdenum, the $\pi$-plane projection of the yield surface was
represented by an irregular hexagon, see \reffig{fig_yieldsurf_MoBOP}. In the case of tungsten, the
edges of the yield polygon in \reffig{fig_yieldsurf_WBOP}, corresponding to
the $(\bar{1}10)[\bar{1}\bar{1}1]$ and the $(\bar{1}10)[11\bar{1}]$ systems, make an obtuse angle and this
causes that the edges of the yield polygon in \reffig{fig_yieldsurf_MoBOP}, corresponding to slip
systems $(110)[1\bar{1}\bar{1}]$ and $(\bar{1}\bar{1}0)[1\bar{1}1]$, degenerate into a
corner. Consequently, the $\pi$-plane projection of the yield surface in tungsten is represented by
an irregular tetragon shown in \reffig{fig_yieldsurf_WBOP}.

\begin{figure}[!htb]
  \centering
  \includegraphics[width=10cm]{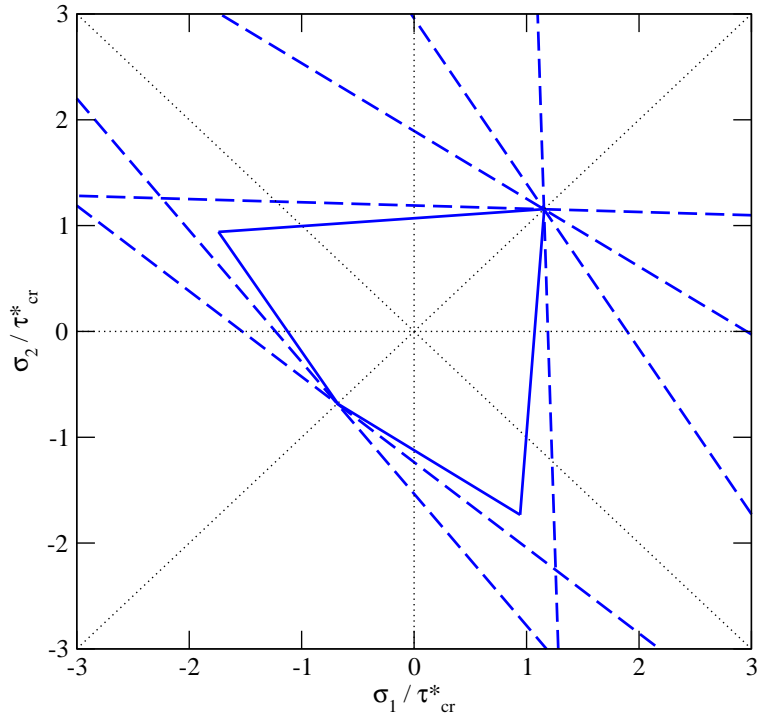}
  \parbox{14cm}{\caption{Yield polygon marking the onset of primary slip (solid line), and the
  critical lines corresponding to operation of non-primary slip systems (dashed lines).}
    \label{fig_yieldsurf_allsys_WBOP}}
\end{figure}

Similarly as in molybdenum, multislip occurs when the loading path reaches one of the corners of the
yield polygon. This is demonstrated in \reffig{fig_yieldsurf_allsys_WBOP}, where the dashed lines
represent the loci of the loading states for which a given non-primary system becomes activated for
slip. Clearly, only the loading directions for which $\sigma_1=\sigma_2$ lead to the onset of
multislip on more than two systems. This means that, as the loading path reaches the yield surface
close to any of these two corners, many slip systems will act simultaneously and significant strain
bursts should be detectable in experiments.

%----------------------------------------------------------------------------------------------------

\subsection{The most operative slip systems in uniaxial loading}

In Section~\ref{sec_predact_MoBOP}, we used the 0~K effective yield criterion for molybdenum to
predict the most operative slip systems under loading in tension and compression. We have shown in
\reffig{fig_sgtria_uniax} that tensile loading along any direction always leads to the primary slip
on the most-highly stressed $(\bar{1}01)[111]$ system, while for compression the nature of slip
depends on the orientation of applied loading. 

\begin{figure}[!htb]
  \centering
  \includegraphics[width=9cm]{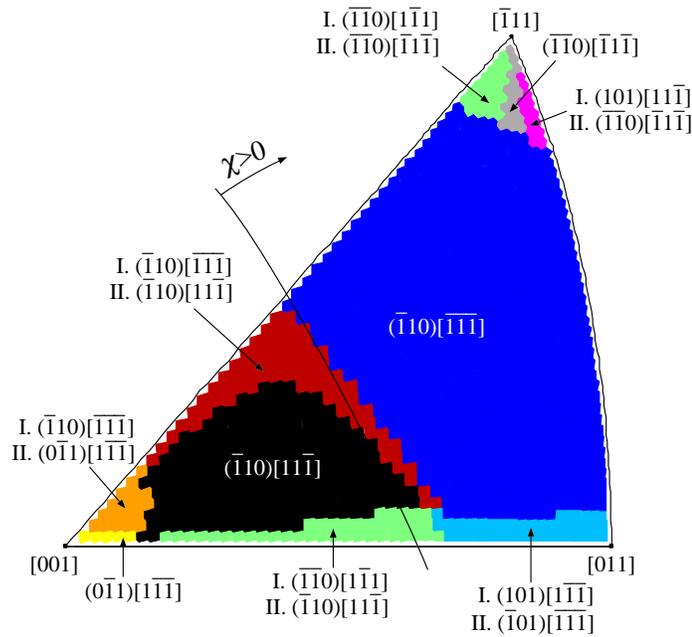} \\
  \parbox{12cm}{\caption{Primary slip systems for loading in \emph{compression} determined from the $0~\K$
    effective yield criterion for tungsten. Two slip systems operate simultaneously if the uniaxial
    yield stress of the secondary system (II.) is within 2\% of that of the primary slip system
    (I.)}
  \label{fig_sgtria_actsys_WBOP}}
\end{figure}

In tungsten, loading in tension along any axis in the stereographic triangle causes single slip on
the most highly stressed $(\bar{1}01)[111]$ system, similarly as in molybdenum. In compression, the
situation is much more complex and the choice of the primary system depends on the orientation of
applied loading (see \reffig{fig_sgtria_actsys_WBOP}). For a large number of loading axes with
MRSSPs corresponding to $\chi>0$, the most operative system with positive Schmid stress is
$(\bar{1}10)[\bar{1}\bar{1}\bar{1}]$. On the other hand, for loading axes corresponding to the
MRSSPs at $\chi<0$, the most operative system changes to $(\bar{1}10)[11\bar{1}]$. As the loading
axis deviates toward one of the corners of the triangle, several slip systems become equally
prominent, and the plastic flow proceeds by multislip on the corresponding systems. Interestingly,
the $(\bar{1}01)[\bar{1}\bar{1}\bar{1}]$ system, which was most prominent in molybdenum, is now deferred
to a tiny elongated region of multislip close to the $[011]$ corner of the triangle.

In \reffig{fig_sgtria_actsys_WBOP}, the Miller indices of the slip systems are written in such a way
that the Schmid stress resolved in each system is always positive. This means that for loading in
tension in the stereographic triangle $[001]-[011]-[\bar{1}11]$ (not shown in
\reffig{fig_sgtria_actsys_WBOP}), only the slip systems 1 to 12 in \reftab{tab_bcc24sys} can become
operative. Under loading in compression, the systems 1 to 12 are sheared in the opposite sense, the
Schmid stress becomes negative, and, therefore, these systems cannot be activated for slip. Instead,
one has to consider the conjugate systems 13 to 24 with opposite slip directions, which leads us to
write $(\bar{1}01)[\bar{1}\bar{1}\bar{1}]$ and not $(\bar{1}01)[111]$.

%----------------------------------------------------------------------------------------------------

\subsection{Tension-compression yield stress asymmetry}

\begin{figure}[!htb]
  \centering
  \includegraphics[width=12.5cm]{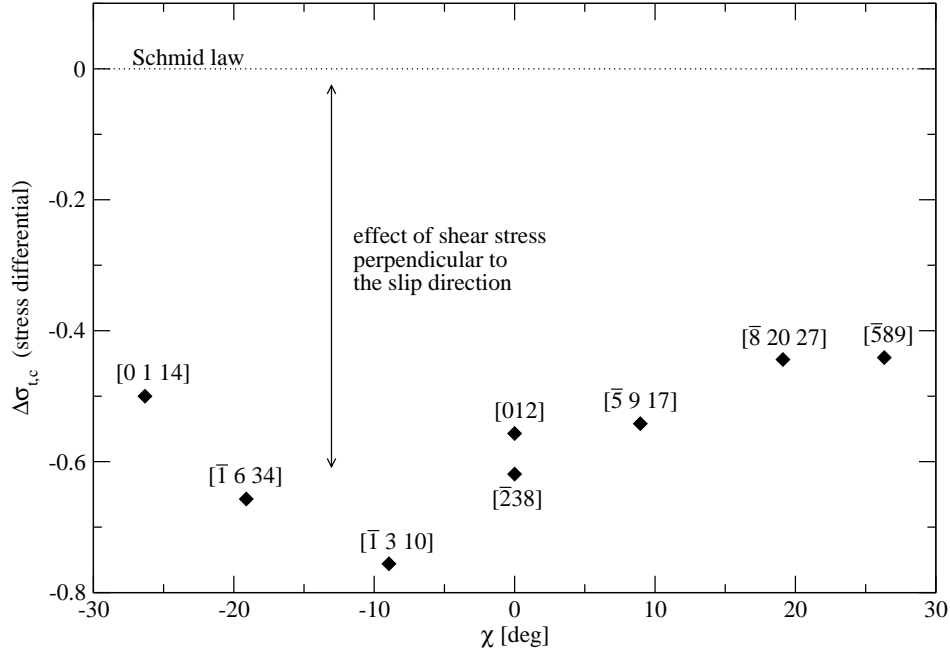}
  \parbox{14cm}{\caption{Variation of the stress differential (\ref{eq_stressdiff}) in tungsten with
      the orientation of the MRSSP for the loading axes from \reffig{fig_CRSS_chi_WBOP}.}
  \label{fig_tc_asym_WBOP}}
\end{figure}

The asymmetry between the yield stress in tension and compression will be again determined by the
stress differential $\Delta\sigma_{t,c}$ that is defined by \refeq{eq_stressdiff}. For any loading
axis, the \emph{uniaxial} yield stress in tension and compression can be calculated directly from
the constructed effective yield criterion. The distribution of $\Delta\sigma_{t,c}$ for tungsten is
plotted in \reffig{fig_tc_asym_WBOP}, where the symbols correspond to the same loading axes as in
\reffig{fig_CRSS_chi_WBOP}.

For any loading axis, we can determine the angle of the MRSSP, $\chi$, and resolve the shear
stresses perpendicular and parallel to the slip direction. If only the shear stress parallel to the
slip direction affected the plastic flow, $\Delta\sigma_{t,c}$ would be antisymmetric with respect
to $\chi$. Since there is no twinning-antitwinning asymmetry in tungsten, this implies that
there would be no asymmetry between the yield stresses in tension and compression, and
$\Delta\sigma_{t,c}$ would be zero for any orientation of the loading axis. This is clearly not the
case, as can be seen in \reffig{fig_tc_asym_WBOP}, and $\Delta\sigma_{t,c}$ is thus nonzero due to
the effect of the shear stress perpendicular to the slip direction. The effect of these non-glide
stresses is so pronounced in tungsten that the region of positive $\Delta\sigma_{t,c}$, that emerged
in molybdenum for orientations close to the $[011]$ corner (see \reffig{fig_sgtria_tcasym}a), is
completely eliminated. Unfortunately, unlike for molybdenum, no comparison with experiments is
currently possible for tungsten due to the lack of measurements performed on a given orientation in
both tension and compression.

%----------------------------------------------------------------------------------------------------
%----------------------------------------------------------------------------------------------------

\section{Construction of the Peierls potential}

The Peierls potential for tungsten can be constructed by following the same procedure that was
introduced in Section~\ref{sec_EffPeierlsPot}. Firstly, the mapping function $m(x,y)$ is written
such that it reproduces the three-fold symmetry of the $[111]$ axis. Here, we will only consider the
Peierls potential with flat saddle-points that was previously shown to reproduce correctly the
experimentally measured temperature dependence of the yield stress for molybdenum. The modified
mapping function that produces the Peierls barrier with flat maximum is obtained by perturbing all
saddle-points of $m(x,y)$ by \refeq{eq_fFD}. In this function, we use the same parameters as for
molybdenum, i.e. $\alpha=0.2$, $\beta=0.12$, and $r_0=a_0/3\sqrt{3}$, where $a_0$ is the periodicity
of the lattice in the $\gdir{112}$ direction in the $\gplane{110}$ plane. This function is applied
to every saddle point of $m(x,y)$, which is emphasized by using the operator notation
$\hat{f}m(x,y)$.

In order to evaluate the height of the Peierls potential, $\Delta V$, we consider loading by pure
shear stress parallel to the slip direction in the $(\bar{1}01)$ plane. The actual MEP for an
elementary jump of the dislocation between two adjacent minima on this plane is calculated using the
NEB method and provides the Peierls barrier $V(\xi)$, where $\xi$ is the curved transition
coordinate. The height of the Peierls potential, $\Delta V$, is obtained from the condition
(\ref{eq_sigmaP}) in which the Peierls stress is now $\sigma_P=0.028C_{44}$. This condition is
satisfied for
\begin{equation}
  \Delta V=0.1369~\eV/\A \ .
  \label{eq_dV_WBOP}
\end{equation}

The next step in the construction of the Peierls potential concerns the fitting of the function
$K_\sigma(\chi)$ in \refeq{eq_Vsigma}, where $V_\sigma(\chi,\theta)$ represents the distortion of
the three-fold symmetric basis of the potential by non-glide shear stresses parallel to the slip
direction. Because the Schmid law now applies for loading by pure shear stress parallel to the
slip direction, $K_\sigma(\chi)=0$, and, therefore, $V_\sigma(\chi,\theta)$ does not enter the
formula (\ref{eq_V3}) for the Peierls potential. Consequently, the Peierls potential for tungsten is
\emph{not} a function of the shear stress parallel to the slip direction.

Hence, we only need to consider the distortion of the Peierls potential by the shear stress
perpendicular to the slip direction that is represented by the function $V_\tau(\chi,\theta)$, given
by \refeq{eq_Vtau}. To determine the parameters of the function $K_\tau(\chi)$, we again consider
loading by $\tau=\pm0.01C_{44}$ for which the dislocation moves on the $(\bar{1}01)$ plane, and,
therefore, the Peierls stress in \refeq{eq_sigmaP} is simply $\sigma_P=\CRSS\cos\chi$. In this
equation, the CRSS is determined from the $\tau^*$ criterion for tungsten, i.e. by means of
\refeq{eq_CRSS@V3} for parameters $a_1$, $a_2$, $a_3$ and $\tau^*_{cr}$ from
\reftab{tab_tstar_params_W}. For a given angle $\chi$, the corresponding value of $K_\tau$ is
obtained such that \refeq{eq_sigmaP} is satisfied. This requirement yields
\begin{equation}
  K_\tau(\chi) = -0.413 - 0.216\chi + 0.782\chi^2 \ .
  \label{eq_Ka3_WBOP}
\end{equation}
In the case of tungsten, the Peierls potential (\ref{eq_V3}) then reduces to
\begin{equation}
  V(x,y) = [\Delta V + V_\tau(\chi,\theta)]\, \hat{f} m(x,y) \ ,
\end{equation}
where $\Delta V$ is given by \refeq{eq_dV_WBOP}, and the functional form of $V_\tau(\chi,\theta)$ by
\refeq{eq_Vtau} with $K_\tau(\chi)$ from \refeq{eq_Ka3_WBOP}.

For positive shear stress perpendicular to the slip direction, $\tau$, the potential develops a
low-energy path along the $(\bar{1}01)$ plane and the dislocation thus preferentially moves on this
plane. At negative $\tau$, this transition is suppressed by a large Peierls barrier and the
dislocation moves instead by elementary steps on the $(0\bar{1}1)$ or the $(\bar{1}10)$ planes for
which the Peierls barrier is lower.

\begin{figure}[!htb]
  \centering
  \includegraphics[width=12cm]{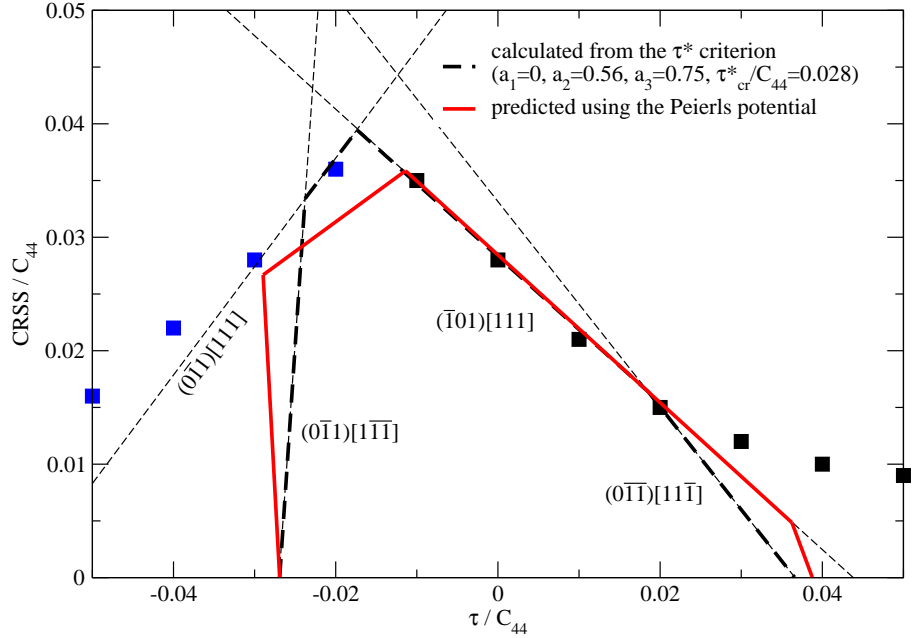}
  \parbox{14cm}{\caption{Critical lines calculated from the $\tau^*$ criterion and their comparison
      with the prediction of the constructed Peierls potential.}
    \label{fig_chi0_fit_WBOP_compare}}
\end{figure}

Similarly as in molybdenum, we can now employ the constructed Peierls potential and predict the
dependence of the CRSS on $\tau$ directly from \refeq{eq_sigmaP}. As explained in detail in
Section~\ref{sec_testpot}, this is achieved by considering a set of uniaxial loadings that
correspond in the $\CRSS-\tau$ projection to straight loading paths emanating from the origin, with
slopes $\eta=\tau/\sigma$. For a given loading, one can use the $\tau^*$ criterion to identify the
four slip systems whose MRSSPs fall in $-30\deg<\chi<+30\deg$, and the Schmid stress is
positive. For each of these systems, the CRSS is then calculated numerically from
\refeq{eq_sigmaP}. This yields a set of four critical points along the given straight loading path
in the $\CRSS-\tau$ projection that correspond to the onset of slip on these four systems. Repeating
this calculation for a number of loading paths with different slopes $\eta$, and connecting the
critical points for individual systems, generates the critical lines plotted as solid lines in
\reffig{fig_chi0_fit_WBOP_compare}. These lines correspond to the loci of the stress states that
activate individual slip systems. Obviously, the critical line corresponding to the
$(\bar{1}01)[111]$ system is in good agreement with the prediction of the $\tau^*$ criterion, because
the data for the $(\bar{1}01)$ slip were used to fit the parameters in the function
$K_\tau(\chi)$. However, the possibility of slip on other systems was not \apriori{} considered, and,
therefore, the agreement with the trend dictated for these systems by the $\tau^*$ criterion is a
natural consequence of the constructed Peierls potential.

%----------------------------------------------------------------------------------------------------
%----------------------------------------------------------------------------------------------------

\section{Thermally activated plastic flow of tungsten}

%----------------------------------------------------------------------------------------------------

\subsection{Parameters entering the activation enthalpy}

In order to incorporate the temperature and strain rate dependence of the yield stress, we will
again first calculate the stress dependence of the activation enthalpy (for details, see
Section~\ref{sec_expt_hollang}). At low temperatures, this is given by the model of the dislocation
bow-out (Section~\ref{sec_bowout}), while the high-temperature regime is treated by the model of
elastic interaction of fully developed kinks (Section~\ref{sec_low_stresses}). Similarly as in
molybdenum, we have to determine first the line tension of a straight screw dislocation such that
the energy of two isolated kinks, $2H_k$, equals the experimentally deduced value. For tungsten, the
measurements of \citet{brunner:00} imply $2H_k=2.06~\eV$, which is identical to the value obtained
from the approximate formula, $2H_k\approx0.1\mu b^3$, proposed by \citet{conrad:63}. If
\refeq{eq_Hk_0stress} is to give the kink energy, the line tension of a straight screw dislocation
has to be $V_0=p\mu b^2$ where $p=0.287$. With the $V_0$ fixed, the stress dependence of the
activation enthalpy can be readily calculated from \refeqs{eq_Hb_stress} and \ref{eq_Hkp_final}.

The temperature and strain rate dependence of the yield stress can be included by means of the
Arrhenius law $\dot\gamma=\dot\gamma_0\exp[-H(\sigma)/kT]$, where $H(\sigma)$ is the
stress-dependent activation enthalpy calculated for a given orientation of applied loading. A
straightforward way of determining $\dot\gamma_0$ is from experiments, by considering the
temperature $T_k$ at which the thermal component of the yield stress vanishes. The activation
enthalpy then reaches its maximum, $2H_k$, and the Arrhenius law gives
$q\equiv\ln(\dot\gamma_0/\dot\gamma)=2H_k/kT_k$, where $k$ is the Boltzmann constant. From the
strain rate sensitivity experiments of \citet{brunner:00}, performed at the nominal plastic strain
rate $\dot\gamma=8.5\times10^{-4}~\s^{-1}$, the temperature at which the thermal component of the
yield stress vanishes is $T_k=760~\K$. Substituting these values into the equation above yields
$\dot\gamma_0=3.7\times10^{10}~\s^{-1}$ and $q=31.4$. Although this value is in good agreement with
that obtained by fitting the experimental data of \citet{brunner:00} at temperatures close to $T_k$,
their measurements also exhibit an abrupt change of $q$ at 600~K, from the value 31.4 (at
$T>600~\K$) to 23.6 (at $T<600~\K$). If we recall that $q=\ln(\rho_m b v_m/\dot\gamma)$ and assume
that the velocity of mobile dislocations, $v_m$, is constant, this discontinuity implies a sudden
change of $\rho_m$, the density of mobile dislocations, by three orders of magnitude. The
temperature dependence of mobile dislocation density in tungsten is likely to exhibit a transient
regime similar to that shown for intermetallic compounds Ni$_3$(Al,Hf) by \citet{matterstock:99}. In
these materials, however, the increase of $\rho_m$ at the transient temperature is only by a factor
of five. At present, the origin of the above-mentioned abrupt change of $\dot\gamma_0$ and $q$ in
tungsten thus remains unresolved, and, therefore, one should consider the estimated $\dot\gamma_0$
only as an effective value.

%----------------------------------------------------------------------------------------------------

\subsection{Stress dependence of the activation enthalpy and volume}

To allow further correlations with the experiments of \citet{brunner:00}, performed under tension
for an orientation close to the center of the stereographic triangle, we will consider tensile
loading along the $[\bar{1}49]$ axis. The MRSSP of the most highly stressed $(\bar{1}01)[111]$
system is directly the $(\bar{1}01)$ plane ($\chi=0$), and the ratio of the shear stress
perpendicular to the shear stress parallel to the slip direction, resolved in this plane, is
$\eta=\tau/\sigma=0.51$. This is a key information that determines how the shape of the Peierls
potential evolves during loading.

\begin{figure}[!htb]
  \centering
  \includegraphics[width=12cm]{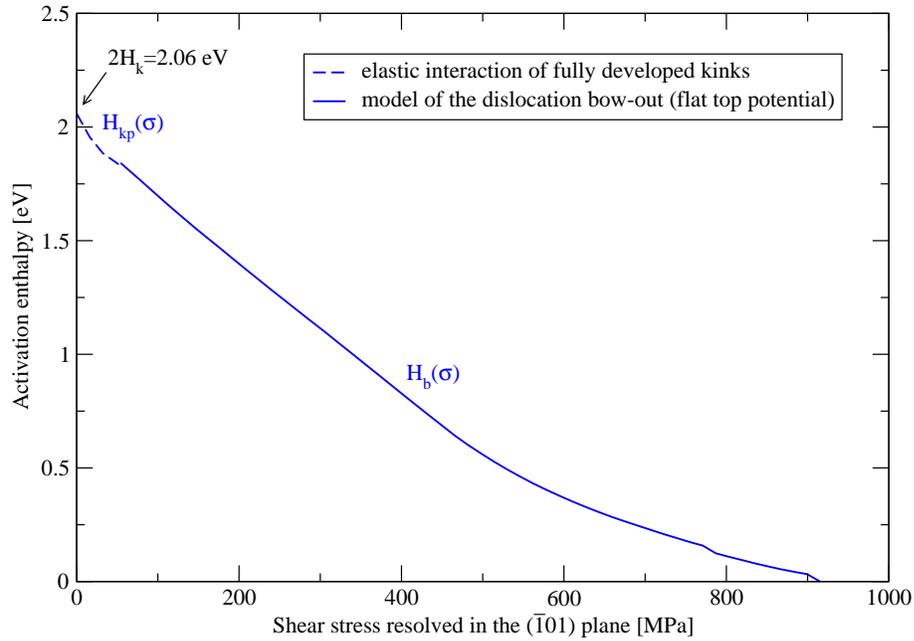}
  \parbox{14cm}{\caption{Stress dependence of the activation enthalpy for loading in tension along
  $[\bar{1}49]$. Here, $2H_k=2.06~\eV$ is the energy of two isolated kinks.}
    \label{fig_actene_brunner_flattop}}
\end{figure}

The activation enthalpy at high and low applied stresses can be obtained from the model of the
dislocation bow-out (Section~\ref{sec_bowout}) and from the model of elastic interaction of fully
developed kinks (Section~\ref{sec_low_stresses}), respectively. This calculation is the same as that
performed in Section~\ref{sec_macroyield_sine} for molybdenum. To ensure agreement between the
theoretically calculated Peierls stress and the yield stress extrapolated from experimental data to
0~K, we have to scale all stresses by a factor of 3.7. This scaling accounts for interactions
between dislocations that are not included in the atomistic simulations and may be also justified by
the model presented in Appendix~\ref{chap_interact}. The obtained stress dependence of the
activation enthalpy is shown in \reffig{fig_actene_brunner_flattop}.

\begin{figure}[!htb]
  \centering 
  \includegraphics[width=12cm]{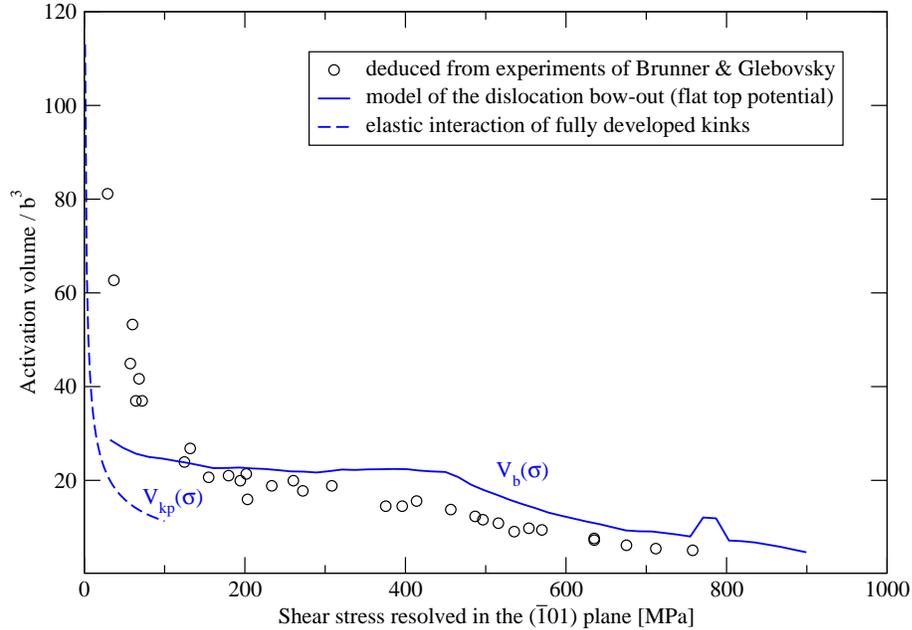}
  \parbox{14.5cm}{\caption{Stress dependence of the activation volume for loading in tension along
      $[\bar{1}49]$. The experimental data are deduced from Fig.~7 of \citet{brunner:00}.}
    \label{fig_actvol_brunner_flattop}}
\end{figure}

In \reffig{fig_actvol_brunner_flattop}, we plot the stress dependence of the activation volume that
was deduced from the experimentally measured temperature dependence of the activation volume
\citep{brunner:00} with help of the temperature dependence of the yield stress for tungsten by
\citet{hollang:01}. The calculated activation volume, obtained as a negative derivative of the
activation enthalpy with respect to stress, is plotted for the two models as full and dashed
curves. The plateau at 400~K and the local maximum close to 800~K are both reminiscent of the same
features shown for molybdenum in \reffig{fig_actvol_hollang_-149_flattop} and are manifestations of
the flat maximum of the Peierls barrier.

%----------------------------------------------------------------------------------------------------

\subsection{Temperature dependence of the yield stress}

The calculation of the temperature dependence of the yield stress was made for the experimental
plastic strain rate $\dot\gamma=8.5\times10^{-4}~\s^{-1}$. The low-temperature region of this
dependence was calculated using the model of the dislocation bow-out, while high temperatures were
treated within the model of elastic interaction of fully developed kinks. At all temperatures
considered, the $(\bar{1}01)[111]$ system has the lowest activation enthalpy and thus the plastic
deformation occurs by single slip on this system.

\begin{figure}[!htb]
  \centering
  \includegraphics[width=12cm]{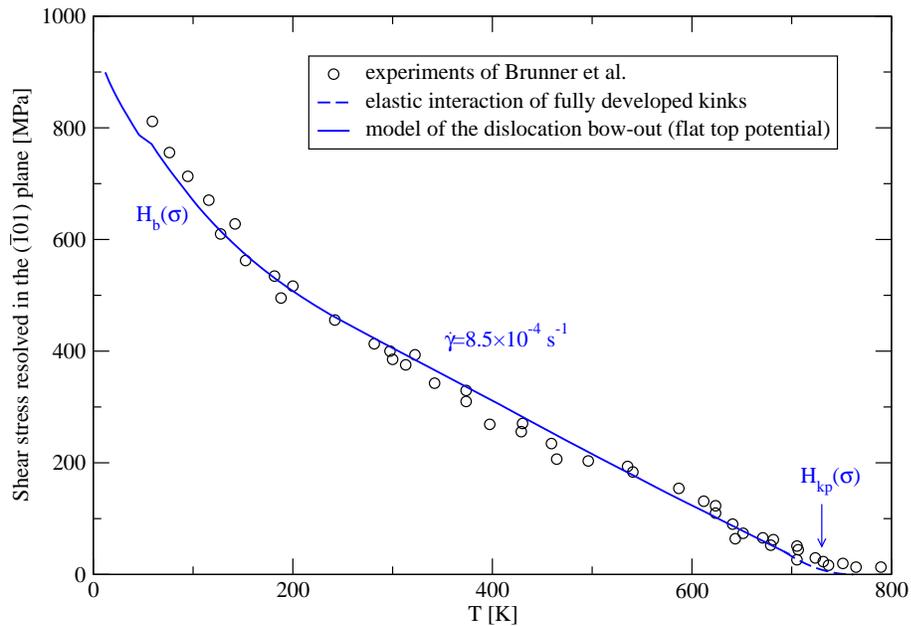}
  \parbox{14cm}{\caption{Calculated temperature dependence of the yield stress (strain rate
      $\dot\gamma=8.5\times10^{-4}~\s^{-1}$) compared with experiments of \citet{brunner:00}.}
    \label{fig_yieldt_brunner_flattop}}
\end{figure}

The theoretically calculated temperature dependence of the yield stress, obtained for the primary
$(\bar{1}01)[111]$ slip system, corresponds to the curve in \reffig{fig_yieldt_brunner_flattop}. It
is important to emphasize, that this dependence has been calculated by considering that the
dislocation moves at all temperatures by elementary jumps on the $(\bar{1}01)$ plane. The very good
agreement of the theoretical predictions with experiments suggests that dislocations may indeed move
by discrete $\gplane{110}$ jumps at all temperatures. Within this model, the change of slope of the
dependence in \reffig{fig_yieldt_brunner_flattop} at 200~K is not caused by any transformation of
the dislocation core, which has been raised as a possible explanation \citep{seeger:04}, but it is a
manifestation of the flat maximum of the constructed Peierls barrier.

%----------------------------------------------------------------------------------------------------
%----------------------------------------------------------------------------------------------------

\section{Temperature and strain rate dependent effective yield criterion}

\begin{figure}[!b]
  \centering
  \includegraphics[width=12cm]{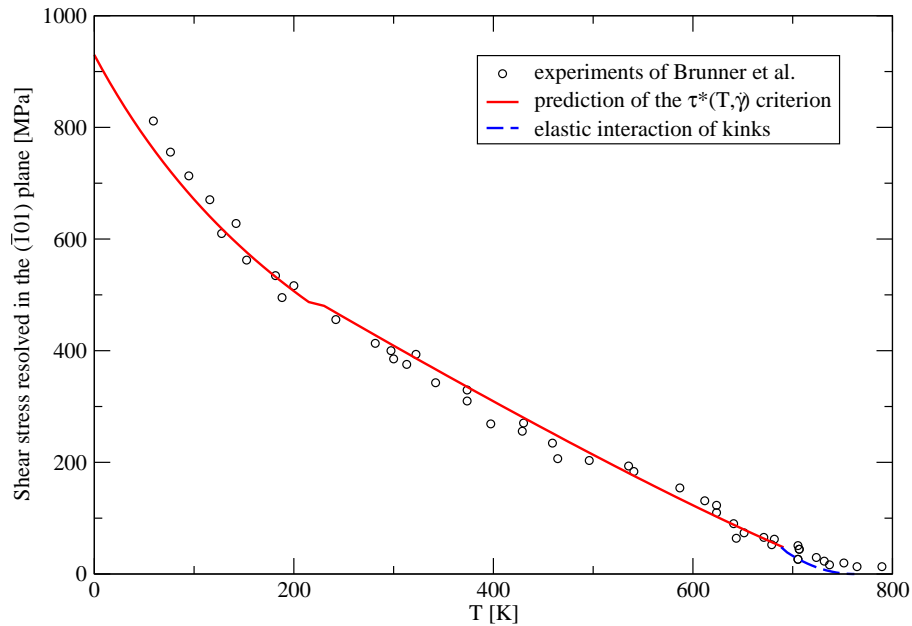}
  \parbox{14cm}{\caption{Temperature dependence of the yield stress calculated from the approximate
  expression (\ref{eq_eallcases_full_sigma-T}) for loading in tension along $[\bar{1}49]$. The
  experimental data are from \citet{brunner:00}.}
    \label{fig_yieldt-tstarC_brunner_flattop}}
\end{figure}

For practical calculations, it is convenient to approximate the stress dependence of the activation
enthalpy by a closed-form expression that replaces the tedious numerical calculations. This
approximation has been thoroughly explained in Chapter~\ref{chap_yieldtemp}. Firstly, we considered
the loading by pure shear parallel to the slip direction that provided the restricted version of
this approximation; see Section~\ref{sec_yieldtemp_restr}. Subsequently, the full form has been obtained by
adding the effect of the shear stress perpendicular to the slip direction; see
Section~\ref{sec_yieldtemp_full}. The same approach can now be used for tungsten, which reduces merely to
fitting the coefficients of the adjustable functions $a,b,a',b'$ to the data calculated for
different $\chi$ and $\eta$ from the model of the dislocation bow-out. The functional forms of
$H_b(\sigma)$, $\sigma(T,\dot\gamma)$ and $\tau^*_{cr}(T,\dot\gamma)$ all remain the same as those
given by \refeqs{eq_eallcases_full_both}, \ref{eq_eallcases_full_sigma-T} and
\ref{eq_eallcases_full_tstarC-T}, respectively. For future reference, we list the coefficients
entering the polynomials $a,b,a',b'$ for tungsten in Appendix~\ref{apx_thermalpar}.

In order to verify that the constructed approximation reproduces correctly the experimental
measurements, consider the loading in tension along $[\bar{1}49]$. The temperature dependence of the
yield stress can then be obtained directly from \refeq{eq_eallcases_full_sigma-T}, where the
polynomials $a,b,a',b'$ are determined for $\chi=0$ and $\eta=0.51$. The obtained approximation of
$\sigma(T)$ is plotted in \reffig{fig_yieldt-tstarC_brunner_flattop} as the full curve. At high
temperatures, the yield stress is determined by the elastic interaction of fully developed kinks
(\ref{eq_lowstress_sigma-T}) that corresponds in \reffig{fig_yieldt-tstarC_brunner_flattop} to the
dashed curve. One can see that the low-temperature approximation is in good agreement with the
experiment at all temperatures for which the dislocation motion is thermally activated. Hence, for
practical purposes, the model of elastic interaction of fully developed kinks is not needed, and one
may safely use \refeq{eq_eallcases_full_sigma-T} at all temperatures up to $T_k=760~\MPa$.

%----------------------------------------------------------------------------------------------------
%----------------------------------------------------------------------------------------------------

\section{Simplified ``engineering'' plastic flow rules}

The simplified plastic flow rules that explicitly involve the effect of non-glide stresses can now
be obtained in the same way as shown in Section~\ref{sec_gdot_engng} for molybdenum. We seek a
reasonably accurate formula that replaces both the low and the high stress approximation of the
plastic strain rate given by \refeqs{eq_gdot_sigma1} and \ref{eq_gdot_sigma2}. The functional form
of this simplified expression is given by the same \refeq{eq_gdot_sigma_eng} that we have
constructed originally for molybdenum. In this expression, the adjustable parameters for
tungsten, determined by fitting, are given in \reftab{tab_parABCD_W}, the critical yield stress
$\tau^*_{cr}=1215~\MPa$ and the pre-exponential factor in the Arrhenius law
$\dot\gamma_0=3.7\times10^{10}~\s^{-1}$.

\begin{table}[!htb]
  \centering
  \parbox{10cm}{\caption{Parameters $A,B,C,D$ for tungsten used in 
      \refeqs{eq_gdot_sigma_eng}, \ref{eq_sigma_T_eng} and \ref{eq_tstarC_engng}.}
  \label{tab_parABCD_W}} \\[1em]
  \begin{tabular}{cccc}
    \hline
    $A$ & $B$ & $C$ & $D$ \\
    \hline
    1.40 & 0.47 & 0.70 & 0.73 \\
    \hline
  \end{tabular} \ .
\end{table}

\begin{figure}[!htb]
  \centering
  \includegraphics[width=12cm]{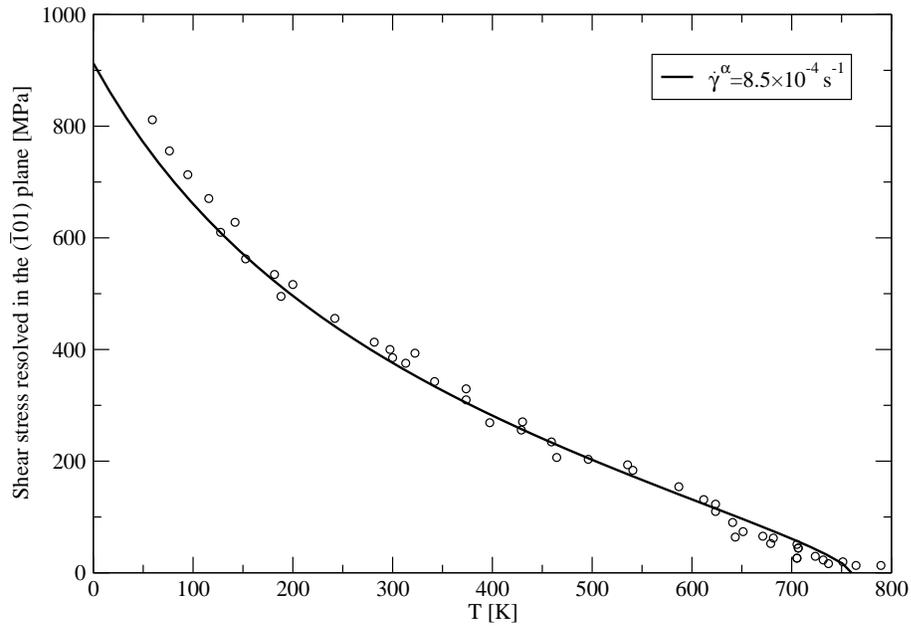}  
  \parbox{14cm}{\caption{Temperature dependence of the yield stress for center-triangle orientation
  calculated from the engineering formula (\ref{eq_sigma_T_eng}) and the parameters for tungsten
  listed in this section. The experimental data (circles) are due to \citet{brunner:00} and taken from
  Fig.\,11.18 of \citet{hollang:01}.}
    \label{fig_yieldt-engng_brunner_flattop}}
\end{figure}

The Arrhenius law (\ref{eq_gdot_sigma_eng}) with the parameters $A,B,C,D$ listed in
\reftab{tab_parABCD_W} correctly reproduces the effect of non-glide stresses that enter $\dot\gamma$
via the magnitude of $\tau^*$. This can be demonstrated by calculating the temperature dependence of
the yield stress from \refeq{eq_sigma_T_eng} that is shown for loading in tension along
$[\bar{1}49]$ in \reffig{fig_yieldt-engng_brunner_flattop}. The overall agreement with the
experimental data of \citet{brunner:00} is very satisfactory and proves that the simplified
expressions given by \refeqs{eq_gdot_sigma_eng}, \ref{eq_sigma_T_eng} and \ref{eq_tstarC_engng} can
indeed be used to describe the plastic flow of single crystals of tungsten at the temperatures for which
the plastic flow is governed by the thermally activated motion of screw dislocations.

  \chapter{Conclusions}
\label{chap_conclusion}

\begin{flushright}
  Perfection is achieved, not when there is nothing more to add,\\
  but when there is nothing left to take away. \\
  \emph{Antoine de Saint-Exup\'ery}
\end{flushright}

Body-centered cubic refractory metals are crystallographically one of the simplest materials, and yet
their plastic deformation exhibits a wealth of features that are not common in close-packed
metals. The observed plastic deformation is always a result of many moving dislocations continuously
produced by dislocation sources and a complex interplay of many dislocation processes. In contrast,
atomic-level theoretical studies focus closely on the behavior of individual dislocations without
invoking the complexity of the system, because it is generally recognized that $1/2\gdir{111}$ screw
dislocations govern the plastic deformation of bcc metals. This approach can provide essential
information about the fundamental processes governing the plastic flow of bcc metals that can be
later incorporated into mesoscopic and continuum models. Experimental observations have provided
firm evidence that the plastic flow of bcc metals at low temperatures occurs by thermally activated
motion of screw dislocations overcoming large lattice friction. However, no general framework has
been introduced so far that would unify the microscopic behavior of isolated screw dislocations
under stress with the macroscopic response determined by operation of many slip systems under a
general loading.

The principal aim of this work has been to develop physically-based rules governing the plastic flow
during macroscopic deformation of bcc transition metals that would explicitly involve the effects of
temperature, strain rate, and the applied stress. This development is based on a multiscale model in
which the fundamental microscopic information about the behavior of individual screw dislocations
under stress is obtained from $0~\K$ atomistic studies. These simulations show that only the
deviatoric component of the stress tensor plays a role in the dislocation glide, but, in contrast to
the common wisdom, they reveal that \emph{both} the shear stresses parallel and perpendicular to the slip
direction affect the plastic flow in bcc metals. Although the latter stresses cannot directly move
the dislocation, because they do not exert any Peach-Koehler force, they affect the structure of the
dislocation core such that the glide of the dislocation occurs at lower or higher shear stress
parallel to the slip direction, depending on the accompanying shear stress perpendicular to the slip
direction. The atomic-level studies supply the dependence of the critical resolved shear stress
(CRSS) at which the dislocation moves, i.e. the Peierls stress, on: (i) the orientation of the maximum
resolved shear stress plane (MRSSP), and (ii) the shear stress perpendicular to the slip
direction. These dependencies can be condensed into a $0~\K$ effective yield criterion that
explicitly involves the effect of non-glide stresses, which are shear stresses parallel to the slip
direction in a plane other than the slip plane and shear stresses perpendicular to the slip
direction. For a given character of applied loading, this criterion can be used not only to
determine the magnitude of the yield stress, but it also determines the order in which individual
$\gplane{110}\gdir{111}$ slip systems become activated for slip.

As the next step, the effective yield criterion, formulated on the basis of atomistic studies, was
utilized to construct the Peierls potential that reflects the variation of the energy of a screw
dislocation moving through the lattice. The Peierls potential is constructed as a two-dimensional
function of the position of the intersection of the screw dislocation with the $\gplane{111}$ plane
perpendicular to the dislocation line. It is based on a three-fold symmetric mapping function that
reproduces the crystal symmetry. The transition path of the dislocation between two potential minima
was determined using the Nudged Elastic Band method and the shape of the potential adjusted such
that it yields the Peierls stress that is in agreement with the prediction of the $0~\K$ effective
yield criterion. The novel concept introduced in this study is that the Peierls potential is a
function of applied shear stresses, both parallel and perpendicular to the slip direction, which
then reflects the calculated dependence of the Peierls stress on these applied stresses. The effects
of temperature and strain rate are included through the mesoscopic model of formation and
propagation of pairs of kinks. The rate of this process is given by an Arrhenius law with the
activation enthalpy determined either from the model of the dislocation bow-out (high stresses and
low temperatures) or the model of elastic interaction of fully developed kinks (low stresses and
high temperatures). Owing to the stress dependence of the Peierls potential, the activation enthalpy
is also a function of non-glide stresses as well as of the shear stress driving the dislocation
motion. This is again a novel approach, in contrast with previous studies in which the activation
enthalpy was always only a function of the shear stress driving the glide, i.e. the Schmid stress.

To allow comparisons with experiments in which many interacting dislocations give rise to the
plastic flow, we proposed a mesoscopic model based on a collective motion of interacting parallel
screw and mixed dislocations emitted from a Frank-Read source. This model suggests that the yield
stress becomes by a factor of 2 to 3 lower than the Peierls stress of individual screw dislocations
if one considers interactions between dislocations. This provides a physical explanation for the
ubiquitous discrepancy between the theoretically calculated Peierls stresses and the measured yield
stresses extrapolated to 0~K. Utilizing the above scaling of the yield stress, the calculated
temperature dependence of the yield stress agrees very closely with experimental measurements. The
stress dependence of the activation enthalpy is further approximated by an analytical formula that
can be utilized to obtain a closed-form expression for the plastic strain rate and, most
importantly, to formulate the temperature and strain rate dependent effective yield criterion. This
criterion is shown to reproduce correctly the effect of non-glide stresses at low temperatures and
their gradual decay with increasing temperature.

The multiscale model developed in this Thesis provides a plausible explanation of many features of
the low-temperature plastic deformation of bcc metals, both at the level of individual dislocations
and on the continuum level of crystal plasticity. In particular, atomistic simulations revealed a
strong twinning-antitwinning asymmetry of the shear stress parallel to the slip direction in bcc
molybdenum, but no such asymmetry has been found in bcc tungsten. However, in both metals, there is
a strong effect of the shear stresses perpendicular to the slip direction that modify the structure
of the dislocation core. If the core is constricted by these stresses in the most highly stressed
$\gplane{110}$ plane in the zone of the slip direction, the slip typically proceeds on one of the
lower-stressed $\gplane{110}$ planes. This may explain the origin of the so-called anomalous slip,
i.e. deformation involving the slip system with a relatively low Schmid factor becomes the major
deformation mode, which has been observed in many high-purity refractory bcc metals at low
temperatures. This observation implies that the anomalous slip is induced by non-glide stresses that
extend the core into one of the planes with a low Schmid factor and compress it in the plane of a
high Schmid factor. Under uniaxial loading, the anomalous slip takes place only in compression. Very
importantly, the experimentally observed slip of $1/2[111]$ dislocations on the $(0\bar{1}1)$ plane
in tension should not be confused with the anomalous slip, because, for loading in tension, the
Schmid factor for the $(0\bar{1}1)[111]$ system is comparable with that for the most highly stressed
$(\bar{1}01)[111]$ system.

The proposed yield criterion reproduces successfully the asymmetry between the yield stress in
tension and compression. In the case of bcc molybdenum, the yield criterion correctly predicts the
experimental observations of \citet{seeger:00} that the yield stress in tension becomes higher than
the yield stress in compression for orientations close to the $[011]$ corner of the stereographic
triangle. It is important to emphasize that the constructed yield criterion predicts the
tension-compression asymmetry, even when the MRSSP coincides exactly with the $(\bar{1}01)$
plane. This conclusion is in full agreement with the experimental observations of
\citet{seeger:00}. Finally, the effective yield criterion that follows from the proposed multiscale
model, based on the behavior of individual screw dislocations at $0~\K$, is expressed as an
analytical function of both temperature and plastic strain rate. This temperature and strain rate
dependent effective yield criterion is shown to reproduce correctly the experimental measurements in
the regime where the plastic deformation is governed by the nucleation and glide of screw
dislocations.

  \chapter{Future research}
\label{chap_future}

\begin{flushright}
  They are ill discoverers that think there is no land, \\
  when they can see nothing but sea. \\
  \emph{Sir Francis Bacon}
\end{flushright}

\noindent
{\it Development of BOPs and plastic flow rules for other bcc metals}

The multiscale approach to the development of the temperature and strain rate dependent yield
criterion formulated in this Thesis for molybdenum and its subsequent testing on tungsten implies
that this approach is likely to be applicable for any bcc metal. A similar development is planned in
the future for tantalum, niobium and vanadium for which the construction of the BOP is currently
under way. Moreover, the BOP for ferromagnetic bcc iron employing the Stoner model of band magnetism
\citep{stoner:38, stoner:39} is now under development at the University of Oxford and the UKAEA
Culham Research Centre in the UK \citep{liu:05}. These BOPs represent accurate empirical potentials
that can be used to study the microscopic details of the plastic flow of bcc metals at low
temperatures and will provide the fundamental
information for the development of yield criteria for these materials.\\

\newpage
\noindent
{\it Calculation of the Peierls potential using the Nudged Elastic Band method}

The Peierls potential constructed in this Thesis is based on an assumed shape of the mapping
function with adjustable parameters fitted such that it yields the Peierls stress that is in
agreement with the $0~\K$ yield criterion constructed on the basis of the results of atomistic
modeling of an isolated screw dislocation. This potential could, in principle, be calculated
atomistically with the help of the Nudged Elastic Band method. In these calculations, the two fixed
images would correspond to atomic blocks in which the dislocation is relaxed in two adjacent lattice
sites. Each image along the elastic band is represented by $3N$ coordinates of the $N$ atoms in the
simulated block. The final configuration corresponds to a chain of atomic blocks involving the
dislocation in metastable positions. These positions and the differences between the energies of
movable and fixed images then provide an energy barrier that has the same shape as the Peierls
barrier and thus helps to formulate the Peierls potential more
accurately.\\

\noindent
{\it Temperature dependence of the density of mobile dislocations}

In the formulation of the macroscopic plastic flow rules, we have adopted the usual approximation
that the density of mobile dislocations in each slip system does not vary significantly during
loading and can thus be considered constant. This is a reasonable approximation at low temperatures,
where the density of mobile dislocation increases mainly in the microyield regime but remains
virtually constant at stresses close to the yield stress. At higher temperatures, the mobile
dislocation density in intermetallic compound Ni$_3$Al typically exhibits a transient behavior
\citep{matterstock:99} during which the density of dislocations rapidly increases. In order to
verify the assumptions used in this work, it is important to investigate experimentally at what
temperatures this transient occurs in refractory metals. If the transient temperature falls into the
low-temperature regime, the rate equation may need a correction for the temperature dependence of
the pre-exponential factor, $\dot\gamma_0$, in the Arrhenius law.\\

\noindent
{\it Mean field model of a group of interacting dislocations}

On the macroscopic level, one of the most puzzling features is the change of shape of the
temperature dependence of the yield stress measured for molybdenum \citep{hollang:97,aono:83}. This
so-called ``hump'' \citep{seeger:00} has been asserted to be due to a hypothetical change of the
mechanism of slip between the nucleation of kinks on $\gplane{110}$ planes at low temperatures and
the formation of kinks on $\gplane{112}$ planes at intermediate and high temperatures. At present,
atomistic simulation do not support this assertion because the slip of the dislocation on a
$\gplane{112}$ plane can always be explained as a simultaneous operation of two or more
$\gplane{110}\gdir{111}$ systems. Provided that the dislocations move in groups, the above-mentioned
shape irregularity could instead occur as a result of a ``phase transition'' between the
low-temperature coherent dynamics in which the jumps of the dislocations are correlated, to the
high-temperature incoherent dynamics in which the individual jumps are mutually uncorrelated
events. This possibility should be studied in the future by investigating the steady-state behavior
of a mean field model representing the chain of oscillators (dislocations) with long-range coupling,
subjected to periodically varying substrate potential (i.e. the Peierls potential) and applied
stress.

  \appendix

  \chapter{Interactions between dislocations}
\label{chap_interact}

In this Appendix, we propose a mesoscopic model that provides an explanation of the discrepancy
between experimentally measured yield stresses of bcc metals at low temperatures and typical Peierls
stresses determined by atomistic simulations of \emph{isolated} screw dislocations. The model
involves a Frank-Read type source emitting dislocations that become pure screws at a certain
distance from the source and, owing to their high Peierls stress, control the operation of the
source.  However, due to the mutual interaction between emitted dislocations the group consisting of
both non-screw and screw dislocations can move at an applied stress that is about a factor of two to
three lower than the stress needed for the glide of individual screw dislocations.

%----------------------------------------------------------------------------------------------------
%----------------------------------------------------------------------------------------------------

\section{Discrepancy between the Peierls stress and experimentally measured yield stresses}

The vast majority of atomistic studies of the core structure and glide of $1/2\langle 111 \rangle$
screw dislocations in bcc metals were carried out using molecular statics techniques and thus they
correspond to 0~K.  A problem encountered universally in all the calculations of the critical
resolved shear stress (CRSS), i.e. the Peierls stress, at which the screw dislocation starts to
glide, is that it is by a factor of two to three larger than the CRSS obtained by extrapolating
experimental measurements to 0~K.  The following are a few examples. 

\citet{basinski:81} measured the flow stress of potassium in the temperature range 1.5~K to 30~K and
extrapolated to 0~K to get 0.002 to 0.003 where $\mu=(C_{11}-C_{12}+C_{44})/3$ is the
$\gplane{110}\gdir{111}$ shear modulus and $C_{11}$, $C_{12}$, $C_{44}$ are elastic moduli.
Similar values were found by \citet{pichl:97c, pichl:97b}.  The values of the CRSS when the MRSSP is
a $\{110\}$ plane, calculated using a pair potential derived on the basis of the theory of weak
pseudopotentials \citep{dagens:75}, is 0.007 to 0.009 \citep{basinski:81, duesbery:84}.  More
recently, \citet{woodward:02} calculated the CRSS in molybdenum using the many-body potentials
derived from the generalized pseudopotential theory \citep{moriarty:90} and a DFT-based method.
When the MRSSP was a $\{110\}$ plane, they found the CRSS to be between 0.018 and 0.020.  A similar
value of the CRSS, 0.019, was found in calculations employing the tight-binding based bond-order
potential for molybdenum \citep{mrovec:04, groger:05}. Experimental measurements of
\citet{hollang:01}, extrapolated to 0~K, give for the CRSS in molybdenum 0.006.  A similar problem
was encountered by \citet{wen:00} who used the Embedded Atom Method (EAM) potential for iron and the
Nudged Elastic Band method to analyze the activation enthalpies for kink-pair nucleation on screw
dislocations.  The calculated yield stress at 0~K was about 0.013 while the experimental values,
reported by \citet{aono:81} are 0.005 to 0.006. This variety of studies suggests that the reason for
the discrepancy between calculated and measured CRSS cannot be sought in the inadequacy of the
description of atomic interactions, which has often been raised as a possible explanation.  In fact
the ubiquitous higher value of the calculated CRSS, found independently of the description of atomic
interactions, suggests that the origin of this discrepancy cannot be sought on the atomic scale of
the motion of individual dislocations but rather on mesoscopic scale where a large number of
elastically interacting dislocations glide at the same time.  In this context it should be noticed
that the only atomic level simulation that predicts the yield stress close to that measured
experimentally considered a planar dislocation network of $1/2[111]$ and $1/2[\bar{1}\bar{1}1]$
screw dislocations with $[001]$ screw junctions \citep{bulatov:02}.  Such a network moved in the
$(\bar{1}10)$ plane at the stress about $50\%$ lower than the Peierls stress of an isolated
dislocation.

A number of in-situ TEM observations of dislocation sources in bcc transition metals showed that in
thin foils straight screw dislocations formed near the source and moved very slowly as a group,
presumably due to a high Peierls stress \citep{vesely:68, louchet:75, matsui:76, saka:76,
takeuchi:77, louchet:79, louchet:79a, garrat-reed:79, louchet:web}.  Hence they fully
control the rate at which the source produces dislocations.  In the foils used in TEM the applied
stresses are very low but, owing to the same properties of screw dislocations, a similar control of
the sources by sessile screw dislocations can be expected in the bulk at stresses leading to the
macroscopic yielding.  However, at higher stresses, dislocations move faster and do not become pure
screws immediately after leaving the source but at a distance from the source.  Indeed, even in situ
observations at higher stresses do not show straight screw dislocations emanating directly from the
sources \citep{vesely:priv06}.

In the following text, we propose a mesoscopic model involving a Frank-Read type source
\citep{hirth:82, hull:01} emitting dislocations of generally mixed character that become pure screw
dislocations at a certain distance from the source and, owing to their high Peierls stress, still
control its operation.  However, there are a number of non-screw dislocations between the screws and
the source, which can move easily.  These dislocations exert a stress on the screw dislocations and
these can then move at an applied CRSS that is about a factor of two to three lower than the CRSS
needed for the glide of individual screw dislocations.  This explains why the CRSS, obtained from
the measurements of the yield and flow stresses extrapolated to 0~K, is usually about a factor of
two to three lower than the Peierls stress of isolated screw dislocations.

%----------------------------------------------------------------------------------------------------

\section{Model of dislocation nucleation and motion}

\begin{figure}[!htb]
  \centering
  \includegraphics[width=7cm]{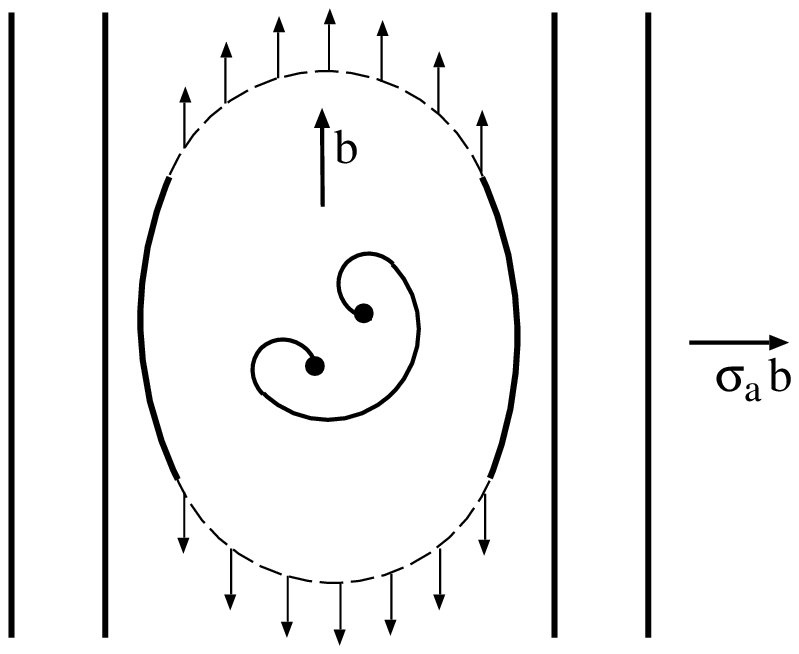}
  \parbox{12cm}{\caption{Schematic operation of a dislocation source
      in bcc metals.  When the curved non-screw segments migrate away,
      they leave behind a new pair of screw dislocations.}
  \label{fig_FRsource}}
\end{figure}

Let us consider a Frank-Read source \citep{hirth:82, hull:01} that produces dislocations in a bcc
metal.  It emits, as always, dislocation loops that have a mixed character and expand easily away
from the source since their Peierls stress is low.  However, at a certain distance from the source,
a significant part of the expanding loop attains the screw orientation and becomes much more
difficult to move owing to the very high Peierls stress.  The rest of the loop, having a mixed
character, continues to move which leads to further extension of the screw segments.  As a result,
the source becomes surrounded by arrays of slowly moving screw dislocations, as depicted
schematically in \reffig{fig_FRsource}.  Further operation of the source is hindered by their back
stress and effectively controlled by the ability of the screw dislocations to glide.
 
The operation of the source is driven by the applied stress, $\sigma_a$, which acts by the
Peach-Koehler force \citep{hirth:82} $\sigma_a b$ per unit length of the dislocation that bows out.
This dislocation obviously has a mixed character.  Let us consider now that there are $N_s$ screw
dislocations at distances $x_i$ from the source and $N_m$ dislocations, generally of mixed
character, positioned between the source and the screw dislocations.  We approximate the latter as
straight lines of the same orientation as the screws, positioned at distances $y_k$ from the source,
but with a negligible Peierls stress when compared with that of the screws.  In the framework of the
isotropic elastic theory of dislocations the condition for the source to operate is then
\begin{equation}
  \sigma_a b \geq \frac{\tau}{R} + \frac{\mu b^2}{2\pi\alpha} \sum_{i=1}^{N_s} \frac{1}{x_i} +
  \frac{\mu b^2}{2\pi\beta} \sum_{k=1}^{N_m} \frac{1}{y_k} \ ,
  \label{eq_source}
\end{equation}
where $\tau$ is the line tension of the emitted dislocations, $b$ their Burgers vector, $R$ the
half-length of the source, $\mu$ the shear modulus and $\alpha$, $\beta$ constants of the order of
unity.  The first term is the force arising from the line tension that pulls the nucleating loop
back and the second and third terms are forces produced by the stress fields of screw and non-screw
dislocations, respectively, present ahead of and/or behind the source.  In the following we neglect
the interaction between dislocations ahead and behind the source as they are far apart.  Moreover,
the dislocation sources are frequently single-ended \citep{hull:01}.  Hence we analyze only
dislocations ahead of the source, i.e. dislocations towards which the source bows out.

It should be noted here that the screw dislocations in the array ahead of the source are not pressed
against any obstacle and thus they do not form a pile-up.  Within the approximations defined above,
the $i$th screw dislocation will move provided
\begin{equation}
  \sigma_a + \frac{\mu b}{2\pi\alpha} \mathop{\sum_{j=1}^{N_s}}_{j\not=i} \frac{1}{x_i-x_j} + 
  \frac{\mu b}{2\pi\beta} \sum_{k=1}^{N_m} \frac{1}{x_i-y_k} + \frac{\mu b}{2\pi\beta} \frac{1}{x_i} \geq
  \sigma_P \ ,
  \label{eq_screw}
\end{equation}
where $\sigma_P$ is the Peierls stress of screw dislocations.  The second and third terms are
stresses arising from screw and non-screw dislocations, respectively, which have been produced by
the source, and the fourth term is the stress arising from the dislocation associated with the
source that is also treated as a straight line of the same type as all the other mixed dislocations.
Since the Peierls stress of non-screw dislocations is negligible, the $l$th non-screw dislocation can
move provided
\begin{equation}
  \sigma_a + \frac{\mu b}{2\pi\alpha} \sum_{j=1}^{N_s} \frac{1}{y_l-x_j} + 
  \frac{\mu b}{2\pi\beta} \mathop{\sum_{k=1}^{N_m}}_{k\not=l} \frac{1}{y_l-y_k} + 
  \frac{\mu b}{2\pi\beta} \frac{1}{y_l} \geq 0 \ .
  \label{eq_mixed}
\end{equation}

Now, the question asked is how large stress, $\sigma_a$, needs to be applied so that the screw
dislocations can move so far away from the source that they either reach a surface or encounter
dislocations of opposite sign from another source and annihilate.  In both cases the source then
keeps producing new dislocations indefinitely.  In the former case these dislocations keep vanishing
at the surface and the latter case leads to the propagation of slip through the sample. In order to
investigate the problem formulated above we performed the following self-consistent simulations
that were carried out for certain fixed values of the Peierls stress, $\sigma_P$, and applied
stress $\sigma_a$.  First we choose a half-length of the source, $R$, and a distance from the
source, $y_{max}$, beyond which the expanding loop always attains the screw character.  The first
mixed dislocation, emitted by the source, becomes screw when reaching the distance $y_{max}$ from the
source and then moves to a distance $x_1$, determined by \refeq{eq_screw}.  Provided that
the source can operate, i.e. the inequality (\ref{eq_source}) is satisfied, another dislocation is
emitted from the source.  The position of this dislocation is determined by \refeq{eq_mixed} if it
does not reach $y_{max}$ and by \refeq{eq_screw} if it does.  Subsequently, the position of the
first dislocation, $x_1$, is updated to satisfy \refeq{eq_screw}, which allows also the second
dislocation to move.  In this way a new position of the first dislocation, $x_1$, and the position
of the second dislocation, either $y_1$ if smaller than $y_{max}$ or $x_2$ if larger than $y_{max}$,
are found self-consistently.  This self-consistent process is then repeated for the third, fourth,
etc., dislocations until the source cannot emit a new dislocation, i.e. when the inequality
(\ref{eq_source}) cannot be satisfied.  The result of this calculation is the number of screw
dislocations, $N_s$, and mixed dislocations, $N_m$, as well as their positions ahead of the source
when it becomes blocked by the back-stress from all the emitted dislocations.  The first screw
dislocation is then at a position $x_1=x_{max}$ and further operation of the source can proceed only
if this screw dislocation is removed.  This can occur if it is either attracted to a free surface or
annihilates after encountering the dislocation of opposite sign produced by another source.  The
source can then continue operating in a steady-state manner producing a large number of dislocations
that mediate the macroscopic plastic flow.

%----------------------------------------------------------------------------------------------------
%----------------------------------------------------------------------------------------------------

\section{Simulation of an array of interacting dislocations}

In the following numerical simulations, the applied stress $\sigma_a$ was set equal to $0.3\sigma_P$
and $0.5\sigma_P$, respectively, in order to investigate whether the source can operate at stress
levels corresponding to experimental yield stresses extrapolated to 0~K.  Three values of the
Peierls stress, $\sigma_P$, have been considered that fall into the range found in atomistic studies
of transition metals \citep{woodward:02, mrovec:04, groger:05, wen:00}, namely $0.01\mu$, $0.02\mu$,
and $0.03\mu$.  Three different positions at which mixed dislocations transform into screw ones have
been considered, namely $y_{max}/b = 500, 1000,$ and 2000.  The dependence on the size of the
source, $R$, was also investigated.  However, this dependence is very weak since $R$ enters only
through the line tension term in \refeq{eq_source} and this is always small compared to the terms
arising from the back-stress of emitted dislocations.  Hence, without the loss of generality, we set
$R = y_{max}$.  The values of parameters $\alpha$ and $\beta$, entering equations (\ref{eq_source})
to (\ref{eq_mixed}) have all been set to one and the usual approximation for the line tension,
$\tau=\mu b^2/2$ \citep{hirth:82} was adopted.

\begin{figure}[!b] 
  \centering
  \includegraphics[width=10cm]{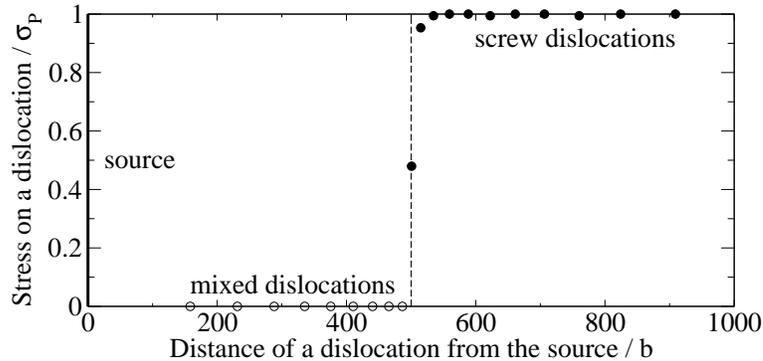}
  \parbox{14cm}{\caption{Positions of mixed (open circles) and screw (full circles) dislocations
      when the source is blocked are shown on the horizontal axis for the case
      $\sigma_a/\sigma_P=0.5$, $\sigma_P/\mu=0.02$, and $y_{max}/b=500$. In this case,
      $x_{max}/y_{max}=1.8$ (see \reftab{tab_xmax}). The stress evaluated at the positions of these
      dislocations is shown on the vertical axis.}
  \label{fig_dstresses}}
\end{figure}

Results of such simulation are presented in detail for $\sigma_a = 0.5\sigma_P$, $\sigma_P=0.02\mu$
and $y_{max}=500b$ in \reffig{fig_dstresses}, where positions of the dislocations ahead of the source
and stresses acting on them are shown. In this case $x_{max}/y_{max}=1.8$. It should be noted that
the stress exerted on the majority of screw dislocations is practically equal to the Peierls stress.
The distances $x_{max}$ found for the above-mentioned two values of $\sigma_a/\sigma_P$, three
values of $\sigma_P/\mu$ and three values of $y_{max}/b$, are summarized in \reftab{tab_xmax}.
These results suggest that, for a given applied stress, the ratio is almost constant, independent of
$y_{max}$, and only weakly dependent on the magnitude of the Peierls stress $\sigma_P$.  At
$\sigma_a/\sigma_P=0.3$, most of the dislocations are mixed and $x_{max}/y_{max} \approx 1.3$.  With
increasing stress, more emitted dislocations become screw and, at $\sigma_a/\sigma_P=0.5$,
$x_{max}/y_{max} \approx 2$, which implies that the numbers of mixed and screw dislocations ahead of
the source are very similar.  Very importantly, the stress exerted on most of the screw dislocations
is practically equal to the Peierls stress..

\begin{table}[!htb]
  \centering
  \parbox{13cm}{\caption{The distance which the leading screw dislocation advances from the source,
    $x_{max}$, as a function of the distance $y_{max}$ from the source at which dislocations become
    screw, applied stress $\sigma_a/\sigma_P$, and the Peierls stress of the screw dislocations
    $\sigma_P/\mu$.}
  \label{tab_xmax}} \\[1em]
  \begin{tabular}{cc|ccc}  
    \hline
    \multicolumn{2}{c}{} & \multicolumn{3}{c}{$y_{max}/b$} \\ \cline{3-5}
    \multicolumn{2}{c}{} & 500 & 1000 & 2000 \\
    \hline
    \multicolumn{5}{c}{$\sigma_a/\sigma_P=0.3$} \\
    \hline
                      & $\sigma_P/\mu=0.01$ & 1.0 & 1.1 & 1.2 \\ \cline{2-5}
    $x_{max}/y_{max}$ & $\sigma_P/\mu=0.02$ & 1.2 & 1.3 & 1.3 \\ \cline{2-5}
                      & $\sigma_P/\mu=0.03$ & 1.2 & 1.2 & 1.2 \\ \cline{2-5}
    \hline
    \multicolumn{5}{c}{$\sigma_a/\sigma_P=0.5$} \\
    \hline
                      & $\sigma_P/\mu=0.01$ & 1.6 & 1.8 & 2.0 \\ \cline{2-5}
    $x_{max}/y_{max}$ & $\sigma_P/\mu=0.02$ & 1.8 & 2.0 & 2.0 \\ \cline{2-5}
                      & $\sigma_P/\mu=0.03$ & 1.9 & 2.0 & 2.0 \\ \cline{2-5}
    \hline
  \end{tabular}
\end{table}

%----------------------------------------------------------------------------------------------------
%----------------------------------------------------------------------------------------------------

\section{Discussion}

\begin{figure}[!htb]
  \centering
  \includegraphics[width=10cm]{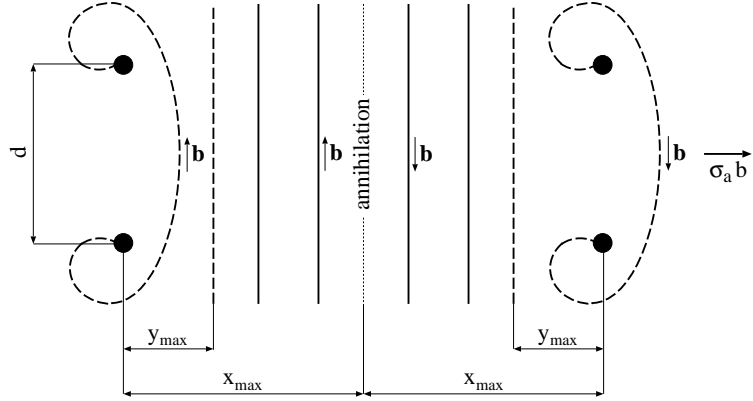}
  \parbox{14cm}{\caption{Schematic illustration of the operation of two dislocation
      sources. Dislocations of opposite Burgers vectors mutually annihilate when reaching the
      distance $x_{max}$ from the source. The circles are pinning points of the source, dashed and
      solid lines are mixed and screw dislocations, respectively. The pinning points are commonly
      forest dislocations intersecting the slip plane and their density is approximately
      $\rho=4/2dx_{max}$, where $d$ is the size of the source.}
  \label{fig_twosources}}
\end{figure}

The mesoscopic model presented in this appendix suggests that in bcc metals the dislocation
generation may be sustained at an applied stress that is by a factor of two or three smaller than
the Peierls stress of $1/2\langle{111}\rangle$ screw dislocations that control the plastic
deformation of these metals.  The distinguishing characteristic of this model is that it does not
consider the glide of a single screw dislocation but movement of a large group of dislocations
produced by a Frank-Read type source.  In general, this source produces dislocation loops of mixed
character that transform into pure screws at a distance $y_{max}$ from the source.  Hence, the group
of dislocations consists of screw dislocations at distances larger than $y_{max}$ and non-screw
dislocations near the source.  It is then the combination of the applied stress with the stress
produced by the dislocations in the group that acts on the screw dislocations and is practically
equal to their Peierls stress.  However, once the leading screw dislocation reaches the distance
$x_{max}$ from the source the operation of the source is blocked. Nonetheless, it can continue
operating if a dislocation of opposite sign, originating from another source, annihilates the
leading screw dislocation, as shown in \reffig{fig_twosources}.  This requires that the average
separation of sources be about $2x_{max}$.  Since the pinning points of the sources for a given slip
system are produced by intersections with dislocations in other slip systems, their separation is
related to the dislocation density in these systems.  For example, in a deformed molybdenum crystal
this density is of the order of $10^{12}\,{\rm m}^{-2}$ \citep{kaspar:00} which implies separation
of dislocations between $3000b$ and $4000b$, for the lattice parameter of Mo equal to $3.15\,\A$.
These values are in the range of $2x_{max}$ for applied stresses that are between one-third and
one-half of the atomistically calculated Peierls stress for the sources of the size compatible with
the above-mentioned density of dislocations.

The implication of the present study is that the values of the Peierls stress of screw dislocations
in bcc metals found in atomistic studies cannot be compared directly with the critical resolved
shear stresses obtained by extrapolating experimental measurements to 0~K.  The experiments do not
determine the stress needed for the glide of individual screw dislocations but, instead, the stress
needed for the operation of sources that are hindered by the sessile screw dislocations.  These
sources can operate at stresses lower than the Peierls stress owing to the collective motion of
screw and mixed dislocations produced by them, as described earlier.  

The model presented here in no way implies a need for some more sophisticated and less comprehensive
simulation schemes. On the contrary, single dislocation studies provide a great insight and
understanding of the onset of plastic deformation in these materials. For the comparison with
experiments, the theoretically calculated Peierls stresses may be scaled down by a factor of two or
three to meet the experimental data. However, in the framework of the proposed model, this scaling
should always be understood as \emph{a posteriori} accounting for interactions between dislocations that
are not included in the atomistic model.

  \chapter{Euler angles and transformations between slip systems}
\label{sec_eulerang}

Consider two Cartesian coordinate systems $\alpha_1$ and $\alpha_2$. The transformation from the
former to the latter system generally requires knowledge of the full ($3\times3$) transformation
matrix. However, according to the Euler's rotation theorem, the same transformation can be
conveniently described using only three angles that define rotations about the three coordinate
axes. The choice of the Euler angles is not unique; the so-called $x$-convention, or the $zxz$
rotation, is one of the most common schemes. In this convention, the rotation from system $\alpha_1$
to $\alpha_2$ is given by Euler angles ($\phi$,$\theta$,$\psi$), where the first rotation is by an
angle $\phi$ about the $z$-axis, the second by an angle $\theta$ about the $x$-axis, and the third
by an angle $\psi$ again about the $z$-axis\footnote{For a detailed explanation of the concept of
  Euler angles, its various alternatives and references, see {\tt
    http://mathworld.wolfram.com/EulerAngles.html}}. The transformation matrix reads
\begin{equation}
  \mat{A} = \left[
    \begin{array}{ccc}
      a_{11} & a_{12} & a_{13} \\
      a_{21} & a_{22} & a_{23} \\
      a_{31} & a_{32} & a_{33} 
   \end{array}
  \right] \ ,
  \label{eq_Atransf}
\end{equation}
where the components $a_{ij}$ are defined in terms of the given Euler angles $\phi$, $\theta$ and
$\psi$ as follows:
\begin{eqnarray} 
  \nonumber
  a_{11} &=& \cos\psi \cos\phi - \cos\theta \sin\phi \sin\psi \\ \nonumber
  a_{12} &=& \cos\psi \sin\phi + \cos\theta \cos\phi \sin\psi \\ \nonumber
  a_{13} &=& \sin\psi \sin\theta \\ \nonumber
  a_{21} &=& -\sin\psi \cos\phi - \cos\theta \sin\phi \cos\psi \\
  a_{22} &=& -\sin\psi \sin\phi + \cos\theta \cos\phi \cos\psi \\ \nonumber
  a_{23} &=& \cos\psi \sin\theta \\ \nonumber
  a_{31} &=& \sin\theta \sin\phi \\ \nonumber
  a_{32} &=& -\sin\theta \cos\phi \\ \nonumber
  a_{33} &=& \cos\theta
\end{eqnarray}

\begin{figure}[!p]
  \centering
  \includegraphics[width=12cm]{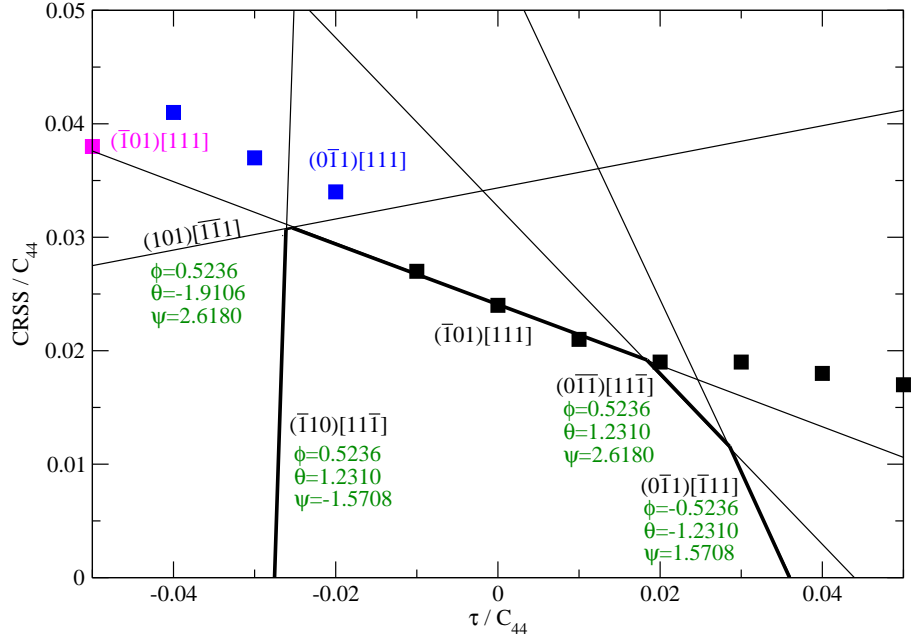} \\
  a) reference system $(\bar{1}01)[111]$ \\[1em]
  \includegraphics[width=12cm]{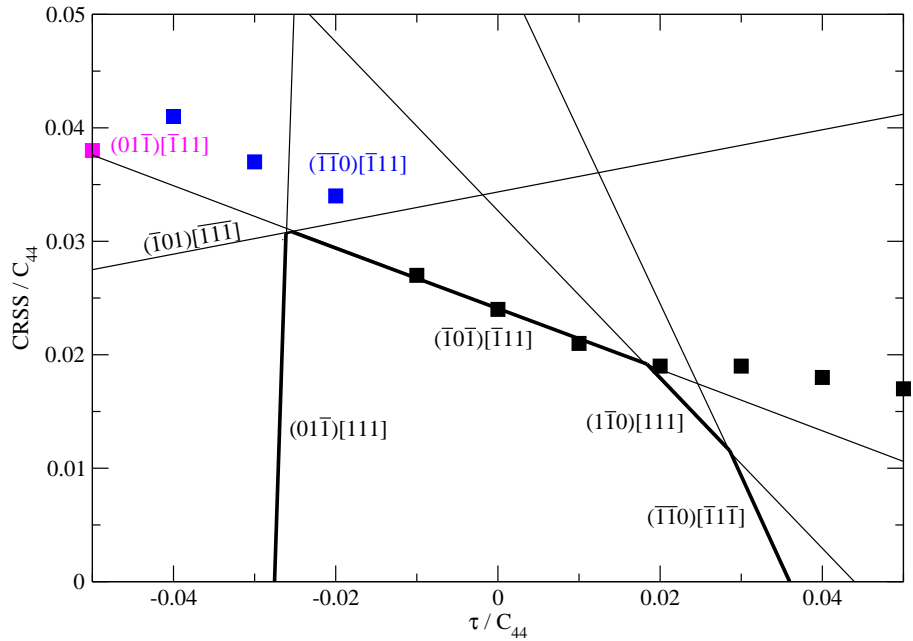} \\
  b) reference system $(\bar{1}0\bar{1})[\bar{1}11]$
  \caption{Projection of the MRSSP at $\chi=0$ for two different reference systems in which the
  angle of $\chi$ is measured. The numbers below the designations of slip systems in a) give the
  Euler angles between an individual system and the corresponding reference system.}
  \label{fig_chi0_fit_MoBOP_apx}
\end{figure}

In the following, we will show how the concept of Euler angles can be used to obtain the
$\CRSS-\tau$ dependencies for real single crystals that correspond to other reference systems than
$(\bar{1}01)[111]$. The fact that all $\gplane{110}\gdir{111}$ systems in bcc crystals are mutually
equivalent allowed us to formulate the 0~K effective yield criterion (\ref{eq_tstar_tensor}) such
that it holds for all slip systems. As a result, we arrived at a series of critical lines in the
$\CRSS-\tau$ projection of each MRSSP, defining the onset of slip on the corresponding slip
systems. For example, the obtained dependence for the MRSSP $(\bar{1}01)$ corresponding to $\chi=0$
is shown in \reffig{fig_chi0_fit_MoBOP_apx}a, where the dashed lines are loci of critical loadings
for which the slip on individual systems takes place.

Since all 24 slip systems in \reftab{tab_bcc24sys} are physically equivalent, the $\CRSS-\tau$ plot
in \reffig{fig_chi0_fit_MoBOP_apx}a, correspoding to the reference system $(\bar{1}01)[111]$, must
be applicable also to other reference systems, provided that the orientation of the MRSSP measured
in these systems is $\chi=0$. The only complication is that whenever one considers other reference
system than $(\bar{1}01)[111]$, the indices of all systems attached to the critical lines in
\reffig{fig_chi0_fit_MoBOP_apx}a have to be recalculated. In order to illustrate this calculation,
let us obtain the $\CRSS-\tau$ plot for the $(\bar{1}0\bar{1})[\bar{1}11]$ reference system. We will
first calculate the Euler angles for rotations from the $(\bar{1}01)[111]$ reference system to each
system for which \reffig{fig_chi0_fit_MoBOP_apx}a shows the critical line. The obtained values of
$(\phi,\theta,\psi)$ for each such rotation are listed under the designation of each system and
provide their orientations relative to the $(\bar{1}01)[111]$ reference system. Now, consider the
new reference system, $(\bar{1}0\bar{1})[\bar{1}11]$, and calculate the indices of the systems that
will be attached to the critical lines in \reffig{fig_chi0_fit_MoBOP_apx}b. Since the two reference
systems, i.e. $(\bar{1}01)[111]$ and $(\bar{1}0\bar{1})[\bar{1}11]$ are equivalent, we can use the
Euler angles from \reffig{fig_chi0_fit_MoBOP_apx}a to obtain the indices of the slip systems
corresponding to the $(\bar{1}0\bar{1})[\bar{1}11]$ reference system that are in the same relative
orientation as the systems in \reffig{fig_chi0_fit_MoBOP_apx}a are to the $(\bar{1}01)[111]$
reference system. This transformation can be written as
\begin{equation}
  \mat{v}_0 = \mat{A}_0^{\rm T} ( \mat{A}^{\rm T} \mat{v} ) \ ,
  \label{eq_mat_v0}
\end{equation}
where $\mat{A}$ is the transformation matrix (\ref{eq_Atransf}) determined using the Euler angles in
\reffig{fig_chi0_fit_MoBOP_apx}a, and $\mat{A}_0$ is the transformation matrix from the
$[100]-[010]-[001]$ cube coordinate system to the new reference system in which the $y$-axis
coincides with the normal to the $(\bar{1}0\bar{1})$ plane and the $z$-axis with the $[\bar{1}11]$
slip direction. For each system in \reffig{fig_chi0_fit_MoBOP_apx}b, the sought index of the slip
plane is obtained in $\mat{v}_0$ by setting $\mat{v}=(0,1,0)$ (i.e. the $y$-axis), and that of the
new slip direction using $\mat{v}=(0,0,1)$ (i.e. the $z$-axis). The complete assignment of slip
systems to individual critical lines, valid for the new $(\bar{1}0\bar{1})[\bar{1}11]$ reference
system, is shown in \reffig{fig_chi0_fit_MoBOP_apx}b. It can be easily shown that the Euler angles
corresponding to rotations from the $(\bar{1}0\bar{1})[\bar{1}11]$ reference system to other systems
in \reffig{fig_chi0_fit_MoBOP_apx}b are identical to those given in
\reffig{fig_chi0_fit_MoBOP_apx}a, which proves that the two plots are equivalent. Finally, the three
possible slip planes corresponding to the atomistic data (squares) in
\reffig{fig_chi0_fit_MoBOP_apx}b are obtained directly from \reffig{fig_8slipdir} for the
$(\bar{1}0\bar{1})[\bar{1}11]$ reference system.

  \chapter{Derivation of the symmetry-mapping function}
\label{sec_mapterm}

In the following text, we present in detail the contruction of the mapping function, $m(x,y)$, of
the Peierls potential which is a two-dimensional surface defined in the $(111)$ plane such that the
$x$-axis coincides with the $[\bar{1}2\bar{1}]$ direction and the $y$-axis with the $[10\bar{1}]$
direction. The fundamental requirement for the construction of this function is that it has to obey
the three-fold symmetry of the $(111)$ plane. This can be conveniently accomplished by writing the
mapping function as a product of three periodic functions, called hereafter \emph{basis
  functions}, that are defined along three characteristic directions in the $xy$ plane. Note,
that generally no restriction is imposed on the shape of these functions, and thus the mapping term
constructed here represents only one of many possible choices.

\begin{figure}[!htb]
  \centering
  \includegraphics[width=8cm]{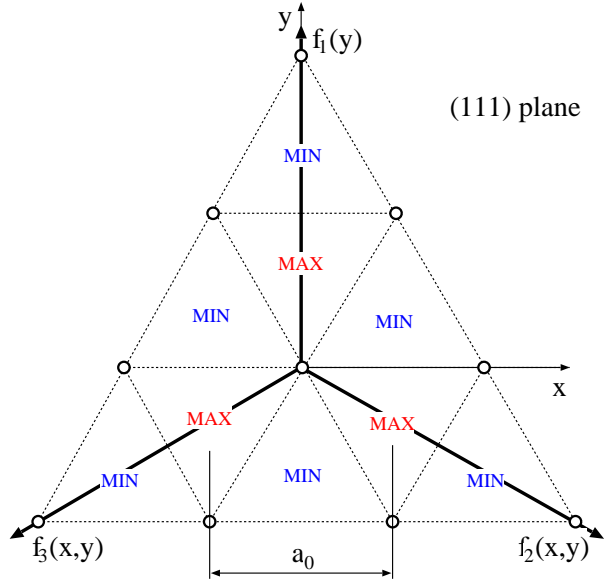}
  \parbox{12cm}{\caption{Orientation of the directions along which the three mapping functions,
  $f_1(y)$, $f_2(x,y)$, $f_3(x,y)$, are defined. The circles correspond to the positions of atoms,
  MIN are potential minima, and MAX potential maxima.}
    \label{fig_mapterm_csys}}
\end{figure}

The coordinate system $(x,y)$ defined above will be centered at an atomic position; the three
directions along which we define the mapping functions $f_1(y)$, $f_2(x,y)$, $f_3(x,y)$ are
shown in \reffig{fig_mapterm_csys}. For simplicity, we describe the basis functions by
sinusoidal waves with periods $a_0\sqrt{3}$, where $a_0=a\sqrt{2/3}$ is the period of the lattice
along the $x$-axis, and $a$ is the $\langle{100}\rangle$ lattice parameter. The function $f_1(y)$ is
then written as
\begin{equation}
  f_1(y) = \sin \left( \frac{2\pi}{a_0\sqrt{3}} y \right) \ .
  \label{eq_f1fun}
\end{equation}
Owing to the lattice symmetry which requires the three basis functions to be identical, the
functional form of $f_2(x,y)$ and $f_3(x,y)$ can now be written in a straightforward manner. If we
denote $r$ the direction along which $f_2$ is defined, the above symmetry requirement and
\refeq{eq_f1fun} demand that $f_2(r)=\sin \left( 2\pi r/ a_0\sqrt{3} \right)$. From
\reffig{fig_mapterm_csys}, the direction $r$ can be written in terms of $x$ and $y$ as $r = x\cos
30\deg - y\sin 30\deg = x\sqrt{3}/2-y/2$. The basis function $f_2$ is then
\begin{equation}
  f_2(x,y) = \sin \frac{\pi}{a_0} \left( x-\frac{y}{\sqrt{3}} \right) \ .
  \label{eq_f2fun}
\end{equation}
The last function, $f_3$, can be found in a similar fashion as $f_2$. The direction $r$ along which
$f_3$ is applied is $r=-x\sqrt{3}/2 - y/2$. After substituting $r$ in $f_3(r)=\sin \left( 2\pi
r/ a_0\sqrt{3} \right)$, the last basis function can be written as
\begin{equation}
  f_3(x,y) = \sin \frac{\pi}{a_0} \left( -x-\frac{y}{\sqrt{3}} \right) \ .
  \label{eq_f3fun}
\end{equation}

\begin{figure}[!htb]
  \centering
  \includegraphics[width=7cm]{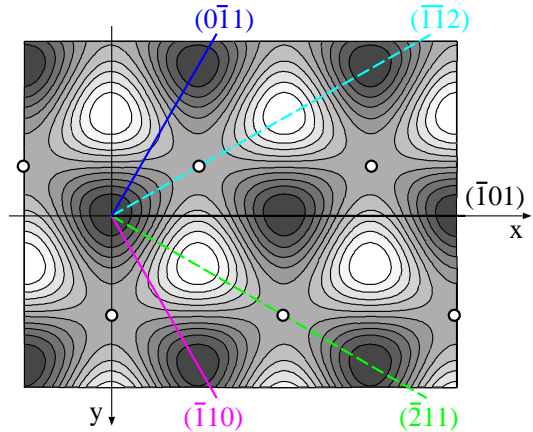}
  \parbox{10cm}{\caption{Contour plot of the mapping function (\ref{eq_mapterm}) of the Peierls
      potential.}
    \label{fig_mapterm_contour}}
\end{figure}

The mapping function can now be defined as a product of the three basis functions $f_1f_2f_3$ that
varies between $\pm 3\sqrt{3}/8$. For practical calculations, we further require that the minimum of
$m$ is zero and its height one, which can be easily accomplished by scaling and shifting the product
$f_1f_2f_3$. The obtained mapping function then reads:
\begin{equation}
  m(x,y) = \frac{1}{2} + \frac{4}{3\sqrt{3}} \sin{ \left( \frac{2\pi}{a_0\sqrt{3}}y \right) } \,
  \sin{ \frac{\pi}{a_0} \left( \frac{y}{\sqrt{3}}-x \right) } \,
  \sin{ \frac{\pi}{a_0} \left( \frac{y}{\sqrt{3}}+x \right) } \ .
  \label{eq_mapterm_noshift}
\end{equation}
In the final step, we will displace the origin of $m$ into one of the potential minima designated in
\reffig{fig_mapterm_csys} as MIN. One possibility is to change the variables in
\refeq{eq_mapterm_noshift} by substituting $x \rightarrow x+a_0/2$, $y \rightarrow
y+a_0\sqrt{3}/6$, which then yields
\begin{equation}
  m(x,y) = \frac{1}{2} + \frac{4}{3\sqrt{3}} \sin{ \frac{\pi}{3a_0} \left( 2y\sqrt{3}+a_0 \right)  } \,
  \sin{ \frac{\pi}{a_0} \left( \frac{y}{\sqrt{3}}-x-\frac{a_0}{3} \right) } \,
  \sin{ \frac{\pi}{a_0} \left( \frac{y}{\sqrt{3}}+x+\frac{2a_0}{3} \right) } \ .
  \label{eq_mapterm}
\end{equation}
This is the final form of the mapping function, shown in \reffig{fig_mapterm_contour}, that is used
in our construction of the Peierls potentials for molybdenum and tungsten. In this Thesis, the
contour plots of the mapping function and the Peierls potential are always depicted with the
$y$-axis pointing downwards. For this particular choice of sinusoidal basis functions, we arrived at
the mapping function that is identical to the unstressed Peierls potential of \citet{edagawa:97}.

  \chapter{Fitted parameters of $\mat{\tau^*_{cr}(T,\dot\gamma)}$}
\label{apx_thermalpar}

In this appendix, we list the values of the adjustable constants entering the functions $a,a',b,b'$
that are involved in the approximations of the stress dependence of the activation enthalpy in
Section~\ref{sec_yieldtemp_restr} (restricted model) and in Section~\ref{sec_yieldtemp_full} (full
model). The restricted model is valid only for loading by pure shear stress parallel to the slip
direction, $\sigma$, acting in the MRSSP defined by the angle $\chi$. The full model adds also the
effect of the shear stress perpendicular to the slip direction, $\tau$, quantified by the ratio
$\eta=\tau/\sigma$. The angles $\chi$ are in radians.

%------------------------------------------------------------------------------------------
%------------------------------------------------------------------------------------------

\section{Parameters for bcc molybdenum}

%------------------------------------------------------------------------------------------

\subsection{Restricted model}

For $\sigma/C_{44}\leq0.003$:

\begin{table}[!htb]
  \centering
  \begin{tabular}{ccc}
    \multicolumn{3}{c}{$a(\chi) = a_0 + a_1\chi + a_2\chi^2$} \\
    \hline
    $a_0$ & $a_1$ & $a_2$ \\
    \hline
    71.0798 & -12.3631 & -30.9276 \\
    \hline
  \end{tabular}
  \hskip3cm
  \begin{tabular}{ccc}
    \multicolumn{3}{c}{$b(\chi) = b_0 + b_1\chi + b_2\chi^2$} \\
    \hline
    $b_0$ & $b_1$ & $b_2$ \\
    \hline
    0.8331 & -0.0163 & -0.0052 \\
    \hline
  \end{tabular}
\end{table}

\newpage
\noindent
For $\sigma/C_{44}\geq0.003$:

\begin{table}[!htb]
  \centering
  \begin{tabular}{ccc}
    \multicolumn{3}{c}{$a'(\chi) = a'_0 + a'_1\chi + a'_2\chi^2$} \\
    \hline
    $a'_0$ & $a'_1$ & $a'_2$ \\
    \hline
    1.3632 & -0.0642 & -0.0078 \\
    \hline
  \end{tabular}
  \hskip3cm
  \begin{tabular}{ccc}
    \multicolumn{3}{c}{$b'(\chi) = b'_0 + b'_1\chi + b'_2\chi^2$} \\
    \hline
    $b'_0$ & $b'_1$ & $b'_2$ \\
    \hline
    0.0026 & 0.0002 & 0.0014 \\
    \hline
  \end{tabular}
\end{table}

%------------------------------------------------------------------------------------------

\subsection{Full model}

For $\sigma/C_{44}\leq0.003$:
\begin{eqnarray}
  \nonumber
  & a(\chi,\eta) = a_0(\eta) + a_1(\eta)\chi + a_2(\eta)\chi^2 &\\
  & \left[
    \begin{array}{c}
      a_0(\eta) \\ a_1(\eta) \\ a_2(\eta)
    \end{array}
  \right] = \left[
    \begin{array}{rrr}
       71.0798 &  17.1061 &   1.6924 \\
      -12.3631 & -19.7773 &  -2.8968 \\
      -30.9276 & -32.1963 &  -1.7529 \\
    \end{array}
  \right] \left[
    \begin{array}{c}
      1 \\ \eta \\ \eta^2
    \end{array} \right] &
\end{eqnarray}

\begin{eqnarray} 
  \nonumber
  & b(\chi,\eta) = b_0(\eta) + b_1(\eta)\chi + b_2(\eta)\chi^2 &\\
  & \left[
    \begin{array}{c}
      b_0(\eta) \\ b_1(\eta) \\ b_2(\eta)
    \end{array}
  \right] = \left[
    \begin{array}{rrr}
       0.8331 &   0.0236 &  -0.0005 \\
      -0.0163 &  -0.0254 &   0.0010 \\
      -0.0052 &  -0.0415 &   0.0049 \\
    \end{array}
  \right] \left[
    \begin{array}{c}
      1 \\ \eta \\ \eta^2
    \end{array} \right] &
\end{eqnarray}
\\

\noindent
For $\sigma/C_{44}\geq0.003$:
\begin{eqnarray}
  \nonumber
  & a'(\chi,\eta) = a'_0(\eta) + a'_1(\eta)\chi + a'_2(\eta)\chi^2 &\\
  & \left[
    \begin{array}{c}
      a'_0(\eta) \\ a'_1(\eta) \\ a'_2(\eta)
    \end{array}
  \right] = \left[
    \begin{array}{rrr}
       1.3632 &   0.0962 &  -0.0099 \\
      -0.0642 &  -0.1121 &   0.0071 \\
      -0.0078 &  -0.1020 &   0.0892 \\
    \end{array}
  \right] \left[
    \begin{array}{c}
      1 \\ \eta \\ \eta^2
    \end{array} \right] &
\end{eqnarray}

\newpage
\begin{eqnarray} 
  \nonumber
  & b'(\chi,\eta) = b'_0(\eta) + b'_1(\eta)\chi + b'_2(\eta)\chi^2 &\\
  & \left[
    \begin{array}{c}
      b'_0(\eta) \\ b'_1(\eta) \\ b'_2(\eta)
    \end{array}
  \right] = \left[
    \begin{array}{rrr}
      0.0026 &  -0.0001 &  0 \\
      0.0002 &   0.0001 &  0 \\
      0.0014 &   0.0003 &  0.0002 \\
    \end{array}
  \right] \left[
    \begin{array}{c}
      1 \\ \eta \\ \eta^2
    \end{array} \right] &
\end{eqnarray}

%------------------------------------------------------------------------------------------
%------------------------------------------------------------------------------------------

\section{Parameters for bcc tungsten}

%------------------------------------------------------------------------------------------

\subsection{Restricted model}

For $\sigma/C_{44}\leq0.003$:

\begin{table}[!htb]
  \centering
  \begin{tabular}{ccc}
    \multicolumn{3}{c}{$a(\chi) = a_0 + a_1\chi + a_2\chi^2$} \\
    \hline
    $a_0$ & $a_1$ & $a_2$ \\
    \hline
    82.6712 & 0 & -35.4748 \\
    \hline
  \end{tabular}
  \hskip3cm
  \begin{tabular}{ccc}
    \multicolumn{3}{c}{$b(\chi) = b_0 + b_1\chi + b_2\chi^2$} \\
    \hline
    $b_0$ & $b_1$ & $b_2$ \\
    \hline
    0.8441 & 0 & -0.0013 \\
    \hline
  \end{tabular}
\end{table}

\noindent
For $\sigma/C_{44}\geq0.003$:

\begin{table}[!htb]
  \centering
  \begin{tabular}{ccc}
    \multicolumn{3}{c}{$a'(\chi) = a'_0 + a'_1\chi + a'_2\chi^2$} \\
    \hline
    $a'_0$ & $a'_1$ & $a'_2$ \\
    \hline
    1.3500 & 0 & 0.0257 \\
    \hline
  \end{tabular}
  \hskip3cm
  \begin{tabular}{ccc}
    \multicolumn{3}{c}{$b'(\chi) = b'_0 + b'_1\chi + b'_2\chi^2$} \\
    \hline
    $b'_0$ & $b'_1$ & $b'_2$ \\
    \hline
    0.0023 & 0 & 0.0013 \\
    \hline
  \end{tabular}
\end{table}

%------------------------------------------------------------------------------------------

\subsection{Full model}

For $\sigma/C_{44}\leq0.003$:
\begin{eqnarray}
  \nonumber
  & a(\chi,\eta) = a_0(\eta) + a_1(\eta)\chi + a_2(\eta)\chi^2 &\\
  & \left[
    \begin{array}{c}
      a_0(\eta) \\ a_1(\eta) \\ a_2(\eta)
    \end{array}
  \right] = \left[
    \begin{array}{rrr}
      82.6712 &  51.8676 &  14.0694 \\
            0 &  26.6283 &  11.8285 \\
     -35.4748 & -108.8250 & -38.7510 \\
    \end{array}
  \right] \left[
    \begin{array}{c}
      1 \\ \eta \\ \eta^2
    \end{array} \right] &
\end{eqnarray}

\newpage
\begin{eqnarray} 
  \nonumber
  & b(\chi,\eta) = b_0(\eta) + b_1(\eta)\chi + b_2(\eta)\chi^2 &\\
  & \left[
    \begin{array}{c}
      b_0(\eta) \\ b_1(\eta) \\ b_2(\eta)
    \end{array}
  \right] = \left[
    \begin{array}{rrr}
       0.8441 &  0.0616 & -0.0018 \\
            0 &  0.0326 & -0.0003 \\
      -0.0013 & -0.1118 &  0.0116 \\
    \end{array}
  \right] \left[
    \begin{array}{c}
      1 \\ \eta \\ \eta^2
    \end{array} \right] &
\end{eqnarray}
\\

\noindent
For $\sigma/C_{44}\geq0.003$:
\begin{eqnarray}
  \nonumber
  & a'(\chi,\eta) = a'_0(\eta) + a'_1(\eta)\chi + a'_2(\eta)\chi^2 &\\
  & \left[
    \begin{array}{c}
      a'_0(\eta) \\ a'_1(\eta) \\ a'_2(\eta)
    \end{array}
  \right] = \left[
    \begin{array}{rrr}
       1.3500 &  0.2027 & -0.0496 \\
            0 &  0.1160 & -0.0454 \\
       0.0257 & -0.2456 &  0.1395 \\
    \end{array}
  \right] \left[
    \begin{array}{c}
      1 \\ \eta \\ \eta^2
    \end{array} \right] &
\end{eqnarray}

\begin{eqnarray} 
  \nonumber
  & b'(\chi,\eta) = b'_0(\eta) + b'_1(\eta)\chi + b'_2(\eta)\chi^2 &\\
  & \left[
    \begin{array}{c}
      b'_0(\eta) \\ b'_1(\eta) \\ b'_2(\eta)
    \end{array}
  \right] = \left[
    \begin{array}{rrr}
      0.0023 & -0.0004 &  0 \\
           0 & -0.0002 &  0 \\
      0.0013 &  0.0006 &  0 \\
    \end{array}
  \right] \left[
    \begin{array}{c}
      1 \\ \eta \\ \eta^2
    \end{array} \right] &
\end{eqnarray}

  \addcontentsline{toc}{chapter}{Bibliography}
  \bibliography{bibliography}

\begin{thebibliography}{}

\bibitem[Ackermann et~al., 1983]{ackermann:83}
Ackermann, F., Mughrabi, H., and Seeger, A. (1983).
\newblock Temperature and strain-rate dependence of the flow stress of
  ultrapure niobium single crystals in cyclic deformation.
\newblock {\em Acta Metall.}, 31(9):1353--1366.

\bibitem[Allen and Tildesley, 1987]{allen:87}
Allen, M.~P. and Tildesley, D.~J. (1987).
\newblock {\em Computer simulation of liquids}.
\newblock Oxford University Press.

\bibitem[Allen et~al., 1956]{allen:56}
Allen, N.~P., Hopkins, B.~E., and McLennan, J.~E. (1956).
\newblock The tensile properties of single crystals of high-purity iron at
  temperatures from 100 to -253 {C}.
\newblock {\em Proc. R. Soc. Lond. A}, 234:221--246.

\bibitem[Andersen et~al., 1985]{andersen:85}
Andersen, O.~K., Jepsen, O., and Gl\"otzel, D. (1985).
\newblock Canonical description of the band structure of metals.
\newblock In Bassani, F., Fumi, F., and Tosi, M.~P., editors, {\em Highlights
  of Condensed Matter Theory}, page~59. North Holland, Amsterdam.

\bibitem[Andersen et~al., 1994]{andersen:94}
Andersen, O.~K., Jepsen, O., and Krieg, G. (1994).
\newblock In Kumar, V., Anderson, O.~K., and Mookerjee, A., editors, {\em
  Lectures on Methods of Electronic Structure Calculations}, page~63. World
  Scientific, Singapore.

\bibitem[Aoki et~al., 2007]{aoki:07}
Aoki, M., Nguyen-Manh, D., Pettifor, D.~G., and Vitek, V. (2007).
\newblock Atom-based bond-order potentials for modelling mechanical properties
  of metals.
\newblock {\em Progress of Mat. Sci.}, 52:154--195.

\bibitem[Aono et~al., 1989]{aono:89}
Aono, Y., Kuramoto, E., Brunner, D., and Diehl, J. (1989).
\newblock Plastic behavior of high-purity molybdenum single crystals in tension
  and compression.
\newblock In {\em Strength of metals and alloys (ICSMA8): {P}roceedings of the
  8th international conference}, pages 271--276. Pergamon Press.

\bibitem[Aono et~al., 1981]{aono:81}
Aono, Y., Kuramoto, E., and Kitajima, K. (1981).
\newblock Plastic deformation of high-purity iron single crystals.
\newblock {\em Reports of Res. Inst. for Appl. Mech.}, 29(92):127--189.

\bibitem[Aono et~al., 1983]{aono:83}
Aono, Y., Kuramoto, E., and Kitajima, K. (1983).
\newblock Fundamental plastic behaviors in high-purity {BCC} metals ({Nb}, {Mo}
  and {Fe}).
\newblock In {\em Strength of metals and alloys (ICSMA6): proceedings of the
  6th international conference}, pages 9--14. Pergamon Press.

\bibitem[Argon and Maloof, 1966]{argon:66}
Argon, A.~S. and Maloof, S.~R. (1966).
\newblock Plastic deformation of tungsten single crystals at low temperatures.
\newblock {\em Acta Metall.}, 14:1449--1462.

\bibitem[Arsenault, 1964]{arsenault:64}
Arsenault, R.~J. (1964).
\newblock Low-temperature creep of alpha iron.
\newblock {\em Acta Metall.}, 12:547--554.

\bibitem[Basinski et~al., 1981]{basinski:81}
Basinski, Z.~S., Duesbery, M.~S., and Murty, G.~S. (1981).
\newblock The orientation and temperature dependence of plastic flow in
  potassium.
\newblock {\em Acta Metall.}, 29(5):801--807.

\bibitem[Basinski et~al., 1971]{basinski:71}
Basinski, Z.~S., Duesbery, M.~S., and Taylor, R. (1971).
\newblock Influence of shear stress on screw dislocations in a model sodium
  lattice.
\newblock {\em Can. J. Phys.}, 49(16):2160.

\bibitem[Baskes, 1992]{baskes:92}
Baskes, M. (1992).
\newblock Modified embedded-atom potentials for cubic materials and impurities.
\newblock {\em Phys. Rev. B}, 46:2727--2742.

\bibitem[Bassani, 1994]{bassani:94}
Bassani, J.~L. (1994).
\newblock Plastic flow of crystals.
\newblock In {\em Advances in applied mechanics}, volume~30, pages 191--258.
  Academic Press.

\bibitem[Bolton and Taylor, 1972]{bolton:72}
Bolton, C.~J. and Taylor, G. (1972).
\newblock Anomalous slip in high-purity niobium single crystals deformed at 77
  {K} in tension.
\newblock {\em Phil. Mag.}, 26(6):1359--1376.

\bibitem[Bowen and Taylor, 1977]{bowen:77}
Bowen, D.~K. and Taylor, G. (1977).
\newblock The deformation behaviour of dilute niobium-nitrogen alloys.
\newblock {\em Acta Metall.}, 25:417--436.

\bibitem[Brunner, 2004]{brunner:04}
Brunner, D. (2004).
\newblock Peculiarities of work hardening of high-purity tungsten single
  crystals below 800 {K}.
\newblock {\em Mat. Sci. Eng. A}, 387-389:167--170.

\bibitem[Brunner and Glebovsky, 2000]{brunner:00}
Brunner, D. and Glebovsky, V. (2000).
\newblock Analysis of flow-stress measurements of high-purity tungsten single
  crystals.
\newblock {\em Mater. Lett.}, 44:144--152.

\bibitem[Bulatov et~al., 1998]{bulatov:98}
Bulatov, V.~V., Abraham, F.~F., Kubin, L., Devincre, B., and Yip, S. (1998).
\newblock Connecting atomistic and mesoscale simulations of crystal plasticity.
\newblock {\em Nature}, 391:669--671.

\bibitem[Bulatov and Cai, 2002]{bulatov:02}
Bulatov, V.~V. and Cai, W. (2002).
\newblock Nodal effects in dislocation mobility.
\newblock {\em Phys. Rev. Lett.}, 89(11):115501.

\bibitem[Caillard and Martin, 2003]{caillard:03}
Caillard, D. and Martin, J.~L. (2003).
\newblock {\em Thermally activated mechanisms in crystal plasticity}.
\newblock Pergamon Press.

\bibitem[Carlsson, 1990a]{carlsson:90}
Carlsson, A.~E. (1990a).
\newblock Beyond pair potentials in elemental transition metals and
  semiconductors.
\newblock In Ehrenreich, H. and Turnbull, D., editors, {\em Solid State
  Physics}, volume~43, page~1. Academic Press, New York.

\bibitem[Carlsson, 1990b]{carlsson:90a}
Carlsson, A.~E. (1990b).
\newblock Derivation of angular forces for semiconductors and transition
  metals.
\newblock In Nieminen, R.~M., Puska, M.~J., and Manninen, M.~J., editors, {\em
  Many-Atom Interactions in Solids}, page 257. Springer, Heidelberg.

\bibitem[Cawkwell, 2005]{cawkwell:05a}
Cawkwell, M.~J. (2005).
\newblock {\em Interatomic bonding and plastic deformation in iridium and
  molybdenum disilicide}.
\newblock PhD thesis, University of Pennsylvania.

\bibitem[Cawkwell et~al., 2006]{cawkwell:06}
Cawkwell, M.~J., Nguyen-Manh, D., Pettifor, D.~J., and Vitek, V. (2006).
\newblock Construction, assessment, and application of a bond-order potential
  for iridium.
\newblock {\em Phys. Rev. B}, 73:064104.

\bibitem[Celli et~al., 1963]{celli:63}
Celli, V., Kabler, M., Ninomiya, T., and Thomson, R. (1963).
\newblock Theory of dislocation mobility in semiconductors.
\newblock {\em Phys. Rev.}, 131(1):58--72.

\bibitem[Chaussidon et~al., 2006]{chaussidon:06}
Chaussidon, J., Fivel, M., and Rodney, D. (2006).
\newblock The glide of screw dislocations in bcc {Fe}: {A}tomistic static and
  dynamic simulations.
\newblock {\em Acta Mater.}, 54:3407--3416.

\bibitem[Christian, 1983]{christian:83}
Christian, J.~W. (1983).
\newblock Some surprising features of the plastic deformation of body-centered
  cubic metals and alloys.
\newblock {\em Metall. Trans. A}, 14:1237--1256.

\bibitem[Conrad and Hayes, 1963]{conrad:63}
Conrad, H. and Hayes, W. (1963).
\newblock Thermally-activated deformation of the bcc metals at low
  temperatures.
\newblock {\em Trans. ASM}, 56:249--262.

\bibitem[Creten et~al., 1977]{creten:77}
Creten, R., Bressers, J., and De~Meester, P. (1977).
\newblock Anomalous slip in high-purity vanadium crystals deformed in
  compression.
\newblock {\em Mat. Sci. Eng.}, 29(1):51--53.

\bibitem[Dagens et~al., 1975]{dagens:75}
Dagens, L., Rasolt, M., and Taylor, R. (1975).
\newblock Change densities and interionic potentials in simple metals:
  {N}onlinear effects. {II}.
\newblock {\em Phys. Rev. B}, 11(8):2726--2734.

\bibitem[Dorn and Rajnak, 1964]{dorn:64}
Dorn, J.~E. and Rajnak, S. (1964).
\newblock Nucleation of kink pairs and the {P}eierls' mechanism of plastic
  deformation.
\newblock {\em Trans. AIME}, 230:1052--1064.

\bibitem[Drautz et~al., 2005]{drautz:05}
Drautz, R., Murdick, D.~A., Nguyen-Manh, D., Zhou, X.~W., Wadley, H. N.~G., and
  Pettifor, D.~G. (2005).
\newblock Analytic bond-order potential for predicting structural trends across
  the sp-valent elements.
\newblock {\em Phys. Rev. B}, 72(14):144105.

\bibitem[Duesbery and Richardson, 1991]{duesbery:91}
Duesbery, M.~D. and Richardson, G.~Y. (1991).
\newblock The dislocation core in crystalline materials.
\newblock {\em CRC Critical Reviews in Solid State and Materials Science},
  17:1.

\bibitem[Duesbery, 1969]{duesbery:69}
Duesbery, M.~S. (1969).
\newblock The influence of core structure on dislocation mobility.
\newblock {\em Phil. Mag.}, 19(159):501--526.

\bibitem[Duesbery, 1984]{duesbery:84}
Duesbery, M.~S. (1984).
\newblock On non-glide stresses and their influence on the screw dislocation
  core in body-centered cubic metals. 1. {T}he {P}eierls stress.
\newblock {\em Proc. R. Soc. Lond. A}, 392(1802):145--173.

\bibitem[Duesbery, 1989]{duesbery:89}
Duesbery, M.~S. (1989).
\newblock The dislocation core and plasticity.
\newblock In Nabarro, F. R.~N., editor, {\em Dislocations in Solids}, volume~8,
  pages 66--173. Elsevier.

\bibitem[Duesbery and Basinski, 1993]{duesbery:93}
Duesbery, M.~S. and Basinski, Z.~S. (1993).
\newblock The flow stress of potassium.
\newblock {\em Acta Metall. Mater.}, 41(2):643--647.

\bibitem[Duesbery and Foxall, 1969]{duesbery:69b}
Duesbery, M.~S. and Foxall, R.~A. (1969).
\newblock A detailed study of deformation of high-purity niobium single
  crystals.
\newblock {\em Phil. Mag.}, 20(166):719.

\bibitem[Duesbery and Vitek, 1998]{duesbery:98}
Duesbery, M.~S. and Vitek, V. (1998).
\newblock Plastic anisotropy in bcc transition metals.
\newblock {\em Acta Mater.}, 46(5):1481--1492.

\bibitem[Duesbery et~al., 2002]{duesbery:02}
Duesbery, M.~S., Vitek, V., and Cserti, J. (2002).
\newblock Non-{S}chmid plastic behaviour in {BCC} metals and alloys.
\newblock In Humphreys, C.~J., editor, {\em Understanding materials: A
  Festschrift for Sir Peter Hirch}, pages 165--192. Maney Publishing, Leeds.

\bibitem[Edagawa et~al., 1997]{edagawa:97}
Edagawa, K., Suzuki, T., and Takeuchi, S. (1997).
\newblock Motion of a screw dislocation in a two-dimensional {P}eierls
  potential.
\newblock {\em Phys. Rev. B}, 55(10):6180--6187.

\bibitem[Elam, 1926]{elam:26}
Elam, C.~F. (1926).
\newblock Tensile tests of large gold, silver and copper crystals.
\newblock {\em Proc. Roy. Soc. A}, 112(760):289--296.

\bibitem[Eshelby and Stroh, 1951]{eshelby:51}
Eshelby, J.~D. and Stroh, A.~N. (1951).
\newblock Dislocations in thin plates.
\newblock {\em Philos. Mag.}, 42(335):1401--1405.

\bibitem[Frederiksen and Jacobsen, 2003]{frederiksen:03}
Frederiksen, S.~L. and Jacobsen, K.~W. (2003).
\newblock Density functional theory studies of screw dislocation core
  structures in bcc metals.
\newblock {\em Phil. Mag.}, 83(3):365--375.

\bibitem[Friedel, 1969]{friedel:69}
Friedel, J. (1969).
\newblock {\em The Physics of Metals}, page 340.
\newblock Cambridge University Press.

\bibitem[Garrat-Reed and Taylor, 1979]{garrat-reed:79}
Garrat-Reed, A.~J. and Taylor, G. (1979).
\newblock Optical and electron microscopy of niobium crystals deformed below
  room temperature.
\newblock {\em Phil. Mag. A}, 39(5):597--646.

\bibitem[Girshick, 1997]{girshick:97}
Girshick, A. (1997).
\newblock {\em Atomistic studies of dislocations in titanium and
  titanium-aluminum compound}.
\newblock PhD thesis, University of Pennsylvania.

\bibitem[Girshick et~al., 1998]{girshick:98}
Girshick, A., Bratkovsky, A.~M., Pettifor, D.~G., and Vitek, V. (1998).
\newblock Atomistic simulation of titanium - {I}. {A} bond-order potential.
\newblock {\em Phil. Mag. A}, 77(4):981--997.

\bibitem[Goodwin et~al., 1989]{goodwin:89}
Goodwin, L., Skinner, A.~J., and Pettifor, D.~G. (1989).
\newblock Generating transferable tight-binding parameters - application to
  silicon.
\newblock {\em Europhys. Lett.}, 9(7):701--706.

\bibitem[Gr\"oger and Vitek, 2005]{groger:05}
Gr\"oger, R. and Vitek, V. (2005).
\newblock Breakdown of the {S}chmid law in bcc molybdenum related to the effect
  of shear stress perpendicular to the slip direction.
\newblock {\em Mat. Sci. Forum}, 482:123--126.

\bibitem[Guiu, 1969]{guiu:69}
Guiu, F. (1969).
\newblock Slip asymmetry in molybdenum single crystals deformed in direct
  shear.
\newblock {\em Scr. Metall.}, 3:449--454.

\bibitem[Henkelman et~al., 2000a]{henkelman:00}
Henkelman, G., J\'ohannesson, G., and J\'onsson, H. (2000a).
\newblock Methods for finding saddle points and minimum energy paths.
\newblock In Schwartz, S.~D., editor, {\em Progress on theoretical chemistry
  and physics}, pages 269--300. Kluwer.

\bibitem[Henkelman and J\'onsson, 2000]{henkelman:00b}
Henkelman, G. and J\'onsson, H. (2000).
\newblock Improved tangent estimate in the nudged elastic band method for
  finding minimum energy paths and saddle points.
\newblock {\em J. Chem. Phys.}, 113(22):9978--9985.

\bibitem[Henkelman et~al., 2000b]{henkelman:00c}
Henkelman, G., Uberuaga, B.~P., and J\'onsson, H. (2000b).
\newblock A climbing image nudged elastic band method for finding saddle points
  and minimum energy paths.
\newblock {\em J. Chem. Phys.}, 113(22):9901--9904.

\bibitem[Hill and Rice, 1972]{hill:72}
Hill, J.~R. and Rice, R. (1972).
\newblock Constitutive analysis of elastic-plastic crystals at arbitrary
  strain.
\newblock {\em J. Mech. Phys. Sol.}, 20:401--413.

\bibitem[Hill, 1965]{hill:65}
Hill, R. (1965).
\newblock Continuum micro-mechanics of elastoplastic polycrystals.
\newblock {\em J. Mech. Phys. Sol.}, 13:89--101.

\bibitem[Hill and Havner, 1982]{hill:82}
Hill, R. and Havner, K.~S. (1982).
\newblock Perspectives in the mechanics of elastoplastic crystals.
\newblock {\em J. Mech. Phys. Sol.}, 30:5--22.

\bibitem[Hirsch, 1960]{hirsch:60}
Hirsch, P.~B. (1960).
\newblock In {\em 5th Int. Conf. on Crystallography}, Cambridge.

\bibitem[Hirsch, 1980]{hirsch:80}
Hirsch, P.~B. (1980).
\newblock Direct observations of dislocations by transmission electron
  microscopy: recollections of the period 1946-56.
\newblock {\em Proc. R. Soc. Lond. A}, 371:160--164.

\bibitem[Hirsch et~al., 1956]{hirsch:56}
Hirsch, P.~B., Horne, R.~W., and Whelan, M.~J. (1956).
\newblock Direct observations of the arrangement and motion of dislocations in
  aluminum.
\newblock {\em Phil. Mag.}, 1:677.

\bibitem[Hirth and Lothe, 1982]{hirth:82}
Hirth, J.~P. and Lothe, J. (1982).
\newblock {\em Theory of dislocations}.
\newblock J.Wiley \& Sons, 2 edition.

\bibitem[Hirth and Nix, 1969]{hirth:69}
Hirth, J.~P. and Nix, W.~D. (1969).
\newblock An analysis of the thermodynamics of dislocation glide.
\newblock {\em Phys. Stat. Sol.}, 35:177--187.

\bibitem[Hollang, 2001]{hollang:01a}
Hollang, L. (2001).
\newblock Mechanical properties.
\newblock In Waseda, Y. and Isshiki, M., editors, {\em Purification process and
  characterization of ultra high purity metals}. Springer-Verlag.

\bibitem[Hollang et~al., 2001]{hollang:01}
Hollang, L., Brunner, D., and Seeger, A. (2001).
\newblock Work hardening and flow stress of ultrapure molybdenum single
  crystals.
\newblock {\em Mat. Sci. Eng. A}, 319-321:233--236.

\bibitem[Hollang et~al., 1997]{hollang:97}
Hollang, L., Hommel, M., and Seeger, A. (1997).
\newblock The flow stress of ultra-high-purity molybdenum single crystals.
\newblock {\em Phys. Stat. Sol. (a)}, 160(2):329--354.

\bibitem[Horsfield et~al., 1996]{horsfield:96}
Horsfield, A.~P., Bratkovsky, A.~M., Fearn, M., Pettifor, D.~G., and Aoki, M.
  (1996).
\newblock Bond-{O}rder potentials: {T}heory and implementation.
\newblock {\em Phys. Rev. B}, 53(19):12694--12712.

\bibitem[Hull and Bacon, 2001]{hull:01}
Hull, D. and Bacon, D.~J. (2001).
\newblock {\em Introduction to dislocations}.
\newblock Butterworth-Heinemann, Oxford, 4th edition.

\bibitem[Ismail-Beigi and Arias, 2000]{ismail-beigi:00}
Ismail-Beigi, S. and Arias, T.~A. (2000).
\newblock Ab initio study of screw dislocations in {Mo} and {Ta}: {A} new
  picture of plasticity in bcc transition metals.
\newblock {\em Phys. Rev. Lett.}, 84(7):1499--1502.

\bibitem[Ito and Vitek, 2001]{ito:01}
Ito, K. and Vitek, V. (2001).
\newblock Atomistic study of non-{S}chmid effects in the plastic yielding of
  bcc metals.
\newblock {\em Phil. Mag. A}, 81(5):1387--1407.

\bibitem[Jeffcoat et~al., 1976]{jeffcoat:76}
Jeffcoat, P.~J., Mordike, B.~L., and Rogausch, K.~D. (1976).
\newblock Anomalous slip in {M}o-5 at.\% {N}b and {M}o-5 at.\% {Re} alloy
  single crystals.
\newblock {\em Phil. Mag.}, 34(4):583--592.

\bibitem[Jeffrey, 2000]{jeffrey:00}
Jeffrey, A. (2000).
\newblock {\em Handbook of mathematical formulas and integrals}.
\newblock Academic Press, 2nd edition.

\bibitem[Jonsson, 2003]{jonsson:03}
Jonsson, A. (2003).
\newblock Discrete dislocation dynamics by an {O(N)} algorithm.
\newblock {\em Comp. Mat. Sci.}, 27:271--288.

\bibitem[J\'onsson et~al., 1998]{jonsson:98}
J\'onsson, H., Mills, G., and Jacobsen, K.~W. (1998).
\newblock {\em Nudged elastic band method for finding minimum energy paths of
  transitions}.
\newblock Classical and Quantum Dynamics in Condensed Phase Simulations. World
  Scientific.

\bibitem[Kaspar et~al., 2000]{kaspar:00}
Kaspar, J., Luft, A., and Skrotzki, W. (2000).
\newblock Deformation modes and structure evolution in laser-shock-loaded
  molybdenum single crystals of high purity.
\newblock {\em Cryst. Res. Technol.}, 35(4):437--448.

\bibitem[Kaun et~al., 1968]{kaun:68}
Kaun, L., Luft, A., Richter, J., and Schulze, D. (1968).
\newblock Slip line pattern and active slip systems of tungsten and molybdenum
  single crystals weakly deformed in tension at room temperature.
\newblock {\em Phys. Stat. Sol.}, 26(2):485.

\bibitem[Keh, 1965]{keh:65}
Keh, A.~S. (1965).
\newblock Work hardening and deformation sub-structure in iron single crystals
  deformed in tension at 298 {K}.
\newblock {\em Phil. Mag.}, 12:9--30.

\bibitem[Kitajima et~al., 1981]{kitajima:81}
Kitajima, K., Aono, Y., and Kuramoto, E. (1981).
\newblock Slip systems and orientation dependence of yield stress in high
  purity molybdenum single crystals at 4.2 {K} and 77 {K}.
\newblock {\em Scripta Metall.}, 15:919--924.

\bibitem[Kocks et~al., 1975]{kocks:75}
Kocks, U.~F., Argon, A.~S., and Ashby, M.~F. (1975).
\newblock Thermodynamics and kinetics of slip.
\newblock {\em Progress in Materials Science}, 19:1--291.

\bibitem[Kubin et~al., 1998]{kubin:98}
Kubin, L., Devincre, B., and Tang, M. (1998).
\newblock Mesoscopic modelling and simulation of plasticity in fcc and bcc
  crystals: {D}islocation intersections and mobility.
\newblock {\em J. Comp.-Aid. Mat. Design}, 5:31--54.

\bibitem[Lassila et~al., 2002]{lassila:02}
Lassila, D.~H., LeBlanc, M.~M., and Kay, G.~J. (2002).
\newblock Uniaxial stress deformation experiment for validation of {3-D}
  dislocation dynamics simulations.
\newblock {\em J. Eng. Mat. Tech.}, 124:290--296.

\bibitem[Lassila et~al., 2003]{lassila:03}
Lassila, D.~H., LeBlanc, M.~M., and Rhee, M. (2003).
\newblock Single crystal deformation experiments for validation of dislocation
  dynamics simulations.
\newblock In {\em MRS Symp. Proc.}, volume 779, pages W2.9.1--2.9.12.

\bibitem[Lau and Dorn, 1970]{lau:70}
Lau, S.~S. and Dorn, J.~E. (1970).
\newblock Asymmetric slip in {M}o single crystals.
\newblock {\em Phys. Stat. Sol. A}, 2:825--836.

\bibitem[Liu et~al., 2005]{liu:05}
Liu, G., Nguyen-Manh, D., Liu, B.-G., and Pettifor, D.~G. (2005).
\newblock Magnetic properties of point defects in iron within the
  tight-binding-bond {S}toner model.
\newblock {\em Phys. Rev. B}, 71:174115.

\bibitem[Louchet, 2003]{louchet:web}
Louchet, F. (2003).
\newblock {\tt http://www.gpm2.inpg.fr/axes/plast/{M}icro{P}last/ddd/{TEM}}.

\bibitem[Louchet and Kubin, 1975]{louchet:75}
Louchet, F. and Kubin, L.~P. (1975).
\newblock Dislocation substructures in the anomalous slip plane of single
  crystal niobium strained at 50 {K}.
\newblock {\em Acta Metall.}, 23:17--21.

\bibitem[Louchet and Kubin, 1979]{louchet:79a}
Louchet, F. and Kubin, L.~P. (1979).
\newblock Dislocation processes in bcc metals.
\newblock {\em Phys. Stat. Sol. A}, 56:169.

\bibitem[Louchet et~al., 1979]{louchet:79}
Louchet, F., Kubin, L.~P., and Vesely, D. (1979).
\newblock In situ deformation of b.c.c. crystals at low temperatures in a
  high-voltage electron microscope. {D}islocation mechanisms and strain-rate
  equation.
\newblock {\em Phil. Mag. A}, 39(4):433--454.

\bibitem[Maragakis et~al., 2002]{maragakis:02}
Maragakis, P., Andreev, S.~A., Brumer, Y., Reichman, D.~R., and Kaxiras, E.
  (2002).
\newblock Adaptive nudged elastic band approach for transition state
  calculation.
\newblock {\em J. Chem. Phys.}, 117(10):4651--4658.

\bibitem[Marian et~al., 2004]{marian:04}
Marian, J., W., C., and Bulatov, V.~V. (2004).
\newblock Dynamic transitions from smooth to rough to twinning in dislocation
  motion.
\newblock {\em Nature Mat.}, 3:2004.

\bibitem[Matsui and Kimura, 1976]{matsui:76}
Matsui, H. and Kimura, H. (1976).
\newblock Anomalous {$\{110\}$} slip in high-purity molybdenum single crystals
  and its comparison with that in {V(a)} metals.
\newblock {\em Mat. Sci. Eng.}, 24:247--256.

\bibitem[Matsui et~al., 1982]{matsui:82}
Matsui, H., Kimura, H., Saka, H., Noda, K., and Imura, T. (1982).
\newblock Anomalous slip induced by the surface effect in molybdenum
  single-crystal foils deformed in a high voltage electron microscope.
\newblock {\em Mat. Sci. Eng.}, 53:263--272.

\bibitem[Matterstock et~al., 1999]{matterstock:99}
Matterstock, B., Martin, J.~L., Bonneville, J., and Kruml, T. (1999).
\newblock Direct measurement of dislocation exhaustion rates during plastic
  deformation of ni$_3$al compounds.
\newblock {\em Mat. Res. Soc. Symp. Proc.}, 552:KK5.17.1--6.

\bibitem[McKee and Page, 1993]{mckee:93}
McKee, M.~L. and Page, M. (1993).
\newblock Computing reaction pathways on molecular potential energy surfaces.
\newblock In Lipkowitz, K.~B. and Boyd, D.~B., editors, {\em Reviews in
  Computational Chemistry}, volume~4. VCH Publishers.

\bibitem[Mendis et~al., 2006]{mendis:06}
Mendis, B.~G., Mishin, Y., Hartley, C.~S., and Hemker, K.~J. (2006).
\newblock Use of the nye tensor in analyzing hrem images of bcc screw
  dislocations.
\newblock {\em Phil. Mag.}, 86(29-31):4607--4640.

\bibitem[Meyers et~al., 2002]{meyers:02}
Meyers, M.~A., Benson, D.~J., V\"ohringer, O., Kad, B.~K., Xue, Q., and Fu,
  H.-H. (2002).
\newblock Constitutive description of dynamic deformation: physically-based
  mechanisms.
\newblock {\em Mat. Sci. Eng. A}, 322:194--216.

\bibitem[Mitchell et~al., 1963]{mitchell:63}
Mitchell, T.~E., Foxall, R.~A., and Hirsch, P.~B. (1963).
\newblock Work-hardening in niobium single crystals.
\newblock {\em Phil. Mag.}, 8:1895--1920.

\bibitem[Moriarty, 1988]{moriarty:88}
Moriarty, J.~A. (1988).
\newblock Density-functional formulation of the generalized pseudopotential
  theory: {III}. {T}ransition-metal interatomic potentials.
\newblock {\em Phys. Rev. B}, 38(5):3199--3231.

\bibitem[Moriarty, 1990]{moriarty:90}
Moriarty, J.~A. (1990).
\newblock Analytic representation of multi-ion interatomic potentials in
  transition metals.
\newblock {\em Phys. Rev. B}, 42(3):1609--1628.

\bibitem[Moriarty et~al., 2002]{moriarty:02}
Moriarty, J.~A., Vitek, V., Bulatov, V.~V., and Yip, S. (2002).
\newblock Atomistic simulations of dislocations and defects.
\newblock {\em J. Comp.-Aid. Mat. Design}, 9:99.

\bibitem[Mrovec, 2002]{mrovec:02}
Mrovec, M. (2002).
\newblock {\em Bond order potentials for bcc transition metals and molybdenum
  silicides}.
\newblock PhD thesis, University of Pennsylvania.

\bibitem[Mrovec et~al., 2007]{mrovec:07}
Mrovec, M., Gr\"{o}ger, R., Bailey, A.~G., Nguyen-Manh, D., Els\"{a}sser, C.,
  and Vitek, V. (2007).
\newblock Bond-order potential for simulations of extended defects in tungsten.
\newblock {\em Phys. Rev. B}, 75:104119.

\bibitem[Mrovec et~al., 2004]{mrovec:04}
Mrovec, M., Nguyen-Manh, D., Pettifor, D.~G., and Vitek, V. (2004).
\newblock Bond-order potential for molybdenum: {A}pplication to dislocation
  behavior.
\newblock {\em Phys. Rev. B}, 69:094115.

\bibitem[Murdick et~al., 2006]{murdick:06}
Murdick, D.~A., Zhou, X.~W., Wadley, H. N.~G., Nguyen-Manh, D., Drautz, R., and
  Pettifor, D.~G. (2006).
\newblock Analytic bond-order potential for the gallium arsenide system.
\newblock {\em Phys. Rev. B}, 73(4):045206.

\bibitem[Nabarro, 1947]{nabarro:47}
Nabarro, F. R.~N. (1947).
\newblock Dislocations in a simple cubic lattice.
\newblock {\em Proc. Phys. Soc.}, 59(2):256--272.

\bibitem[Nawaz and Mordike, 1975]{nawaz:75}
Nawaz, M. H.~A. and Mordike, B.~L. (1975).
\newblock Slip geometry of tantalum and tantalum alloys.
\newblock {\em Phys. Stat. Sol. A}, 32:449--458.

\bibitem[Nguyen-Manh et~al., 2000]{nguyen-manh:00}
Nguyen-Manh, D., Pettifor, D.~G., and Vitek, V. (2000).
\newblock Analytic environment-dependent tight-binding bond integrals:
  {A}pplication to {M}o{S}i$_2$.
\newblock {\em Phys. Rev. Lett.}, 85(19):4136--4139.

\bibitem[Orowan, 1934]{orowan:34}
Orowan, E. (1934).
\newblock {\em Z. Phys.}, 89:634.

\bibitem[Peach and Koehler, 1950]{peach:50}
Peach, M. and Koehler, J.~S. (1950).
\newblock The forces exerted on dislocations and the stress fields produced by
  them.
\newblock {\em Phys. Rev.}, 80(3):436--439.

\bibitem[Peierls, 1940]{peierls:40}
Peierls, R. (1940).
\newblock The size of a dislocation.
\newblock {\em Proc. Phys. Soc.}, 52(1):34--37.

\bibitem[Pettifor, 1995]{pettifor:95}
Pettifor, D.~G. (1995).
\newblock {\em Bonding and structure of molecules and solids}.
\newblock Oxford University Press.

\bibitem[Pettifor et~al., 2004]{pettifor:04}
Pettifor, D.~G., Finnis, M.~W., Nguyen-Manh, D., Murdick, D.~A., Zhou, X.~W.,
  and Wadley, H. N.~G. (2004).
\newblock Analytic bond-order potentials for multicomponent systems.
\newblock {\em Mat. Sci. Eng. A}, 365(1-2):2--13.

\bibitem[Pettifor and Oleinik, 2002]{pettifor:02}
Pettifor, D.~G. and Oleinik, I.~I. (2002).
\newblock Analytic bond-order potential for open and close-packed phases.
\newblock {\em Phys. Rev. B}, 65(17):172103.

\bibitem[Pichl and Krystian, 1997a]{pichl:97b}
Pichl, W. and Krystian, M. (1997a).
\newblock The flow stress of high purity alkali metals.
\newblock {\em Phys. Stat. Sol. A}, 160(2):373--383.

\bibitem[Pichl and Krystian, 1997b]{pichl:97c}
Pichl, W. and Krystian, M. (1997b).
\newblock The non-uniform temperature dependence of the flow stress of
  potassium.
\newblock {\em Phil. Mag. Lett.}, 75(2):75--82.

\bibitem[Pichl and Krystian, 1997c]{pichl:97}
Pichl, W. and Krystian, M. (1997c).
\newblock The plasticity of potassium.
\newblock {\em Mat. Sci. Eng.}, A234-236:426--429.

\bibitem[Polanyi, 1934]{polanyi:34}
Polanyi, M. (1934).
\newblock {\em Z. Phys.}, 89:660.

\bibitem[Qin and Bassani, 1992a]{qin:92b}
Qin, Q. and Bassani, J.~L. (1992a).
\newblock Non-associated plastic flow in single crystals.
\newblock {\em J. Mech. Phys. Sol.}, 40(4):835--862.

\bibitem[Qin and Bassani, 1992b]{qin:92}
Qin, Q. and Bassani, J.~L. (1992b).
\newblock Non-{S}chmid yield behavior in single crystals.
\newblock {\em J. Mech. Phys. Sol.}, 40(4):813--833.

\bibitem[Reed and Arsenault, 1976]{reed:76}
Reed, R.~E. and Arsenault, R.~J. (1976).
\newblock Further observations of anomalous slip in niobium single crystals.
\newblock {\em Scripta Metall.}, 10:1003--1006.

\bibitem[Rice, 1971]{rice:71}
Rice, J.~R. (1971).
\newblock Inelastic constitutive relations for solids: {A}n internal-variable
  theory and its application to metal plasticity.
\newblock {\em J. Mech. Phys. Sol.}, 19:433--455.

\bibitem[Saka et~al., 1976]{saka:76}
Saka, H., Matsui, H., Noda, K., Kimura, H., and Imura, T. (1976).
\newblock Direct observation of {$(\bar{1}01)$} anomalous slip in molybdenum by
  {HVEM}.
\newblock {\em Scripta Metall.}, 10:59--62.

\bibitem[Schmid and Boas, 1935]{schmid:35}
Schmid, E. and Boas, W. (1935).
\newblock {\em Kristallplastizit\"at}.
\newblock Springer, Berlin.

\bibitem[Schoeck, 1965]{schoeck:65}
Schoeck, G. (1965).
\newblock The activation energy of dislocation movement.
\newblock {\em Phys. Stat. Sol.}, 8:499--507.

\bibitem[Seeger, 1956]{seeger:56}
Seeger, A. (1956).
\newblock On the theory of the low-temperature internal friction peak observed
  in metals.
\newblock {\em Phil. Mag.}, 1(7):651--662.

\bibitem[Seeger, 2004]{seeger:04}
Seeger, A. (2004).
\newblock Experimental evidence for the $\{110\} \leftrightarrow \{112\}$
  transformation of the screw-dislocation cores in body-centred cubic metals.
\newblock {\em Phys. Stat. Sol. (a)}, 201(4):21--24.

\bibitem[Seeger and Hollang, 2000]{seeger:00}
Seeger, A. and Hollang, L. (2000).
\newblock The flow-stress asymmetry of ultra-pure molybdenum single crystals.
\newblock {\em Mater. Trans., JIM}, 41(1):141--151.

\bibitem[Shields et~al., 1975]{shields:75}
Shields, J.~A., Goods, S.~H., Gibala, R., and Mitchell, T.~E. (1975).
\newblock Deformation of high purity tantalum single crystals at 4.2 {K}.
\newblock {\em Mat. Sci. Eng.}, 20:71--81.

\bibitem[Sigle, 1999]{sigle:99}
Sigle, W. (1999).
\newblock High-resolution electron microscopy and molecular dynamics study of
  the a/2[111] screw dislocations in molybdenum.
\newblock {\em Phil. Mag. A}, 79(5):1009--1020.

\bibitem[Slater and Koster, 1954]{slater:54}
Slater, J.~C. and Koster, G.~F. (1954).
\newblock Simplified lcao method for the periodic potential problem.
\newblock {\em Phys. Rev.}, 94(6):1498--1524.

\bibitem[Stoner, 1938]{stoner:38}
Stoner, E.~C. (1938).
\newblock Collective electron ferromagnetism.
\newblock {\em Proc. Roy. Soc. London A}, 165(922):372--414.

\bibitem[Stoner, 1939]{stoner:39}
Stoner, E.~C. (1939).
\newblock Collective electron ferromagnetism: {II.} {E}nergy and specific heat.
\newblock {\em Proc. Roy. Soc. London A}, 169(938):339--371.

\bibitem[Strogatz, 2003]{strogatz:03}
Strogatz, S. (2003).
\newblock {\em Sync: {A}n emergenging science of spontaneous order}.
\newblock Hyperion.

\bibitem[Sutton et~al., 1988]{sutton:88}
Sutton, A.~P., Finnis, M.~W., Pettifor, D.~G., and Ohta, Y. (1988).
\newblock The tight-binding bond model.
\newblock {\em J. Phys. C}, 21:35--66.

\bibitem[Suzuki et~al., 1995]{suzuki:95}
Suzuki, T., Koizumi, H., and Kirchner, H. O.~K. (1995).
\newblock Plastic flow stress of bcc transition metals and the {P}eierls
  potential.
\newblock {\em Acta Metall. Mater.}, 43(6):2177--2187.

\bibitem[Tabata et~al., 1976]{tabata:76}
Tabata, T., Mori, H., and Fujita, H. (1976).
\newblock In-situ deformation of tungsten single crystals with [100] tensile
  axis in an ultra-high voltage electron microscope.
\newblock {\em J. Phys. Soc. Japan}, 40(4):1103--1111.

\bibitem[Takeuchi and Maeda, 1977]{takeuchi:77}
Takeuchi, S. and Maeda, K. (1977).
\newblock Slip in high purity tantalum between 0.7 and 40 {K}.
\newblock {\em Acta Metall.}, 25:1485--1490.

\bibitem[Taylor et~al., 1973]{taylor:73}
Taylor, G., Bajaj, R., and Carlson, O.~N. (1973).
\newblock Anomalous slip in high-purity vanadium crystals.
\newblock {\em Phil. Mag.}, 28(5):1035--1042.

\bibitem[Taylor and Saka, 1991]{taylor:91}
Taylor, G. and Saka, M. (1991).
\newblock Some observations on slip in niobium and {Nb-Ti} alloy deformed in
  situ in a {HVEM}.
\newblock {\em Phil. Mag. A}, 64(6):1345--1354.

\bibitem[Taylor, 1928]{taylor:28}
Taylor, G.~I. (1928).
\newblock The deformation of crystals of $\beta$-brass.
\newblock {\em Proc. R. Soc. Lond. A}, 118(779):1--24.

\bibitem[Taylor, 1934]{taylor:34}
Taylor, G.~I. (1934).
\newblock The mechanism of plastic deformation of crystals. {P}art {I}.
  {T}heoretical.
\newblock {\em Proc. Roy. Soc. A}, 145(855):362--387.

\bibitem[Taylor and Elam, 1925]{taylor:25}
Taylor, G.~I. and Elam, C.~F. (1925).
\newblock The plastic extension and fracture of aluminium crystals.
\newblock {\em Proc. Roy. Soc. A}, 108:28--51.

\bibitem[Taylor and Elam, 1926]{taylor:26}
Taylor, G.~I. and Elam, C.~F. (1926).
\newblock The distortion of iron crystals.
\newblock {\em Proc. Roy. Soc. A}, 112(761):337--361.

\bibitem[Taylor and Farren, 1926]{taylor:26a}
Taylor, G.~I. and Farren, W.~S. (1926).
\newblock The distortion of crystals of aluminium under compression - {P}art
  {I}.
\newblock {\em Proc. Roy. Soc. A}, 111:529--551.

\bibitem[Vesely, 1968]{vesely:68}
Vesely, D. (1968).
\newblock The study of deformation of thin foils of {M}o under the electron
  microscope.
\newblock {\em Phys. Stat. Sol. A}, 29:675--683.

\bibitem[Vesely, 2006]{vesely:priv06}
Vesely, D. (2006).
\newblock private communication.

\bibitem[Vineyard, 1957]{vineyard:57}
Vineyard, G.~H. (1957).
\newblock Frequency factors and isotope effects in solid state rate processes.
\newblock {\em J. Phys. Chem. Solids}, 3:121--127.

\bibitem[Vitek, 1968]{vitek:68}
Vitek, V. (1968).
\newblock Intrinsic stacking faults in body-centred cubic crystals.
\newblock {\em Phil. Mag. A}, 18(154):773.

\bibitem[Vitek, 1992]{vitek:92}
Vitek, V. (1992).
\newblock Structure of dislocation cores in metallic materials and its impact
  on their plastic behaviour.
\newblock {\em Prog. Mater. Sci.}, 36:1--27.

\bibitem[Vitek, 2004]{vitek:04a}
Vitek, V. (2004).
\newblock Core structure of screw dislocations in body-centred cubic metals:
  relation to symmetry and interatomic bonding.
\newblock {\em Philos. Mag.}, 84(3-5):415--428.

\bibitem[Vitek et~al., 2004]{vitek:04}
Vitek, V., Mrovec, M., Gr\"oger, R., Bassani, J.~L., Racherla, V., and Yin, L.
  (2004).
\newblock Effects of non-glide stresses on the plastic flow of single and
  polycrystals of molybdenum.
\newblock {\em Mat. Sci. Eng. A}, 387-389:138--142.

\bibitem[Vitek et~al., 1970]{vitek:70}
Vitek, V., Perrin, R.~C., and Bowen, D.~K. (1970).
\newblock Core structure of 1/2$\langle111\rangle$ screw dislocations in bcc
  crystals.
\newblock {\em Philos. Mag.}, 21(173):1049.

\bibitem[{von Neumann}, 1885]{neumann:1885}
{von Neumann}, J. (1885).
\newblock {\em Vorlesungen \"uber die {T}heorie der {E}lasticit\"at}.
\newblock Leipzig.

\bibitem[Voyiadjis and Abed, 2005]{voyiadjis:05}
Voyiadjis, G.~Z. and Abed, F.~H. (2005).
\newblock Microstructural based models for bcc and fcc metals with temperature
  and strain rate dependency.
\newblock {\em Mech. Mater.}, 37:355--378.

\bibitem[Wasserb\"ach and Nov\'ak, 1985]{wasserbach:85}
Wasserb\"ach, W. and Nov\'ak, V. (1985).
\newblock Optical investigation of anomalous slip-line patterns in high purity
  niobium and tantalum single crystals after tensile deformation at 77 {K}.
\newblock {\em Mat. Sci. Eng.}, 73:197--202.

\bibitem[Webb et~al., 1974]{webb:74}
Webb, G.~L., Gibala, R., and Mitchell, T.~E. (1974).
\newblock Effect of normal stress on yield asymmetry in high purity tantalum
  crystals.
\newblock {\em Metall. Trans.}, 5(7):1581--1584.

\bibitem[Wen and Ngan, 2000]{wen:00}
Wen, M. and Ngan, A. H.~W. (2000).
\newblock Atomistic simulation of kink-pairs of screw dislocations in
  body-centred cubic iron.
\newblock {\em Acta Mater.}, 48:4255--4265.

\bibitem[Woodward and Rao, 2001]{woodward:01}
Woodward, C. and Rao, S.~I. (2001).
\newblock Ab-initio simulation of isolated screw dislocations in bcc {Mo} and
  {Ta}.
\newblock {\em Phil. Mag. A}, 81(5):1305--1316.

\bibitem[Woodward and Rao, 2002]{woodward:02}
Woodward, C. and Rao, S.~I. (2002).
\newblock Flexible ab initio boundary conditions: {S}imulating isolated
  dislocations in bcc {Mo} and {Ta}.
\newblock {\em Phys. Rev. Lett.}, 88(21):216402.

\bibitem[Xu and Moriarty, 1996]{xu:96}
Xu, W. and Moriarty, J.~A. (1996).
\newblock Atomistic simulation of ideal shear strength, point defects, and
  screw dislocations in bcc transition metals: {M}o as a prototype.
\newblock {\em Phys. Rev. B}, 54(10):6941--6951.

\bibitem[Xu and Moriarty, 1998]{xu:98}
Xu, W. and Moriarty, J.~A. (1998).
\newblock Accurate atomistic simulations of the {P}eierls barrier and kink-pair
  formation energy for $\langle{111}\rangle$ screw dislocations in bcc {Mo}.
\newblock {\em Comp. Mater. Sci.}, 9:348--356.

\bibitem[Yang et~al., 2001]{yang:01a}
Yang, L.~H., S\"oderlind, P., and Moriarty, J.~A. (2001).
\newblock Atomistic simulation of pressure-dependent screw dislocation
  properties in bcc tantalum.
\newblock {\em Mat. Sci. Eng. A}, 309-310:102--107.

\bibitem[Zbib et~al., 2000]{zbib:00}
Zbib, H.~M., de~la Rubia, T.~D., Rhee, M., and Hirth, J.~P. (2000).
\newblock {3D} dislocation dynamics: stress-strain behavior and hardening
  mechanisms in fcc and bcc metals.
\newblock {\em J. Nucl. Mat.}, 276:154--165.

\bibitem[Zerilli, 2004]{zerilli:04}
Zerilli, F.~J. (2004).
\newblock Dislocation mechanics-based constitutive equations.
\newblock {\em Metall. and Mater. Trans. A}, 35:2547--2555.

\bibitem[Zhou et~al., 1998]{zhou:98}
Zhou, S.~J., Preston, D.~L., Lomdahl, P.~S., and Beazley, D.~M. (1998).
\newblock Large-scale molecular dynamics simulations of dislocation
  intersection in copper.
\newblock {\em Science}, 279:1525--1527.

\bibitem[Zinkle et~al., 2002]{zinkle:02}
Zinkle, S.~J., Victoria, M., and Abe, K. (2002).
\newblock Scientific and engineering advances from fusion materials {R}\&{D}.
\newblock {\em J. Nuclear Mat.}, 307-311:31--42.

\bibitem[Znam, 2001]{znam:01}
Znam, S. (2001).
\newblock {\em Bond-order potentials for atomistic studies of dislocations and
  other extended defects in TiAl}.
\newblock PhD thesis, University of Pennsylvania.

\bibitem[Znam et~al., 2003]{znam:03}
Znam, S., Nguyen-Manh, D., Pettifor, D.~G., and Vitek, V. (2003).
\newblock Atomistic modelling of {TiAl} - {I}. {B}ond-order potentials with
  environmental dependence.
\newblock {\em Phil. Mag.}, 83(4):415--438.

\end{thebibliography}

\end{document}